\documentclass[iop,numberedappendix]{emulateapj-rtx4}
\usepackage{graphicx}
\usepackage{epstopdf}
\usepackage{apjfonts}
\usepackage{lscape}

\newif\ifpdffig
\pdffigtrue 
\ifpdffig
\pdfoutput=1
\fi

\def\puncspace{\ifmmode\,\else{\ifcat.\C{\if.\C\else\if,\C\else\if?\C\else%
\if:\C\else\if;\C\else\if-\C\else\if)\C\else\if/\C\else\if]\C\else\if'\C%
\else\space\fi\fi\fi\fi\fi\fi\fi\fi\fi\fi}%
\else\if\empty\C\else\if\space\C\else\space\fi\fi\fi}\fi}
\def\SP{\let\\=\empty\futurelet\C\puncspace }

\def\deg{$^\circ$\ }
\def\kms{km s$^{-1}$}
\def\h1{$h^{-1}$}

\def\deg{\ifmmode^\circ\kern- .38em\;\else$^\circ\kern- .38em\;$\fi}

\def\amin{\ifmmode^\prime\kern- .38em\;\else$^\prime\kern- .38em\;$\fi}
\def\asec{\ifmmode^{\prime\prime}\kern- .38em\;\else$^{\prime\prime}
    \kern- .38em\;$\fi}
\def\s.{\kern+ .1em\lower 0.0ex\hbox{$\buildrel ^{\prime\prime} \over 
    {\rm .} \kern- .08em$}} 
\def\m.{\kern+ .1em\lower 0.0ex\hbox{$\buildrel ^{\prime} \over 
    {\rm .} \kern- .08em$}} 
\def\grad{\kern+ .1em\lower 0.0ex\hbox{$\buildrel {\rightharpoonup} \over 
    {\nabla} \kern- .08em$}} 
\def\div{\kern+ .1em\lower 0.0ex\hbox{$\buildrel {\rightharpoonup} \over 
    {\nabla} \kern- .08em$}\cdot} 
\def\ltsim{\lower 0.5ex\hbox{$\; \buildrel < \over \sim \;$}} 
\def\gtsim{\lower 0.5ex\hbox{$\; \buildrel > \over \sim \;$}} 
\def\gteq{\lower 0.5ex\hbox{$\; \buildrel > \over = \;$}} 

\def\xx{\enspace\enspace}

 \def\apj{{ApJ}}
\def\apjs{{ApJS}}
\def\aj{{AJ}}
\def\aaas{{A\&AS}}
\def\pasp{{PASP}}
\def\mnras{{MNRAS}}

\def\includepdf{\includegraphics}

\begin{document}

\title{The DEEP2 Galaxy Redshift Survey: Design, Observations,
Data Reduction, and Redshifts\altaffilmark{1}} 
\author{Jeffrey A. Newman\altaffilmark{2},
 Michael C. Cooper\altaffilmark{3,\dag,\ddag},
 Marc Davis\altaffilmark{4},
 S. M. Faber\altaffilmark{5},
 Alison L. Coil\altaffilmark{6,$\nabla$},
 Puragra Guhathakurta\altaffilmark{5}, 
 David C. Koo\altaffilmark{5},  
 Andrew C. Phillips\altaffilmark{5},
 Charlie Conroy\altaffilmark{7},
 Aaron A. Dutton\altaffilmark{8},
 Douglas P. Finkbeiner\altaffilmark{7},
 Brian F. Gerke\altaffilmark{9},
 David J. Rosario\altaffilmark{10}, 
 Benjamin J. Weiner\altaffilmark{11}, 
 C. N. A. Willmer\altaffilmark{11},
 Renbin Yan\altaffilmark{12},
 Justin J. Harker\altaffilmark{5},
 Susan A. Kassin\altaffilmark{13,$\flat$}, 
 N. P. Konidaris\altaffilmark{14},
 Kamson Lai\altaffilmark{5},
 Darren S. Madgwick\altaffilmark{5},
 K. G. Noeske\altaffilmark{15},
 Gregory D. Wirth\altaffilmark{16},
 A. J. Connolly\altaffilmark{17},
 N. Kaiser\altaffilmark{18},
 Evan N. Kirby\altaffilmark{14,\dag},
 Brian C. Lemaux\altaffilmark{19},
 Lihwai Lin\altaffilmark{20},
 Jennifer M. Lotz\altaffilmark{15},
 G. A. Luppino\altaffilmark{18},
 C. Marinoni\altaffilmark{21},
 Daniel J. Matthews\altaffilmark{2},
 Anne Metevier\altaffilmark{22}, 
 Ricardo P. Schiavon\altaffilmark{23}
}

\altaffiltext{1}{Based on observations taken at the W. M. Keck
  Observatory, which is operated jointly by the University of
  California and the California Institute of Technology, and on
  observations made with the NASA/ESO {\it{Hubble Space Telescope}},
  obtained from the data archives at the Space Telescope Science
  Institute, which is operated by the Association of Universities for
  Research in Astronomy, Inc., under NASA contract NAS5-26555, and
  from the Canadian Astronomy Data Centre.}
\altaffiltext{2}{Department of Physics and Astronomy, University of Pittsburgh,
  Pittsburgh, PA 15260, {janewman@pitt.edu, djm70@pitt.edu} }
 \altaffiltext{3}{Center for Galaxy Evolution, Department of Physics
 and Astronomy, University of California, Irvine, 4129 Frederick
 Reines Hall, Irvine, CA 92697, USA; {m.cooper@uci.edu}}
\altaffiltext{\dag}{Hubble Fellow}
\altaffiltext{\ddag} {Spitzer Fellow}
\altaffiltext{4}{Departments of 
Astronomy \& Physics, University of California, 601
  Campbell Hall, Berkeley, CA 94720, {mdavis@berkeley.edu}}
\altaffiltext{5}{UCO/Lick Observatory, University of California, 1156
  High Street, Santa Cruz CA, 95064, {faber@ucolick.org, koo@ucolick.org,
  raja@ucolick.org, jharker@ucolick.org, klai@ucolick.org, phillips@ucolick.org}}
\altaffiltext{6}{Department of Physics, University of California, San Diego,
La Jolla, CA 92093, {acoil@ucsd.edu}} 
\altaffiltext{$\nabla$}{Alfred P.\ Sloan Foundation Fellow}
\altaffiltext{7} {Harvard-Smithsonian Center for Astrophysics, 
Harvard University, 60 Garden St., Cambridge, 
MA 02138, {dfinkbeiner@cfa.harvard.edu, cconroy@cfa.harvard.edu}}
\altaffiltext{8}{Max Planck Institute for Astronomy, K\"onigstuhl 17, 69117, Heidelberg,
Germany, {dutton@mpia.de}}
\altaffiltext{9} {Lawrence Berkeley National Laboratory, 
1 Cyclotron Rd., MS 90R4000, Berkeley, CA 94720 {bfgerke@lbl.gov}}
\altaffiltext{10} {Max-Planck-Institut fŸr extraterrestrische Physik, Giessenbachstra§e, 85748 Garching
Germany, {rosario@mpe.mpg.de}}
\altaffiltext{11}{Steward Observatory, University of Arizona, 
933 N.~Cherry Ave.,Tucson, AZ 85721-0065, 
{bjw@as.arizona.edu, cnaw@as.arizona.edu}}
\altaffiltext{12} {Center for Cosmology and Particle Physics,
  Department of Physics, New York University, 4 Washington Place, New
  York, NY 10003, {ry9@nyu.edu}}
\altaffiltext{13} {Astrophysics Science Division, Goddard Space Flight
  Center, Code 665, Greenbelt, MD 20771, USA, {susan.kassin@nasa.gov}}
\altaffiltext{$\flat$}{NASA Postdoctoral Program Fellow}
\altaffiltext{14} {Astronomy Department, California Institute of
  Technology, 1200 E.\ California Blvd., Pasadena, CA 91125,
  {npk@astro.caltech.edu, enk@astro.caltech.edu}}
\altaffiltext{15} {Space Telescope Science Institute, 3700 San Martin
  Drive, Baltimore, MD 21218, {lotz@stsci.edu, noeske@stsci.edu}}
\altaffiltext{16} {Keck Observatory, Kamuela, HI 96743, 
{wirth@keck.hawaii.edu}}
\altaffiltext{17} {Department of Astronomy, University of
 Washington, Box 2351580, Seattle, WA 98195-1580, {
 ajc@astro.washington.edu}}
\altaffiltext{18} {Institute for Astronomy, 2680 Woodlawn Drive 
 Honolulu, HI, 96822-1897, {kaiser@ifa.hawaii.edu, ger@ifa.hawaii.edu}}
\altaffiltext{19} {Laboratoire d'Astrophysique de Marseille, Marseilles,
France, {brian.lemaux@oamp.fr}}
\altaffiltext{20} {Institute of Astronomy \& Astrophysics, Academia Sinica, Taipei 106,
Taiwan, 
{lihwailin@asiaa.sinica.edu.tw}}
\altaffiltext{21} {Centre de Physique Theorique de Marseilles, Marseilles,
France, {marinoni@cpt.univ-mrs.fr}}
\altaffiltext{22} {Dept. of Physics and Astronomy, Sonoma State University, 1801 E. Cotati Ave., Rohnert Park, CA 94928,
{ajmetevier@gmail.com}}
\altaffiltext{23} {Gemini Observatory, 670 N. A'ohoku Pl., Hilo HI 96720, 
{rschiavo@gemini.edu}}

\begin{abstract}

  In this paper, we describe the design and data analysis of the DEEP2
  Galaxy Redshift Survey, the densest and largest high-precision
  redshift survey of galaxies at $z \sim 1$ completed to date.  The
  survey was designed to conduct a comprehensive census of massive
  galaxies, their properties, environments, and large-scale structure
  down to absolute magnitude $M_B = -20$ at $z \sim 1$ via $\sim 90$
  nights of observation on the Keck telescope. The survey covers an
  area of 2.8 deg$^2$ divided into four separate fields observed to a
  limiting apparent magnitude of $R_{\rm AB}=24.1$.  Objects with $z
  \ltsim 0.7$ are readily identifiable using $BRI$ photometry and
  rejected in three of the four DEEP2 fields, allowing galaxies with
  $z > 0.7$ to be targeted $\sim 2.5$ times more efficiently than in a
  purely magnitude-limited sample.  Approximately sixty percent of
  eligible targets are chosen for spectroscopy, yielding nearly 53,000
  spectra and more than 38,000 reliable redshift measurements. Most of
  the targets which fail to yield secure redshifts are blue objects
  that lie beyond $z \sim 1.45$, where the [O {\scriptsize II}] 3727
  \AA\ doublet lies in the infrared.  The DEIMOS 1200-line/mm grating
  used for the survey delivers high spectral resolution ($R \sim
  6000$), accurate and secure redshifts, and unique internal kinematic
  information.  Extensive ancillary data are available in the DEEP2
  fields, particularly in the Extended Groth Strip, which has evolved
  into one of the richest multiwavelength regions on the sky.

  This paper is intended as a handbook for users of the DEEP2 Data
  Release 4, which includes all DEEP2 spectra and redshifts, as well
  as for the DEEP2 DEIMOS data reduction pipelines.  Extensive details
  are provided on object selection, mask design, biases in target
  selection and redshift measurements, the \emph{spec2d}
  two-dimensional data-reduction pipeline, the \emph{spec1d} automated
  redshift pipeline, and the \emph{zspec} visual redshift verification
  process, along with examples of instrumental signatures or other
  artifacts that in some cases remain after data reduction.  Redshift
  errors and catastrophic failure rates are assessed through more than
  2000 objects with duplicate observations. Sky subtraction is
  essentially photon-limited even under bright OH sky lines; we
  describe the strategies that permitted this, based on high image
  stability, accurate wavelength solutions, and powerful b-spline
  modeling methods.  Summary data are given that demonstrate the
  superiority of DEEP2 over other deep redshift surveys at $z\sim 1$
  in terms of galaxy numbers, redshift accuracy, sample number
  density, and amount of spectral information.  We also provide an
  overview of the scientific highlights of the DEEP2 survey thus far.
\end{abstract}


\section{Introduction}
\label{introduction}

Spectroscopic redshift surveys have been a major gateway to
understanding galaxy evolution.  Locating each galaxy in space and
time, they provide the raw material for compiling a census of galaxy
properties as a function of cosmic epoch and position in the cosmic
web. Clustering measures derived from redshift surveys of sufficiently
large volume -- such as the correlation function, $\xi(r_p,\Pi)$
diagram, and measures of the local overdensity of galaxies -- probe
the gravitational growth of structure.  Pair counts contain
information on merger rates, while satellite motions probe dark-halo
masses and dynamics. When the galaxy sampling is sufficiently dense,
groups of galaxies become well defined and allow the estimation of the
number density of groups (constraining cosmology), as well as the
determination of the environments in which individual galaxies are
found. The spectra themselves can reveal a wealth of detail on
emission-line strengths, stellar populations, star-formation rates,
AGN activity, gas and stellar metallicities, internal motions, and
dynamical masses.  Finally, spectroscopic redshifts provide the
fundamental data needed for calibrating photometric redshifts, which
are the only way to estimate distances for the hundreds of thousands
of fainter galaxies that are beyond the reach of current redshift
surveys.

Pioneering low-redshift surveys out to $z \sim 0.2$, such as the Sloan
Digital Sky Survey (SDSS, York et al.~2000; Abajazian et al.~2003) and
the Two Degree Field Survey (2dF, Colless et al.~2001) have
demonstrated the value of mammoth samples of galaxies containing
hundreds of thousands of objects.  Huge samples enable finer slicing
in redshift and galaxy-parameter space, reveal rare phases of galactic
evolution, and provide sufficient weight to measure {\it statistical
  scatter} about the various galactic scaling laws.

Redshift surveys of more distant galaxies at $z\sim1$ progressed more
slowly, primarily as they have targeted galaxies more than 100 times
fainter than SDSS or 2dF.  The first substantial surveys were
conducted in the 1990's, measuring redshifts of several hundred
objects at intermediate redshift; e.g., the LDSS survey (Colless et
al.~1990), the ESO-Sculptor Survey (Bellanger et al.~1995, Arnouts et
al.~1997), the CNOC and CNOC2 surveys (Yee et al.~1996, 2000), and the
Hawaii Deep Fields Survey (Cowie et al.~1996).

Deeper surveys also began taking shape at roughly the same time.  The
pioneering Canada-France Redshift Survey (CFRS; Lilly et al.~1995)
garnered $\sim600$ redshifts and was the first to provide a dense,
statistical sample of galaxies out to $z\sim1$.  Though conducted with
only a 3.6-m diameter telescope, it had a fairly deep magnitude limit
of $I_{\rm AB}= 22.5$ and yielded a median redshift of $z = 0.56$.
Augmented by Hubble imaging, it established landmark norms for the
luminosities and sizes of galaxies and the star-formation history of
the Universe to $z\sim1$.

The Keck telescopes were also used for deeper surveys. Cowie and
collaborators (Cowie et al.~1996) used roughly 400 redshifts down to
$I\sim23$ in the Hawaii Deep Fields to elucidate the phenomenon of
``downsizing'', that the peak of star formation occurred first in the
most massive galaxies and swept down later to smaller galaxies.  This
concept has since become central to understanding galaxy
evolution. Later surveys by the same group have probed the nature of
distant X-ray sources (e.g., Barger et al.~2005,~2008).

The Caltech Faint Galaxy Redshift Survey (CFGRS, Cohen et al.~2000)
exploited the Keck 10 m telescope and went roughly one magnitude
fainter than CFRS ($R = 23$-24), assembling several hundred redshifts
with median $z=0.70$ in two fields (one of them containing the Hubble
Deep Field-N). Notable discoveries included the extreme amount of
clustering found even at $z\sim1$, the relatively normal morphologies
of most field galaxies observed (i.e., the rarity of true peculiars
and mergers), the measurement of the shape of the luminosity function
to deeper magnitudes, and the exploration of the nature of distant
mid-IR and radio sources (van den Bergh et al.~2000; Hogg, Cohen, \&
Blanford 2000; Cohen 2002).

The largest of this first generation of deep redshift surveys was
DEEP1.  This survey was conducted using the Keck I telescope to a
magnitude limit of $R \sim 24$ in and around the HDF-N and the Groth
Survey Strip (GSS), a 127 arcmin$^2$ region observed with WFPC2 on
$HST$ in both $I$ and $V$ (Groth et al.~1994; Vogt et al.~2005).
DEEP1 was conceived and executed as a pilot survey for the DEEP2
Galaxy Redshift Survey described in this paper.  Over 720 redshifts
were measured in total (Phillips et al.~1997, Weiner et al.~2005)
having a median value of $\sim$0.65, and bulge and disk parameters
were fitted to the Groth Strip $HST$ images (Simard et
al.~2002). Science highlights include an early survey that helped to
establish the nature of Lyman-break galaxies (Lowenthal et al.~1997),
the detection of a strong color bimodality out to $z\sim 1$ (Im et
al.~2002, Weiner et al.~2005), the measurement of fundamental plane
evolution for distant field spheroidal galaxies (Gebhardt et
al.~2003), the detection of ``downsizing'' in the star formation rate
of compact galaxies (Guzman et al.~1997), measurements of ISM
metallicity evolution to $z\sim 1$ (Kobulnicky et al.~2003), an
extensive analysis of selection effects on disk-galaxy radii and
surface brightnesses (Simard et al.~2002), and the discovery of the
prevelance of already-red bulges in typical disk galaxies at $z\sim1$
(Koo et al.~2005; Weiner et al.~2005).  A distinguishing feature of
DEEP1 was the use of relatively high spectral resolution ($R\sim
3000$, FWHM = 3-4 \AA), which resolved redshift ambiguities by
splitting the [O {\scriptsize II}] $\lambda$3727 doublet and also
yielded linewidths and rotation curves. These were used to extend the
Tully-Fisher relation to $z\sim1$ (Vogt et al.~1996, 1997), a strategy
that was later used in the Team Keck Redshift Survey (GOODS-N; Wirth
et al.~2004) to yield an even larger sample of linewidths (Weiner et
al.~2006a,b).

DEEP1 was a valuable learning experience.  In addition to proving the
scientific value of high spectral resolution, it confirmed that faint
spectral features could be more easily seen in the dark stretches of
spectrum {\it between} the OH lines. On the other hand, achieving
photon-limited sky subtraction under the newly concentrated OH lines
places even higher requirements on flat-fielding accuracy, wavelength
calibration, and night-sky spectral modeling.  Furthermore, pushing
beyond the {\it de facto} redshift limit that then existed near $z\sim
1$ (the so-called ``redshift desert'') would require pursuing [O
{\scriptsize II}] $\lambda$3727 to well beyond 8000 \AA. A new
spectrograph on one of the largest optical telescopes would be
necessary to achieve these goals.

The present paper describes the design and execution of the DEEP2
Galaxy Redshift Survey (commonly referred to as DEEP2 in the remainder
of this paper), which was conducted by a collaboration of observers
primarily based at UC Berkeley and UC Santa Cruz, together with
astronomers from the University of Hawaii and other institutions.  The
survey was executed with the DEIMOS multi-object spectrograph (Faber
et al.~2003) on Keck 2; aspects of the survey have previously been
described in Davis et al.~(2003); Davis, Gerke \& Newman (2005); Coil
et al.~(2004b); and Willmer et al.~(2006).  The survey obtained
roughly 53,000 spectra, coming close to the target of $\sim60,000$
galaxies in the original survey design.

DEEP2 was conceived as a legacy survey for the astronomical community,
and all data are now public.  The present paper is intended as a
handbook for users to understand and utilize the data products in Data
Release 4 (hereafter DR4), which is available on the web at
\url{http://deep.berkeley.edu/DR4/}.  Brief descriptions of the data
and methods are given there, but the present account is more complete
and presents new details on the final data reduction techniques,
redshift measurement procedures, the current generation of redshifts,
and summary data tables.  Additional measurements that were not part
of previous DEEP2 data products are also included.  All data in this
paper are available in the form of FITS BINTABLE files at the DR4
website.

The outline of this paper is as follows.
Section~\ref{designconsiderations} sets forth the overarching goals
and tradeoffs that determined the design of the DEEP2 survey.
Section~\ref{scienceresults} summarizes some of the major science
results so far, giving examples of how various aspects of the design
have translated into specific science gains.
Section~\ref{surveyparameters} gives the basic DEEP2 survey parameters
and field properties and describes the extensive existing ancillary
data that are available in the DEEP2 fields.  DEEP2 is put into
context in Section~\ref{othersurveys}, which compares DEEP2 to other
leading $z\sim1$ surveys (VVDS-deep, VVDS-wide, zCOSMOS, and PRIMUS)
using several different measures of survey size and information
content.

Section~\ref{objectselection} describes the parent galaxy photometric
catalogs and selection cuts that were used to determine potential
targets for DEEP2 spectroscopy.  The slitmask design process is
covered in Section~\ref{maskdesign}, which describes the DEIMOS
detector and slitmask geometry, as well as the algorithms used to
place objects on masks.  Section~\ref{targetbiases} is an exhaustive
discussion of all known selection effects resulting from the slitmask
design process, caused either by loss of objects from the initial
target pool or by systematics in the rate at which we select pool
objects of given properties for masks.

Sections~\ref{exposures} and \ref{reduceddata} describe the
spectroscopic exposures and their reduction, including detailed
information on the \emph{spec2d} pipeline that produces the reduced
1-d and 2-d spectra.  This pipeline has been employed by many DEIMOS
users; this paper provides its definitive description.  The features
of the DEIMOS spectrograph and \emph{spec2d} pipeline that enable
photon-limited sky subtraction accuracy even under the OH lines are
described in the Appendix.  The range in quality of DEEP2 data is
discussed in Section~\ref{reduceddata}, where examples of typical data, bad
data, and interesting spectra are given.

Section~\ref{redshiftmeasurements} describes the redshift measurement
process, including the automated \emph{spec1d} redshift pipeline that
produces a menu of redshift possibilities and the subsequent visual
inspection process to select among them via the interactive user GUI
\emph{zspec}.  Final results on redshift success, quality codes,
completeness, biases, and redshift accuracy are also given here.
Section~\ref{multiplicity} investigates the impact of multiple
galaxies masquerading as single galaxies in the ground-based
\emph{pcat} photometry. Section~\ref{datatables} is a guide to DR4 and
the two main data tables in this paper, which summarize the mask data
and final redshifts and quality codes.

Section~\ref{selectionfunction} is provided to enable correlation
function measurements with DR4 data and to aid theorists who wish to
make mock DEEP2 catalogs from model galaxy data.  It describes a set
of files being distributed along with DR4, which describe the
selection and redshift success probability for a DEEP2 target as a
function of location on the sky. Finally, Section~\ref{redshifttrends}
summarizes the properties of the DEEP2 dataset via plots of numbers,
colors, magnitudes, etc., as a function of redshift.

Depending on their interests, a reader may wish to focus on certain
sections of this paper.  For instance:

\begin{itemize}

\item {\bf Users of DEEP2 Data Release 4 data:} Key sections are
  \S\ref{surveyparameters} (survey parameters), \S\ref{targetbiases}
  (assessment of biases in targeting), \S\ref{samplespectra} (presenting
  sample spectra), \S\ref{redshiftresults} and \S\ref{redshiftresults2}
  (redshift results and completeness), \S\ref{datatables} (describing
  the released data tables), \S\ref{selectionfunction} (describing the
  released 2-d selection functions) and \S\ref{redshifttrends}
  (describing trends in the sample with $z$).

\item {\bf Individuals interested in the design of DEEP2 and
    comparison to other deep surveys:} Key sections are
  \S\ref{designconsiderations} (design goals), \S\ref{scienceresults}
  (science results), \S\ref{surveyparameters} (survey parameters),
  \S\ref{othersurveys} (comparison to other surveys),
  \S\ref{targetbiases} (assessment of biases in targeting),
  \S\ref{multiplicity} (evaluating how frequently spectroscopic
  targets consisted of multiple unresolved galaxies), and
  \S\ref{redshifttrends} (describing trends in the sample with $z$).

\item {\bf Users of the DEEP2/DEIMOS data reduction pipelines:} Key
  sections are \S\ref{reduceddata} (on the 2-d data reduction pipeline
  and spectral extractions) and \S\ref{redshiftmeasurements} (on the
  redshift determination pipeline).

\item {\bf Individuals interested in details of DEEP2 target selection
    and techniques:} Key sections are \S\ref{objectselection}
  (describing the target selection criteria), \S\ref{maskdesign}
  (slitmask design algorithms), and \S\ref{exposures} (observational
  strategy).
\end{itemize}

Throughout this paper, unless specified otherwise we utilize
magnitudes on the AB system and assume a $\Lambda$CDM concordance
cosmology with $\Omega_m = 0.3$, $\Omega_{\Lambda}$=0.7, and $H_0 =
100$ $h$ km s$^{-1}$ Mpc$^{-1}$. The DR4 data tables also use these
quantities.

\section{Considerations Guiding the Design of the DEEP2 Survey}
\label{designconsiderations}

DEEP2 was originally envisioned as a tool for simultaneously studying
galaxy evolution and the growth of large-scale structure since $z\sim
1$.  However, while the survey was being designed it was realized that
such data could also be used to constrain the nature of dark energy
(i.e., its equation of state) by counting the abundance of groups and
clusters as a function of velocity dispersion and redshift (Newman et
al.~2002; Newman \& Davis 2000, 2002; Gerke et al.~2005).  This was
one of the first methods proposed for studying dark energy by counting
massive objects (see also Haiman, Mohr, \& Holder 2001).  The velocity
and redshift distribution of clusters depends on both the growth rate
of large-scale structure and on the volume element (i.e., the amount
of volume per redshift interval per angular area), which are each
related to fundamental cosmological parameters in predictable ways.
Counting the abundance of intermediate-mass groups to high accuracy
therefore became a third major goal guiding the design of DEEP2.

It was clear from the start that one of the most important DEEP2 data
products would be {\it precision counts}, to be compared with various
types of model predictions. Obtaining high-precision counts requires:
\begin{itemize}
\item Many galaxies,
\item An accurately known and simple 
      selection function -- DEEP2 is  
      basically magnitude-limited with nearly uniform 
      sampling on the sky and in redshift space  at $z > 0.75$ -- and,
\item Low sample (commonly referred to as ``cosmic'') variance.
\end{itemize}
These aims -- {\it large numbers, simple selection function, and low
  cosmic variance} -- were the central guiding principles for DEEP2.
They all had to be achieved, however, while using as little total
telescope time as possible.
				   
Optimizing the measurement of environment and clustering statistics
played a major role in setting the sizes and shapes of the DEEP2
survey fields.  Galaxy properties vary systematically with
environment, and the length-scale of this influence can shed important
light on the physical mechanisms causing these effects: for example,
differentiating between processes associated with the parent dark
matter halo of a galaxy from those that act on larger scales. Since
the typical comoving radius of galaxy groups and clusters is of order
$\sim 1$ $h^{-1}$ Mpc, measuring the local overdensity of other
galaxies around a given object (the most common measure of environment
used today) requires counting neighbors to a separation at least this
large, corresponding to 1\m.5 at $ z \sim 1$.  As a result, even when
corrections for lost area are applied, neighbor counts become
inaccurate for galaxies too near to survey boundaries (cf.\ Cooper et
al.~2005); minimizing the number of lost objects in environment
studies requires:
\begin{itemize}
\item Fields that are at least ten times wider in each direction than
  the radius within which environments are measured, corresponding to
  a minimum dimension of $\sim$15\amin\ on the sky for a survey at
  $z\sim 1$.
\end{itemize}   
This restriction also ensures that fields will be well-sized for
characterizing groups and clusters, in support of our third science
goal.

The measurement of correlation functions drives the choice of a
minimum field dimension in similar ways, as the dominant error terms
on large scales increase rapidly for pair separations larger than half
the minimum field dimension (e.g., Bernstein 1994).  It is also
necessary to survey a large volume to minimize the effects of cosmic
variance on correlation function estimates.  Typical values of the
scale length of galaxy clustering, $r_0$, for galaxy populations of
interest are generally $\lesssim 5$ $h^{-1}$ co-moving Mpc at $z \sim
1$ (Coil et al.~2004a, 2008), corresponding to an angular separation
of 7\m.4.  Measuring correlation functions well therefore requires:
\begin{itemize}
\item Contiguous volumes spanning several times $r_0$ in one
  dimension, with the other two dimensions spanning many times $r_0$.
\end{itemize}
A field size of 0.5\deg\ $\times$ 2\deg\ was accordingly selected for
three out of the four DEEP2 fields, corresponding to 4 $r_0$ by 16
$r_0$ at $z = 1$.  Rectangular fields rather than square fields were
chosen to increase the statistical power of the sample: the opposite
ends of a strip will be more statistically independent than regions
closer to each other, reducing cosmic variance (Newman \& Davis 2002).

Since environment measures involve the pairwise counting of objects,
the statistical weight of such data increases as the {\it square} of
the volume density of galaxies sampled, all other things being
equal.\footnote{This is straightforward to understand when considering
  the use of counts within some aperture (cylinder, sphere, etc.) as a
  measure of local overdensity. When the number density, $n$, is
  increased, the fractional uncertainty in each object's neighbor
  count (i.e., its environment estimate) goes down as $1 \over
  \sqrt{n}$, but the number of objects in the sample also increases as
  ${n}$.  Hence, the standard deviation of the mean overdensity
  determined for a given sample scales as 1/$n$, corresponding to an
  $n^2$ increase in inverse variance/statistical weight.}  Essentially
all environmental studies in the Poisson-dominated regime should
benefit from this scaling, including those which employ a wide variety
of galaxy environment estimators, the measurement of correlation
functions and cross-correlations for dilute samples or at small
scales, pair counts, comparisons of group and field populations, and
studies incorporating the dynamical masses of groups.  The DEEP2
targeting strategy is therefore designed to obtain:
\begin{itemize}
\item A high number density of galaxies (or equivalently, number of
  objects per unit area and redshift) down to the faintest feasible
  magnitude limit.
\end{itemize}

However, under the constraint of finite telescope time, the need for
large, densely sampled fields to enable accurate correlation function
and environment measurements conflicts with the desirability of
covering many independent regions on the sky to minimize (and allow
assessment of) cosmic variance.  Since its subregions will be
correlated due to large-scale modes of the power spectrum, the sample
variance in a single large field will always be greater than in a
larger number of widely-separated fields on the sky covering the same
total area.  Our preferences for large field sizes but small cosmic
variance are most efficiently reconciled by choosing:
\begin{itemize}
\item Several widely separated areas on the sky, each one of which
  exceeding the minimum width and length needed for correlation and
  environment studies.
\end{itemize}
The final choice to survey four fields was driven by cosmic variance
calculations utilizing the QUICKCV code of Newman \& Davis
(2002)\footnote{The QUICKCV code is publicly available at
  \url{http://www.phyast.pitt.edu/$\sim$janewman/research.html}.} and
by the instability of standard deviation measurements with fewer
samples. Surveying multiple fields spaced well apart in RA also makes
telescope scheduling easier.

Obtaining a dense sample of galaxies can be achieved by selecting down
to a very faint magnitude limit; but that would require prohibitive
observation time.  Instead, it is desirable to go as faint as one must
to obtain enough targets to fill slitmasks, but no fainter.  Since the
DEIMOS spectrograph generally has 10 slots available for user
slitmasks, observation times of order 1 hour per mask are optimal.  In
order to obtain redshifts for a majority of objects in an hour's
observation time, we desire:
\begin{itemize}
\item A target population brighter than $R_{\rm AB} \sim 24$.
\end{itemize}
In order to include a large enough sample to optimize the filling of
slitmasks, the final DEEP2 sample extends down to $R_{\rm AB} = 24.1$.

A single DEIMOS slitmask covers an area on the sky of roughly 5\m.3
$\times$ 16\m.7 (cf.\ \S\ref{maskdesign}), of which more than 30\% is
unsuitable for galaxy target slitlets due to vignetting, chip gaps, or
variation in wavelength coverage.  DEIMOS masks can generally include
$\sim 130-150$ objects, corresponding to a surface density of 2.2--2.6
galaxies per square arcminute, comparable to the number counts of
galaxies down to $R_{\rm AB} \sim 21.4$.  However, targeting just
those objects would result in a relatively bright and quite dilute
sample of galaxies, weighted towards more luminous galaxies and lower
redshifts than were of interest for DEEP2.  Furthermore, many objects
of interest would be lost as we are unable to observe multiple objects
at the same displacement in the spatial direction on a slitmask, lest
spectra overlap, reducing our sampling of overdense regions.  Given
our desire to maximize the number density of high-redshift galaxies,
we therefore require:
\begin{itemize}
\item Rejection of lower-redshift ($z < 0.7$) galaxies, and coverage
  of each point on the sky by at least two slitmasks.
\end{itemize}
In the Extended Groth Strip (EGS, DEEP2 Field 1), however, we do not
reject low-$z$ galaxies, both to test our selection methods and to
take advantage of the rich multiwavelength data in that field; as a
consequence, in that field we require coverage of each point on the
sky by at least four slitmasks to compensate.  To keep the number of
slitmasks per field roughly constant, the width of the area covered is
half as large in the EGS as for the remaining fields (i.e., the survey
area goal in this field was 0.25\deg\ $\times$ 2\deg, rather than
0.5\deg\ $\times$ 2\deg\ as in the other DEEP2 fields).

The field area of 0.5\deg\ $\times$ 2\deg\ is roughly 60 times the
area covered by a single DEIMOS slitmask; double-coverage thus
requires 120 slitmasks per field.  Assuming an average of 135 targets
per slitmask, this led to a survey goal of $\sim$65,000 spectra.

With the field geometry and sample sizes determined, the main
parameters remaining were spectral dispersion and spectral coverage.
These also involve tradeoffs.  For a fixed detector size, low
dispersion captures more spectral features and thereby enhances
spectral information and (potentially) redshift completeness.
However, as noted above, high dispersion yields more information per
angstrom by resolving internal galaxy motions and yielding more
accurate redshifts.  It splits the [O {\scriptsize II}] $\lambda$3727
doublet, converting that feature into a unique, robust redshift
indicator even when it is the only feature visible.  At high
resolution, OH sky lines are concentrated into a few pixels, leaving
the remainder of the pixels dark and thereby shortening required
exposure times (see \S\ref{dataquality}).  High dispersion also
reduces flat-fielding errors due to ``fringing'' (see
Appendix~\ref{fringing}).  In conventional grating spectrographs, it
increases anamorphic demagnification, which narrows slitwidth images
still further and reduces the displacement of spectra along the
spectral direction due to variations in slitlet position, maximizing
the spectral region common to all objects.  Finally, high dispersion
allows the use of wider slits for the same net spectral resolution,
thereby capturing more galaxy light.

Evidently, we would like to have both high dispersion and broad
spectral coverage.  The DEIMOS CCD detector/camera system was designed
to do this by providing a full 8192 pixels along the spectral
direction, with minimal chip gap.  Our ideal choice balancing
resolution and coverage would have used an 830-line/mm grating, which
covers 3900 \AA\ of spectrum with DEIMOS.  However, this grating
suffers significantly from ghosting, which we feared would interfere
with accurate sky subtraction.  We therefore conservatively chose the
1200-line/mm grating, which captures 2600 \AA\ of spectrum over 2000
separate resolution elements.  When centered at 7800 \AA, this
spectral range yields at least one strong spectral feature for
essentially all galaxies from $z = 0$ to $ z = 1.4$.  Redshift
ambiguities are rare and, if present, can largely be resolved in the
future by using photometric redshifts (Kirby et al.~2007).  We
therefore selected:
\begin{itemize} 
\item DEIMOS' highest-resolution grating,
      1200 line mm$^{-1}$, typically covering 
      6500 \AA\ to 9100 \AA\  and 
      yielding a 
      spectral resolution $R \equiv \lambda/\Delta\lambda \sim 6000$ 
      with a 1\asec-wide slit.\footnote{Here $\Delta\lambda$ 
is the FWHM of an OH sky line.}
\end{itemize} 

Practical considerations influenced many of the final design decisions
for DEEP2.  The detailed layout of slitmasks on the sky was governed
by the DEIMOS slitmask geometry (cf.\ \S\ref{maskdesign}).  The final
fields were chosen to have low Galactic reddening utilizing the
Schlegel, Finkbeiner, \& Davis (1998) dust map and to be well spaced in
RA to permit good-weather observing at Keck.  Field 1 (the Extended
Groth Strip) was specially selected for its excellent multiwavelength
supplemental data (as described in \S\ref{surveyparameters}).

Finally, detailed attention was paid to ensuring good sky-subtraction
accuracy even under the bright OH lines.  These lines would occupy
only a small portion of the spectrum (due to the use of high
dispersion), but any sky-subtraction errors would spread to other
wavelengths whenever spectra were smoothed.  Requirements for
excellent sky subtraction include (1) a highly accurate and stable
calibration of the wavelength corresponding to every pixel on the
detector, (2) a constant PSF for OH lines along each slitlet, and (3)
highly reproducible CCD flat-fielding, which was to be attained by
maintaining {\it exactly the same wavelength on each pixel} between
the afternoon flat-field calibration and the night-time observation.
These requirements in turn require a very stable spectrograph, uniform
image quality over the field of view, careful calibration and
observing techniques, and high-precision data reduction methods.
Further details are provided in the Appendix, which discusses CCD
fringing, sky subtraction, the DEIMOS flexure compensation system,
image stability, and wavelength calibration methods.

\section{Science Highlights of DEEP2}
\label{scienceresults}

The design resulting from the above considerations has enabled a wide
variety of scientific investigations; more than 75 refereed papers
have appeared as of this writing.  An early summary appeared in the
May 1, 2007, special issue of {\it Astrophysical Journal Letters}
(devoted to the AEGIS collaboration, described in
\S\ref{surveyparameters}); Table~\ref{table.previouspapers} of this
paper presents an updated list of selected highlights. Amongst our
major findings are:

\begin{itemize}
\item DEEP2 confirmed and buttressed the conclusion of Bell et
  al.~(2004) that the number of red-and-dead, quenched galaxies has at
  least doubled since $z\sim1$, implying that some galaxies arrived on
  the red sequence relatively recently (Willmer et al.~2006; Faber et
  al.~2007).  Stellar populations imply late quenching (Schiavon et
  al.~(2007), and the mean restframe $U-V$ color of galaxies evolves
  slowly due to these late arivals (Harker et al.~2006).
\end{itemize}

\begin{itemize}
\item Star-formation rates have been measured as a function of~stellar
  mass using both 24-$\micron$ fluxes and GALEX photometry (Noeske et
  al.~2007a; Salim et al.~2009).  In star-forming galaxies, the star
  formation rate follows stellar mass closely at all epochs out to
  $z\sim1$, displaying a ``star-forming main sequence'' that can be
  modeled by assuming that larger galaxies start to form stars earlier
  and shut down sooner (Noeske et al.~2007b).  The RMS scatter about
  this relation is of order a factor of two.  Most star-formation at
  $z \sim 1$ is taking place in normal galaxies, not in
  merger-triggered starbursts (Noeske et al.~2007b, Harker 2008).  Red
  spheroidal galaxies are truly quenched, and the strong [O
  {\scriptsize II}] emission in some is due to AGN and/or LINER
  activity (Konidaris 2008, Yan et al.~2006).
\end{itemize}

\begin{itemize}
\item The fraction of quenched galaxies is larger in high-density
  environments (Cooper et al.~2006) and is higher in groups versus the
  field (Gerke et al.~2007). However, the decline in global star
  formation rate since $z \sim 1$ is strong in all environments, and
  cannot be driven by environmental effects (Cooper et al.~2008). The
  fraction of quenched galaxies in high-density environments decreases
  with lookback time and essentially vanishes at $z \gtrsim 1.3$,
  implying that massive galaxies started shutting down in bulk near $z
  \gtrsim 2$. Environment correlates much more strongly with galaxy
  color than luminosity; a color-density relation persists to $z > 1$
  (Cooper et al.~2006, 2007, 2010). Many massive star-forming galaxies
  were still found in high-density regions at $z \sim 1$ but have no
  analogs today; they must have quenched in the interim time (Cooper
  et al.~2006, 2008). Post-starburst galaxies are found in similar
  environments as quenched galaxies at $z~0.8$, but similar to
  star-forming galaxies at $z~0.1$; given the overall movement of
  galaxies towards denser environments at lower $z$, the absolute (as
  opposed to relative) overdensity around quenching galaxies may have
  not have evolved with redshift (Yan et al. 2009).  Finally, massive
  early-type galaxies in high-density regions are $\gtrsim \! 25\%$
  larger in size relative to their counterparts in low-density
  environs at $z \sim 1$, strongly suggesting that minor mergers may
  be important in the size evolution of massive ellipticals (Cooper et
  al.~2012a).
\end{itemize}

\begin{itemize}
\item The roots of the Tully-Fisher relation are visible out to $z\sim
  1$ (Kassin et al.~2007).  Disturbed and merging galaxies have low
  rotation velocities but high line-widths that will place them on the
  TF relation after settling (Covington et al.~2009).  The zero point
  of the stellar-mass TF relation has evolved very little since
  $z\sim1$, in agreement with LCDM based models (Dutton et al.~2011),
  and the Faber-Jackson relation for spheroids likely arose from the
  TF relation via mergers and quenching.
\end{itemize}

\begin{itemize}
\item The number of disky galaxies has declined and the number of
  bulge-dominated galaxies has increased since $z\sim 1$ (Lotz et
  al.~2008).  However, X-ray detected AGN are not preferentially found
  in major mergers, and the fraction of merging galaxies is not
  appreciably higher at $z\sim1$ than now (Pierce et al.~2007; Lin et
  al.~2004, 2008; Lotz et al.~2008).  Many distant galaxies previously
  classed as peculiar or merging are more likely normal disk galaxies
  in the process of settling.
\end{itemize}

\begin{itemize}
\item The clustering properties of a wide variety of galaxy and AGN
  samples have been measured.  The auto-correlation function of
  typical DEEP2 galaxies implies halo masses near 10$^{12}$
  M$_{\sun}$.  The clustering amplitude is a stronger function of
  color (Coil et al.~2008) than luminosity (Coil et al.~2006b),
  matching the results from environment studies.  Red galaxies cluster
  more strongly than blue ones, with intermediate-color "green valley"
  galaxies preferentially found on the outskirts of the same massive
  halos that host red galaxies (Coil et al.~2008). The halo mass
  associated with star formation quenching has been quantified both
  through the correlation function (Coil et al.~2008) and through the
  motions of satellite galaxies (Conroy et al.~2007).  Coil et
  al.~(2006a) use the clustering of galaxies with group centers to
  separate out the ``one-halo'' and ``two-halo'' contributions to the
  correlation function.  Using the cross-correlation of AGN and DEEP2
  galaxies, Coil et al.~(2007) find that quasars cluster like blue,
  star-forming galaxies at $z=1$, while lower-accretion X-ray detected
  AGN cluster similarly to red, quiescent galaxies and are more likely
  to reside in galaxy groups (Coil et al.~2009).  Comparison between
  X-ray and optically-selected AGNs indicates that the fractions of
  obscured AGN and Compton-thick AGNs at $z \sim 0.6$ are at least as
  large as those fractions in the local universe (Yan et al.~2011).

\end{itemize}

\begin{itemize}
\item Ubiquitous outflowing winds have been detected in normal
  star-forming galaxies for the first time, at $z \sim 1.3$ using [Mg
  {\scriptsize II}] absorption (Weiner et al.~2009).  Outflow speeds
  are higher in more massive galaxies and are comparable to the escape
  velocity at all masses.  At $z$=0.1-0.5, outflows are detected in
  both blue and red galaxies but are more frequent in galaxies with
  high IR luminosity or recently truncated star formation (Sato et
  al.~2009).  Overall, these results suggest that stellar feedback is
  an important process in regulating the gas content and star
  formation rates of star-forming galaxies and that galactic-scale
  outflows play an important role in the quenching and migration of
  blue-cloud galaxies to the red sequence.
\end{itemize}

\begin{itemize}
\item The cause of AGN activity and the role of AGN in galaxy
  quenching remain major puzzles.  X-ray AGN are found in massive
  galaxies at the top of the blue cloud, on the red sequence, and in
  the green valley (Nandra et al.~2007), as expected if AGN activity
  is associated with black-hole growth and bulge-building.  However,
  X-ray AGNs are not especially associated with mergers (Pierce et
  al.~2007) or with high star formation rates (Georgakakis et
  al.~2008).  A surprise is the large number of luminous, hard,
  obscured X-ray sources in morphologically normal red spheroidals and
  post-starburst galaxes (Georgakakis et al.~2008), well after star
  formation should nominally have stopped.  If AGN feedback causes
  galaxy quenching, these persistent hard AGN imply that considerable
  nuclear gas and dust must remain in the central regions, where it
  continues to feed the black hole.
\end{itemize}

To summarize these findings: the roots of most present-day galaxy
properties are visible far back in time to the edge of DEEP2 ($z=1.4$,
more than 8 billion years in the past); galaxy evolution over this
time has on balance been a fairly regular and predictable process; and
a good predictor of a galaxy's properties at any era is its stellar
mass.  An overarching hypothesis that may unite all of these findings
is that the most important physical driver controlling a galaxy's
evolution at $z<1.4$ is the growth of its dark halo mass versus
time. On the other hand, several of the above findings point to close
couplings between AGNs, starbursts, quenching, and galactic winds, and
the causal connections among these seems more complex than the simple
halo-driven model would predict.  These issues are the great frontier
for future work.

The present data in DR4 do not exhaust the information from DEEP2.
Many additional quantities have been derived from survey data, and
have either been published separately (see
Table~\ref{table.previouspapers}, or may be obtained by contacting the
individual scientists involved.  These include measurements of
emission-line equivalent widths (from B.~Weiner, R.~Yan, N.~Konidaris,
and J.~Harker); improved velocity widths for emission and absorption
lines (from B.~Weiner and S.~Kassin); local environmental densities
(from M.~Cooper); group identifications and group memberships (from
B.~Gerke); stellar masses (from K.~Bundy and C.~N.~A.~Willmer); ISM
metallicities (from A.~Phillips); D4000 and Balmer-line absorption
strengths (from J.~Harker and R.~Yan), optical AGN IDs (from R.~Yan),
morphologies (from J.~Lotz), pair catalogs (from L.~Lin), satellite
kinematics (from C.~Conroy), star-formation rates (from K.~Noeske and
S.~Salim), and indicators for poststarbust galaxies (from R.~Yan).

\section{Survey Overview and Field Selection}
\label{surveyparameters}

In Section \ref{designconsiderations}, we presented the goals which
were used to fix the basic properties of the DEEP2 survey.  In the
remainder of this paper, we describe the design and execution of the
project and the resulting datasets in more detail.  This first section
serves as a basic introduction to the DEEP2 survey.

To first order, DEEP2 is a magnitude-limited spectroscopic galaxy
redshift survey with limiting magnitude of $R_{\rm AB} =
24.1$.\footnote{In some cases, spectra of fainter objects which
  serendipitously fell on DEIMOS slitlets were obtained and extracted
  by our data reduction pipelines; such objects are not presented
  here, as their selection is less well-characterized than the main
  DEEP2 sample, though they may be included in a later data release.}
Instrument and exposure parameters are summarized in
Table~\ref{table.instrumentparams}.  The survey was designed to be
executed in four separate 0.5\deg\ (or 0.25\deg\ in the case of the
Extended Groth Strip) $\times$ 2\deg\ rectangular fields widely spaced
in Right Ascension, with an average of 120 DEIMOS slitmasks per field.
The spectral setup used the 1200-line/mm high-resolution DEIMOS
grating with a spectral resolution of $R \sim 6000$ and a central
wavelength of 7800 \AA; the typical exposure time was 1 hour per mask.
The survey is primarily sensitive to galaxies below a redshift of $z
\sim 1.45$, past which the [O {\scriptsize II}] 3727 \AA\ doublet
moves beyond the red limit of our typical spectral coverage. The total
number of spectra obtained is 52,989, and the total number of objects
with secure (classes with $>95\%$ repeatability) redshifts is 38,348.
Due to the total allocation of telescope time being somewhat smaller
than originally envisioned, the total area of sky covered by the final
survey sample is roughly 3.0 square degrees (rather than 3.5), and 411
(rather than 480) slitmasks were observed.

Objects are pre-selected in DEEP2 Fields 2, 3, and 4 using broad-band
CFHT 12K $BRI$ photometry to remove foreground galaxies below $z \sim
0.7$ (cf.\ Section~\ref{pre-selection}); this selection is essentially
complete for galaxies at $z>0.75$.  This preselection removes 60\% of
galaxies brighter than $R_{\rm AB}=24.1$, thus multiplying the
efficiency of the survey for studying galaxies at $z \sim 1$ by nearly
$2.5\times$.  Field 1 is the Extended Groth Strip (EGS, Groth et
al.~1994) and is treated differently owing to the wealth of ancillary
data there.  In that field, we do not reject low-redshift galaxies on
the basis of their colors.

To select objects, we first define a target pool of candidate galaxies
in each field, consisting of all objects between $R_{\rm AB} = 18.5$
and $R_{\rm AB} = 24.1$ except those with exceptionally low surface
brightness or large ($>80\%$) Bayesian probability of being a
star.\footnote{Three sets of weak deweighting functions are applied to
  suppress very faint galaxies, very nearby galaxies, and very blue
  galaxies, as explained in \S\ref{sample}.}  The photometric
catalogs ("{\it pcat}"s) used for this selection are described in
\S\ref{photometry}.  We then apply a color pre-selection to this pool
to remove $z<0.7$ objects (in Fields 2, 3, and 4). The remaining
galaxies are used to design the masks; in a typical field, roughly
60\% of potential targets receive slitlets.
 
The chosen depth of $R_{\rm AB} = 24.1$ is suitable for several
reasons.  At this depth, DEEP2 probes down to $L^*$ in the luminosity
function at $z \sim 1.2$.  The surface density of 20,000
objects/\sq\deg\ (after high-$z$ preselection) is large enough to pack
slitmasks efficiently while leaving a modest excess for flexibility.
Finally, the majority of galaxies at this magnitude limit still yield
secure redshifts in one hour of observation time.

Information on the four DEEP2 fields is given in
Table~\ref{table.fields}, including coordinates, average reddening,
field sizes, number of slitmasks planned and observed, and number of
target galaxies and redshifts.  By ranging in RA from 14$^{\rm h}$
20$^{\rm m}$ to 2$^{\rm h}$ 30$^{\rm m}$, with spacings of 3-7 hours
between fields, the four DEEP2 fields are well suited for observing in
what is historically a good-weather period at Keck, ranging from March
through October.  Masks were designed over 0.5\deg\ wide by 2\deg\
long regions in Fields 2, 3, and 4, but not all of these masks were
ultimately observed.  As noted, Field 1 (EGS) is special: it hosts a
wide variety of deep multiwavelength data from X-rays to radio.  It is
also one of the darkest and most dust-free regions of the sky.  Since
no $z > 0.7$ preselection was done in Field 1, the surface density of
targets on the sky there is roughly twice that in the other DEEP2
fields; in order to sample high-redshift objects at the same number
density as in other fields, only half as much area could be covered
using the same number of slitmasks.  As a result, the spectroscopy in
EGS covers an area that is only $\sim$16\amin\ wide (but still 2\deg\
long).  All masks in EGS were observed.

\begin{figure}[t]
\ifpdffig
\includepdf[scale=0.4]{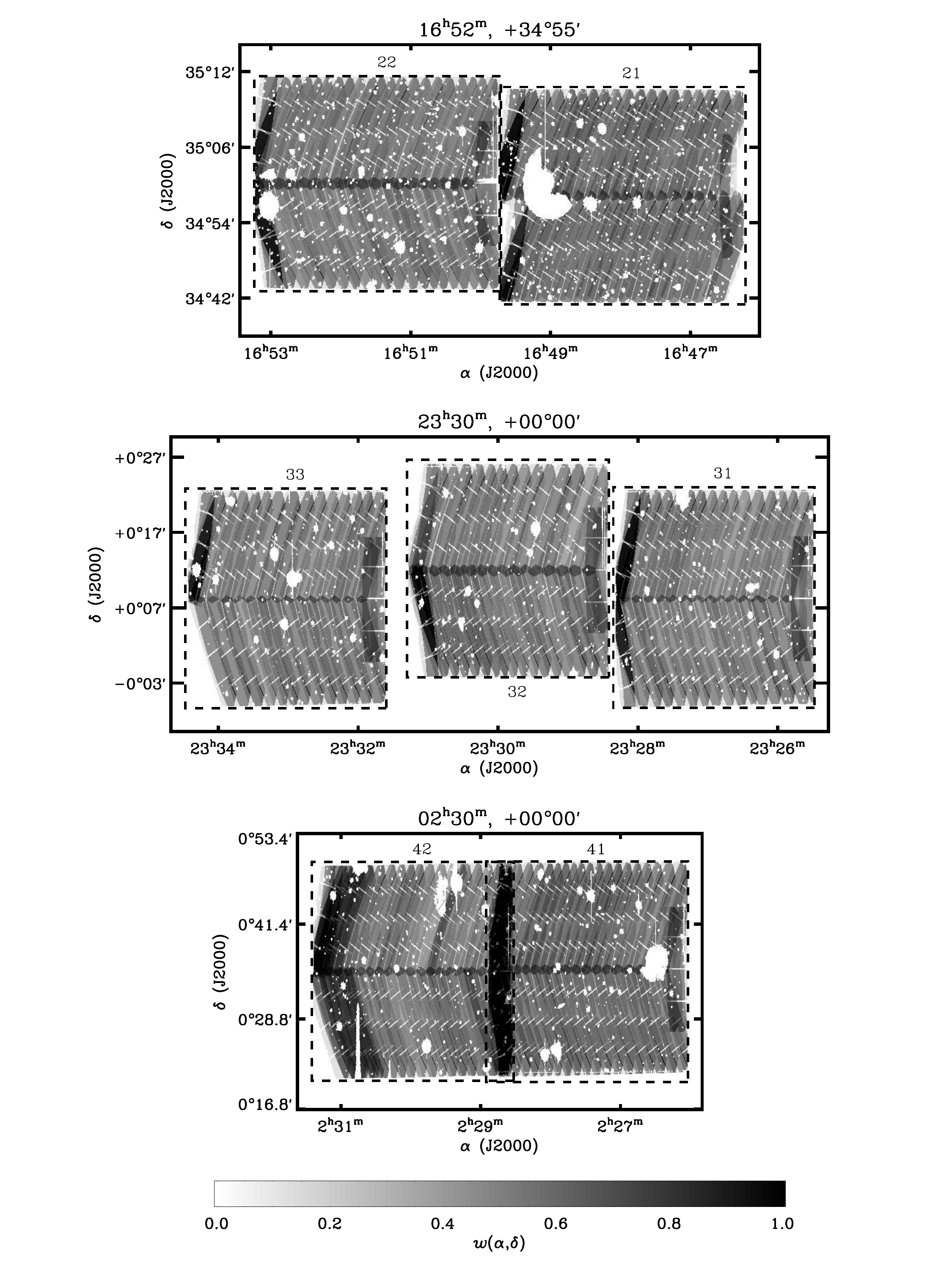}
\else
\includegraphics[scale=0.4]{cooper_deep2_wfn.eps}
\fi
\caption{2-d redshift completeness maps, $w(\alpha,\delta)$, in DEEP2
  Fields 2, 3, and 4, updated from Cooper et al.~(2006).  The
  boundaries for each pointing of the DEEP2 CFHT 12K $BRI$ photometry
  are indicated by the dashed lines, labelled by the pointing number
  (e.g., pointing 21 is the first pointing in DEEP2 Field 2).  The
  grey scale at each point represents the probability that a galaxy in
  that mask meeting the DEEP2 sample selection criteria was targeted
  for spectroscopy and that a $Q=3$ or $Q=4$ redshift was measured.
  Most of the slitmasks for each pointing are in two rows of
  approximately 20 masks each.  The small white lines correspond to
  gaps between the DEIMOS CCDs.  The darker regions show areas where
  masks overlap and objects are therefore observed with higher
  probability.  At the intersection of the top and bottom rows,
  objects may be observed twice, allowing verification of redshift
  repeatability, etc.; however, the vertical masks filling in the
  ``fishtails'' at the east end of each pointing do not contain
  duplicate objects.  }
\label{cooper.deep2.wfn.eps}
\end{figure}

The field layouts and final slitmask coverage in Fields 2, 3, and 4
are shown in Figure~\ref{cooper.deep2.wfn.eps}.  The boundary of each
individual CFHT 12k $BRI$ {\it pcat} photometric pointing is shown by
dashed lines; individual fields are labelled by their field and
pointing number (e.g., Pointing 12 is the second CFHT pointing in
Field 1).  The chevron pattern permits slitlets to be aligned with
atmospheric dispersion both east and west of the meridian (the odd
pattern in Field 2 was a result of this requirement, combined with the
high priority assigned to observations in Field 1); see Section
\ref{maskdesign} for details.  The grey scale at each point represents
the probability that a galaxy in that mask meeting the DEEP2 sample
selection criteria was targeted for spectroscopy and that a secure
($Q=3$ or $Q=4$, cf.\ \S\ref{redshiftmeasurements}) redshift was
measured.  This spatial selection function can be downloaded by users
who wish to perform clustering measurements or make theoretical models
of the survey (see \S\ref{selectionfunction}).  Certain areas were
redesigned following the first semester of survey operations, which
caused a few regions to be observed twice with different masks (darker
stripes).  All DEEP2 masks, including these regions of double
coverage, are included in DR4. The density of masks is especially high
where CFHT 12K pointings overlap in Field 4.  This duplication of
coverage occurred because slitmasks were designed separately pointing
by pointing in Fields 2--4 (but not EGS).  Although photometry was
obtained for three CFHT pointings in each of fields 2-4, due to
limited time for spectroscopic observations field 23 (which had
inferior photometry) and most of field 43 were not observed with
DEIMOS.  The photometric ({\it pcat}) catalogs for these fields are
still included in the data release accompanying this paper.

\begin{figure}
\ifpdffig
\includepdf[scale=0.4]{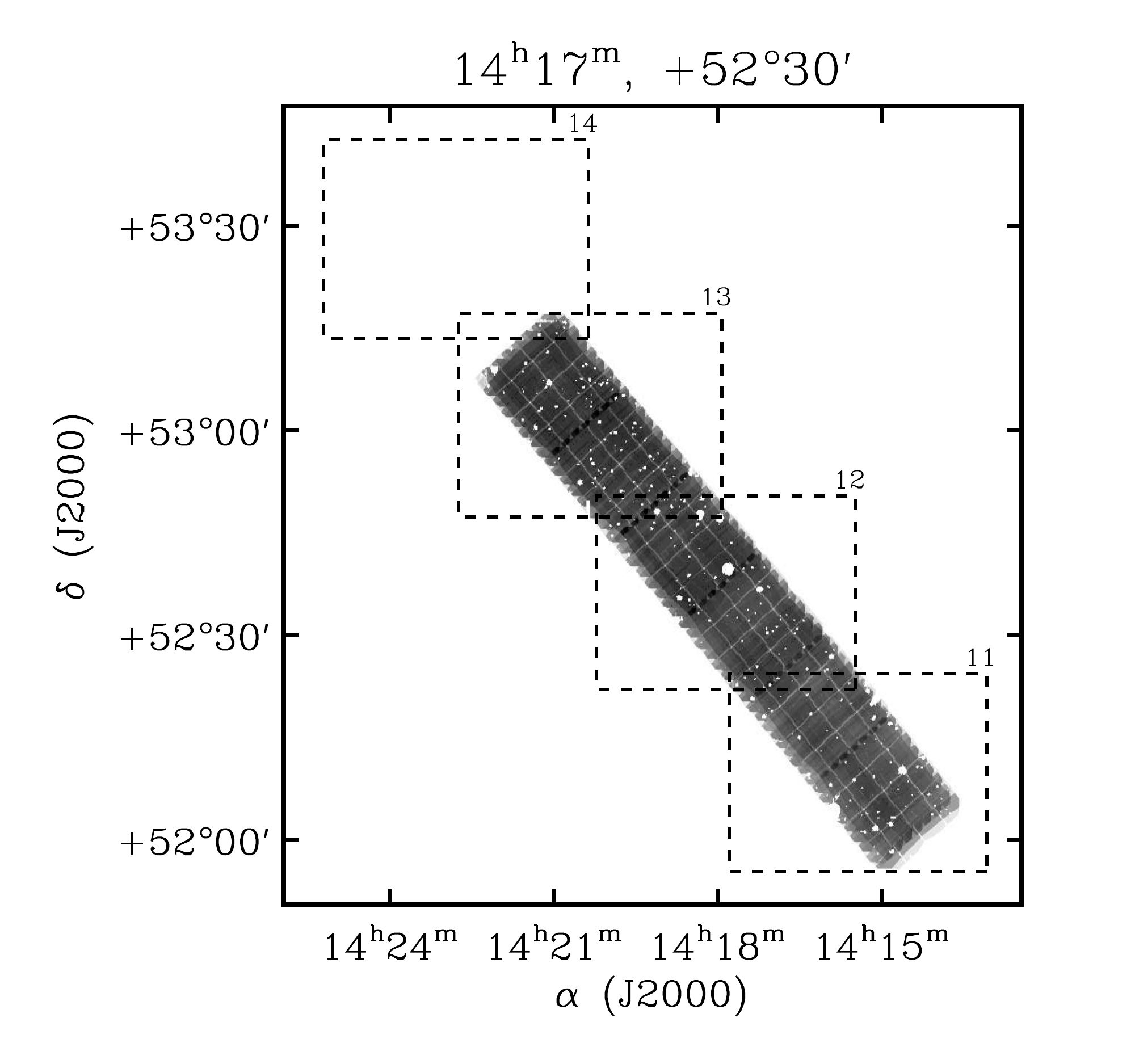}
\else
\includegraphics[scale=0.4]{cooper_deep2_egs_wfn.eps}
\fi
\caption{As Figure~\ref{cooper.deep2.wfn.eps}, but for the Extended
  Groth Strip (DEEP2 Field 1).  Pointing boundaries for the CFHT 12K
  $BRI$ photometry catalog are again shown as dashed lines.  Masks
  were designed in eight blocks along the long direction of the Strip.
  Each block has 15 masks, eight perpendicular to the strip and 7
  parallel.  Masks overlap extensively but unlike in Fields 2, 3, and
  4 there are very few duplicate observations built into the design.
  DEEP2 pointing 14 is omitted here, as mask design in that region
  followed different algorithms to account for its poorer photometry.
  Pointing 14 is therefore not included in any DEEP2 large-scale
  structure analyses.  }
\label{cooper.deep2.egs.wfn.eps}
\end{figure}

A similar plot of combined selection and redshift success probability
is given for Field 1 (the Extended Groth Strip) in
Figure~\ref{cooper.deep2.egs.wfn.eps}.  The layout of this field is
long and narrow both to allow dense coverage and in order to follow
the geometry of preexisting space-based data.  The strip is divided
into eight blocks, each containing 15 masks.  Within each block, eight
masks run perpendicular to the strip and seven run parallel to it.
Unlike Fields 2, 3, and 4, a single {\it pcat} was created by merging
the photometry from all four pointings before the masks were defined.
There are thus no regions where separately-designed masks overlap,
unlike in Field 4.

\begin{figure}
\ifpdffig
\includepdf[scale=0.4]{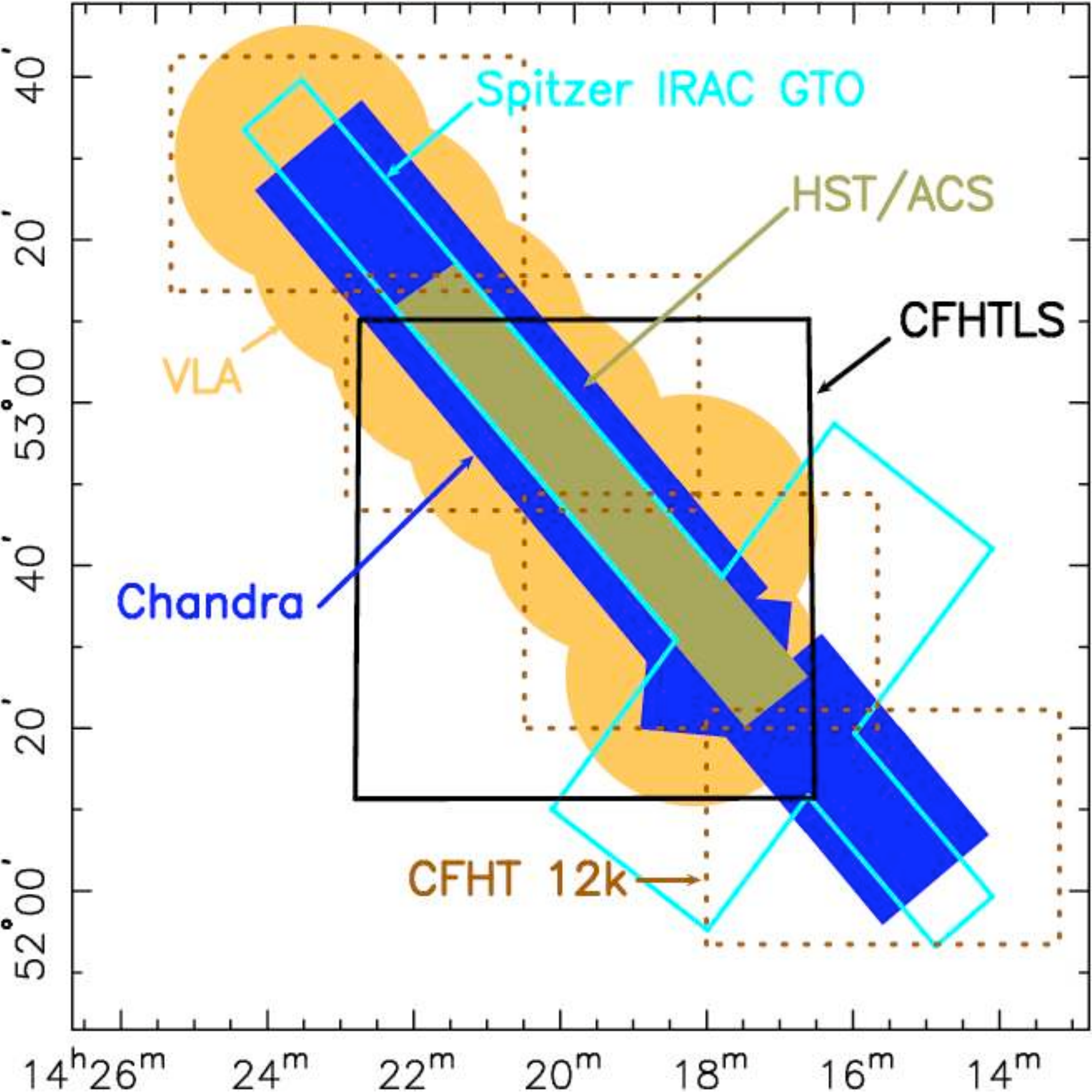}
\else
\includegraphics[scale=0.4]{willmer7nsf_egs1.eps}
\fi
\caption{Initial data in the Extended Groth Strip (DEEP2 Field
  1). Details of the individual data sets are given in
  Table~\ref{table.otherdata}.  The CFHT 12K imaging provides the
  $BRI$ {\it pcat} photometry used for DEEP2 target selection. The
  {\it HST}/ACS mosaic is one of the largest two-color $V+I$ mosaics
  on the sky.  The Guaranteed Time Observation {\it Spitzer}/IRAC data
  and the {\it Chandra} 200 ksec data are the deepest/widest of their
  kind, and the CFHT Legacy Survey provides valuable synoptic
  variability data.  EGS has also been imaged deeply with the VLA at
  20 cm and with the GMRT (not shown) at 50 cm.  }
\label{willmer.7.nsf_egs1.epsi}
\end{figure}

A summary of the existing data in all DEEP2 fields is given in
Table~\ref{table.otherdata}, and skymaps of the major multiwavelength
EGS surveys are shown in Figures~\ref{willmer.7.nsf_egs1.epsi} and
\ref{willmer.7.nsf_egs1.epsi}.  The data in EGS are particularly deep
and rich.  EGS is one of only two deep-wide {\it Spitzer} MIPS and
IRAC regions (Fazio et al.~2004, Papovich et al.~2004, Dickinson 2007
[FIDEL survey]) and is receiving ultra-deep imaging at 3.6 $\mu$m and
4.8 $\mu$m in the SEDS survey, a Warm {\it Spitzer} mission.  It has
one of only two large two-color {\it HST} ACS mosaics (Davis et al.~2007, Lotz et
al.~2008), the other being GEMS in ECDFS (Rix et al.~2004).  It has
the deepest {\it GALEX} imaging on the sky and the deepest wide-area
{\it Chandra} mosaic (800 ksec, Laird et al.~2008).  It is a CFHT
Legacy Deep Survey field, is slated for deep Herschel imaging, and
many other deep optical, IR, sub-mm, and radio data have been taken or
planned.  This includes the follow-on DEEP3 survey (Cooper et al.\
2011, 2012a), which is acquiring $\sim$8,000 new spectra down to
$R_{\rm AB} = 25.5$ and will double the redshift sampling density over
the area covered by the {\it HST} ACS mosaic (see
Table~\ref{table.otherdata}).  Most recently, EGS has become one of
five fields targeted by the CANDELS (Cosmic Assembly Near-infrared
Deep Extragalactic Legacy Survey) Multi-Cycle Treasury program on {\it
  HST} (Grogin et al.~2011; Koekemoer et al.~2011); it is being imaged
deeply with both the ACS and WFC3 instruments as part of this project.

\begin{figure}[t]
\ifpdffig
\includepdf[scale=0.475,angle=0]{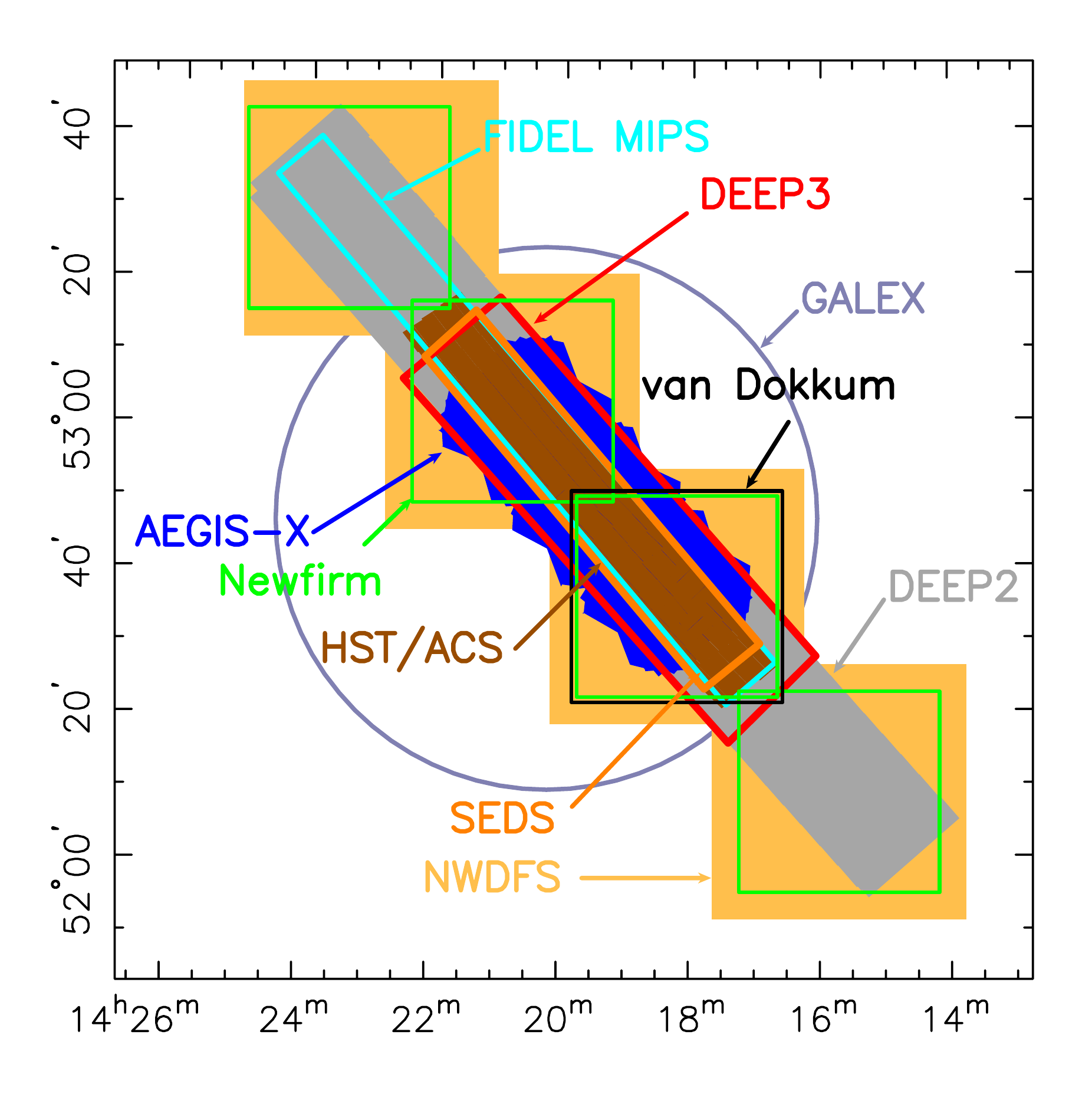}
\else
\includegraphics[scale=0.5,angle=270]{willmer2.ps}
\fi
\caption{Data in the Extended Groth Strip are continually expanding.
  Details of the individual data sets are given in
  Table~\ref{table.otherdata}.  The FIDEL {\it Spitzer}/MIPS, AEGIS-X
  {\it Chandra/ACIS}, and {\it GALEX}/FUV+NUV imaging are the deepest
  exposures of their size on the sky.  The ongoing DEEP3 survey
  (Cooper et al.\ 2012, in prep) will increase the number of DEEP2
  spectra by 50\% and triple the weight of environmental data in the
  upper three fields of EGS.  The Warm {\it Spitzer} Deep
  Extragalactic Survey (SEDS) will provide a total of 10 hours of
  integration time per pointing in IRAC Channels 1 and 2 (centered at
  3.6 and 4.5 $\mu$m).  The NEWFIRM survey will provide $JK$
  photometry to $\sim$24 AB mag, while the NEWFIRM Intermediate Band
  Survey (PI: Pieter van Dokkum; cf.\ van Dokkum et al.~2009) will
  measure photoz's to roughly $K_{\rm AB} \sim 23.4$.  In addition
  (not shown), EGS is a deep SCUBA-2 Legacy field and a deep field for
  the {\it Herschel} HERMES survey with the PACS and SPIRE
  instruments.  }
\label{willmer.7.nsf_egs2.epsi}
\end{figure}

To exploit the great richness of data in EGS, a broader research
collaboration has been formed called AEGIS (the All-wavelength
Extended Groth Strip International Survey).  This collaboration is
combining efforts from more than a dozen teams who have obtained data
in this field ranging from X-ray to radio wavelengths.  More
information on AEGIS may be found at the AEGIS website
(\url{http://aegis.ucolick.org}) and in the May 1, 2007, special issue
of {\it Astrophysical Journal Letters}.

With Field 1 designated to be the Extended Groth Strip, Fields 2, 3,
and 4 were selected by finding regions free of bright stars, with low
reddenings based on the IRAS dust map (Schlegel, Finkbeiner \& Davis
1998), spaced roughly 4 hours apart in RA, and weighted toward higher
RA where Keck weather is better.  Their distribution also avoids the
prime Galactic observing season and balances the number of nights
required in each semester, in order to ease telescope scheduling. The
chosen fields all have declinations such that they are observable for
more than 6 hours with an airmass less than 1.5, to ensure that
differential refraction across a slitmask is below $0\s.2$ (DEIMOS
does not have an atmospheric dispersion compensator).  Fields 3 and 4
are also in the multiply observed Equatorial Strip (Stripe 82) of the
Sloan Digital Sky Survey (SDSS, York et al.~2000, Adelman-McCarthy et
al.~2007, Ivezic et al.~2007), and are visible from both northern and
southern hemispheres.  The recently-released deep SDSS photometry in
this stripe should yield useful photometric redshifts down to
DEEP2-like depths.

The original design of the DEEP2 survey consisted of 480 slitmasks.
The expected number of 135 slitlets per mask would then target
$\sim$65,000 galaxies, of which 52,000 were expected to yield secure
redshifts assuming an 80\% success rate.  One can therefore crudely
think of DEEP2 as having roughly twice the number of objects and three
times the volume as the Las Campanas Redshift survey (Shectman et
al.~1996), but at $z\sim1$ rather than at $z\sim0.1$.  The design
volume of $9 \times 10^6 h^{-3}$ Mpc$^3$ would be expected to contain
only a handful of rich clusters but is large enough to count both
galaxies and groups of galaxies to good accuracy.  At one hour per
mask and eight masks per night, such a survey would take 60 clear
nights of Keck time.  Ninety nights were allocated via a Keck LMAP
(Large Multi-Year Approved Project) proposal with Marc Davis as PI,
which left a cushion of 50\% for bad weather and equipment
malfunction.

The above plan was followed closely, but the yield of reliable
redshifts per galaxy targeted was only 71\%, not 80\%, owing to the
fact that more galaxies than expected were beyond the effective
redshift limit of $z\sim1.4$ (\S\ref{redshiftmeasurements}). Weather
also did not fully cooperate, with the result that only 411 out of the
480 original slitmasks were observed, yielding a total of $\sim$53,000
spectra. The final area and numbers of slitmasks and targets observed
in each field are given in Table~\ref{table.fields}, and sky maps of
the regions covered are shown in Figure~\ref{cooper.deep2.wfn.eps} and
Figure~\ref{cooper.deep2.egs.wfn.eps}.  The final survey observed 86\%
percent of the proposed number of slitmasks and 88\% of the proposed
number of galaxies.

\section{ DEEP2 in Comparison to Other \boldmath$z \sim 1$ Surveys}
\label{othersurveys}

In this section and Table~\ref{table.othersurveys}, we compare the
properties of DEEP2 (considering both Fields 2-4 and EGS separately)
to other large $z \sim 1$ surveys to date in a variety of ways.  The
other projects considered include the Team Keck Redshift Survey in
GOODS-North (TKRS, Wirth et al.~2004), the VVDS-deep survey in CDFS
and in one other field (Le Fevre et al.~2005), the VVDS-wide survey in
four fields (Garilli et al.~2008), zCOSMOS-bright in the COSMOS field
(Lilly et al.~2007), and PRIMUS, which covers 7 fields including
several of the above (Coil et al.~2011).\footnote{Although they
  include some objects at comparable redshifts, the AGES (Hickox et
  al.~2009; Kochanek et al.~2011), BOSS (Aihara et al.~2011), and
  WiggleZ (Drinkwater et al.~2010) redshift surveys are not included
  since their magnitude limits are 3-4 magnitudes brighter than these
  other surveys .}  The first five are conventional spectroscopic
surveys, but PRIMUS is a low-resolution prism survey yielding less
precise redshifts but targeting many more galaxies.  DEEP2 and TKRS
have released all data, but the properties of the others have been
deduced from stated plans and/or partial published data.  Details of
the data and sources used for these comparisons are given in the
footnotes to Table~\ref{table.othersurveys}.

Given the significant redshift failure rate in all distant galaxy
surveys, it is important to have a consistent definition of ``redshift
success.''  ``Reliable'', "secure", or "robust" redshifts are defined
in this paper to be those with $\ge$95\% probability of being correct,
according to their authors (for all surveys described, this means
quality codes 3 and 4 [or equivalent, e.g. including codes 13, 14, 23,
and 24 for VVDS], but not 1 or 2.) In comparing surveys we use the
sizes of the presently published datasets for all surveys except
zCOSMOS, for which we use the design values.

We now proceed to compare the power of the various $z \sim 1$ surveys
using a variety of metrics.  Numerical results are summarize in
Table~\ref{table.othersurveys}.

$\bullet$ {\it Redshift efficency.}  A useful concept is ``redshift
efficiency,'' for which we offer two definitions.  The first is the
overall redshift efficiency, $z_{\rm effic1}$, which is defined as is
the fraction of {\it all} slitlets (including stars) that eventually
yield reliable {\it galaxy} (or QSO) redshifts. Values of $z_{\rm
  effic1}$ are shown for the various surveys in
Table~\ref{table.othersurveys}.  Apart from the two VVDS surveys,
there is considerable degree of homogeneity in this quantity, with
efficiencies hovering between 56-72\%.  The efficiencies of VVDS-deep
and VVDS-wide are only 28-40\%, however, partly because these surveys
do not exclude stars, which hence take up a larger fraction of the
slitlets than in other surveys, but mostly because of our high
standard for redshift reliability, $\ge$95\% repeatability, which
excludes many VVDS redshifts.

The second definition of efficiency is the galaxy-only efficiency,
$z_{\rm effic2}$, which is defined as the fraction of {\it
  galaxy-only} slitlets that yield reliable galaxy
redshifts.\footnote{To calculate this number, we assume that the
  fraction of objects with uncertain redshifts that are stars is the
  same as their fraction amongst objects with secure
  redshifts/identifications.}  The percentages rise by 3-4\%, to
59-76\%, for the non-VVDS surveys and are now 42-43\% for the two VVDS
surveys.  The efficiency of VVDS-wide rises the most due to the high
fraction of stars amongst its targets.

\begin{figure}
\ifpdffig
\includepdf[scale=0.375]{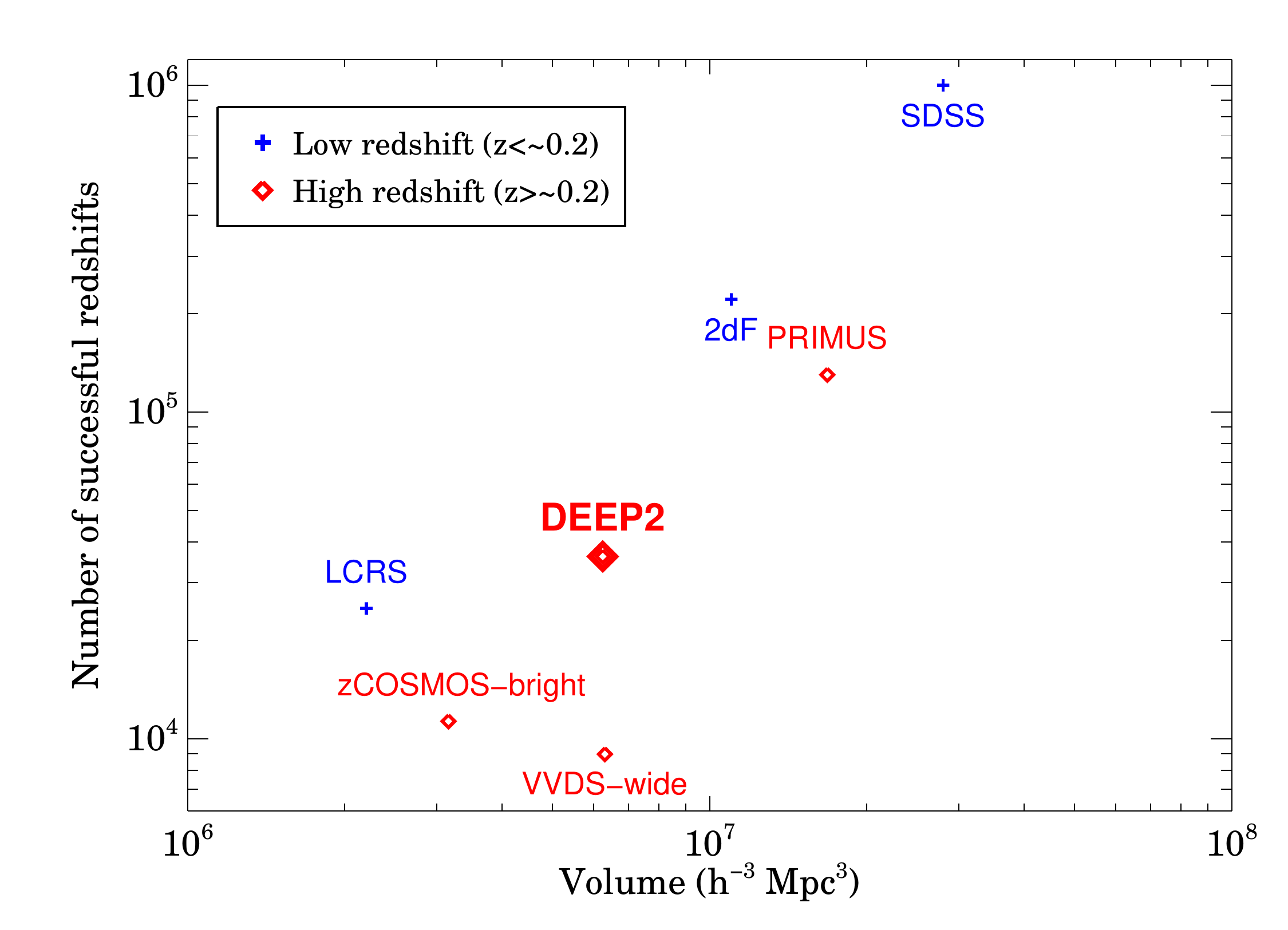}
\else
\includegraphics[scale=0.375]{volnum.eps}
\fi
\caption{Number of redshifts vs.~volume sampled for major
  spectroscopic redshift surveys out to $z \sim 1$.  For all surveys
  except zCOSMOS-bright, we use the number of redshifts and areas
  covered to date (as given in Table~\ref{table.othersurveys} for the
  distant surveys).  For zCOSMOS-bright, which is still in progress,
  design data are used.  Numbers in all cases use reliable redshifts
  only (those with probabilities of being correct $\ge$95\%).  Volumes
  covered are computed from the field areas and magnitude limits of
  each survey as described in \S\ref{surveyparameters}.  The TKRS
  survey is omitted from this figure (although included in
  Table~\ref{table.othersurveys}) due to the relatively small sample
  size and volume.  PRIMUS has low-resolution spectra, and provides
  coarser redshift information for brighter samples than the other
  higher-$z$ surveys shown. If it is set aside, DEEP2 leads amongst
  distant surveys in both volume surveyed and number of reliable
  redshifts.  }
\label{volnum.eps} 
\end{figure}

$\bullet$ {\it Volume sampled vs.~number of objects.}
Figure~\ref{volnum.eps} compares the number of reliable redshifts
versus volume sampled for both high- and low-redshift surveys.  The
volume used is computed from the areas in
Table~\ref{table.othersurveys}, bounded by the redshifts corresponding
to the 2.5 percentile and 97.5 percentile point within a given survey
(or by $z=0.2$ and $z=1.2$, the redshift range the team uses for
science, in the case of PRIMUS).  For surveys other than DEEP2 and
TKRS (where the actual percentiles are used), these redshift limits
are computed from analytic redshift distributions of the form $n(z)
\propto z^2 e^{-z_0}$, using fits to $z_0$ as a function of limiting
magnitude determined as described in Coil et al.~(2004b), but using the
full DEEP2 dataset for calibration.  This results in values of $z_0$
of 0.26, for DEEP2 and TKRS, 0.224 for VVDS-wide and zCOSMOS-bright,
and 0.283 for VVDS-deep.  Among the contemporaneous distant
spectroscopic surveys (i.e., excluding PRIMUS), DEEP2 has the best
combination of total number and volume sampled, with 2.5 times as many
redshifts as the next competitor, zCOSMOS-bright, and roughly twice
its effective volume.  PRIMUS surpasses DEEP2 and all other high
redshift surveys in this space, targeting a larger number of galaxies
over a larger volume (but with lower spectral resolution).

\begin{figure}
\ifpdffig
\includepdf[scale=0.375]{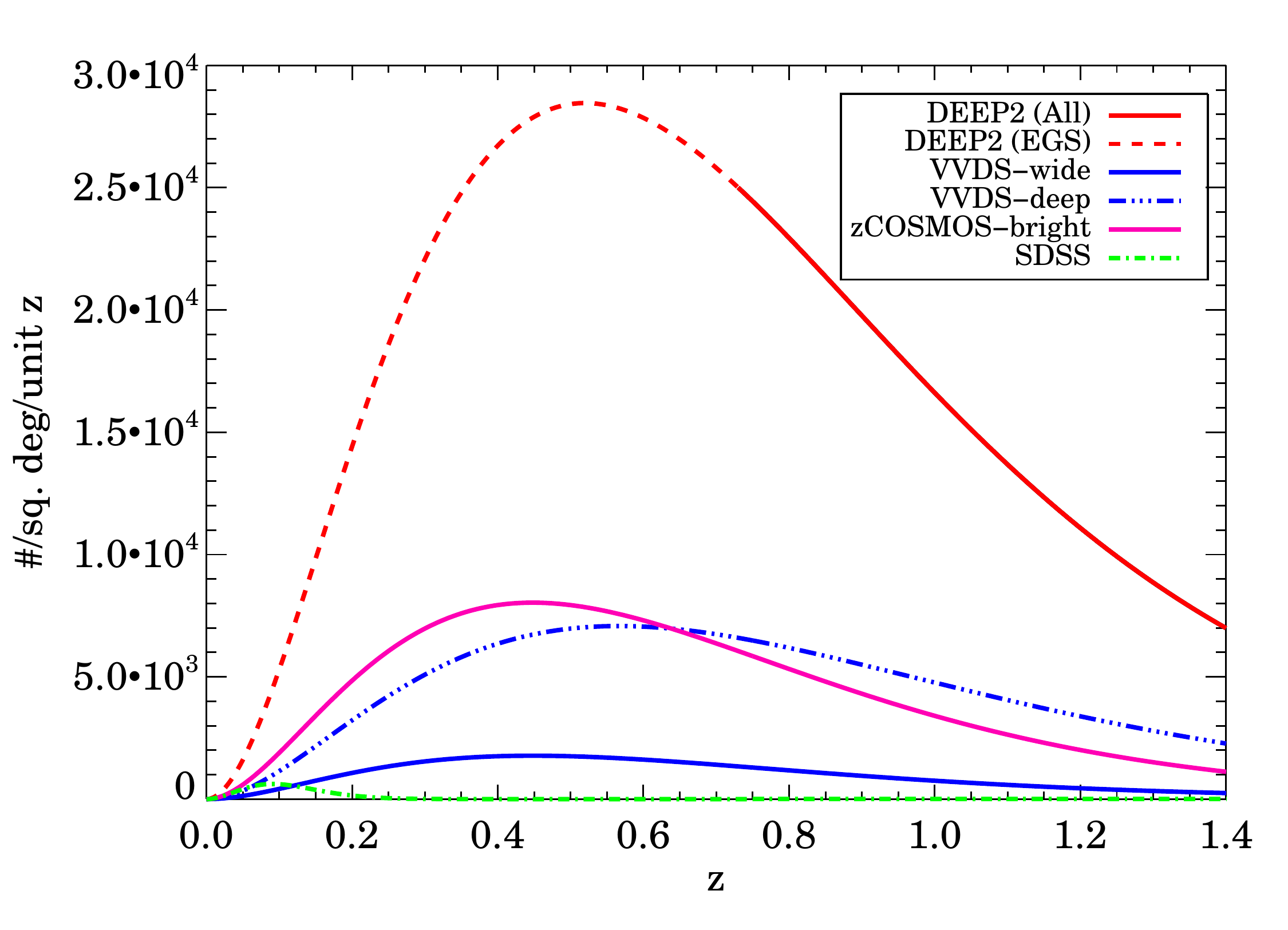}
\else
\includegraphics[scale=0.375]{densitycompare.eps}
\fi
\caption{Sample number densities of various spectroscopic surveys,
  expressed as the number of galaxies per square degree per unit
  redshift interval (eliminating any dependence on cosmological
  assumptions).  Curves are based on the stated magnitude limits, the
  number of reliable galaxy redshifts per square degree ($\ge$95\%,
  see Table~\ref{table.othersurveys}), and the redshift distribution
  model described in \S\ref{surveyparameters}, applied to each
  survey.  Note that the statistical weight of a survey for
  environmental and clustering purposes increases as the {\it square}
  of the co-moving number density if area and magnitude limit are held
  fixed.  Because of its comparably high number density at $z \ltsim
  1$, DEEP2 is significantly better suited for environmental studies
  at intermediate redshifts than other deep surveys.  The PRIMUS
  survey is not shown here, as in that sample spectral resolution,
  rather than sample number density, limits the ability to measure
  Mpc-scale environments for individual objects.  Its number density
  at peak ($z\sim 0.4-0.6$) is roughly 40\% that of DEEP2, or
  approximately 50\% larger than that of zCOSMOS-bright or VVDS-deep.
}
\label{densitycompare.eps}
\end{figure}

\begin{figure}
\ifpdffig
\includepdf[scale=0.375]{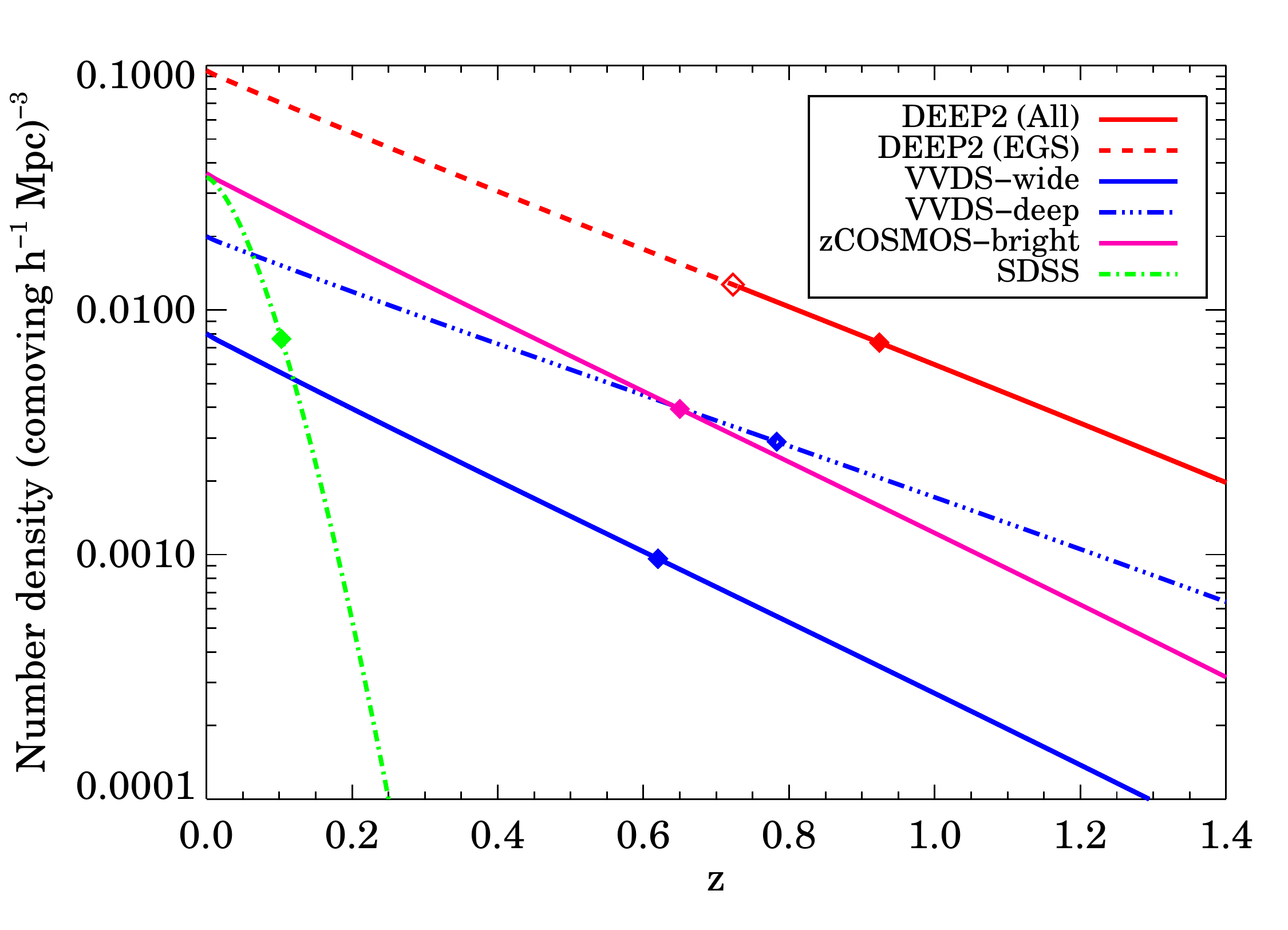}
\else
\includegraphics[scale=0.375]{densitycompare_comoving_log_new.eps}
\fi
\caption{Number densities of various spectroscopic surveys, expressed
  as number of galaxies per comoving Mpc$^{-3}$ (assuming our standard
  $\Lambda$CDM cosmology).  Curves are based on the same models and
  data used in Figure~\ref{densitycompare.eps}.  The tick marks
  indicate the estimated median redshifts of each survey; the open
  tick mark is for EGS.  Since the weight of a survey for
  environmental and clustering data scales as the {\it square} of the
  number density, the square of the relative heights of the tick marks
  ranks the surveys according to the strength of their clustering
  information per unit volume at their peak redshifts.  DEEP2 and its
  EGS subset are the most densely sampled surveys shown, followed by
  SDSS, zCOSMOS-bright, VVDS-deep, and VVDS-wide.  The number
  densities of all distant surveys considered decline roughly
  exponentially with redshift. As in
  Figure \ref{densitycompare.eps}, the PRIMUS sample is not shown
  here as its number density is not the limiting factor for
  small-scale environment measures; it is intermediate between the
  DEEP2 and zCOSMOS-bright surveys.  }
\label{densitycompare_comoving_log.new.eps}
\end{figure}

$\bullet$ {\it Co-moving number density.}
Figure~\ref{densitycompare.eps} uses the same model to compute
cosmology-independent sample densities, expressed as the number of
galaxies with secure redshifts per square degree per unit redshift
interval; we also include the corresponding curve for the Sloan
Digital Sky Survey (for which we take $n(z)\propto z^2
e^{(-z/0.075)^1.5}$) for
comparison. Figure~\ref{densitycompare_comoving_log.new.eps} converts
this to the co-moving number density of objects, assuming our standard
cosmology.  The diamonds mark the median redshifts of the various
surveys.  The sample density is important since the statistical weight
of a survey for environmental and small-scale clustering measurements
increases as the {\it square} of this density if sky area is held
constant (see below).
Figure~\ref{densitycompare_comoving_log.new.eps} shows that DEEP2/EGS
is $\sim 4$ times denser than zCOSMOS-bright and VVDS-deep in the
redshift range $z = 0.6$-1.0 and $\sim 20$ times denser than
VVDS-wide.  Both the higher secure-redshift rate and denser targeting
of DEEP2 contribute to this difference.  We omit PRIMUS from these
figures and the environment figure of merit calculation (below) as the
ability to determine local environment for individual objects in the
PRIMUS sample is limited by redshift errors ($\sigma_z/(1+z) \gtsim
0.005 \gtsim 15$ $h^{-1}$ Mpc comoving) rather than sample number
density; its number density is intermediate between that of
zCOSMOS-bright or VVDS-deep and that of DEEP2.

$\bullet$ {\it Statistical weight for environment measures.}  The
figure of merit that describes a survey's total statistical weight for
environmental purposes scales as $N^2/A$, where $N$ is the number of
reliable redshifts and $A$ is the area covered (assuming matching
redshift distributions).\footnote{On scales where Poisson variance,
  rather than cosmic variance, dominates, the statistical weight of a
  survey for environment measures scales is proportional to the number
  of pairs of galaxies it contains on the relevant scale.  For a fixed
  redshift distribution, the number of pairs will be proportional to
  the number of galaxies with secure redshifts, $N$, times the number
  of companions per galaxy.  This latter is proportional to the number
  density of galaxies, and hence (for constant $z$ distribution) the
  surface density $N/A$.  Hence, the total number of pairs scales as
  $N^2/A$.}  This metric is given for the various surveys in
Table~\ref{table.othersurveys} (``Env metric'').  The combined
environmental weight of DEEP2+EGS will be 4 times that of
zCOSMOS-bright when that survey is fully completed, and is 14-35 times
the weight of VVDS-deep and VVDS-wide in their present states.

$\bullet$ {\it Spectral power.}  Other measures of survey power
include spectral resolution and the total number of independent
spectral resolution elements sampled.  With a spectral resolution that
is 10 and 26 times higher than zCOSMOS-bright and VVDS respectively,
DEEP2 is the only distant survey that can measure the internal
kinematics of galaxies, satellite motions, and small group velocity
dispersions.  Each DEEP2/DEIMOS spectrum has $\sim 2000$ independent
spectral resolution elements, versus $\sim 320$ for zCOSMOS-bright,
$\sim $120 for VVDS-wide and VVDS-deep, and $\sim 20$ for PRIMUS.
Multiplying by the number of design targets (excluding stars) gives
the total number of spectral elements present in galaxy spectra.  By
this metric, DEEP2+EGS has 16 times more information than
zCOSMOS-bright, 10 and 33 times more information than VVDS-wide and
VVDS-deep, and 185 times more information than PRIMUS.  These figures
are generous, as we have here assumed design numbers for total survey
size and that all spectra in all surveys are equally likely to yield
reliable redshifts once stars are excluded.

$\bullet$ {\it Galaxy counts.}  A final metric of survey power is the
accuracy of galaxy counts.  Two sources of noise contribute, Poisson
statistics in the number of galaxies counted and sample (or "cosmic"
variance), which is determined by the number of fields, their areas,
and geometries.  Using the publicly-available QUICKCV code of Newman
\& Davis (2002), we find that the cosmic variance in the count of a
population of galaxies with $r_0 = 4$ $h^{-1}$ Mpc (comoving) and
$\gamma=1.8$ at $z=0.75$-$0.95$ will be 7.6\%, as opposed to 30\% for
TKRS, 14\% for the completed area of VVDS-deep, 6.6\% for VVDS-wide,
12\% for the full zCOSMOS-bright area, and 5.1\% for PRIMUS (here we
assume square field geometry and equal area per field for all surveys
but DEEP2). As a result, the noise in DEEP2 galaxy counts
(e.g. measurements of the abundance of any particular population) will
be smaller than in other distant $z\sim1$ grating spectroscopic
surveys (though not than in PRIMUS); the larger sample size yields
smaller Poisson errors, while the relatively long, narrow field
geometry utilized gives smaller cosmic variance than nearly-square
fields of the same area do (Newman \& Davis 2002).

To conclude, by all of these measures of survey power, DEEP2+EGS is
more powerful than any contemporaneous, high-resolution spectroscopic
survey at $z\sim1$; in many cases, by a factor of 4-30$\times$.  This
power enables a wide variety of unique science.

We have excluded PRIMUS in many of these comparisons; it has largely
followed after DEEP2 in time, and operates in a very different domain.
Due to the very low resolution spectroscopy employed, redshift
uncertainties are relatively large, $\sim 1800-3300$ km s$^{-1}$
(depending on redshift, due to the wavelength dependence of resolution
in prism spectroscopy).  The large redshift errors make it unsuitable
for measuring environment on $\sim$Mpc scales for individual objects
(see Cooper et al.~2005), which is why we have not computed an
environmental figure of merit for it in
Table~\ref{table.othersurveys}.  However, with nearly 80,000 redshifts
over 7 different fields, it has, by a significant factor, the lowest
cosmic variance of any of these surveys and will be able to count rare
objects and measure mass and luminosity functions for bright objects
with unparalleled accuracy.  It will also be effective at measuring
clustering extending to relatively large scales; this allows
measurement of the average overdensity of a population, an important
measure of environment complementary to the measurement of
overdensities for individual objects.  However, information on
redshift-space distortions will be lost (compared to these other
surveys) due to the larger redshift errors.

\section{Object Selection}
\label{objectselection}

\subsection{Photometric Catalog and \boldmath$p_{gal}$ Probabilities}
\label{photometry}

The photometric catalogs used for selecting DEEP2 targets (the {\it
  pcat} catalogs, described in detail in Coil et al.~2004b) are
derived from Canada-France-Hawaii Telescope (CFHT) images taken with
the 12K $\times$ 8K mosaic camera (Cuillandre et al.~2001) in the $B$,
$R$, and $I$ bands.  These images were taken as part of a major
weak-lensing survey by Nick Kaiser and Gerry Luppino, and we were very
fortunate to piggy-back on their efforts.  The field of view of the
camera is 28\amin \ $\times$ 42\amin, as illustrated by the pointing
boundaries in Figures~\ref{cooper.deep2.wfn.eps}
and~\ref{cooper.deep2.egs.wfn.eps}.

The $R$-band images have the highest signal-to-noise and are used to
define the object catalog, which is complete to $R_{\rm AB} > 24.15$
in all pointings.  Objects were identified using the {\it imcat}
software package written by N.~Kaiser and described in Kaiser, Squires
\& Broadhurst (1995).  This package was also used to calculate other
image parameters used in object selection, such as object sizes and
photometry. In Fields 2, 3, and 4, the calibrated {\it pcat} catalogs
for separate pointings are only considerated separately, and slitmasks
in these fields were designed independently (though pointings overlap
in Field 4, which means that one object can appear in two {\it
  pcat}'s).  In Field 1 (EGS), the {\it pcat}'s were merged to make a
single catalog for the whole area before the masks were designed
(retaining the higher-quality photometry in overlap regions), such
that each galaxy can appear only once.  Each object is assigned an
8-digit object number; the first digit of the object number indicates
the field the object was found in, the second digit indicates the CFHT
pointing number, and the remaining digits provide a unique identifier,
counting upwards from zero.  As an example, object 32001226 is the
1,227th object in the {\it imcat} catalog for Field 3, Pointing 2.

\begin{figure}
\ifpdffig
\includepdf[scale=0.4]{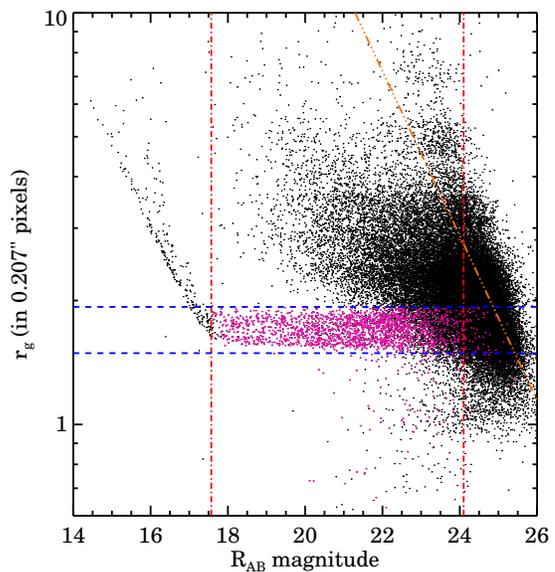}
\vspace*{-1.0in}
\else
\includegraphics[scale=0.4]{sizemag.eps}
\fi
\caption{Size-magnitude diagram resulting from {\it pcat} photometry
  in CFHT pointing 32, which has median seeing for the CFHT data set
  (0\s.84).  Here $r_g$ is the RMS radius in pixels of a circular
  Gaussian fit to an object's image (1 CFHT pixel = $0\s.207$).  The
  stellar locus is visible as the narrow horizontal band below and to
  the left of galaxies.  The left vertical red line is where stars
  begin to saturate, and the right vertical line is the magnitude
  limit of the DEEP2 survey, $R_{\r, AB} = 24.1$.  The blue horizontal
  lines indicate the 95\% size range for for bright stellar sources in
  this pointing; it is determined individually for each CFHT pointing.
  The orange slanting line marks a line of constant surface
  brightness, 26.5 mag/\sq\asec, above which objects are excluded from
  DEEP2 targeting.  Most exluded objects prove to be multiple objects
  incorrectly identified as single in the {\it pcat} photometry (see
  \S\ref{multiplicity}).  Stars begin to blend with galaxies near $R
  \sim 22.5$.  A Bayesian probability for star-galaxy separation, $0 <
  p_{gal} < 1$, is calculated for all objects with size below the
  maximum of the stellar band by combining information on the
  magnitude and color distributions of stars and galaxies in DEEP2
  (cf.\ Figure~\ref{egsstars.eps}); objects with $p_{gal} < 0.2$ are
  colored purple in this plot. \S\ref{photometry} provides more
  details on the algorithms used.  }
\label{sizemag.eps}
\end{figure}

Objects with a high probability of being a star (i.e., a low
probability of being a galaxy, $p_{gal}$, as described in Coil et
al.~2004b) were excluded when defining the pool of objects from which
DEEP2 targets are selected. Figure ~\ref{sizemag.eps} plots object
radius $r_g$ versus apparent $R_{\rm AB}$ magnitude; where $r_g$ is
the 1-$\sigma$ radius in pixels of a circular Gaussian fitted to the
CFHT photometry.  The plot shown is for DEEP2 Field 3, pointing 2
(also referred to as pointing 32), which has median seeing for the
CFHT/DEEP2 data (0\s.84 FWHM); however, apparent object sizes are
analyzed separately for each CFHT pointing.  Stars are identifiable as
the relatively tight, horizontal locus of points below and to the left
of galaxies.  Stellar radii increase at bright magnitudes due to
saturation; the vertical line marks the point at which saturation
becomes detectable in this particular pointing.  Saturation sets in at
$R_{\rm AB} \sim 16.7-18.0$ depending upon pointing, establishing the
bright magnitude limit for DEEP2 targets, $R_{\rm AB} > 18.5$.

\begin{figure}
\ifpdffig
\vspace*{-0.3in}
\includepdf[scale=0.425]{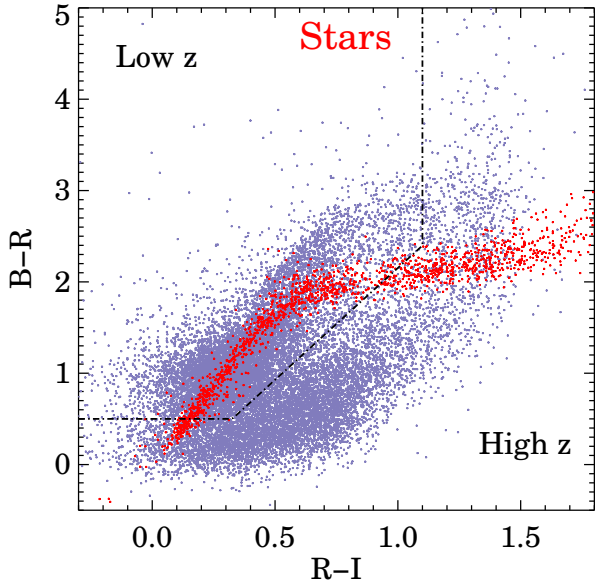}
\vspace*{-1.2in}
\else
\includegraphics[scale=0.425]{egsstars.eps}
\fi
\caption{$BRI$ Color-color diagram for stars and galaxies.  Grey
  points are candidate target galaxies in the Extended Groth Strip;
  i.e., they are objects in the {\it pcat} with $18.5 < R_{\rm AB} <
  24.1$ and $p_{gal} > 0.2$.  Stars, defined here as objects with
  $p_{gal} < 0.2$, are indicated by the red points.  Straight line
  segments show the boundary used for color pre-selection to screen
  out low-redshift galaxies in Fields 2, 3, and 4; candidate objects
  at high redshift lie below and to the right of these lines.
  Specifically, the lines show the locus in color-color space where
  objects are given 10\% weight for selection in Fields 2-4, the
  minimum possible for inclusion in the DEEP2 sample.  The weight
  falls off from a maximum of 1 (a little below and to the right of
  this locus) as a 2-d Gaussian having $\sigma=0.05$ mag in both
  coordinates.  This ``pre-whitening'' of the boundary is done to
  reduce the impact of any systematic color errors within a CFHT 12K
  pointing, which are of order 0.02-0.04 mag (Coil et al.~2004b).  The
  division between objects treated as low- and high-redshift in EGS is
  similar to the lines shown (cf.\ \S\ref{slitmaskgeometry}).  Stars
  are distinguished from galaxies by computing a Bayesian probability,
  $p_{gal}$, based on the distributions of stars and compact galaxies
  in both magnitude and color-color space.  \S\ref{photometry} gives
  more details.  }
\label{egsstars.eps}
\end{figure}

Information from the size-magnitude diagram in
Figure~\ref{sizemag.eps} is used in conjunction with the $B-R$
vs.~$R-I$ color-color diagram to compute $p_{gal}$.  First, objects in
any part of Figure~\ref{sizemag.eps} that have colors unlike any other
stars or bright galaxies are given $p_{gal} = 2$.  Second, the upper
horizontal dashed line denotes the 95\% upper radius range for stars
(determined separately for each CFHT pointing), and all sources above
this line are considered extended and assigned $p_{gal} = 3$.
However, stars have sizes indistinguishable from galaxies at dim
magnitudes, below $R_{\rm AB} \sim 23$.  In this regime, we can use
the fact that stars tend to be brighter than and occupy a different
locus in color-color space from compact galaxies to differentiate the
two classes.  The differences in colors are illustrated in
Figure~\ref{egsstars.eps}, which shows the $BRI$ color-color diagrams
for stars and extended galaxies in a DEEP2 {\it pcat}.  For all
objects having a size consistent with the stars in a given pointing,
the value of $p_{gal}$ indicates the net Bayesian probability that an
object is a galaxy (as opposed to a star) based on both color
information and brightness; it is therefore a value between 0 and 1.
The pink points in Figure~\ref{sizemag.eps} are the objects identified
as likely stars, defined as those having $p_{gal} < 0.2$.

To summarize: $p_{gal} = 3$ indicates an extended source, all of which
are included for target selection.  Objects with peculiar colors have
$p_{gal} = 2$; these are also all taken as candidates (they vast
majority have turned out to be objects with large photometric errors).
Finally, values of $p_{gal}$ from $0$ to 1 are used for compact
objects with a size consistent with stars.  More details are provided
in Coil et al.~(2004b).  All objects with $p_{gal} > 0.2$ are included
as candidate targets for DEEP2, with selection weight proportional to
$p_{gal}$ (to a maximum weight of 1).  Further details on sample
selection are given in Section~\ref{sample}, and possible biases are
described in Section~\ref{targetbiases}.

Note that the trajectory of stellar colors in
Figure~\ref{egsstars.eps} is approximately described by two straight
lines with a ``knee'' separating the two.  Demanding that this knee be
identical in all fields allows us to place all CFHT 12k pointings on a
common color system, with an overall zero point determined via
comparison to stars in the Sloan Digital Sky Survey.  From a variety
of tests, we find that pointing-to-pointing variations in the average
photometric zero points are at the $<0.01$ magnitude level, while
variations within pointings and absolute zero point uncertainties are
$\sim$ 0.02-0.04 mag (Coil et al.~2004b).

\subsection{Target Pool Selection Procedure}
\label{sample}

With $p_{gal}$ determined, a pool of potentially acceptable DEEP2
target galaxies can be defined.  The final slitmasks are designed by
selecting objects from this pool and placing them on slitlets.  All
galaxies selected as targets must fulfill three separate selection
criteria that apply in all 4 DEEP2 fields.

First, the object must be classified as a potential galaxy by our
probabilistic star-galaxy separation procedure.  All extended objects
and objects with peculiar colors are automatically included ($p_{gal}
= 2$ or 3).  Objects whose sizes are consistent with stars are taken
as eligible targets if they have $p_{gal} >0.2$, i.e., the probability
of being a galaxy is greater than 20\%.  The distribution of $p_{gal}$
is extremely bimodal, with values piling up around 0 or 1 (Coil et
al.~2004b); as a result, varying the cut level between 0.2 and 0.8
makes very little difference to the sample.

The next cut is on apparent $R$-band magnitude: to be considered for
DEEP2 targeting, an object must lie in the range $18.5 < R_{\rm AB} <
24.1$.\footnote{The DEEP2 $BRI$ photometric system is an approximately
  AB magnitude system in which the observed magnitude in each filter
  has been zeropointed by requiring that it matches SDSS photometry in
  the closest available band for stars with zero AB color (Coil et
  al.~2004b): e.g. we choose a zero point such that DEEP2 $R$ equals
  SDSS $r$ plus a color term proportional to SDSS $(r-i)$.  To the
  degree to which the SDSS system is on AB, DEEP2 magnitudes will be
  AB magnitudes within the native filter+telescope system; however,
  this assumption is imperfect at the $\sim 0.02$ mag level
  (D.~Eisenstein, private communication).  System response curves
  including the filters and detector may be found at
  \url{http://deep.berkeley.edu/DR1/photo.primer.html}.}  For science
purposes, it would have been more ideal to use a redder band like $I$
to set the survey magnitude limit, but the CFHT $I$ photometry
obtained was less uniform in depth, necessitating the adoption of
$R$.\footnote{Ironically, the originally-planned fields with worst $I$
  ended up being discarded due to the descope from the
  originally-planned coverage.}

Finally, the object must have surface brightness above a standard
limit (indicated by the slanting line in Figure~\ref{sizemag.eps}).
This surface brightness (SB) requirement is defined by the equation:
\begin{equation}
   SB = R_{\rm AB} + 2.5 \log_{10}{[\pi(3r_g)^2]} \leq 26.5,      
\end{equation}
where $r_g$ here is the Gaussian profile radius for that object, given
in arc seconds.  The minimum $r_g$ is set to $0\s.33$, so that objects
with 3$r_g < 1$\asec \ are presumed to have an effective 3$\sigma$
radius of 1\asec \, minimizing the effect of noise in measuring the
sizes of compact objects on this cut.  Visual inspection of the
low-surface-brightness objects which are excluded by this cut
indicates that nearly all are double or multiple sources in the $BRI$
images that were incorrectly identified as single objects in the {\it
  pcat} due to blending.  Their surface brightnesses are low because
their radii are falsely inflated.  Since the data for multiple objects
are ambiguous, they are not good targets for DEEP2, and this cut
properly excludes many of them.  Further discussion of multiple
interlopers is given in \S\ref{multiplicity}.

\subsection{Color Pre-selection in Fields 2, 3, and 4}
\label{pre-selection}

In Fields 2, 3, and 4, the target pool is further refined to remove
galaxies with redshifts likely to be below $z \sim 0.75$.  All but the
bluest galaxies exhibit a significant break in their spectra at
$\lambda \sim 4000$ \AA.  This break causes a strong separation in
color-color space between objects where the break falls in the $B$
band or on the blue side of $R$, versus those where it occurs at
redder wavelengths.  We therefore use galaxies' $B-R$ and $R-I$ colors
to identify the low-$z$ interlopers.  Since the bluest,
flattest-spectrum objects exhibit minimal break (having colors near
$0$ at all redshifts), we always include those objects within our
selection region, in order to ensure that our sample is complete for
galaxies at $z>0.75$.

\begin{figure}
\ifpdffig
\vspace*{-0.25in}
\includepdf[scale=0.4]{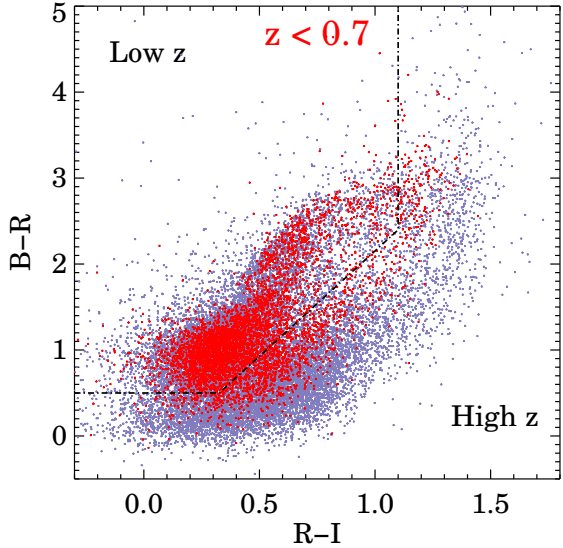}
\vspace*{-1.1in}
\else
\includegraphics[scale=0.4]{egsrangefail.eps}
\fi
\caption {Assessment of the accuracy of the DEEP2 $BRI$ color-color
  pre-selection for rejecting low-redshift ($z < 0.7$) galaxies.  Grey
  points are the same set of DEEP2 candidate galaxies shown in
  Figure~\ref{egsstars.eps}; red points are those EGS galaxies whose
  spectroscopic redshifts are below $z = 0.7$ (only secure, $Q = 3$
  and $Q=4$ redshifts are used). We here take advantage of the fact
  that low-redshift galaxies are not excluded from DEEP2 targeting in
  the Extended Groth Strip.  The boundary is the 10\%-weight locus
  repeated from Figure~\ref{egsstars.eps}.  Nearly all red points lie
  above the boundary, indicating that they would be {\it successfully
    excluded} from DEEP2 by the color-color pre-selection cut.  }
\label{egsrangefail.eps}
\end{figure}

\begin{figure}
\ifpdffig
\vspace*{-0.25in}
\includepdf[scale=0.4]{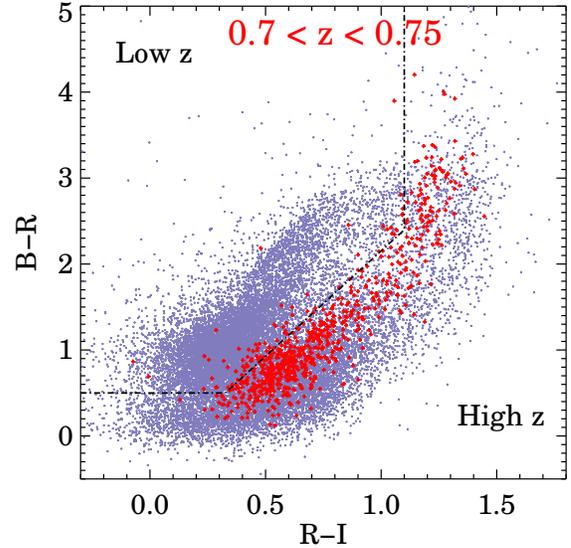}
\vspace*{-1.1in}
\else
\includegraphics[scale=0.4]{egsrangemid.eps}
%
\fi
\caption {A test of the accuracy of the DEEP2 $BRI$ color-color
  pre-selection for galaxies near the transition redshift.  This
  figure is similar to Figure~\ref{egsrangefail.eps}, but colored
  points indicate EGS galaxies with spectroscopic redshifts in the
  range $0.70$--$0.75$.  Most of these lie below the boundary,
  indicating that they would be {\it accepted} by the color-color
  pre-selection cut despite their redshifts below 0.75.  This is
  conservative, ensuring that objects with redshifts above $z = 0.75$
  are not lost.  }
\label{egsrangemid.eps}
\end{figure}

\begin{figure}
\ifpdffig
\vspace*{-0.25in}
\includepdf[scale=0.4]{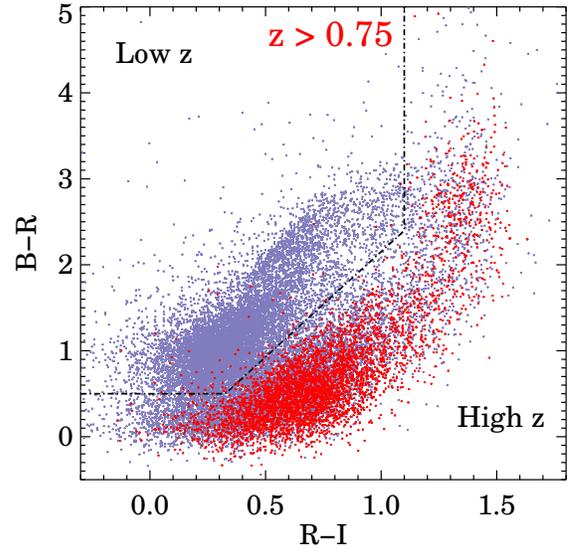}
\vspace*{-1.1in}
\else
\includegraphics[scale=0.4]{egsrangesuccess.eps}
\fi
\caption {The accuracy of color-color pre-selection for known
  high-redshift galaxies.  This figure is similar to
  Figure~\ref{egsrangefail.eps}, but colored points indicate EGS
  galaxies with spectroscopic redshifts above 0.75.  Nearly all of
  these lie below the boundary, indicating that nearly all would be
  {\it accepted} by the color-color cut.  The DEEP2 color selection
  produces a sample which is highly (approaching 100\%) complete for
  high-redshift objects.  }
\label{egsrangesuccess.eps}
\end{figure}

Figure~\ref{egsrangefail.eps}, Figure~\ref{egsrangemid.eps}, and
Figure~\ref{egsrangesuccess.eps} show color-color plots of galaxies in
different redshift regimes in the Extended Groth Strip, where we
obtained spectroscopy of galaxies at all redshifts (i.e., galaxies are
not excluded based on this color cut).  The lines show the nominal
color boundary used to screen out low-redshift galaxies --
essentially, only objects to the right and below this line are chosen
in Fields 2, 3, and 4, while objects of all colors are observed in
EGS.

The boundary used is not perfectly sharp but rather the weight given
an object for selection falls off (from 1) as a two-dimensional
Gaussian with $\sigma = 0.05$ mag in both coordinates; we call the
resulting weight the ``color-weighting factor'' for Fields 2, 3, and 4
in the discussion below.  This ``pre-whitening'' is done to reduce the
impact of possible systematic color errors within a single CFHT 12K
pointing, which are of order 0.02-0.04 mag (Coil et al.~2004b).  The
lines shown are not the nominal (i.e., sharp) boundary but rather show
(for $R-I>0.2$) or approximate (for $R-I<0.2$) the locus of points
where the color-weighting factor equals $0.1 = 10$\%, the minimum for
objects selected for DEEP2 spectroscopy.  Although this weight is used
in target selection, selection probability is a slow function of the
weight\footnote{The exact position of these lines changed slightly
  after 2002, taking account of lessons learned from the first
  semester's data.}.  The boundary in these figures corresponds to
selection criteria for DEEP2 targets of:
\begin{enumerate}
\item $(B-R) < 0.5$; and/or:\\
\item $(R-I) > 1.1$; and/or: \\
\item $(B-R) < 2.45 \times (R-I) - 0.2976$ .
\end{enumerate}
The selection weight is 100\% for objects with $(B-R) < 0.389$; $(R-I)
> 1.211$; and/or $(B-R) < 2.45 \times (R-I) - 0.311$. 

\begin{figure}
\ifpdffig
\vspace*{-0.25in}
\includepdf[scale=0.4]{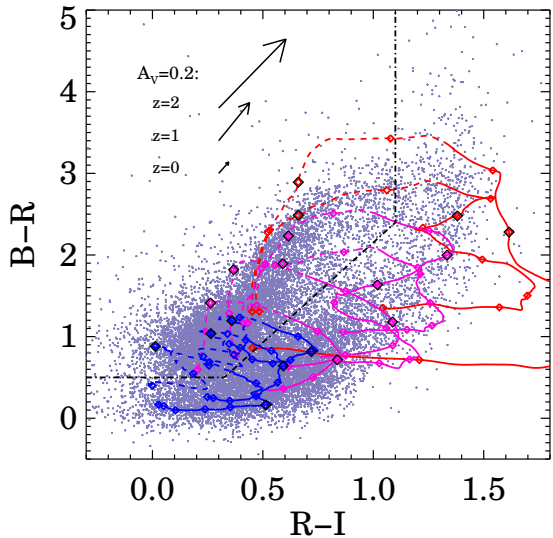}
\vspace*{-1.1in}
\else
\includegraphics[scale=0.4]{allcolor.eps}
\fi
\caption {Model color-color distributions for redshifted spectral
  energy distributions (SEDs) superimposed on DEEP2 targets.  The SEDs
  used are based on spectroscopy of normal nearby galaxies from
  Coleman, Wu, and Weedman (1980) and selected starburst galaxies from
  Kinney et al.~(1996).  Each SED executes a loop as redshift
  increases as the Balmer/4000-\AA\ break moves first through $B-R$
  and then through $R-I$ at higher redshifts.  Each model loop runs
  from $z = 0$ to $z = 2.0$ with diamonds shown at increments of 0.2;
  the line type switches from dashed to solid at $z = 0.7$.  The lines
  are colored to improve visibility, with early-type spectra in red,
  later-types in purple, and starbursts in blue.  The dashed lines are
  the boundaries used to pre-select target galaxies with $z > 0.75$ in
  Fields 2, 3, and 4.  Arrows indicate the impact of $A_V=0.2$ of
  extinction at a variety of redshifts.  Dust reddening tends to move
  galaxies parallel to, rather than across, the DEEP2 color cut.
  These models are illustrative only -- the final positions of the
  boundaries were tuned using EGS data, which contain galaxies at all
  redshifts.  }
\label{allcolor.eps}
\end{figure}

Figure~\ref{allcolor.eps} demonstrates why we would expect such a
color pre-selection to be effective. This figure shows the tracks of
redshifted galaxy SEDs from Coleman, Wu \& Weedman (1980) (augmented
by starburst galaxies from Kinney et al.~1996) in color-color space.
Galaxies move across the color-color boundary rapidly as their Balmer
and 4000-\AA\ breaks move first through $B-R$ and then through $R-I$
near $z = 0.7$.  The resulting valley in the colors can be used to
sort galaxies into two groups below and above $ z \sim 0.75$.

\begin{figure}
\ifpdffig
\vspace*{-0.25in}
\includepdf[scale=0.4]{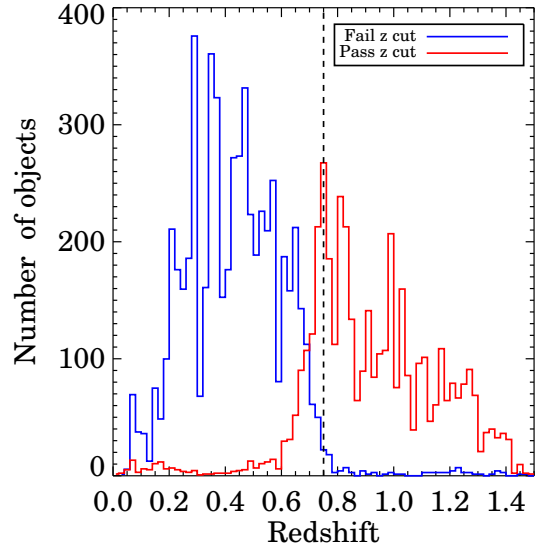}
\vspace*{-1.1in}
\else
\includegraphics[scale=0.4]{zhist_cut.eps}
%
\fi
\caption {Redshift histograms for EGS galaxies with secure ($Q = 3$ or
  $4$) redshifts that either pass or fail the pre-selection color cut.
  We correct here for the selection probability of each object, as
  otherwise the $R$-magnitude weighting scheme used in EGS would
  influence the redshift distributions (see \S\ref{maskdesign}).  The
  50-50\% crossover point has been deliberately placed at $ z \sim
  0.71$ to insure high sample completeness above the nominal target
  redshift cut at $z = 0.75$ (vertical dashed line).  }
\label{zhist_cut.eps}
\end{figure}

\begin{figure}
\ifpdffig
\vspace*{-0.2in}
\includepdf[scale=0.4]{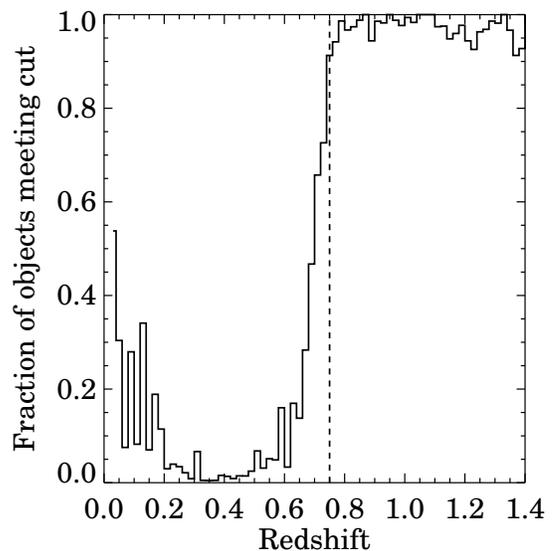}
\vspace*{-1.1in}
\else
\includegraphics[scale=0.4]{fracincut.eps}
%
\fi
\caption {The fraction of EGS galaxies passing the pre-selection color
  cut relative to the total number of galaxies with measured
  redshifts, as a function of $z$.  Numbers are taken from
  Figure~\ref{zhist_cut.eps}.  This figure illustrates the high
  completeness of the color cut, which is 91\% at $z = 0.75$ and
  averages 97.5\% for $0.8 < z < 1.4$.}
\label{fracincut.eps}
\end{figure}

The CWW models are illustrative only -- the actual boundary was tuned
using DEEP1 data and early redshifts in Field 1 (EGS), where galaxies
are targeted regardless of their color.  Figure~\ref{zhist_cut.eps}
shows the final redshift histograms of objects passing and failing the
color cut in EGS.  The vertical dashed line is the desired redshift
cut at $z = 0.75$, and the 50-50\% crossover point is at $z=0.714$.
The crossover has been placed significantly below the target cut in
order to enhance sample completeness at $z>0.75$ (at the expense of
somewhat decreased efficiency at targeted only high-$z$ objects).
Figure~\ref{fracincut.eps} shows the targeted fraction at all
redshifts, which indicates the completeness of our color cut at a
particular $z$.  For the bin centered at $z = 0.75$, the pre-selection
color cut correctly captures 87\% of all objects and incorrectly
rejects 13\%.  For bins above $z = 0.8$, these fractions are constant
at 97.5\% and 2.5\% respectively.  Figures~\ref{egsrangefail.eps} and
\ref{egsrangesuccess.eps} show that the loss of high-$z$ galaxies and
the creeping in of low-$z$ galaxies are primarily due to blue objects
in the lower-left corner of the color-color diagrams, whose colors
tend to blur together at low and high redshift, as well as to objects
with catastrophic photometric errors.

The net effect of the color pre-selection in Fields 2, 3, and 4 is to
remove $\sim 55\%$ of $R < 24.1$ objects from the target list --
almost all unwanted foreground galaxies -- with very minimal loss of
the desired distant galaxies, and thus to increase the survey's
efficiency for studying objects at the target redshift of $z \sim 1$
by a factor of $\sim 2.2$ at the cost of a $\sim 3\%$ loss of
desirable targets.

\newpage
\vspace*{0.1in}

\section{Slitmask Design}
\label{maskdesign}

\subsection{DEIMOS Detector and Slitmask Geometry}
\label{deimosgeometry}

Having described the factors which determine whether an object is
targeted by DEEP2, as well as the factors which contribute to their
selection weight, we now turn to the methods used to design individual
slitmasks from this target pool.

\begin{figure}
\ifpdffig
\includepdf[scale=0.375]{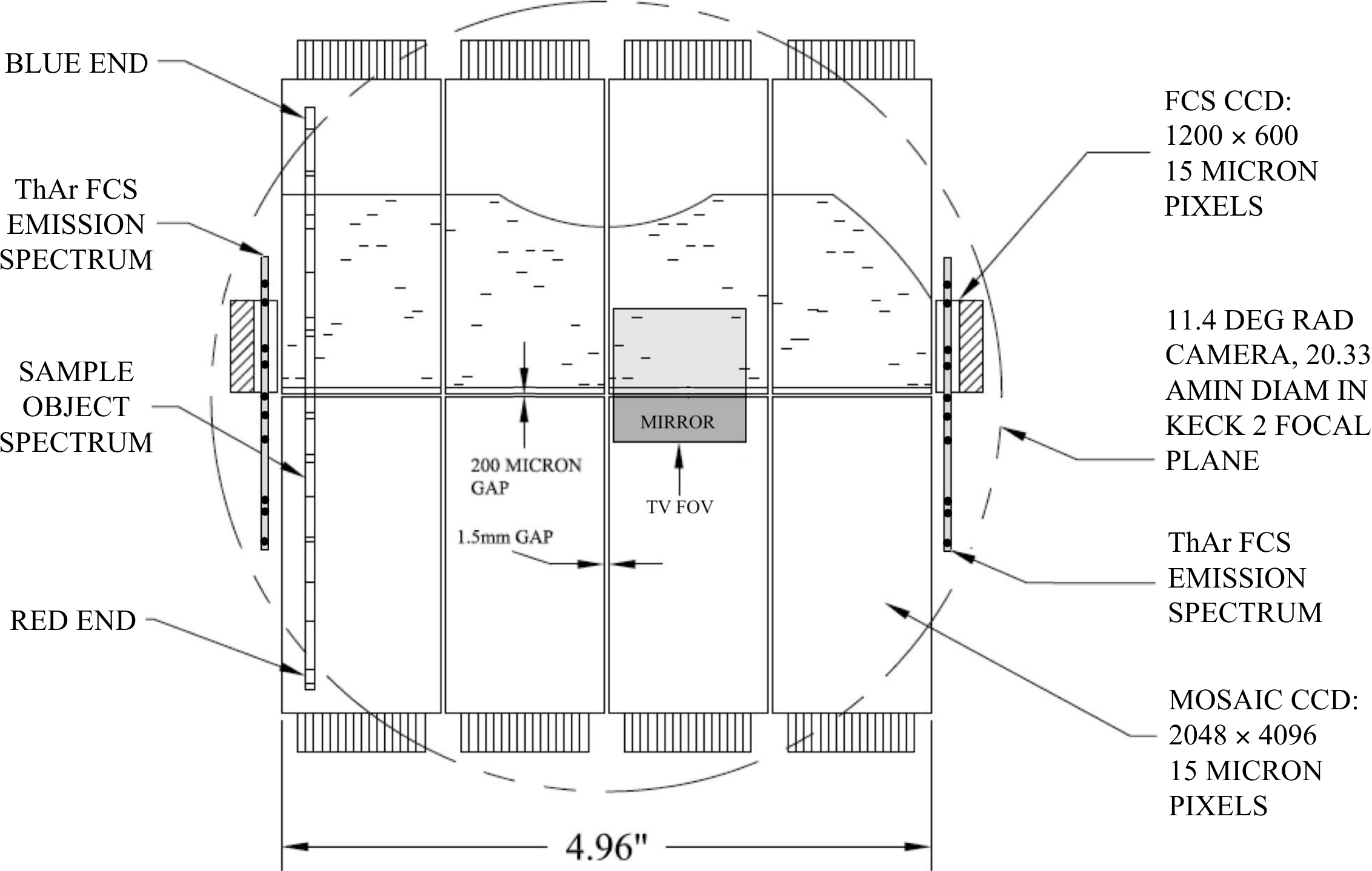}
\vspace*{0.2in}
\else
\includegraphics[scale=0.375,angle=270]{deimosfocalplane.eps}
%
\fi
\caption{Schematic of the DEIMOS focal plane, showing the geometry of
  the 2$\times$4 CCD mosaic.  The dispersion directions of the
  individual spectra run vertically.  Superposed is a slitmask image
  as projected onto the detector in direct imaging mode. The TV guider
  camera stares directly at a fixed of the focal plane.  Light is
  reflected into the camera from the shiny slitmask (upper, light grey
  region) and more efficiently from a reflective mirror (lower, dark
  grey region).  The geometry is precisely known, so that placing a
  given star at a predicted TV guider pixel at the proper spectrograph
  angle should succeed in placing the alignnment stars for that mask
  within their respective 4\asec$\times$4\asec \ alignment boxes.
  Alignment proceeds by taking successive direct images through the
  mask followed by small corrections to the telescope RA/Dec and
  spectrograph position angle.  Flanking the main detector array are
  the flexure compensation CCDs.  These are fed by ThAr emission
  spectra, which are piped into the slitmask focal plane by optical
  fibers (there are actually two fiber spectra at each end, but only
  one is shown).  The ThAr spectra are so rich that at least one
  emission feature falls on the FCS CCDs at all grating tilts.  }
\label{deimosfocalplane.thar.epsi}
\end{figure}

The design and selection of objects is intimately intertwined with the
geometry of the DEIMOS detector array and slitmasks, which are
illustrated in Figure~\ref{deimosfocalplane.thar.epsi}.  Related
instrument parameters are given in Table~\ref{table.instrumentparams}.
The telescope focal plane and slitmask geometry are shown as they are
imaged onto the CCD detector array.  For simplicity, we adopt the same
coordinate system convention used for DEIMOS pipeline software
outputs: ``columns'' run horizontally in the figure, i.e. parallel to
the slitlet direction, while ``rows'' run vertically in the figure,
parallel to the spectral direction.  A typical extracted spectrum will
therefore have 8192 columns, but considerably fewer
rows.\footnote{Note that the reduced data keyword {\it bcol} as
  explained in \S\ref{datatables} refers to readout columns on the
  individual DEIMOS CCDs, which run parallel to the dispersion
  direction, opposite from our standard convention.}

The CCD detector is composed of eight 2K $\times$ 4K MIT Lincoln CCDs
with 15-$\mu$ pixels, for a total size of 8192 $\times$ 8192 pixels
and total physical dimensions 126 mm $\times$ 126 mm.  The CCD
thickness is greater than normal (40 $\mu$ rather than 15 $\mu$),
which renders fringing negligible at wavelengths below 8000 \AA\ and
still yields low fringe amplitudes at longer wavelength (e.g. only
$\pm$2\% at 9000 \AA).  The thickness of these CCDs also boosts total
system quantum efficiency at far red wavelengths, e.g.\ to 16\% at
9000 \AA.\footnote{See throughput plots at \\
  \url{http://www.ucolick.org/$\sim$ripisc/Go1200/through\_go1200\_tilts.gif}
  for details.}  The between CCDs in the spectral direction is only
seven pixels wide, while the gap between the long sides of the chips
is roughly 1 mm, or $8\asec$.  A single spectrum in general falls on
only two CCDs.

The slitmasks are flexible flat aluminum sheets that are 23.6 cm wide,
75.7 cm long, and 0.50 mm thick.  The total length of the mask on the
sky is $16\m.7$, while the maximum usable slit length (subtracting the
CCD gaps) is $16\m.3$.  The masks are milled flat on a
computer-controlled milling machine and spring-loaded onto a
cylindrically curved mandrel in the spectrograph that approximates the
spherically curved telescope focal surface.  Eleven slitmasks can be
held in a jukebox-like cassette at one time.

\subsection{Mask Design Algorithm}
\label{slitmaskgeometry}

The widest part of a DEIMOS slitmask projects to $5\m.2$ across on the
sky, but we place slitlets only within the inner 4\amin\ because the
outline on one side is irregular.  The wavelength range of the portion
of the spectrum from a given slitlet which actually falls on the
DEIMOS CCDs will vary according to slitlet placement.  To maximize the
wavelength range that is common to all objects (given anamporphic
shifts in central wavelength with position on the sky), we select
targets for spectroscopy only within a strip which is curved slightly
``backwards'' on the sky, but whose edges correspond to lines of
constant wavelength on the detector.  To ensure good sky subtraction,
we require that slitlets generally be at least $\sim 3$\asec \ long,
implemented by requiring that successive targets be at least 3\asec\
apart in the long direction of a DEIMOS slitmasks (i.e.\ along a
'row').  Objects may not be placed within $0.\s.3$ of the end of a
slitlet, in order to ensure uniform slit width along each object for
sky subtraction.  We also require $0\s.6$ of dead space in the
slitmask's long direction between slitlet ends to limit
cross-contamination between light from adjacent spectra (this is part
of, rather than adding to, the minimum spacing between objects).  The
average slitlet length resulting from this procedure is 7\asec.

In Fields 2, 3, and 4, the pattern of slitmasks on the sky is
chevron-shaped, as shown in Figure~\ref{cooper.deep2.wfn.eps}.  The
two arms of the chevron are observed either east or west of the
meridian to better align the slitlets with the atmospheric dispersion
direction.  Masks are designed separately for each of these two sets
of masks; where they overlap at the point of the ``V'', some objects
will be multiply-observed, allowing tests of the repeatability of
DEEP2 measurements.  The slanting pattern also keeps the masks within
the boundaries of the 28\amin-wide CFHT 12k pointings, despite the
masks being $>14$\amin\ long.  Mask centers are separated along by
2\amin \ on average, which gives each object roughly two chances to be
included on a slitmask.  The actual step size is modulated by an
adaptive tiling scheme (see below).

The algorithm to place objects on masks in Fields 2, 3, and 4 assigns
each object a weight $W$ between 0 and 1 that is the product of four
factors.  The first factor, $W_{SG}$, comes from the star-galaxy
separation probability $p_{gal}$; it is unity if $p_{gal} = 2$ or 3
but equals $p_{gal}$ if $p_{gal}$ is in the range 0 to 1 (i.e., for
small galaxies or stars); only objects with $p_{gal}>0.2$ are
considered for targeting.  The second factor, $W_c$, which we will
refer to as the ``color-weight'' is a measure of each object's
consistency with the DEEP2 color-color cut.  As explained in
\S\ref{pre-selection}, this factor runs between 0 and 1 and is the
result of smoothing a sharp-edged step function in the color-color
diagram by a 2-d Gaussian of width $\sigma = 0.05$ mag in each
coordinate.  Only objects with color-weight $>0.1$ are considered for
targeting.

The third factor, $W_R$, is a function only of $R$ magnitude and is
designed to flatten the distribution of objects in redshift and
magnitude.  In Fields 2, 3, and 4, this weight is given by the smaller
of the two numbers $(0.75\times10^{-0.4*(24.1-R})$ and 1, which yields
a weight that rolls off smoothly from 1.0 to 0.75 in the last 0.31
magnitude, de-weighting the faintest galaxies moderately.  The final
factor, applied only for DEEP2 objects observed in 2003 or later, is
designed to reduce the number of high-redshift galaxies beyond $z =
1.4$ targeted, as they often fail to yield redshifts in DEEP2
spectroscopy.  The target selection weights for these ``blue corner''
objects (those with $B-R < 0.5$ and $R-I < 0.45$) are multiplied by a
factor $W_{bc}$ that rises linearly from 0.1 for objects with $R-I \le
0.045$ to 1 for objects with $R-I \ge 0.45$.  Hence, in fields 2-4, $W
= W_{SG} \times W_c \times W_R \times W_{bc}$.

In the Extended Groth Strip, $p_{gal}$ is still used for weighting,
but no color pre-selection or blue-corner deweighting are applied,
eliminating the second and fourth weighting factor.  However, ignoring
color information entirely yields a sample which is dominated by faint
foreground galaxies.  Therefore, the $R$-magnitude-based weight, $W_R$
is modified in EGS to counteract this.

The magnitude-based weight factor is calculated differently
accordingly to which of two possible different scenarios applies: \\
$\bullet$ If $R < 21.5$ {\bf or} the galaxy passes the color cut to be
at high redshift,\footnote{This cut uses the nominal color-color
  boundary rather than a smoothed version; so objects with $(B-R) <
  0.5$, $(R-I) >1.1$, and/or $(B-R) < 2.45 \times (R-I) -0.5$ are
  treated as high-redshift in EGS, while those failing all these
  criteria are considered to be at low $z$.}  $W_R$ takes the same
value as it would in Fields 2, 3, and 4, falling from $1$ at $R\ltsim
23.8$ to
0.75 at $R=24.1$. \\
$\bullet$ If $R > 21.5$ {\bf and} the galaxy fails the color cut
(i.e., it is both faint and at low redshift), the weight is the lesser
of the numbers $0.1\times10^{-0.4(24.1-R)}$ and 1.  This yields a
weight that rolls off smoothly from 1.0 at $R = 21.6$ to 0.1 at $R =
24.1$, decreasing the sampling rate for the faintest low-$z$
galaxies. \\
Thus, distant EGS galaxies are weighted very like those in Fields 2,
3, and 4 (except for small differences in the color boundary), and
rare, intrinsically bright low-redshift galaxies are favored in target
selection; so in EGS $W = W{SG} \times W_R$, but unlike in Fields 2-4
$W_R$ depends indirectly on $BRI$ color.  We summarize the effects of
the two weighting schemes on the resulting sample at the end of this
section after describing the actual mask design
process.\footnote{The weighting scheme and target selection in
  EGS is complicated somewhat due to the fact the photometry in
  pointing CFHT 12K pointing 14 was comparatively poor and less
  well-calibrated than in the remainder of the DEEP2 spectroscopic
  fields, which leads to less well-characterized $B-R$ and $R-I$
  colors there.  In this pointing, $p_{gal}$ is still retained as a
  factor, but the $R$-magnitude weighting is the same as in Fields 2,
  3, and 4 regardless of object color, and there is no
  surface-brightness cut or blue-corner de-weighting.  Improved
  photometry for this field will be separately released in an upcoming
  paper, Matthews et al.~(2012).}

In Fields 2, 3, and 4, objects are allocated among the slitmasks in a
two-stage process\footnote{This process begins after the selection of
  four to six bright stars to be used for slitmask alignment.  Each
  star is then used as the center of a 4\asec\ $\times$ 4\asec \
  square alignment box.  To the degree possible, these are placed
  evenly on both ends of a mask for maximum leverage.}.  The first
pass places slitlets only in a central strip that is 2\amin \ wide on
average (though this width can vary due to adaptive tiling,
q.v. below).  This central, curved strip is favored in order to
maximize the wavelength region common to all spectra (with our chosen
grating, the wavelength shift across 2\amin\ is $\sim$200 \AA).

\begin{figure}
\ifpdffig
\hspace*{-0.37in}
\includepdf[scale=0.48]{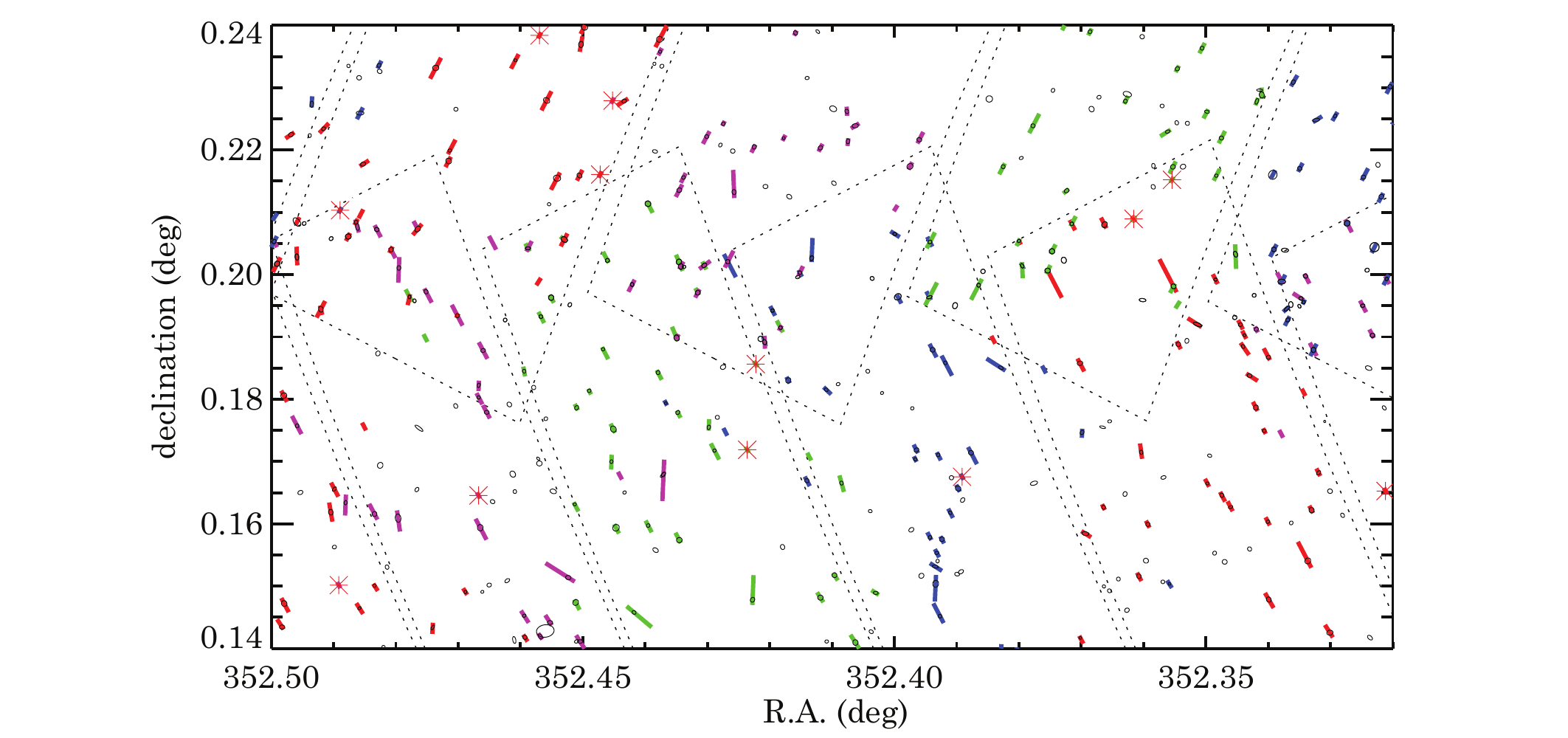}
\else
\includegraphics[scale=0.38,angle=90]{plotmasks_32cut.eps}
\fi
\caption{Blowup of mask designs in CFHT pointing 32 showing a region
  where many masks overlap, near the ``V'' of the chevron pattern.
  Dotted lines show the prime inner $\sim$2\amin\ strip within each
  mask (actual widths vary due to the adaptive tiling algorithm).
  Target objects are indicated by black ellipses with size $3r_g$ and
  axis ratio $(1-e_2)$.  Slitlets on a given mask have the same
  color. Alignment stars are indicated by the red asterisks.  }
\label{plotmasks.32cut.eps}
\end{figure}

Cosmic variance in the number of targets within a 16\amin\ $\times$
2\amin\ region is high, so that some masks will have fewer slitlets
than others if masks are evenly spaced.  To minimize this variation,
we iteratively adjust the positions (and correspondingly the widths of
the central regions) of masks in Fields 2, 3, and 4 such that the
number of targets selected in the first pass is held roughly constant
from mask to mask (see Figure~\ref{plotmasks.32cut.eps}, which
reflects these variations).  This ``adaptive tiling'' strategy helps
to ensure that a uniform fraction of target galaxies is sampled over
the survey region.

At the beginning of the first pass, the list of targets is temporarily
trimmed by generating a random number between 0 and 1 and retaining
only objects for which this random number is less than their weight.
Thus, low-$W$ objects (e.g., likely stars and galaxies just outside
the nominal color cut) are only rarely placed on slits in this first
pass.  The trimmed list is then searched for cases where we can place
two objects (separated by less than 3\asec \ in the long direction) on
a single slitlet that has a PA relative to the long axis of the mask
of less than 30\deg.  Slitlets are allocated first to all such cases,
boosting our ability to study close pairs of galaxies.

A random priority uniformly distributed between 0 and 1, $P_1$, is
then generated for each of the remaining objects in the trimmed list.
Slit allocation then begins at one end of the mask (along its long
axis) and proceeds to the opposite limit.  Each object that can be
observed without precluding the observation of any other is selected
for observation, while in cases of conflicts the object with greatest
$P_1$ is taken.  A new object will be allocated a slitlet only if it
is more than a minimum distance along the long axis of the mask away
from any object already targeted. This minimum buffer distance is the
greater of 3\asec\ or ($0\s.3$ + twice the radius of the new object
projected along the slit, $r_{g,proj}$), where the latter is computed
from the new object's Gaussian radius $r_g$, axis ratio, and PA.  New
objects that are large must therefore be farther from existing objects
and will be allocated longer slitlets as a result.

This procedure (initial selection according to $W$, and nearly random
selection amongst the surviving targets) yields a sample where the
probability of selection is almost perfectly proportional to an
object's weight, making the best possible use of the central portion
of each mask.  In Fields 2, 3, and 4, 77-80\% of slitlets are
allocated in the first pass.

After the first pass through all masks is completed, the remaining
galaxies that overlap a given mask (over the full 16\amin\ $\times$
4\amin\ area) are given new weight values $W_2$ according to their
initial selection weights, with $W_2 = W \over {1-0.55W}$.  This new
weight accounts for the fact that the probability a given object is
selected in pass 1 is very close to $0.55 W$, so the probability it is
not selected is $1-0.55W$.  A new random number uniformly distributed
between 0 and 1, $P_2$, is then generated for each object; all
potential second-pass targets are then ranked in descending order of
$W_2 \over P_2$, providing a priority to each one.  If all objects in
pass 2 were subject to conflicts (i.e., have to compete with other
objects for slitlets), this priority would ensure that the final
probability an object is selected after passes 1 and 2 is proportional
to its initial weight, $W$.

Objects are then allocated slitlets in descending order of their
priority, subject to the requirement that so long as they do not
conflict with any slitlet that has already been added to the mask
(e.g. as an alignment star, in Pass 1, or as an object with higher
priority in Pass 2).  In the end, any target that can be placed on the
mask without causing a conflict is included, so in the second pass,
low-weight objects will frequently be selected if they are cost-free.
This causes the final selection probability of an object not to be
simply proportional to $W$; in fields 2-4, the probability is instead
closely approximated by $p_{selection} = 0.27976 + 0.44717 W - 0.09137
W^2$.  Hence, an object with weight of 0.1 will be selected for
observation 32\% of the time, while one with $W=1$ will be selected
64\% of the time.  Since most eligible targets have a weight near 1,
the overall selection rate is close to the latter value.  The second
pass proceeds from west to east, assigning targets to all the masks in
a row (i.e., all the masks with similar PA) in order.

The choice of targets in the Extended Groth Strip differs in two
respects from Fields 2, 3, and 4.  First, as noted above, there is no
pre-selection based on color, and the $R$-magnitude weighting factor
used is different in order to suppress the number of faint foreground
galaxies. The tile pattern in EGS is also different owing to the
narrower width of the region, 16\amin \ rather than 30\amin.  Masks
are divided into 8 groups with 15 masks per group, each group
approximately square in size (see
Figure~\ref{cooper.deep2.egs.wfn.eps}).  In each square, eight masks
have their long dimensions perpendicular to the long axis of the
strip, and seven masks have their long axes parallel to the strip;
masks parallel to the strip (within a given pointing) are given larger
mask numbers.

In order to maximize the number of galaxies having measured rotation
curves, we first divide all extended, elliptical objects (those with
eccentricity $\epsilon > 0.081$ and size $3 r_g > 1$ arcsec) into two
sets, one consisting of objects with their major axis within 45\deg\
of the long axis of the strip and the other including the remainder.
The two-pass maskmaking algorithm described previously is then run on
the set of objects in the second category, combined with with those
which are too small or round to have a well-defined position angle, to
assign objects to masks with their long direction perpendicular to the
strip.  Then, any galaxies not assigned to a mask, as well as the
extended objects with long axis along the strip, are used to design
the masks with long dimension parallel to the strip, again using our
standard slit assignment algorithm. The crossed orientations again
permit observations both east and west of the meridian, as well as
enabling rotation curve measurements.  However, in EGS the step size
between masks is only 1\amin \ owing to the greater density of targets
on the sky, and here it is not varied adaptively.
 
In EGS, 74\% of objects on slitmasks are selected in the first pass,
which again only places targets in the central 2\amin \ width of a
mask.  The second pass is similar to the second pass in Fields 2, 3,
and 4; again galaxies are selected over the full 4\amin \ active
width.  By limiting the first pass selection to a field of only
2\amin, the spectral coverage of the dispersed light for the galaxies
is made more uniform.  In the EGS, the probability an object is
selected for observation depends on the weight, $W$ (which here is the
product of $W_{SG}$ and the magnitude-dependent weight $W_R$) as
$p_{selection} = 0.33398 + 0.42687 W$; so an object with $W=0.1$ is
selected 38\% of the time, while 76\% of objects with $W=1$ are
targeted.

Once all objects are chosen, the final step is to set the lengths of
the slitlets.  Since no account was taken of the sizes of previously
selected objects when new ones were added, it may happen that two
projected buffer distances, each of size $2r_{g,proj} + 0\s.3$,
overlap for two neighboring objects.  In this case the available space
is evenly allocated between the objects regardless of which one is
bigger.  Much more commonly, when the separation is larger than the
two buffer distances, the available slit lengths are allocated in
proportion to their values of $r_{g,proj}$.  In both cases, half the
dead space of $0\s.6$ between slitlets is subtracted from the extent
of each slitlet in the direction of overlap.

Once slitlets have been placed on targets, the design of all DEEP2
masks are adjusted to include $\sim$8 sky-only slitlets placed in open
spaces on the masks.  Their positions along the long axis of the mask
are chosen to fall in between the objects with largest separations,
while their positions along the short axis are chosen by choosing
empty regions from the CFHT 12K imaging. The sky slitlets are used to
perform fallback, ``non-local'' sky subtraction useful in the presence
of certain rare instrumental anomalies and for the occasional very
short slitlet.  These sky slitlets are only 0\s.7 wide; to produce a
nonlocal sky model, the spectrum from a sky slitlet is convolved with
a kernel that varies with wavelength, fit using an individual
slitlet's estimated sky spectrum.  In practice, this method proved
inferior to local sky subtraction using the object's own sky spectrum
even for short slitlets, and it has not generally been used
extensively.  However, non-local-sky-subtracted versions of both 1-d
and 2-d spectra are included in pipeline-produced data files for
completeness.

As a final step, in order to preserve rotation curve information, the
long axis of each slitlet is oriented as closely as possible along the
major axis of its galaxy (as measured from the CFHT photometry)
provided the following conditions are met: 1) the axis is within
$\pm$60\deg \ of the long axis of the mask; 2) the galaxy has $r_g \ge
1$\asec; 3) the slitlet has not already been tilted to observe another
object; and 4) the galaxy is measurably ellipsoidal with CFHT axis
ratio less than 0.85.  As slitlet tilts, $\theta$, can be up to
30\deg\ away from the long axis of the mask but no more, the mismatch
between slitlet and galaxy PA can be up to 30\deg\ (note also that PAs
measured from high-resolution ACS images can differ considerably from
the PAs from ground-based images, as found by Wirth et al.~2004).

In the central portions of a slitmask, slitlets are still tilted by
$\pm 5$\deg \ compared to the mask direction for objects which do not
fulfill these requirements.  This ensures better wavelength sampling
of the night sky spectrum, improving sky subtraction (optical
distortions provide that sampling for untilted slits at the ends of a
mask, so those slitlets are designed with PA=0 relative to the mask).
The crosswise widths of tilted slitlets are reduced by cos($\theta$)
in order to maintain a fixed projected slitwidth along the dispersion
direction.

Figure~\ref{plotmasks.32cut.eps} shows a blowup of a region covered by
several slitmasks in Field 3, indicating the central first-pass
regions, eligible targets, selected targets, and slitlets with their
final lengths and orientations.  The overlap of slitmasks at the ``V''
of the chevron pattern provides a sample of multiply observed objects
for statistical and reliability assessment purposes.

In the end, 61\% of eligible targets are assigned
slitlets in Fields 2, 3, and 4.  More precisely, this is the
fraction selected out of targets that have
$18.5 < R < 24.1$, $p_{gal} > 0.2$, pass
the surface-brightness cut, and have probability $> 0.1$ 
of meeting the color-color cut.   
Slitmasks in these fields contain $\sim 110-140$ target slitlets (mean
121) that are typically 7\asec \ long but span the range 3-15+\asec.

In EGS, the average number of slitlets per mask is 148.  In pointings
11-13 of EGS, $R$-band de-weighting of low-redshift galaxies boosts
the relative number of high-$z$ galaxies, with the result that roughly
73\% of targets out of a Field 2-4 equivalent sample (as defined
above) are selected in EGS\footnote{It is thus possible to simulate a
  high-$z$ sample from these three EGS pointings by drawing a subset
  of the high-$z$ galaxies at random, compensating for the dependence
  of weight on color in the other fields. The resulting sample will be
  essentially identical to that in Fields 2, 3, and 4 except for
  slightly different statistics in a small corner of color space near
  the color-cut boundary.}  In contrast, due to the deweighting of
faint objects that fail the color cut, only 53\% of low-$z$ galaxies
are selected.  Without this deweighting, only 55\% of the galaxies in
a Field 2-4 equivalent sample would be selected in EGS, making it
sparser in number density than other fields.

\section{Selection Effects in the Final Target Sample}

\label{seleffects}

The final DEEP2 target sample suffers from a variety of known
selection effects.  These effects should in general be taken into
account and/or corrected for when utilizing the DEEP2 dataset.  A
variety of methods for doing so have been implemented in DEEP2 science
papers; see \S\ref{scienceresults} and Table
\ref{table.previouspapers} for examples.

$\bullet$ {\it Galaxy color bias due to the $R$ magnitude limit}
Selection according to a fixed apparent $R$-band magnitude limit
causes the resulting sample to reach different depths (in terms of
absolute magnitude) for galaxies of different intrinsic colors at the
same redshift.  At $z \sim 0.5$, the $R$-band filter corresponds
closely to restframe $B$, so objects are selected down to the same
$B$-band absolute magnitude independent of color.  Below that
redshift, $R$-band selection favors redder galaxies, while above this,
it favors bluer galaxies.  The effect was first illustrated in Willmer
et al.~(2006) and is discussed in more detail in
\S\ref{redshifttrends}.  Though fundamental, this bias is well
understood and can readily be applied to mock catalogs and other
statistical predictions.  The same type of bias will inevitably affect
any magnitude-limited sample which spans a range of redshifts, due to
the broad span of intrinsic galaxy colors.

$\bullet$ {\it Loss of bright star-like objects} The use of $p_{gal}$
discriminates against point sources (i.e., objects with apparent size
consistent with stars; cf.\ Figure~\ref{sizemag.eps}), and thus
potentially discriminates against bright Type I AGNs, i.e., QSOs.
AGNs will be lost only if they are both point-like {\it and} lie on or
near the stellar locus in the color-color diagram.  As a result, Type
I AGN will be included in the DEEP2 sample at some redshifts, but
excluded at others.

$\bullet$ {\it Misclassification of faint stars as galaxies} Although
the differential number counts of objects classified as stars
($p_{gal} < 0.2$) in the DEEP2 {\it pcat}s rise approximately linearly
at bright magnitudes, they flatten out at $R_{AB} \sim 22.5$; it is
likely not coincidental that this is where large numbers of galaxies
begin to populate the stellar regime in Figure~\ref{sizemag.eps}.  It
appears that the $p_{gal}$ values overestimate the probability that
faint objects are galaxies rather than stars; we therefore tend to
include more stars than expected within the DEEP2 target list.  This
is conservative from the standpoint of galaxy selection, as the
contaminant objects are trivial to weed out once spectra are
obtained.\footnote{Overall, 1.8\% of the objects targeted by DEEP2
  turned out to be stars; 69\% of those stars have $R > 22.5$.}
However, as a result of this misclassification, a list of photometric
stars selected purely via their $p_{gal}$ values will be incomplete at
the faint end and would be of limited use for stellar studies.

$\bullet$ {\it Loss of objects due to missing $B$ and $I$ photometry}
The definition of $p_{gal}$ and the DEEP2 color pre-selection both
depend on having measured $B-R$ and $R-I$ colors.\footnote{Note:
  colors are used to select objects in Fields 2, 3, and 4 and (less
  directly) in Pointings 11, 12, and 13, but not in Field 14; see
  \S\ref{pre-selection}.}  Though the $R$-band photometry is complete
to the $R_{\rm AB} = 24.1$ magnitude limit in all fields (Coil et
al.~2004b), very red or blue objects may not be detectable in either
the $B$ or $I$ imaging. However, in the three EGS {\it pcat} pointings
with accurate $B$ and $I$ photometry (11-13), which contain 59,594
objects brighter than $R_{\rm AB} = 24.1$, only 153 galaxies are
missing $B$.  They tend to be faint and red and constitute only 4\% of
distant (red-sequence) galaxies with $R_{\rm AB}>23$ and $R-I>1.0$.
Their loss is not significant.  Similarly, 71 galaxies with $R_{\rm
  AB} < 24.1$ are not detected in $I$; they are predominantly very
blue.

$\bullet$ {\it Loss of small, distant, faint red galaxies}
Discriminating faint stars from galaxies depends on the contrast
between stars and galaxies in the color-color plot
(Figure~\ref{egsstars.eps}) (as well as on the fitting functions used
for the magnitude distributions of each sample).  The blue stellar
locus is quite narrow in this figure, so the loss of blue galaxies
misclassified as stars is fractionally small.  The situation changes
where $B-R > 1.5$ and $R-I > 0.6$, where the surface density of stars
becomes more comparable to that of galaxies.  As shown by
Figure~\ref{allcolor.eps}, this region of the color-color plot is
populated by intrinsically red galaxies beyond $z \sim 0.7$.  Loss of
such galaxies was estimated by Willmer et al.~(2006) using the
higher-resolution {\it HST} ACS images in EGS and partial redshift
data.  We redo that calculation using final EGS redshifts and a larger
database.

\begin{figure}
\ifpdffig
\vspace*{-0.2in}
\includepdf[scale=0.4]{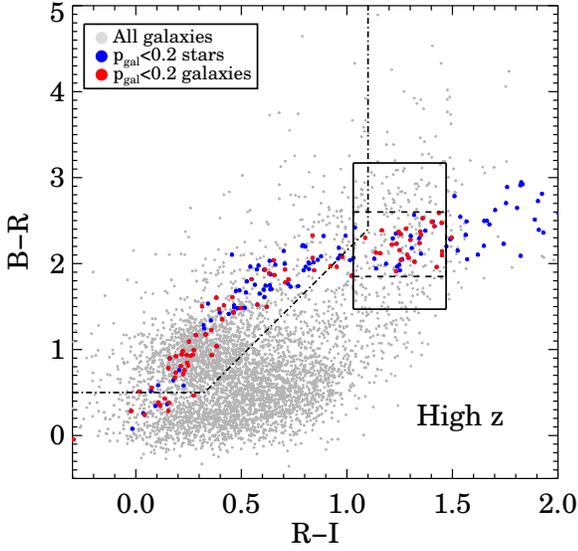}
\vspace*{-1.0in}
\else
\includegraphics[scale=0.4]{faintstars2.eps}
%
\fi
\caption{ Color-color diagram for faint ($R > 23$) objects in the
  Extended Groth Strip with {\it HST}/ACS imaging.  This is the domain
  where we expect confusion between stars and galaxies to be most
  likely, based on Figure~\ref{sizemag.eps}.  Objects classed as stars
  based on the CFHT 12K photometry ($p_{gal} < 0.2$) are the colored
  points.  Objects confirmed as stars in the ACS images are blue;
  those which proved to be galaxies are red.  The red points are
  missed by DEEP2 and are thus a concern.  The loss is worst (as a
  fraction of all objects) in the dashed rectangle, where
  high-redshift red sequence galaxies in the range $z =0.75$-1.0 cross
  the stellar locus.  The loss within that rectangle is 27\%.
  However, roughly half of galaxies in the affected redshift and
  parameter regime lie in the upper and lower rectangles, which are
  not lost, giving a final estimated loss fraction of $\sim$13.5\%
  (see \S\ref{targetbiases}).  }
\label{faintstars2.eps}
\end{figure}

Figure~\ref{faintstars2.eps} plots the color-color diagram for EGS but
restricts to objects in the ACS mosaic fainter than $R = 23.0$, where
misclassification is most likely (cf.\ Figure~\ref{sizemag.eps}).
DEEP2 stellar candidates ($p_{gal} < 0.2$) that are confirmed as stars
in EGS ACS images are plotted in blue; those shown instead to be
galaxies are red.  The latter are missed by DEEP2 and are thus a
concern.  Galaxies are lost in a narrow band in $B-R$, which
corresponds to certain combinations of intrinsic red-sequence color
and redshift (cf.\ Figure~\ref{allcolor.eps}).

The central rectangle in Figure~\ref{faintstars2.eps} delineates the
region in which galaxies are lost. Setting true stars aside (the blue
points) leaves a total of 99 galaxies in the rectangle, of which 27
were misclassified as stars by $p_{gal}$ (red points).  However, only
about half of all $R > 23.1$ red-sequence galaxies at $z = 0.75$-1.1
lie in the rectangle -- the others lie in the rectangles above or
below -- and this fraction is found to be virtually constant at all
redshifts above $z = 0.75$.  Since galaxies outside the central
rectangle are not lost, the net result is that $\sim 0.27 \times 0.50
\sim$0.13=13\% of red-sequence galaxies in the dimmest magnitude bin
are lost, and this loss is uniform beyond $z = 0.75$.  This matches
the estimate of Willmer et al.~(2006).  A loss of this size is not
generally important for analyses of total counts along the red
sequence but might be important (effects as large as 27\%) for studies
in individual restframe color bins {\it within} the red sequence that
have apparent $B-R$ near 2.25.

\begin{figure}
\ifpdffig
\includepdf[scale=0.4]{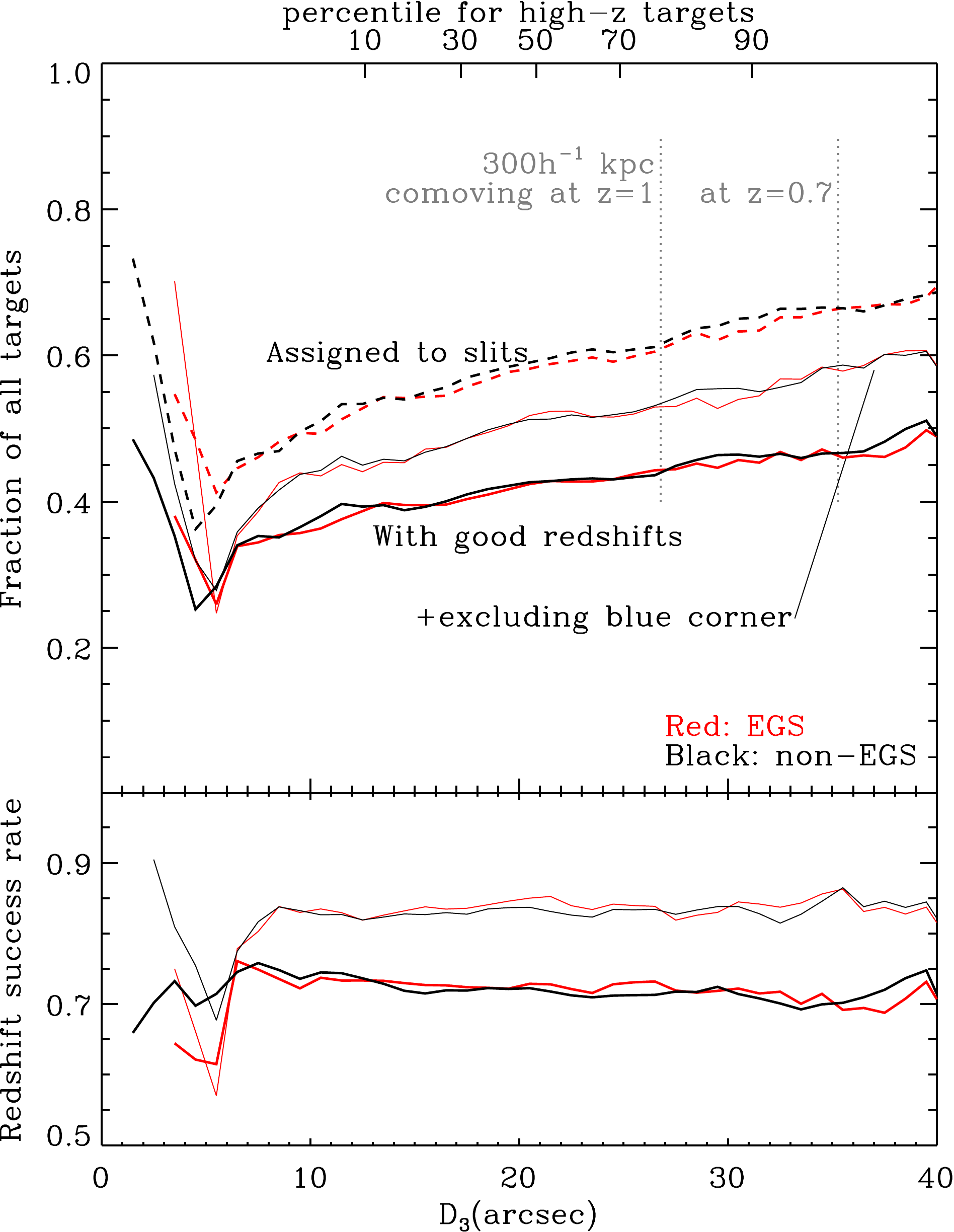}
\vspace*{0.15in}
\else
\includegraphics[scale=0.4]{neigh_figure.eps}
%
\fi
\caption{ Rate of spectroscopic observations and redshift success as a
  function of local density of DEEP2 target galaxies on the sky, as
  measured by the third-nearest-neighbor distance, $D_3$.  Upper
  panel: dashed lines show the probability that a galaxy meeting the
  DEEP2 targeting criteria receives a slitlet.  The heavy solid lines
  give the {\it total} probability that its redshift is reliably
  measured ($Q \ge 3$).  The light solid lines repeat this with all
  galaxies in the blue corner excluded, as most of those do not yield
  reliable redshifts.  The top axis shows the cumulative percentage of
  $z \sim 1$ galaxies with $D_3$ up to a given value.  Vertical lines
  show the apparent sizes of dense cluster cores at two redshifts
  (co-moving).  The sharp increase in targeted fractions at low $D_3$
  arises because close pairs may be placed on a single slitlet.  The
  lower panel separately shows the redshift success rate once a galaxy
  is targeted.  The probability of observation is reduced by almost a
  factor of two in regions of high density, but redshift success rate
  is unaffected.  }
\label{neigh_figure.epsi}
\end{figure}

$\bullet$ {\it Loss of objects at small separations} Because spectra
of galaxies on the same mask cannot be allowed to overlap, the number
of slitlets that can be put down in a small area is limited for
objects that would have similar positions on a mask (along its long
axis); this affects the sampling rate in dense regions.  This effect
is illustrated in Figure~\ref{neigh_figure.epsi}, expanded from Gerke
et al.~(2005), which shows that the sampling rate in the very densest
regions (in terms of nearest neighbor distance on the sky) is reduced
by almost a factor of two from nominal, but that the fraction of
sampled galaxies that yield reliable redshifts is unaffected.  The
upturn of the sampling rate at small separations is due to the
placement of close pairs on a single slitlet.

It should be noted that this effect is much weaker as a function of
three dimensional density: unlike in nearby galaxy surveys like SDSS,
most conflicting objects (and most sets of nearest-neighbors on the
sky) are actually at very different redshifts from each other (a
consequence both of $L_*$ brightening with $z$ and the comparatively
small contrast in luminosity distance between $z \sim 0.7$ and $z\sim
1.4$).  The effect is already substantially smaller in impact than
fiber collisions in nearby-galaxy surveys, since in DEEP2 objects can
conflict with each other along only one dimension, and at scales not
much larger than an individual galaxy (1\asec spans $\sim 10 h^{-1}$
kpc comoving at $z=1$, while the minimum separation between objects
along the mask direction -- i.e. the separation within which they
conflict with each other -- is 3\asec).

Nonetheless, the diminished sampling rate in dense regions is a large
enough effect that it does need to be taken into account for certain
statistics.  For example, Coil, Davis \& Szapudi (2001) investigated
the impact of prototype DEEP2 target selection algorithms on the
$\xi(r_p,\pi)$ diagram; they found that it introduces moderate
distortion into the two-point redshift-space autocorrelation function
at small projected separations, reducing the elongation of contours
along the ``finger of God'' by about 20\%.  The data shown in
Figure~\ref{neigh_figure.epsi} can be used for modeling purposes.  In
all DEEP2 clustering analyses (e.g.\ Coil et al.~2006a, 2008) this
effect is quantified and corrected using mock catalogs that have been
passed through the DEEP2 target selection code.

$\bullet$ {\it Multiple galaxies masquerading as single galaxies} The
final known bias in the DEEP2 target sample is due to the finite
resolution of the CFHT $BRI$ ground-based images, in which separate
galaxies may appear blended.  This blending can distort the magnitudes
and colors of DEEP2 galaxies, and it can also elevate galaxies that
would otherwise be too faint to meet the $R_{\rm AB} = 24.1$ magnitude
limit above that threshold.  Investigating these composite objects
requires redshift information, and so this topic is deferred to
\S\ref{multiplicity} after DEEP2 spectroscopy, data processing, and
redshift measurements have been discussed.

\label{targetbiases}

\section{Spectroscopic Procedures}
\label{exposures}

Instrument and exposure parameters for the DEEP2 survey are summarized
in Table~\ref{table.instrumentparams}.  Each mask is observed for at
least three 20-minute exposures with the DEIMOS 1200 l mm$^{-1}$
gold-coated grating centered at 7800 \AA.  No dithering is performed
between exposures because sky subtraction is natively almost
photon-limited (see \S\ref{dataquality}), and dithering would waste
roughly half of the detector real estate.

DEIMOS' guider camera is rigidly mounted in the slitmask coordinate
frame with known orientation and offset
(Figure~\ref{deimosfocalplane.thar.epsi}).  Locating a guide star at a
precomputed pixel location while the mask is at the proper PA
therefore succeeds in placing all alignment stars within their 4\amin\
square slitlets.  In a direct image of the night sky through the mask,
the locations of alignment stars and of all slitlets, including the
alignment boxes, are visible (the latter are illuminated by the sky
and hence much brighter than the interslit background).

We obtain the necessary direct alignment images efficiently by tilting
the grating to zeroth-order, rather than swapping the grating for a
mirror; the images obtained are then analyzed using custom-designed
software using a graphical interface.  The pointing of the telescope
and the PA of DEIMOS are then adjusted to center all alignment stars
within their boxes, which in general centers all galaxies in their
slitlets to within 0.1\asec (modulo astrometric errors).  Slitmask
alignment generally converges after two direct images and takes
roughly 5 minutes.  It is generally checked once using the same
procedure after the first of the three spectroscopic exposures (see
below).

\begin{figure}
\ifpdffig
\vspace*{-0.25in}
\includepdf[scale=0.4]{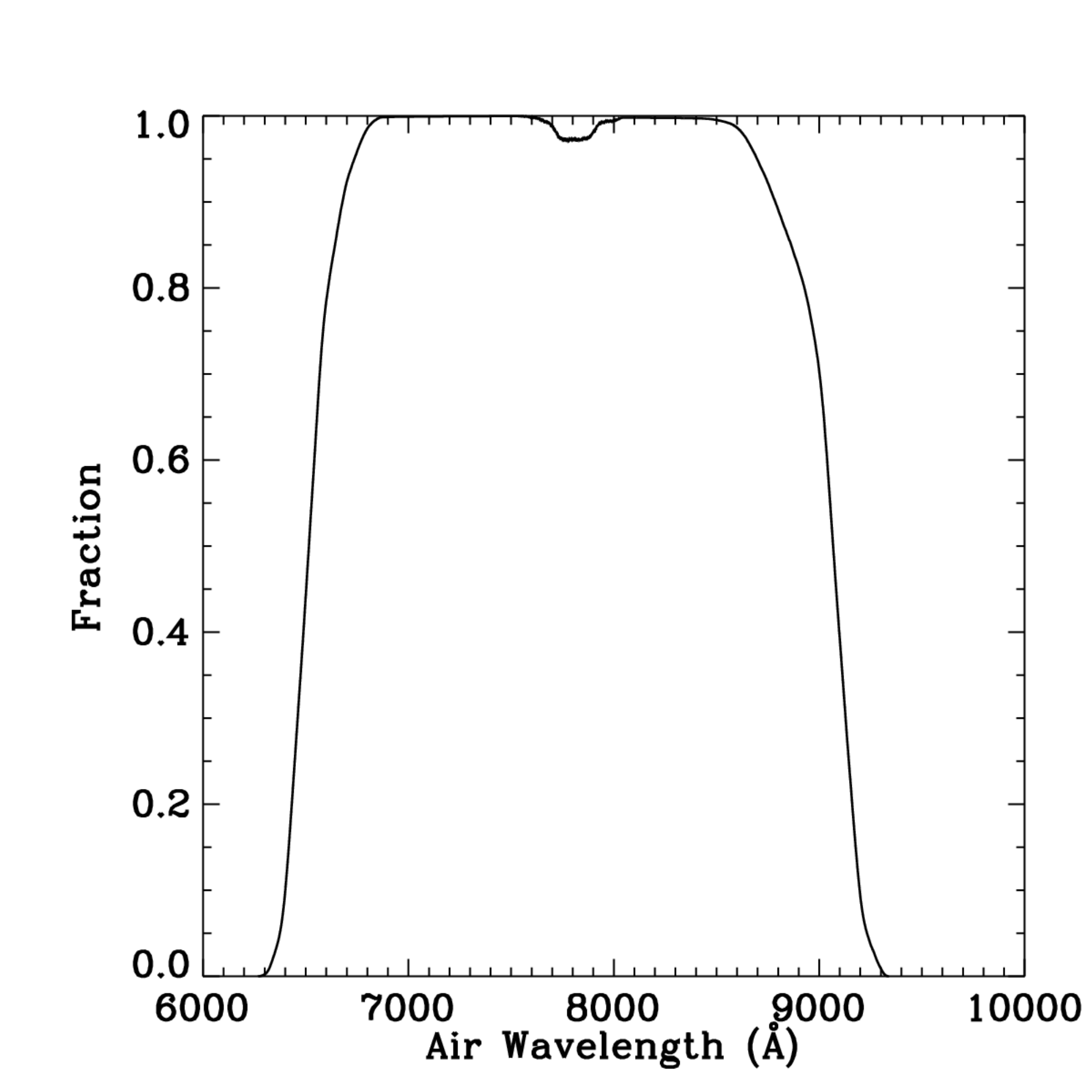}
\else
\includegraphics[scale=0.4]{wavecover.eps}
%
\fi
\caption{The fraction of DEEP2 spectra having coverage at a given
  wavelength. The dip around 7800 \AA\ is due to the gap between blue
  and red chips in the DEIMOS detector; this gap falls at different
  wavelengths for different spectra due to variation in slitlet
  position along the wavelength direction.  }
\label{wavecover.eps}
\end{figure}

The fraction of wavelengths common to all spectra is shown in
Figure~\ref{wavecover.eps}; 50\% coverage limits are 6500 \AA\ to 9100
\AA.  The 1.0\asec \ slitwidth used for all targets yields OH skylines
with a FWHM of $\sim1.3$ \AA, which produces a spectral resolution $R
\equiv \Delta\lambda/\lambda = 5900$ at 7800 \AA.  A custom data
verification program called {\it quicklook}\footnote{See
  \url{http://deep.berkeley.edu/deep2manual/quicklook.html}
  for details.} is run during observing.  This program performs a
coarse reduction of 10\% of the slitlets on each exposure and computes
the cumulative signal-to-noise ($S/N$) for an $R=23.5$ source obtained
by combining all exposures so far.  We continue obtaining exposures
until a minimum $S/N$ threshold is met, which enables productive
observing even under clouds or poor seeing.  The {\it quicklook}
software also checks for missing lamps in arcs, the occasional
slitmask buckling, flexure compensation system misalignments, guiding
errors along the slit, and bad or degrading seeing (based on the FWHM
of the alignment star spectra).  Run on a Sun workstation from circa
2002, it is able to evaluate a single DEIMOS frame in roughly 10
minutes (compared to our 20 minute exposure time).

All slitmasks are observed at airmass below 1.5.  Under optimal
conditions, eight masks can be observed per night, yielding 1000-1200
spectra.

\section{Reduced Spectroscopic Data}
\label{reduceddata}

\subsection{The {\it spec2d} Data Reduction Pipeline}
\label{pipeline}

The DEIMOS {\it spec2d} pipeline (Cooper et al.~2012b) is an IDL-based code that is modeled on and borrows significant code from the SDSS spectral pipeline of David Schlegel and Scott Burles (Schlegel \& Burles 2012). Most of it was written by Douglas Finkbeiner, Marc Davis, Jeffrey Newman, and Michael Cooper, with important contributions from Brian Gerke regarding non-local sky subtraction (see below). The pipeline operates in five separate stages, each of which produces its own output files. The code operates without supervision and usually does an excellent job of wavelength fitting, sky subtraction, and object fitting.\footnote {The code is publicly available, and further information may be found about how it works and how to download it at \url{http://deep.berkeley.edu/DR4} and \url{http://deep.berkeley.edu/spec2d/.}}

The first step in data reduction is creation of a "planfile" (a plain-text file with extension .plan) which may be generated automatically after a night's observing and defines the locations of all files needed to reduce a given mask's data.  The planfile may optionally control various parameters of the reduction such as the chips to be reduced, the arc line list to be used, etc.  The planfile is used by a controlling IDL program, DOMASK, which calls separate routines which perform a series of reduction steps:

{\it Step I:} The first stage of data reduction is to produce {\it calibSlit} files (i.e., a set of FITS BINTABLE format files whose filename begins with "calibSlit") from the flatfield and arc frames taken through a given mask.  This processing is controlled by the IDL procedure DEIMOS\_MASK\_CALIBRATE.  Each of the 8 DEIMOS CCDs is analyzed completely independently through all stages of the 2-d reductions, allowing trivial parallelization; in general, every spectrum will span 2 of these 8 CCDs, which we will generally refer to as 'blue' and 'red'; hence for each object's spectrum there are two separate {\it calibSlit}, {\it spSlit}, and {\it slit} files (q.v.\ below). The {\it calibSlit} files indicate where the slitlets fall on the CCD array for each mask and contain flatfield information and a 2-d wavelength
 solution for each slitlet.   
 
 To produce the {\it calibSlit} files, first the multiple flats are first read in, corrected for pixel-to-pixel response variations, and processed to reject cosmic-rays.  To identify every slitlet, edges that are detected in the flats using an unsharp mask are compared to a table in the DEIMOS header data that describes where each slitlet should appear; a smooth polynomial is fitted to the edges of eacqh curved spectrum to define its upper and lower limits. Each chip's combined flat frame is then mapped into rectangular arrays for each slitlet, which greatly simplifies further processing, by shifting its edges to be parallel, using interpolation (over the spatial direction) to accomodate subpixel shifts.  Hence, we turn the entire flatfield into a set of individual, rectangular flatfields for each slitlet.  Each slitlet's flatfield is first used to measure the ``slit function,'' i.e.\ the relative throughput (maximum 1) of each slitlet as a function of position (i.e.\ row number) along it.\footnote{Throughout this discussion, we will use 'rows' and 'columns' to refer to the rows of a rectified image stored by the pipeline; hence rows run along the wavelength direction, with columns along the spatial direction, perpendicular to the definitions of rows and columns in raw DEIMOS data.}  After normalizing by the slit function and any large-scale variation in light intensity with wavelength and position along a given slitlet, the flats are smoothed on small scales to generate a 2-d fringing-correction map with mean 1.

 {\it Step II:} The next step, also controlled by DEIMOS\_MASK\_CALIBRATE, is to solve for the wavelength of each pixel in the 2-d data array.  First, the entire arc (or arcs, if multiple ones are available\footnote{In general, shifts between successive arc frames are greater than our wavelength-solution tolerance for sky subtraction; a single, KrArNeXe arc yielded best results.}) is read in and corrected for pixel-to-pixel response variations.  Next, a shift in the spatial direction between a given chip's arc data and the original flatfield, $dx$, is solved for by cross-correlating the intensity summed over the central (in the wavelength direction) region of the slit with the light profile predicted for the edges found in Step I.  Then, individual slitlet arc spectra are extracted and rectified in the spatial direction using the same algorithms applied for the flatfield, but using slit edges shifted from the flat by $dx$; the individual spectra are then corrected for the slit function and fringing using the results of Step I.

 Starting with the DEIMOS optical model (produced by Drew Phillips), which gives a rough initial guess, we fit for the relationship between wavelength and pixel in a spectrum in multiple stages, obtaining a wavelength solution that is typically accurate at the 0.007 \AA\ ($\sim$0.02 px) RMS level or better over the entire 2-d spectrum; 0.003 \AA\ RMS is commonly achieved.  The fits used are linear regressions for the coefficients of Legendre polynomials of up to fifth order (for wavelength as a function of pixel number along the central row of the slitlet) or second order (for specifying the local tilt of the constant-$\lambda$ locus as function of central pixel number).  The mean residual about the fit in each row of the slitlet is also stored and incorporated into the wavelength solution (i.e., we apply a row-by-row shift from the smooth polynomial-based solution, e.g. to account for defects or dust along the slit which shift the mean wavelength of a row).  The requirements that accurate OH-line subtraction places on sky subtraction are discussed in Section~\ref{dataquality} and in the Appendix.  We generally require errors below $\sim 0.01$ \AA\ RMS for good sky subtraction; to be useful in attaining wavelength solutions, individual arclines must contain $\sim10,000$ photons in each row.  The result of Steps I and II is a single {\it calibSlit} file containing all necessary calibration information per slitlet per chip (i.e., each spectrum is split into blue and red files).

 {\it Step III:} The third stage of the reductions, controlled by the procedures DEIMOS\_2DREDUCE and DEIMOS\_SPSLIT, uses the {\it calibSlit} files produced in Steps I \& II to flatfield and rectify the on-sky ("science") data for each slitlet. Response correction, slitlet extraction/rectification, and flatfielding is all done using the same procedures as for the arc files.  All CCDs are again treated separately.  An inverse-variance image is also produced based on the photon and read noise (with compensations for all multiplicative corrections applied), as well as maps of bad pixels (which will have zero inverse variance; this map provides flag information on why a given pixel is bad) and questionable ones (e.g., those which may be affected by imperfect vignetting corrections or cosmic rays, which will not have their inverse variance set to zero; again, this map provides flag information for each pixel).

The most important operation in this step is computing the proper sky background level (including bright OH lines) at every location in a 2-d science spectrum as a function of both wavelength and position along the slitlet.  First, a constant wavelength shift (relative to the arc wavelength solution) for each frame is determined using cross-correlation against a high-resolution night-sky spectrum.  Then, a B-spline model for sky intensity as a function of wavelength is fitted to the sky regions of the slitlet (i.e., regions where, based on its location and size, both contamination from DEEP2 targets and slitlet edge effects are expected to be minimal).  A key strategem is tilting all slitlets by at least 5\deg\ with respect to the detector array so that each night sky line is substantially oversampled, making the B-spline fit more accurate; see Appendix~\ref{skysubtractcode}.  The result of Step IV is one {\it spSlit} file per slitlet per chip, which contains the reduced two-dimensional spectra (with associated wavelength, inverse variance, and mask arrays) from each science frame for that slitlet and that frame's bspline fit in alternating HDUs.

{\it Step IV:} The fourth stage, controlled by SPSLIT\_COMBINE and SLITEXAM, combines the separate science exposures for each slitlet into one inverse-variance-weighted mean, sky-subtracted, cosmic ray-cleaned 2-d spectrum (which we will generally refer to below as a ``Combined'' or {\it slit} file).  First, the B-spline model sky for each slitlet is subtracted from its flux. Next, the (generally) three 20-minute exposures are combined, weighting them according to the inverse variance maps from the previous stage.  Cosmic ray rejection is also done at this time based on the time variability of a given pixel.  The result of this process is one {\it slit} file for each slitlet for each chip, containing the processed, combined two-dimensional spectrum, inverse variance, and wavelength solution, along with various diagnostic information (e.g. bad pixel masks).  Each row of each spectrum has its own wavelength solution; a 2-d map of wavelength as a function of position in an image may be obtained with the IDL function LAMBDA\_EVAL.

{\it Step V:} In the fifth and final stage, 1-d spectra are extracted from the Combined 2d spectrum using the routine EXTRACT1D (controlled by the procedure DO\_EXTRACT).  Extraction is done using both "optimal" and boxcar (i.e., fixed-width sum) extraction techniques (see below), as usual processing the blue and red chips separately.  As a first step in both approaches, each 2d spectrum is rectified in the wavelength direction by using whole-pixel shifts (as opposed to interpolation), making the wavelength scale in each row match the central row of the object.

The optimal extraction (or "horne" as it is labeled in the FITS headers) loosely follows the algorithm of Horne et al.~(1986).  However, due to the lower signal-to-noise of some DEEP2 data, instead of using the light profile evaluated at a single location along the slit to weight pixels, a Gaussian model is used for weighting.  First, we measure the light profile along each slit by performing an inverse-variance weighted average of the flux along each row of the slit, excluding regions around bright night sky emission lines.  Next, an overall offset between the positions of objects along each slit predicted from the mask design and their actual observed positions is determined by fitting a Gaussian to each peak in the individual slits' light profiles and comparing their locations to the prediction.  We also determine the quantity that must be added to or subtracted from (in quadrature) the sizes of objects in the CFHT photometry ($r_g$ values) to match the observed widths of each Gaussian peak, allowing compensation of sizes for differences in seeing. The center of the extraction window is determined from either the peak of the spatial profile of the object along the slit or (in low-flux cases where the peak differs by at least 5 pixels from the prediction) using the predicted position with the just-determined shift in the spatial direction applied.  The extraction center is determined independently on the red and the blue sides, ameliorating the small shifts between them expected due to differential refraction.

The weighting kernel $P$ used for optimal extraction is then a Gaussian with peak at the center of the extraction window and $\sigma$ derived from a Gaussian fit to the light profile in the 2-d spectrum in high-S/N cases, or the expected seeing-corrected size as derived from the photometry in the event of low S/N.  In our reductions the extracted flux $f$ at a particular wavelength is:
\begin{equation}
f = {\Sigma_i f_i P_i S_i} \over {\Sigma_i P_i^2 S_i} , 
\end{equation}
where $f_i$ is the flux, $P_i$ is the model profile, and $S_i$ is the inverse variance at a particular row (at fixed wavelength).  Including the inverse variance in the weight in this way yielded slightly improved results over using the Gaussian profile alone.

In the boxcar approach, bad pixels are masked (rather than interpolated over as in the optimal extraction) and the fraction of flux missed within these pixels is determined using the Gaussian model spatial profile along the slit determined for the optimal extraction. The flux from all pixels at a given wavelength is then summed, and that sum is divided by the fraction of flux which should have come from the included rows. Hence, this algorithm differs only in the case of bad pixels from a generic boxcar or tophat weighting. For this reason, the extraction is called a "boxsprof" extraction in the headers of the spec1d FITS files. The half-width of the extraction window used is a multiple of the light-profile model Gaussian width: 1.75$\sigma$ (in the optimal case) or 1.3$\sigma$ (in the boxcar case).

The extractions are saved in a {\it spec1d} FITS file, which contains the blue and red portions of each object's spectrum for each type of extraction, each in its own HDU (i.e.\ 4 HDUs total).  Note that in cases where multiple DEEP2 targets are placed on the same slitlet, there will be only one set of {\it calibSlit/spSlit/slit} files, but separate {\it spec1d} files for each object.

There is no step in the pipeline data reduction process that explicitly flux-calibrates the data using standard star spectra.  However, DEIMOS is a fairly stable instrument, and a {\it relative} flux calibration accurate to roughly $\pm$10\% over the whole spectral range has been determined, based on tests with standard stars and the coadded spectra of galaxies on each chip (Yan et al., in prep).  This error does not include variations in absolute throughput due to clouds, seeing, and long-term drifts in optics reflectivity (the latter at the 30\% level); in applications requiring flux calibration, we tie our observed spectra to the CFHT photometry.  Standard star observations are taken on a regular basis, but these data are not part of the regular reduction process.  Throughput plots (for freshly recoated optics) are available online.\footnote{\url{http://www.ucolick.org/$\sim$ripisc/Go1200/through\_go1200\_tilts.gif}}
    
\subsection{Final Data Quality}
\label{dataquality}

\begin{figure*}
\ifpdffig
\begin{center}
\includepdf[scale=0.7]{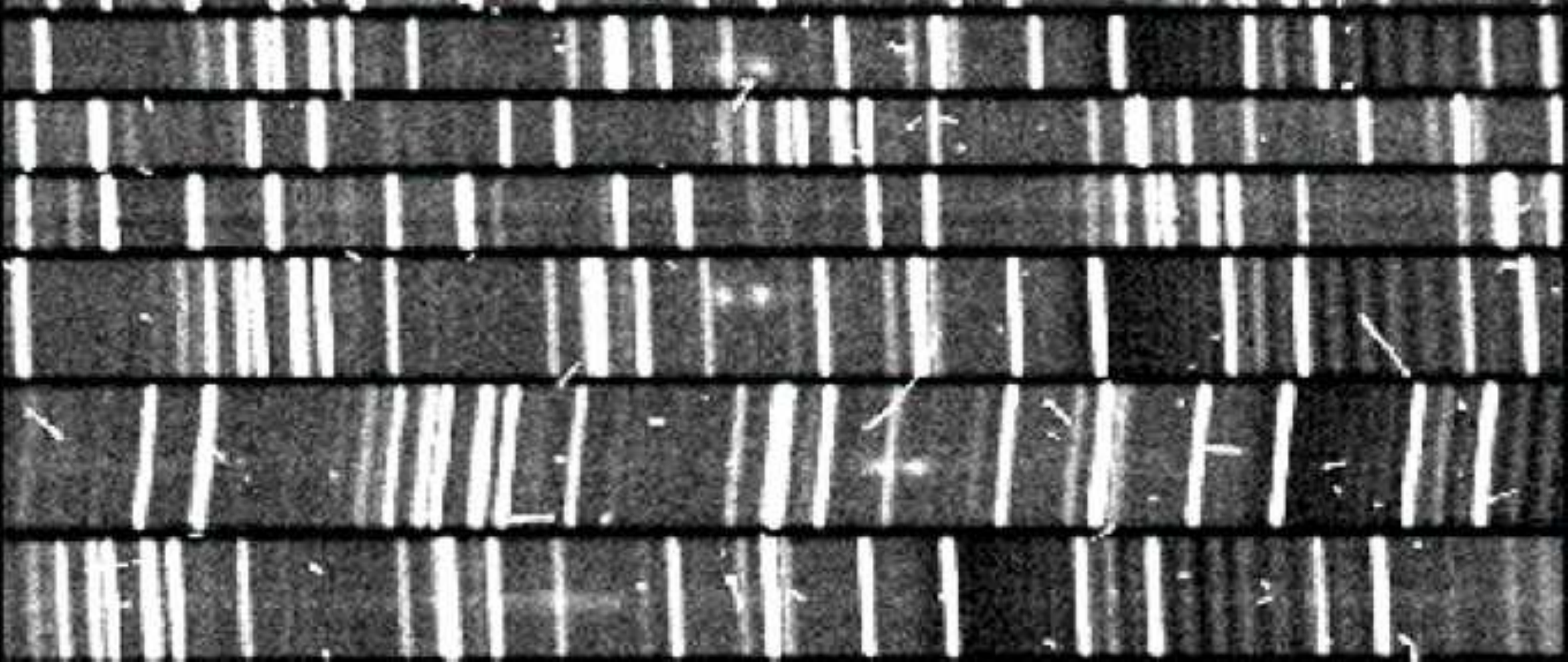}
\end{center}
\else
\begin{center}
\includegraphics[scale=0.7]{newman_rawimage.eps}
\end{center}
\fi
\caption{A section of single raw 2-d 20-min exposure from the DEIMOS spectrograph.
The [O {\scriptsize II}] doublet is visible in four objects,
but cosmic rays and night-sky lines are much brighter.  
} 
\label{davis.matched.rawimage.epsi}
\end{figure*}

\begin{figure*}
\ifpdffig
\begin{center}
\vspace*{0.1in}
\includepdf[scale=0.7]{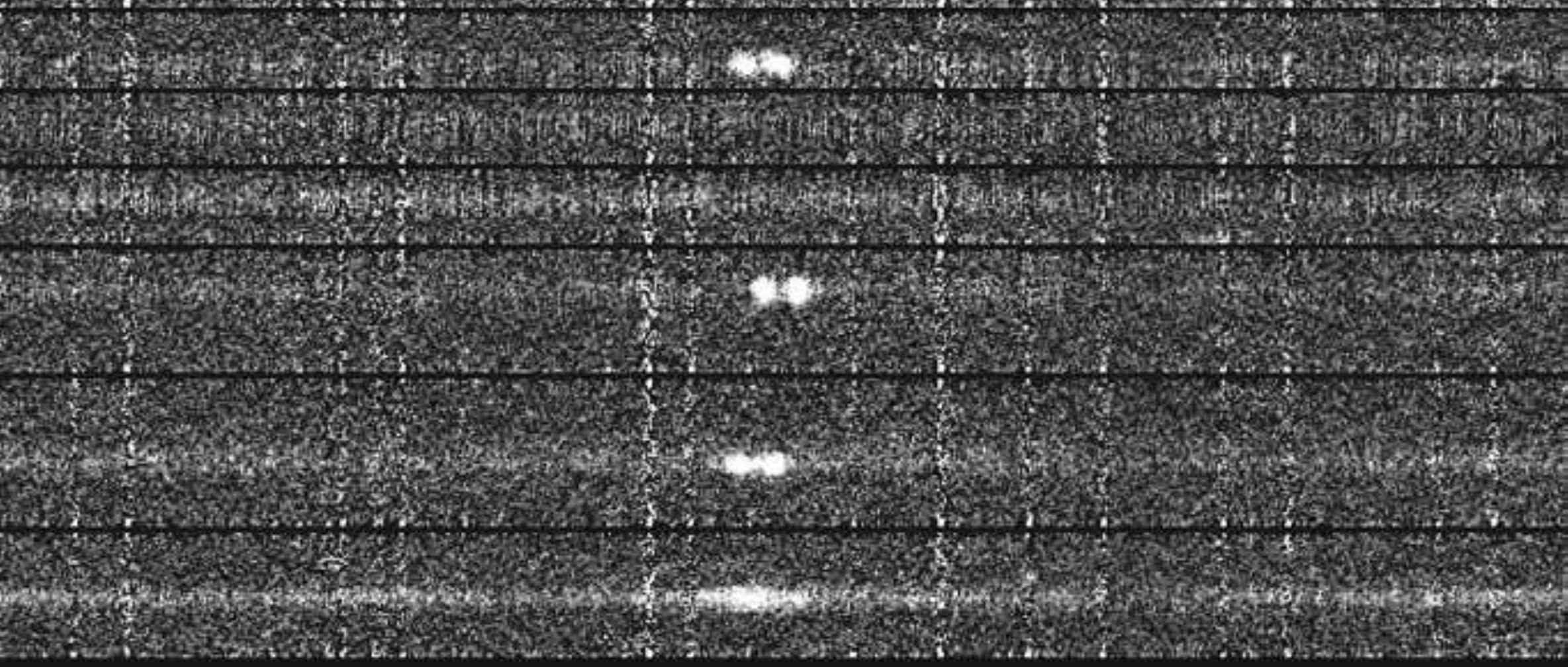}
\end{center}
\else
\begin{center}
\vspace*{0.1in}
\includegraphics[scale=0.7]{newman_rawimage_newreduction.eps}
\end{center}
\fi
\caption{A comparable section of a 2-d image showing the reduced and rectified sum of 3 exposures. The [O {\scriptsize II}] doublets seen in Fig.~\ref{davis.matched.rawimage.epsi} stand out strongly, and the noise under night sky lines is consistent with random noise except at slit ends (due to their sharper PSF compared to the bulk of the slit).  This is an example of a typical reduction and subtraction of night sky.  Some vertical striping is visible in object regions due to undersampling of the sky spectrum on nearly-vertical slits (which were only used in 2002 observations); this correlated noise is absent from the great majority of DEEP2 spectra.  }
\label{davis.matched.reducedimage.epsi}
\end{figure*}

Sample raw and reduced spectroscopic images are shown in Figures~\ref{davis.matched.rawimage.epsi} and \ref{davis.matched.reducedimage.epsi}.  Figure~\ref{davis.matched.rawimage.epsi} shows a blowup of a portion of a single 20-minute sub-exposure (cosmic rays are clearly visible), while Figure~\ref{davis.matched.reducedimage.epsi} shows the fully reduced, CR-cleaned, sky-subtracted, and coadded ensemble of three 20-minute exposures.  This particular region represents a typical reduction, with sky subtraction that is photon-noise-limited as signaled by the random noise pattern under the OH lines (although some aliasing is visible in the sky subtraction for slits which were not tilted significantly compared to the detector; this mask was one of the first observed, and we required a 5 degree minimum slit tilt only afterwards).

\begin{figure}
\ifpdffig
\includepdf[scale=0.4]{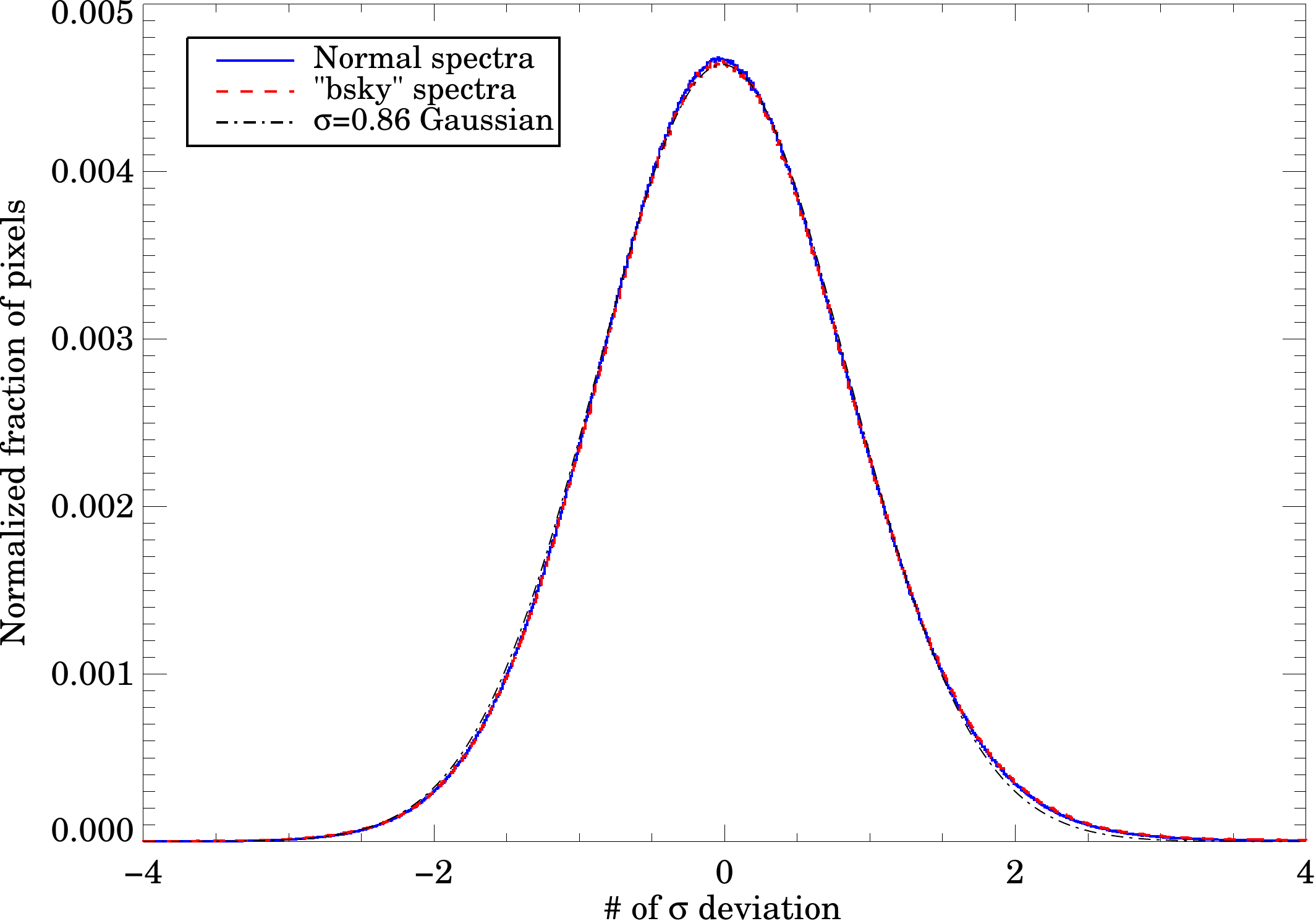}
\else
\includegraphics[scale=0.4,angle=90]{hist_offsky.eps}
\fi
\caption{Distribution of the quantity $\Delta/\sigma$ for sky pixels {\it not} under bright sky lines.  Here $\Delta$ is the difference between the actual photon count in the pixel and the predicted sky count at that location inferred from a local sky median, and $\sigma$ is the predicted total noise based on the read noise and predicted sky count.  Each curve combines the pixels from the sky regions of 800 2-d galaxy spectra.  The blue solid curve is based only on normal spectra; the red dashed line is for spectra with poor sky subtraction in at least part of the range (designated by the ``bsky'' comment, see \S\ref{datatables}).  The curves are nearly identical, and both closely follow a Gaussian.  If the noise were due purely to photon statistics plus read noise and were properly estimated, the curves should follow a Gaussian distribution with rmd width $\sigma$ = 1.  The actual distributions are nearly Gaussian but with width $\sigma$ = 0.86, indicating that we have if anything {\it over}estimated the noise in our spectra.  Regardless of any normalization error, the tails are very nearly Gaussian, indicating that our sky subtraction is well behaved.  }
\label{hist_offsky.epsi}
\end{figure}

\begin{figure}
\ifpdffig
\vspace*{0.2in}
\includepdf[scale=0.4]{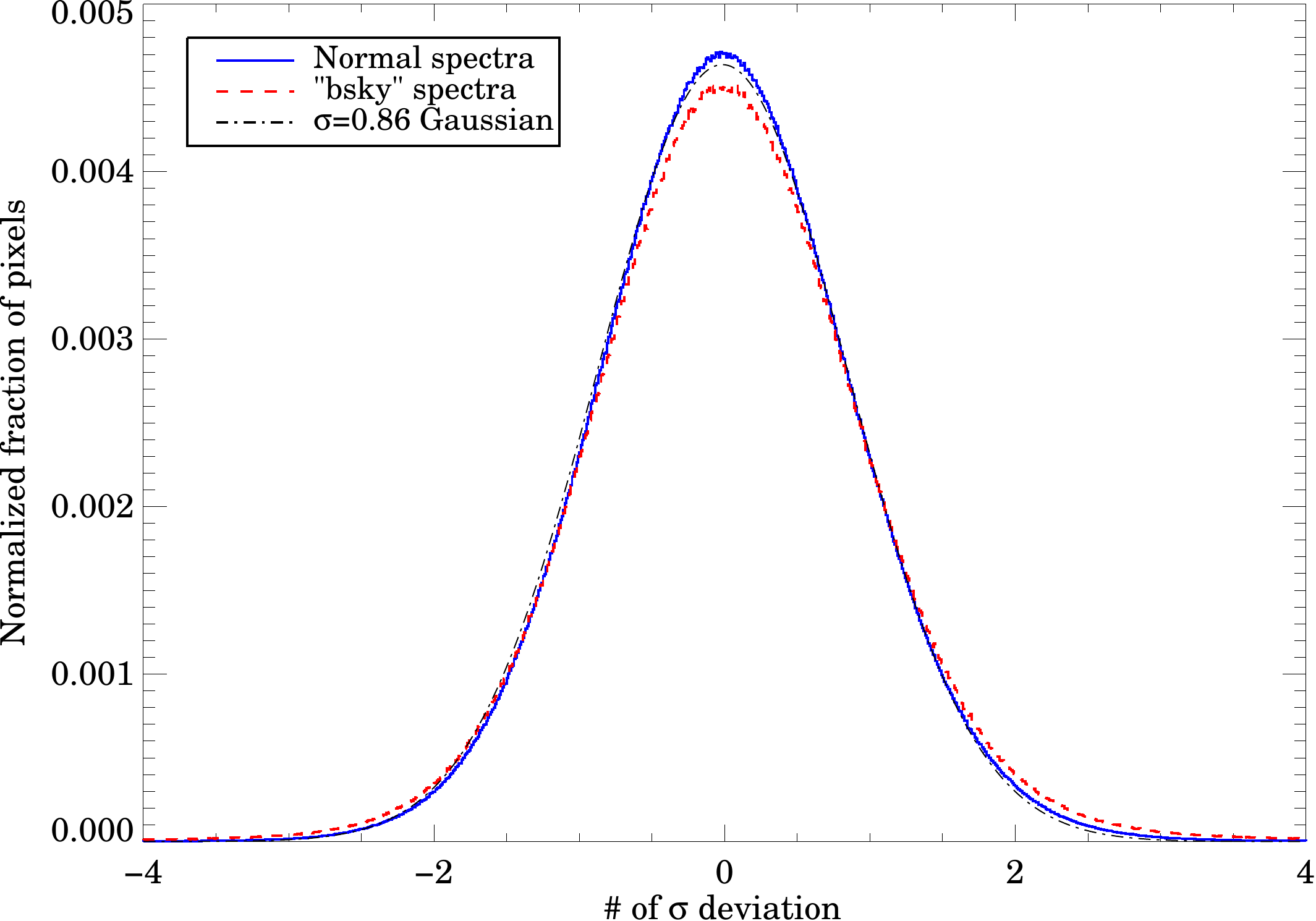}
\else
\includegraphics[scale=0.4,angle=90]{hist_onsky.eps}
\fi
\caption{This figure repeats Figure~\ref{hist_offsky.epsi} but for sky pixels falling on bright OH lines.  Wavelength and  flat-fielding errors would each produce residuals that are larger than the expectation from combined photon and read noise, causing an excess at the tails of the distribution.  Normal and ``bsky'' spectra are again compared to each other.  The normal spectra again closely follow a Gaussian curve with $\sigma = 0.86$.  The ``bsky'' residual distribution is slightly worse, but are still remarkably similar to the results for typical spectra on sky-lines or for the benign regions depicted in Figure~\ref{hist_offsky.epsi}.  These two figures collectively demonstrate that our sky subtraction accuracy is very close to being photon-limited both on and off the OH lines.  } 
\label{hist_onsky.epsi} 
\end{figure}

Poor sky subtraction happens occasionally, and some examples will be shown below.  However, quantitative analysis indicates that the sky subtraction accuracy is close to photon noise-limited even under bright sky lines and under most conditions.  This is illustrated in Figure~\ref{hist_offsky.epsi} and Figure~\ref{hist_onsky.epsi}, which show distributions of the quantity $\delta/\sigma$ for sky pixels, where $\delta$ is the difference between the actual and predicted pixel intensity and $\sigma$ is the read noise combined with the photon noise in that pixel based on the measured sky level at that location (inferred from a local median filtering of the inverse-variance map).  If sky subtraction is noise-limited, this distribution should follow a Gaussian with RMS width unity.  Significant numbers of pixels with bad sky subtraction will depress the middle and enhance the tails.

We consider first Figure~\ref{hist_offsky.epsi}, which treats pixels {\it between} the bright sky lines.  Two sets of 800 spectra were analyzed, one set for which visual inspection confirmed good sky subtraction over the whole spectrum (``normal''), and one set for which visual inspection indicated significant systematic sky-subtraction errors over at least part of the range (see ``bsky'' comment code in Section~\ref{datatables}).  The horizontal axis is the ratio of the actual flux deviation at that pixel divided by the predicted noise due to read noise and photon sky noise at that pixel.  If the noise is random and is properly predicted, the observed curves should be Gaussian with RMS $\sigma = 1$.  Both sets of spectra do indeed populate the same Gaussian, but with RMS width 0.86, not 1.0, indicating that if anything the errors have been {\it over}estimated, It is reassuring to see that the tails of the distribution are not significantly overpopulated.

Figure~\ref{hist_onsky.epsi} repeats this test under more difficult conditions, using the same spectra but now with pixels {\it under} bright OH lines.  Small errors in the wavelength scale or in flat-fielding will cause the residuals to be large and populate the tails of the distribution.  The normal, well-subtracted spectra again closely follow the same Gaussian curve with $\sigma = 0.86$; our pipeline error model appears to be modestly overconservative. The ``bsky'' spectra look slightly worse but are still remarkably similar to the normal spectra here as well as in the benign sky regions in Figure~\ref{hist_onsky.epsi}.  The two figures collectively demonstrate that our sky subtraction accuracy is very close to being photon-limited both on and off the OH lines, even in spectra that the eye sees as having imperfect sky subtraction.

\begin{figure}
\ifpdffig
\vspace*{-0.1in}
\includepdf[scale=0.38]{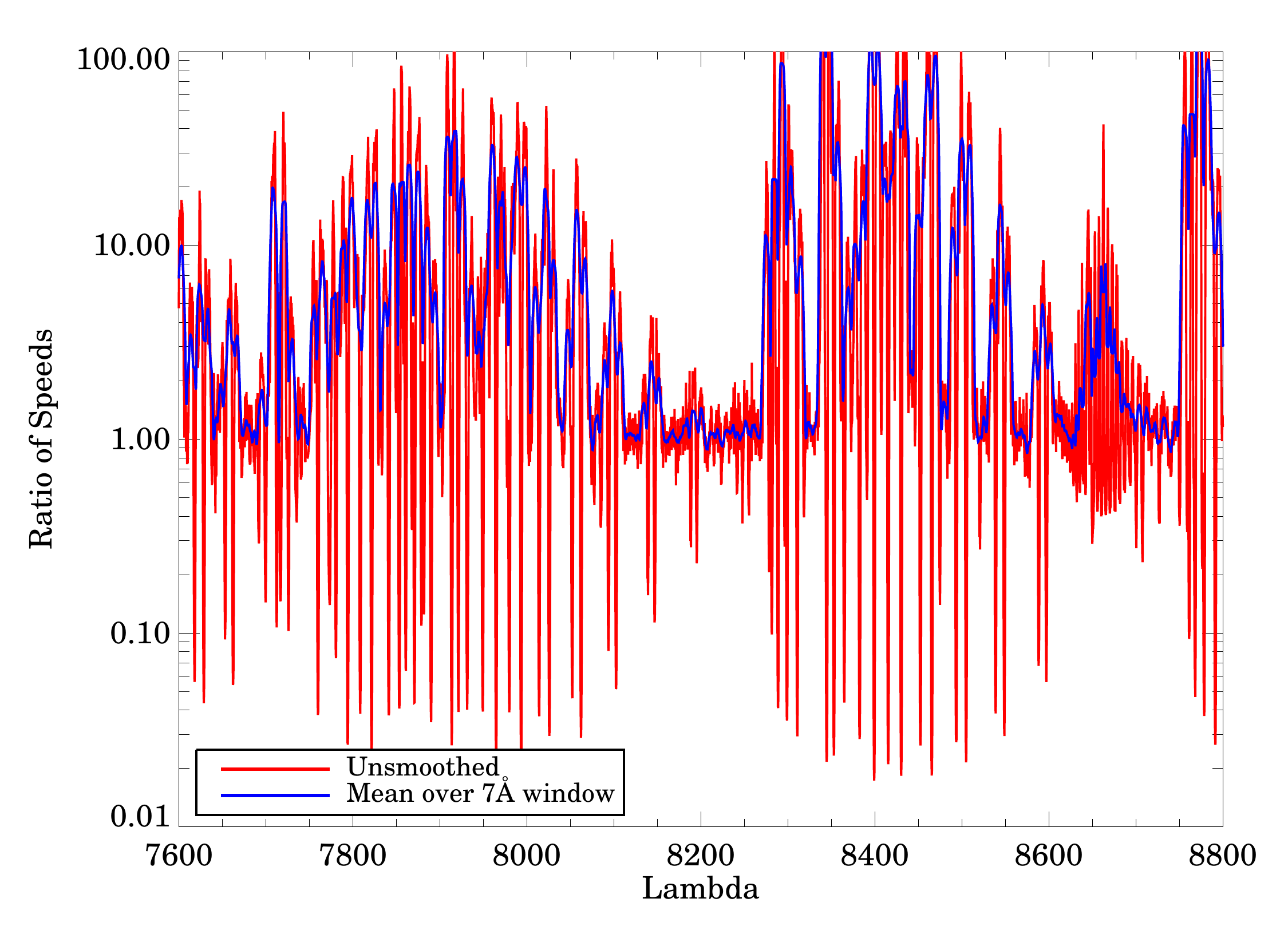}
\else
\includegraphics[scale=0.38,angle=90]{speeds_600.eps}
\fi
\caption{The speed gain (in exposure time at each wavelength) that is obtained by using a spectral resolution $\delta \lambda/\lambda = 6000$ versus 600.  The red curve is based on the full-resolution DEEP2 sky spectrum; the blue curve has been smoothed with a 7 \AA\ boxcar.  In regions that are relatively free of sky lines (e.g., near 8200 \AA), the exposure times are equal.  In regions near sky lines, the high-resolution mode saves a large amount of time by removing sky-line photons from those pixels, while in regions directly on top of sky lines, the low-resolution mode is faster because the sky brightness is diluted there.  The blue curve is the best gauge of total savings, as it provides a running average of time saved over 7 angstroms.  The time savings in regions that are dense with sky lines can exceed a factor of 10.}
\label{speeds_600.eps}
\end{figure}

Figure~\ref{speeds_600.eps} shows the major speed gain that has been obtained by deciding to use high spectral resolution for the DEEP2 survey.  This figure shows the ratio of exposure times needed to obtain the same $S/N$ at each wavelength using a spectral resolution of $R= 600$, which is typical of distant redshift surveys, versus DEEP2's resolution of $R \sim 6000$.  The red curve is the raw ratio at each pixel, while the blue curve is the red curve smoothed by a 7 \AA\ boxcar.  The speed gain varies rapidly with wavelength depending on whether one is on a sky line or not, but the median (unsmoothed) gain from 7600-8800 \AA\ is 1.58.  This factor is the savings in exposure time.  The gain occurs because high resolution confines OH photons to a small fraction of the pixels, leaving most of the pixels much darker than they otherwise would be.
  
\begin{figure*}
\ifpdffig
\begin{center}
\includepdf[scale=0.7]{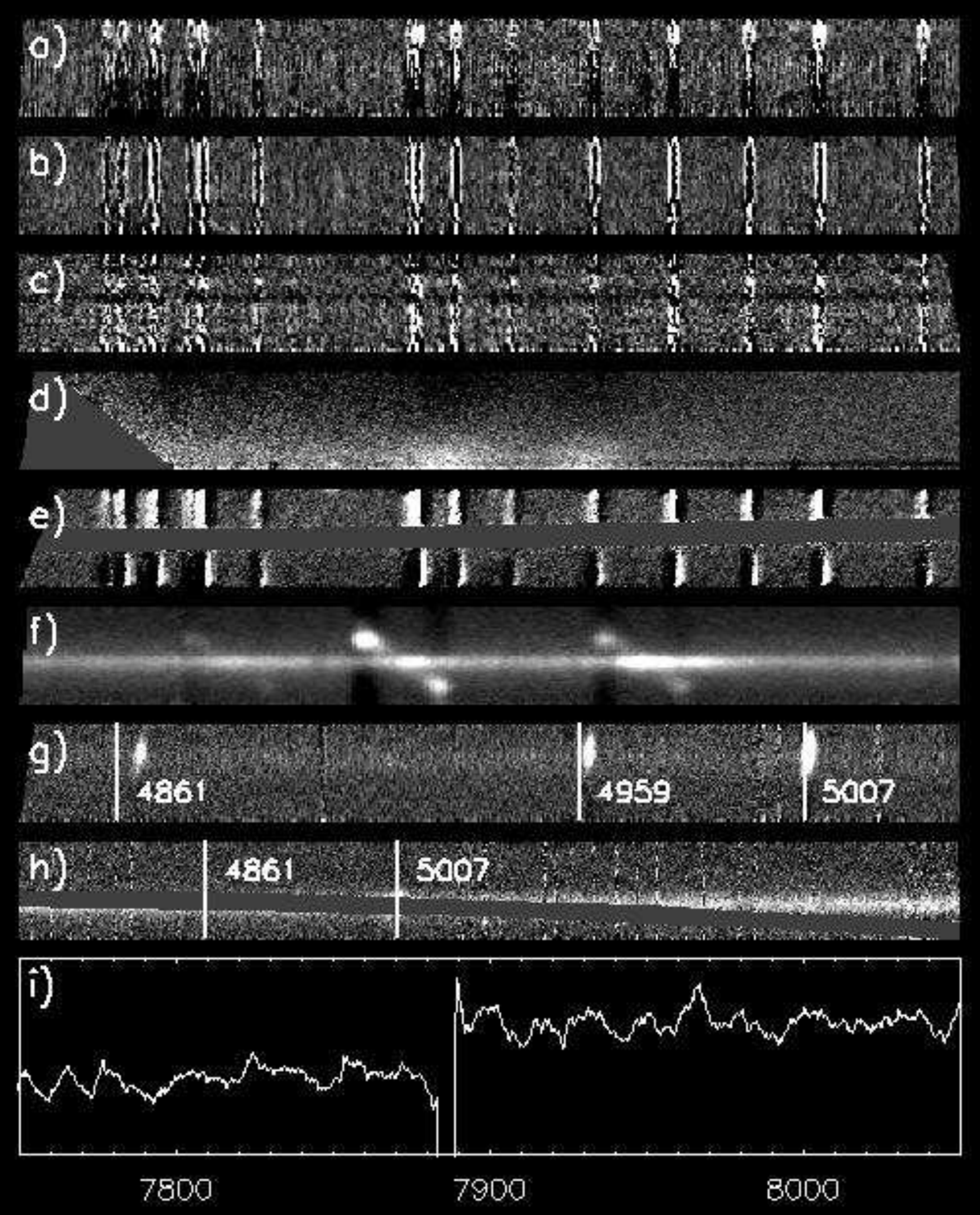}
\end{center}
\else
\begin{center}
\includegraphics[scale=0.7]{bad_redux_2d_v5.eps}
\end{center}
\fi
\caption{Examples of problematic data and the corresponding comment codes: (a) - (b): Bad modeling of OH lines and resulting systematic residuals.  A sky spectrum with the wrong profile width (e.g. due to a different effective slit width on sky vs. object regions due to dust or burrs) causes residuals visible at both the line center and wings.  This is tagged by {\tt bsky} if problems affect substantial fraction of wavelength range.  (c): Dust or burrs along slitlet can cause high-order intensity variations in the slit function too strong to be modeled.  This is tagged by {\tt bsky}.  (d) Sky continuum errors caused by scattered OH light from a neighboring slitlet.  The extra-wide alignment-star boxes are the most common sources.  If this affects many wavelengths, it gets tagged by {\tt bsky}.  (e) Missing columns caused by a set of adjoining bad columns on the CCD.  This has caused an inaccurate wavelength solution.  Wavelength errors as small as 0.015 px can cause positive vs.~negative errors across the line, yielding a 'barberpole' effect at OH lines.  Slitlets with only partial coverage due to a position overhanging a CCD edge often exhibit similar problems.  If many wavelengths are affected, the slitlet is useless and the spectrum is assigned quality code $Q = -2$; if only a small region is affected, the spectrum is tagged with {\tt bcol} and {\tt bsky}.  (f) Emission lines have extended beyond the extraction window into the sky region , so that the sky is locally oversubtracted.  This case would yield a good redshift but get the {\tt iffy} comment code to signify that emission-line EWs are incorrect.  (g) A slitlet with a bad wavelength solution.  Gets $Q = 2$ to indicate need for rereduction.  The problem is so rare there is no specific comment code for it.  (h) A set of bad columns ({\tt bcol}) cross the spectrum, obliterating most of the galaxy, but a hint of $\lambda$5007 peeks above and below.  If one other feature is clear, this spectrum would get $Q = 4$ and comments {\tt bcol} and {\tt iffy}.  (i) Example of a flux discontinuity across the blue-red CCD gap.  Signalled by {\tt disc}.  The DEEP2 object numbers for these spectra are: (a) 11013128, (b) 11033793, (c) 11039367, (d) 11044563, (e) 11043877, (f) 11045853, (g) 13010507, (h) 12029377, (i) 11048040.  }
\label{bad_data.epsi}
\end{figure*}

Figure~\ref{bad_data.epsi} illustrates various data reduction problems and their associated comment codes.  Spectra (a) and (b) show regions with poor sky subtraction in the OH lines along the slitlets, recognizable by the {\it systematic patches} of bright and dark sky values under the OH lines.  Variable wavelength errors sometimes cause these residuals to wrap along the slitlet, yielding a ``barberpole'' pattern.  Appendix~\ref{fringing} derives an allowed image-motion criterion of 0.6 px (rms) in the wavelength direction in order to keep flat-field fringing-induced errors in sky subtraction to tolerable levels.  DEIMOS's actual image motion is two times smaller than this.  Thus, the sky-subtraction errors seen in spectra (a) and (b) should not be due to fringing errors but rather tiny residual sky-line modeling errors, due either to wrong sky-line profile shapes or to small wavelength calibration errors.  Errors in the latter case of only 0.015 px are sufficient to shift the model sky lines enough during ``de-tilting'' to produce noticeable subtraction errors.  Less than 1\% of the total pixel area in the DEEP2 spectra is noticeably affected by such problems.

Spectrum (c) in Figure~\ref{bad_data.epsi} shows an example of burrs or dirt along a slitlet, which perturbs the slit function and the sky-line PSFs.  Spectrum (d) shows scattered light from sky lines in the neighboring slitlet that has contaminated the sky continuum in this slitlet; such light most often comes from the alignment boxes, whose 4\asec slit widths admit more light.  The gap between CCDs has eliminated large chunks of the spectrum in spectrum (e).  Such loss can also be caused by bad CCD columns.  Spectrum (f) illustrates a common problem in which galaxy emission extends into the sky pixels, causing the sky spectrum to be too bright, which creates a dim spot in the extracted spectrum.  Spectrum (g) shows a (rare) bad wavelength solution.  In spectrum (h), a bad column crosses the spectrum, obliterating most of the galaxy, but enough of $\lambda$5007 peeks through to yield a reliable redshift.  Spectrum (i) illustrates a flux discontinuity across the red and blue CCD gap.  These cases usually look fine in the 2-d data and are caused by a slight mismatch in the extraction window between the two CCDs.  The final continuum flux is very sensitive to the exact width and position of the extraction window.

\subsection{Sample Spectra}
\label{samplespectra}

\begin{figure*}
\ifpdffig
\begin{center}
\includepdf[scale=0.7]{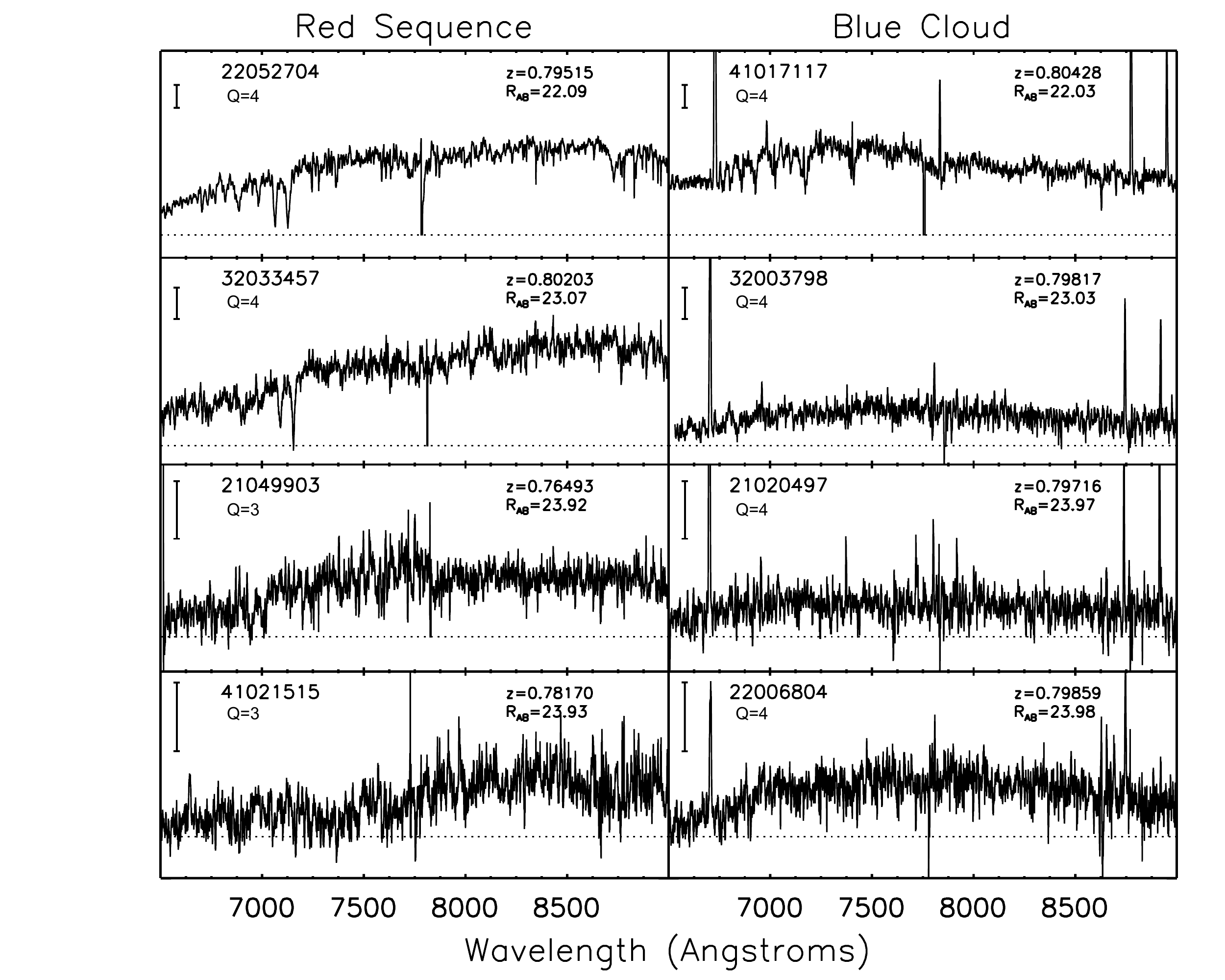}
\end{center}
\else
\begin{center}
\hskip -0.5in \includegraphics[scale=0.7]{oned_showcase.eps}
\end{center}
\fi
\caption{Examples of extracted, 1-d spectra from DEEP2, illustrating a selection of
red and blue galaxies near $z = 0.8$. Objects are ordered
by brightness within the red sequence and blue cloud
to illustrate the declining $S/N$ for fainter
galaxies.   The dip in counts near
the centers is due to the gap between the red and blue
CCDs.  The spectra have been smoothed with
an inverse variance-weighted boxcar of width 15 pixels.
The spectra have not been flux calibrated;
the vertical bar denotes an intensity of 50 DN/pix. 
There is a strong bias to preferentially determine redshifts for faint red
galaxies that have [O {\scriptsize II}] emission, as in object 41021515 at lower left.  } 
\label{sample1dspectra.eps}
\end{figure*}

Typical 1-d spectra for eight galaxies are shown in Figure~\ref{sample1dspectra.eps}, arranged by $R$-band magnitude and color.  The spectra have been smoothed with a 15-pixel boxcar.  The selected galaxies lie near $z = 0.8$ and are sorted into two color bins, four from the blue cloud and four from the red sequence.  Notable features in the 4000 \AA\ region are visible, including [O {\scriptsize II}] $\lambda$3727, the 4000-\AA\ break, Ca H and K, two CN bands near 3800 \AA\ and 4100 \AA, and the higher-order Balmer lines.  The lowest four panels illustrate galaxies near the survey magnitude limit ($R_{AB} = 24.1$).  Emission-line redshifts are still quite easy at this level for galaxies in the blue cloud (right side), but continuum $S/N$ is becoming marginal for red-sequence galaxies, as noted in \S\ref{redshiftmeasurements}.  Velocity broadening is detectable in most spectra; 66\% of all galaxies with redshifts have emission linewidths at least 3$\sigma$ larger than the instrumental signature (while 79\% of these galaxies have at least one line detectable at $>3\sigma$).

\begin{figure}
\ifpdffig
\begin{center}
\vspace*{0.1in}
\includepdf[scale=0.9]{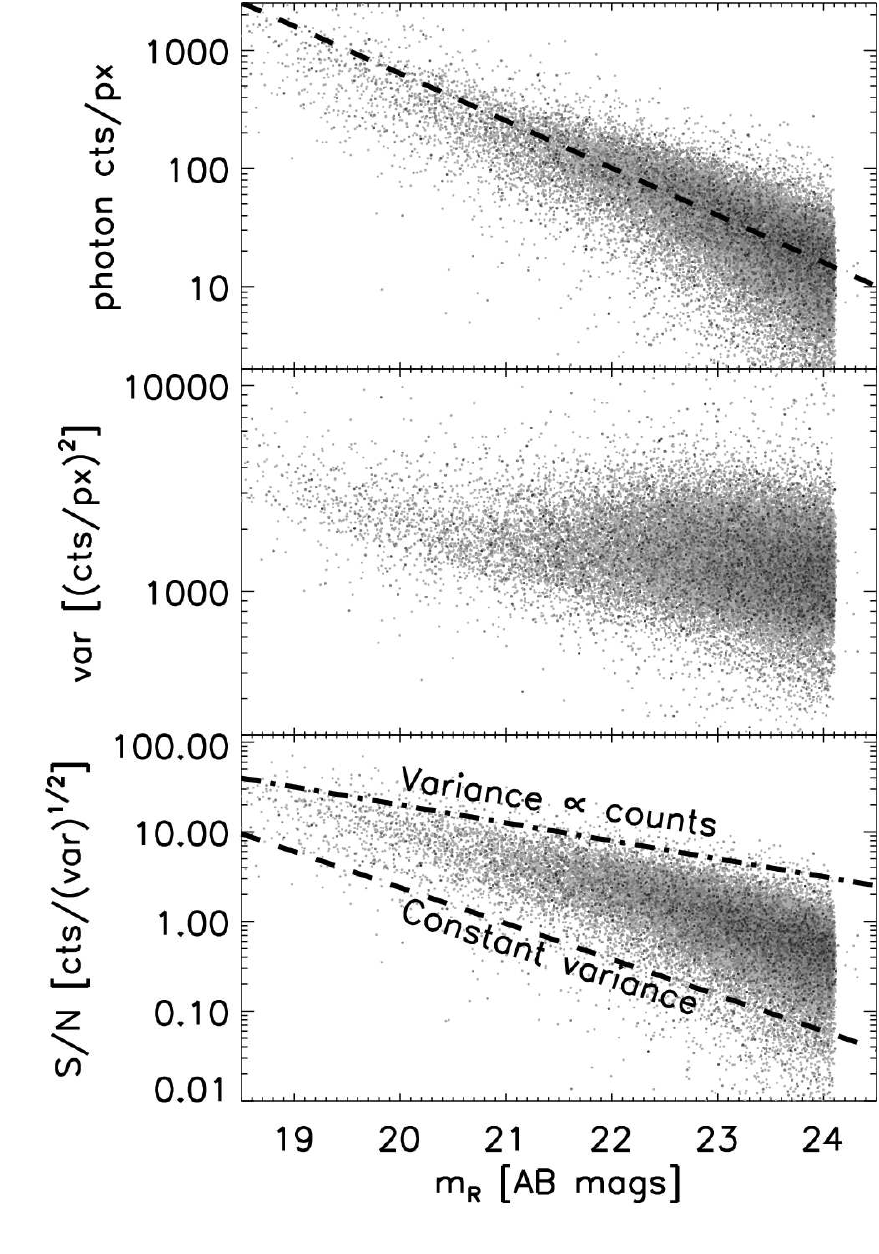}
\vspace*{-0.2in}
\end{center}
\else
\begin{center}
\vspace*{0.1in}
\includegraphics[scale=0.9]{harker_SN.eps}
\end{center}
\fi
\caption{ Mean counts, variance, and $S/N$ of the hour-exposure 1-d spectra as a function of $R$-band magnitude.  This figure can be used to estimate the $S/N$ for a single spectrum or from stacked spectra as a function of magnitude.  Panel (a) shows the median continuum number of photons per pixel for a range of galaxy brightnesses near $z \sim 0.8$.  The median value is 14 photons pix$^{-1}$ hr$^{-1}$ near $R_{\rm AB} = 24.1$ and scales slightly more slowly than a linear relation (shown by the dashed line) because brighter galaxies tend to overfill the slit.  Panel (b) shows the median variance per pixel for the same spectra, predicted from the noise model tested in Figures \ref{hist_offsky.epsi} and \ref{hist_onsky.epsi}.  The variance is nearly flat as a function of magnitude, as it is generally dominated by sky noise.  Panel (c) plots the median continuum $S/N$ per pixel, which is the quantity plotted in panel (a) divided by the square root of the corresponding object's value in panel (b).  The upper line shows the trend expected if noise is dominated by photon statistics from an object, while the lower line shows the expected trend if noise is dominated by sky.  The actual data are bounded by these extremes.  }
\label{harker.S-to-N.eps}
\end{figure}

Figure~\ref{harker.S-to-N.eps} summarizes the mean counts and $S/N$ of the combined one-hour 1-d spectra as a function of $R$-band magnitude.  Panel (a) shows the median continuum number of photons per px for a range of galaxy brightnesses near $z \sim 0.8$.  The median value is 14 photons px$^{-1}$ near $R_{\rm AB} = 24.1$ and scales slightly more slowly than the linear relation (shown by the dashed line) because brighter galaxies tend to overfill the slit.  Much of the scatter is due to size variations -- larger galaxies at fixed magnitude have lower counts per pixel.  Panel (b) shows the median inverse variance per px for the same spectra, which is nearly flat with brightness, being dominated by sky noise (see below).  Panel (c) plots the median continuum $S/N$ per px, which is panel (a) divided by the square root of panel (b).  The upper, dot-dashed line shows the shape of the trend expected if noise is dominated by photon statistics from the object, while the lower dashed line shows the shape expected if noise is dominated by sky.  The actual data are bounded by these extremes.

This figure is useful for predicting the $S/N$ available from single and stacked DEEP2 spectra.  For example, it says that summing together roughly 50 spectra at $R \sim 22.0$ will yield $S/N =25$ per \AA, which is the recommended minimum for absorption-line velocity dispersion and stellar-population work (e.g., Schiavon et al.~2007).
  
More information about sky counts is given in Table~\ref{table.instrumentparams}.  Comparing the sky continuum level between the OH lines with the read noise tells whether an individual 20-minute exposure is sky-limited or read-noise limited.  The typical continuum sky count in the range 7700-8300 \AA\ is 13 photons per 2-d px in 20 minutes while the variance of the read noise is only $2.55^2 = 6.5 $e$^-$ per px.  Thus, read noise is less than sky noise but is not negligible.  This suggests that some gain in continuum $S/N$ could be obtained by using a 600-line grating instead of the 1200-line grating.  Having explored several tens of thousands of galaxies at very high resolution for DEEP2, we have indeed chosen DEIMOS' 600-line grating for the DEEP3 survey in EGS, which will capture galaxies over a wider redshift range, sample more spectral features, and yield somewhat higher continuum $S/N$, with the tradeoff of greater difficulty in measuring linewidths and in identifying [O {\scriptsize II}] amidst sky lines.
 
\begin{figure*}
\ifpdffig
\begin{center}
\includepdf[scale=0.85]{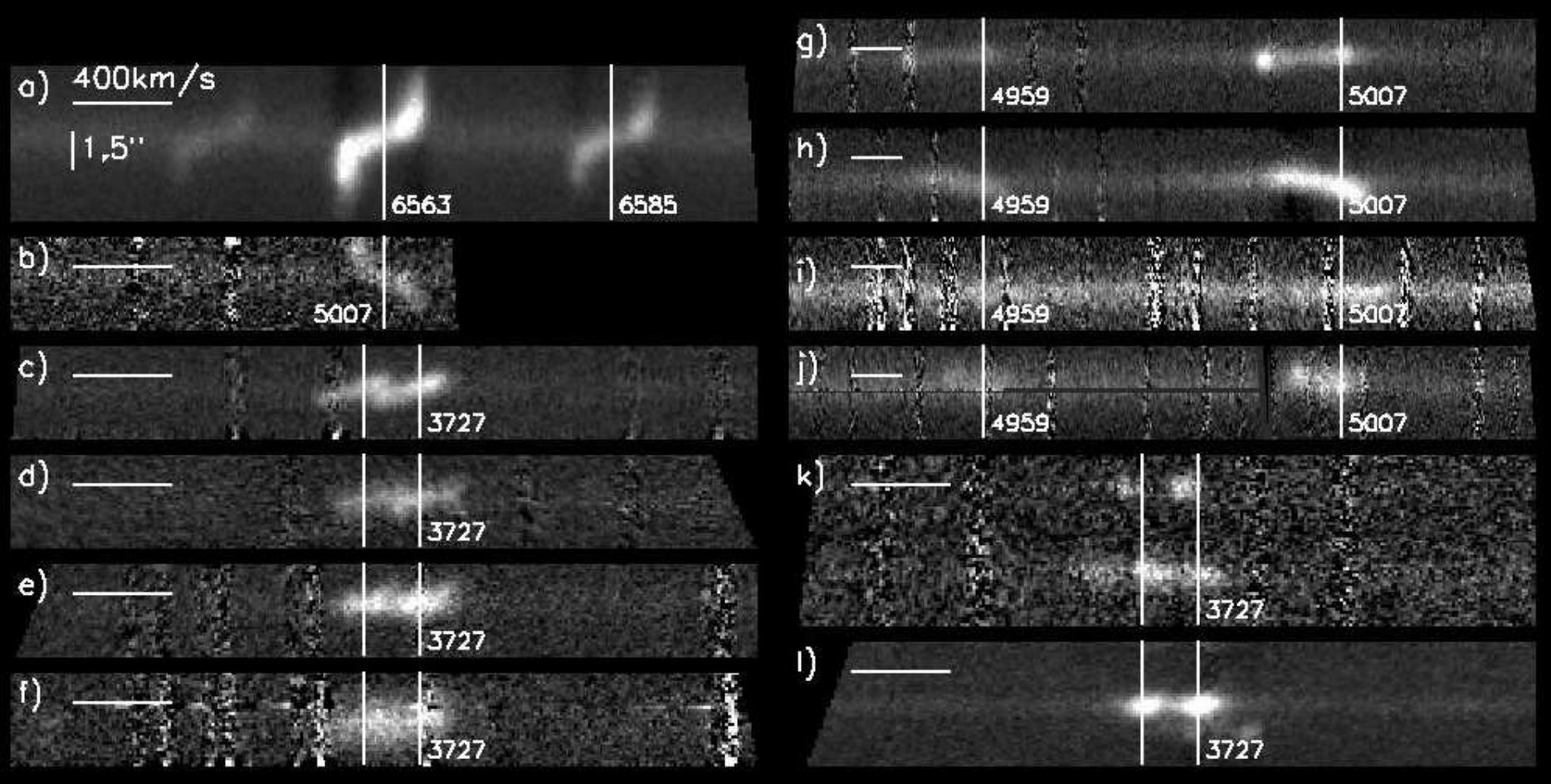}
\end{center}
\else
\begin{center}
\includegraphics[scale=0.85]{interesting_2d_v5.eps}
\end{center}
\fi

\caption{A montage of interesting or unusual 2-d spectra.  The horizontal line corresponds to a restframe velocity of 400 \kms\ in all cases.  Spectra (a)-(f) show rotation curves and/or broadened spectral lines, illustrating the kind of information on internal kinematic motions afforded by our $R \sim 6000$ spectral resolution.  Spectra (g) and (h) show galaxies that exhibit two well separated sets of AGN emission lines, possibly indicative of merging black holes.  Spectra (i) and (j) are examples of significant velocity offsets ($\sim$170 \kms) between the emission velocity and the stellar absorption velocity (marked by the vertical white line); these may also be a result of in-spiraling black holes.  Spectra (k) and (l) show nearby offset companions ({\tt offser}'s), which can be used to measure the speeds of satellite motions as in Conroy et al.~(2005).  The DEEP2 object numbers for these spectra are: (a) 11039144, (b) 11045441, (c), 13033966, (d) 13058074, (e) 12011735, (f) 12004795, (g) 13025437, (h) 11026433, (i) 11028015, (j) 13051207, (k) 11051416, (l) 12025255.  }
\label{interesting_spectra}
\end{figure*}

Figure~\ref{interesting_spectra} illustrates various types of information that are available from the 2-d spectra.  Spectra (a)-(f) illustrate a range of galaxy internal motions, from highly ordered and resolved rotation curves, as in spectrum (a), to unresolved but broadened lines, as in spectrum (f).  The quantity $S_{0.5} \equiv \sqrt{0.5V_{rot}^2 + \sigma^2}$, where $V_{rot}$ is rotation speed and $\sigma$ is linewidth, combines both systematic and random motions into a single line-width broadening parameter (Weiner et al.~2006ab) and proves to be a remarkably useful and robust internal velocity indicator that places the internal speeds of both highly regular and highly disturbed galaxies on the same scale (Kassin et al.~2008).

Spectra (g)-(j) show four candidate dual AGNs (Gerke et al.~2007b, Comerford et al.~2009).  The first two exhibit spatially separated AGNs, while the third shows a large velocity offset between [O {\scriptsize III}] and the stellar absorption lines, as would be expected for an in-spiralling black hole following a galaxy merger.  Several dozen such offset cases have now been found (Comerford et al.~2009).  Panels (j)-(l) show examples of close pairs that are suitable for measuring satellite galaxy motions.  Also measurable from the resolved spectra but not shown are equivalent widths and line-ratio gradients.

\section{Redshift Measurements and Completeness}
\label{redshiftmeasurements}

\subsection{The {\it spec1d} Redshift Pipeline}
\label{redshiftpipeline}

The output of the {\it spec2d} pipeline is a set of extracted  
spectra, one for each object.  Obtaining redshifts and object  
classifications from these spectra is the task of a separate software  
pipeline, {\it spec1d}.  This IDL code base inherits much from the  
SDSS {\it specBS} pipeline produced by David Schlegel, Scott Burles,  
and Douglas Finkbeiner, with modifications and new routines by  
Michael Cooper, Marc Davis, Darren Madgwick, Renbin Yan, and Jeffrey  
Newman.

The algorithms employed are relatively straightforward; a single
routine, REDUCE1D, guides every step of the process.  First, the 1-d
extracted spectra are read in and corrected for telluric absorption
(using templates for the A and B band absorption provided by
C. Steidel scaled as airmass to the 0.55 power), for variation in the
overall response of the DEIMOS spectrograph with wavelength, and for
the difference between air and vacuum wavelengths (all spectra are
provided on an air wavelength scale using the IAU standard method of
Morton (1991). The spectrum is then interpolated onto a uniform grid
in ${\rm log}_{10}(\lambda)$ with spacing $2\times10^{-5}$ for
efficiency; in log(wavelength) space, redshifting corresponds to a
shift in pixel numbers rather than a change in wavelength scale.
Finally, a version of the spectrum that has been median-smoothed using
a 2500-pixel window is subtracted from it; this renders the redshift
fit insensitive to any large-scale, coarse features in the spectrum
(e.g., due to instrumental effects causing variation in throughput or
continuum level with wavelength).

Next, the spectrum is separately fit to find the best galaxy, QSO, and
stellar template matches and redshifts.  The figure of merit used is
$\chi^2(z)=\Sigma (f_{object,i}-f_{template fit,i,z})^2/ \sigma_i^2)$,
where $\Sigma$ denotes summation over all pixels of the spectrum,
$f_{object,i}$ is the flux from the object in pixel $i$, $f_ {template
  fit,i}$ is the flux predicted for the best fitting template or
linear combination of templates at redshift $z$ evaluated at pixel
$i$, and $\sigma_i$ is the predicted uncertainty in the flux
measurement for pixel $i$ (i.e., the inverse variance at that pixel to
the $-1/2$ power).  In the limit where $\sigma_i$ is a constant, this
is equivalent to standard cross-correlation techniques (Tonry \& Davis
1979),\footnote{Davis and Tonry were able to rapidly do the
  cross-correlation by shifting to Fourier space, but their spectra
  had an assumed noise that was homoskedastic, and the strong sky line
  line at 6300 \AA\ was blocked out.  In our case $\sigma_i$ is far
  from constant and we must do the cross-correlation in wavelength
  space.}  but de-weights pixels with poor flux measurements (e.g.,
due to night sky lines).

For galaxy fits, the best-fitting linear combination of three
templates is used for each trial redshift: an early-type galaxy
spectrum (based on the composite luminous red galaxy spectrum of
Eisenstein et al.~ (2003); a model Vega spectrum from the Kurucz
library\footnote{\url{http://kurucz.harvard.edu/stars/vega/veg1000pr25.500000}}
(cf.\ Yoon et al.~2010 and references therein) convolved to a velocity
dispersion of 84.5 \kms\ (corresponding to intrinsic velocity
dispersion and instrumental broadening of 60 \kms\ each); and an
emission-line-only template with line strengths matched to the coadded
spectrum of all blue galaxies and with line FWHMs matching the
predicted width for a DEIMOS observation of an emission line having
rest-frame velocity dispersion = 60 \kms\ but observed at the assumed
trial redshift.  All template spectra have the continuum subtracted
off using the same window size and pixelization as for the data.  Then
we construct the function $\chi^2(z)$ with spacing $2\times 10^{-5}$
in ${\rm log}_{10}(1+z)$, spanning all redshifts from -0.0001 to the
largest redshift at which [O {\scriptsize II}] $\lambda$3727 will fall
within the spectrum ($z=1.47$ for a spectrum extending to 9200 \AA).

The five deepest minima in $\chi^2$ are identified, and  each one is fit  
with a quadratic over the seven pixels centered at the minimum. This is  
used both to obtain an improved estimate of the redshift of each  
minimum and to determine its corresponding redshift error (as the  
distance away from the minimum of the fit curve where $\chi^2$  
changes by 1).  For each of these minima, an emission-line velocity  
dispersion is measured by measuring $\chi^2$ between the  
spectrum and 40 emission-line templates constructed with velocity  
dispersions $\sigma$ evenly spaced from 0 to 360 km s$^{-1}$ (but  
otherwise utilizing the same as the emission-line template used for redshift  
determination).  We then determine the best-fit velocity dispersion  
for a galaxy as the minimum of a spline fit to $\chi^2(\sigma)$  
tabulated in 9 km s$^{-1}$ bins, and the uncertainty in that  
dispersion as half the size of the range in velocity with $\chi^2$ no  
more than 1 larger than the minimum.

For QSO fits, we use a single composite spectrum constructed from the
QSO eigenspectra used by {\it specBS}, and search for the two deepest
$\chi^2$ minima over the redshift range $0.0033 < z < 5$.  For stars,
we choose the three best (i.e. minimum-$\chi^2$) fits with $-0.004 < z
< 0.004$ using any of the SDSS template spectra of stars of type O,
A0, F2, G0, K1, M0V, M1, and L0, as well as the SDSS carbon star
template. Otherwise, the routines proceed in the same way as for
galaxies, but the velocity dispersion is left undetermined.

At the end of this process, for each DEEP2 spectrum we have 10
possible estimates of the redshift plus the best-fitting
spectral-template match for each one (for galaxies, we specify the
linear combination of old stellar population, young stellar
population, and emission lines; for QSOs we only have one template;
while for stars we specify the spectral type of the best match).  The
results for all objects on a mask are compiled into a single "{\it
  zresult}" file, stored in FITS BINTABLE format.  We have learned
from experience that it is impossible to choose amongst them
automatically, unlike in high-signal-to-noise measurements using data
with accurate flux calibrations (e.g., SDSS; cf.\ Abazajian et
al.~2003).  In DEEP2 data the lowest value of $\chi^2$ often, but far
from always, corresponds to the correct redshift. Human intervention
is required to choose amongst these candidate redshifts, or (rarely)
to find redshifts that the automated pipeline missed.

\subsection{Visual Redshift Inspection Process: the {\bf \it zspec} Tool}
\label{visualinspection}

To allow us to check and fit for redshifts interactively, we have
developed an IDL widget-based program called {\it zspec}.  This
program allows the user to examine interactively each of the eight
candidate redshifts proposed by the {\it spec1d} pipeline, or to
provide an alternative redshift solution.  Using this program, all
redshifts from {\it spec1d} have been checked by eye, a process
typically requiring 1-2 minutes per spectrum.

\begin{figure*}
\vspace{190mm}
\ifpdffig 
\vspace*{-8.0in}
\begin{center}
\includepdf[scale=0.3]{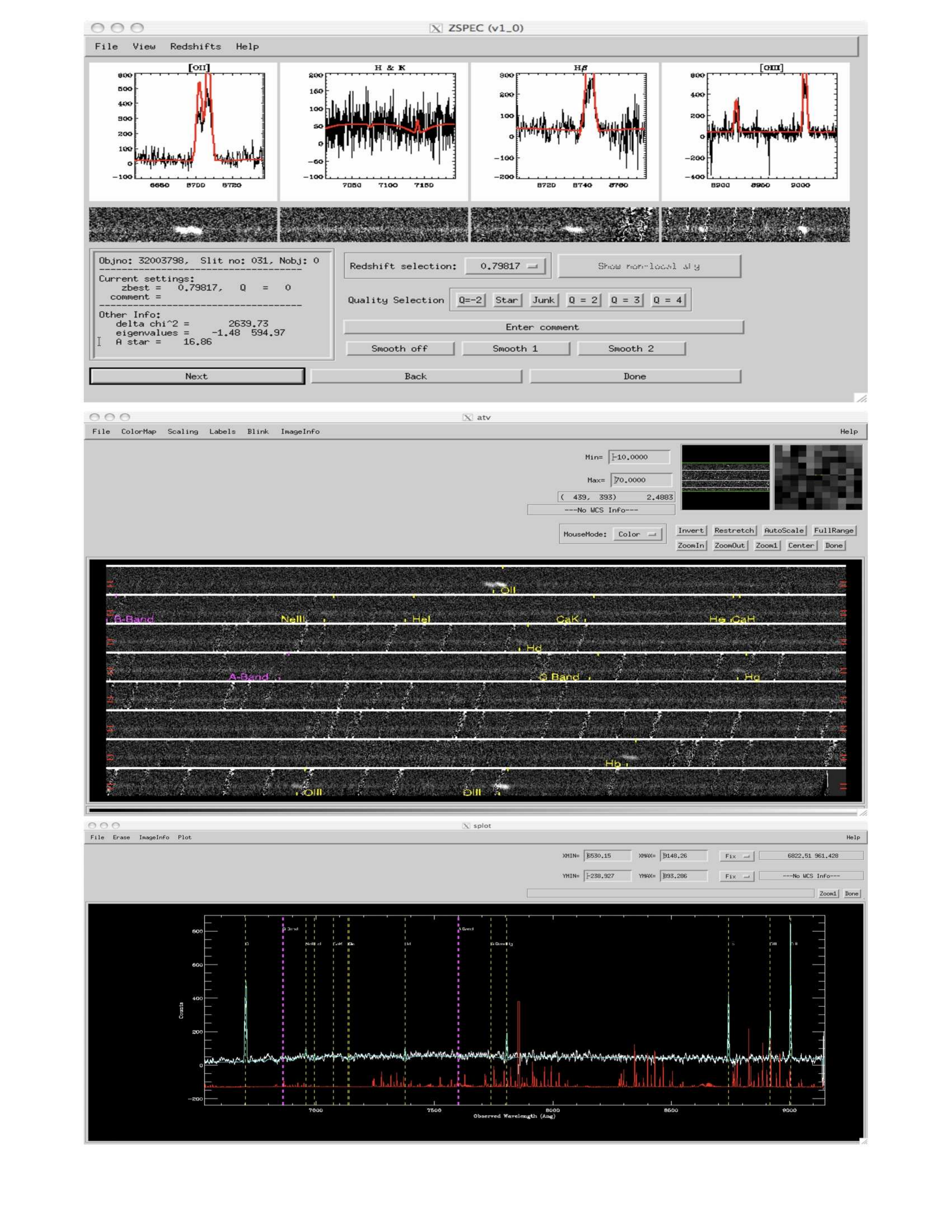}
\vspace*{-0.6in}
\end{center}
\else
\vspace*{-8.0in}
\begin{center}
\includegraphics[scale=0.3]{zspec.eps}
\end{center}
\fi
\caption{\footnotesize Illustration of the {\it zspec} widget being
  used to determine the redshift for a typical star-forming galaxy at
  $z = 0.798$.  Top: the main control panel, showing the four most
  likely spectral features for the chosen redshift in both 1-d and
  2-d.  Buttons allow the user to choose a redshift and quality, enter
  comment codes, control the smoothing of the 1-d spectrum, and force
  fit a redshift if necessary.  Middle: the 2-d spectrum for this
  object (displayed using {\it ATV}, Barth 2001), with tick marks
  at features for the chosen redshift.  An option exists to rectify
  the slit, removing the apparent tilt.  Red tick marks at either side
  indicate the extraction window that was used to produce the 1-d
  spectrum.  Bottom: the extracted 1-d spectrum (here smoothed by
  taking the inverse-variance-weighted mean over a rolling 15 pixel
  window), displayed using {\it SPLOT}, with candidate features marked.  The aqua spectrum shows
  the best-fitting linear combination of templates, while the red
  curve shows the variance in the spectrum (rescaled for convenience).
}
\label{zspec}
\end{figure*}

The three computer display screens for the {\it zspec} IDL widget are
shown in Figure~\ref{zspec}.  When a candidate redshift is selected,
the {\it zspec} control panel (top) displays both 1-dimensional and
2-dimensional spectra of small regions around any of the six most
common absorption or emission features seen in DEEP2 spectra. The middle panel shows the full 2-d spectrum, which is displayed using the {\it ATV} tool developed by Aaron Barth (Barth 2001; middle panel).  The final window, shown at bottom, displays the full extracted 1-d spectrum using the {\it SPLOT} tool developed by David Schlegel; the
user can smooth the 1-d spectrum by varying amounts if so desired. The predicted locations of more
than 20 common absorption and emission features are marked in each of
these windows. The {\it spec1d} pipeline template that is fitted for a
given choice of redshift is plotted both in the full view and
subwindows, and the variance is also plotted in the 1-d spectrum view
to aid the user in identifying false features associated with
sky-subtraction residuals or Poisson noise.  Optionally, a smoothed
version of the 1-d spectrum may be plotted instead of the raw
measurements, and plot ranges may be adjusted arbitrarily to help the
user check redshifts.

In many cases, obvious, resolved multiple emission (e.g., the [O
{\scriptsize II}] doublet or H$\beta$ and [O {\scriptsize III}] )
and/or absorption (e.g., Ca H \& K) features will appear in the upper
inspection windows at the proper positions (compared to the template)
for some choice of redshift, providing immediate confirmation.  In
other cases, the full spectrum in either the 2-d or 1-d windows must
be examined in order to find rarer emission features or to confirm a
redshift using the ensemble of absorption lines in the spectrum of an
early-type or post-starburst galaxy.

Confirmation of multiple features (where a resolved [O {\scriptsize
  II}] doublet counts as two features) is required to assign a
redshift to a given object with a reliable redshift quality code
($Q=3$ or $Q=4$; see \S\ref{redshiftresults}).  The one-dimensional
extracted spectrum and the two-dimensional spectrum are both checked
to ensure that the features are real and not a consequence of perverse
Poisson noise under a sky line or due to much of the spectrum being
obscured by a bad column, for instance. Other redshift quality codes
which the user may assign indicate that an object's spectrum is
completely masked by instrumental artifacts so that the object was
effectively never observed ($Q=-2$), is a star ($Q=-1$), yields no
useful redshift information ($Q=1$), or may potentially yield redshift
information but needs more analysis or re-reduction ($Q=2$); many, but
not all, $Q=2$ cases may be resolved with more inspection.
Section~\ref{datatables} below for a fuller description of the DEEP2
redshift quality codes.

During the $zspec$ process, the user may also fill in one or more
comment codes that signify additional information about the data; the
standard codes are also described in \S\ref{datatables}.  These codes
are useful, but their application has varied among various $zspec$
users, and they are the least homogeneous aspect of the DEEP2
database.  Additional description can be found in \S\ref{datatables}.

In a small minority of cases, none of the 10 $\chi^2$-minimum
redshifts from the {\it spec1d} pipeline are correct for a given
galaxy, but the true redshift is still measurable with human
intervention; these constitute less than 1\% of the reliable redshifts
in the DEEP2 sample.  In those cases, the person inspecting the
redshifts can use the {\it zspec} widget to fit for a redshift using
the same algorithms used by {\it spec1d} (described in
\S\ref{redshiftpipeline}) but searching only a limited range in
redshift around a user-specified value and optionally using only a
limited set of templates and/or spectral features.

The result of the {\it zspec} process for a given mask is a FITS BINTABLE
{\it "zspec"} file, recording the properties of the best redshift choice identified
by the user (template fit, velocity dispersion, etc.), the assigned redshift
quality code $Q$, and the user's comment codes, if any.  These are 
compiled into a catalog with one entry for each spectrum that has been
checked.

The results of the {\it zspec} process are combiled into a single
redshift catalog, labelled the "{\it zcat}".  To make this catalog,
for each mask, we use the {\it zspec} results from the most
experienced {\it zspec} user who has examined it.  The redshifts
($z$), redshift quality codes (ZQUALITY, or $Q$ for short), and
comments provided by this user, together with all derived parameters
determined by the {\it spec1d} pipeline (e.g., the coefficients for
each template, the measured linewidth, etc.) and selected photometric
information from the {\it pcat} files are combined in this
catalog.\footnote{In a few dozen cases, object numbering changed in
  the course of the survey in regions of the Extended Groth Strip
  where multiple fields overlap.  In the {\it zcat} file, the
  integer-format object number, OBJNO, indicates the object number for
  the current best photometry of a given source, whereas the string
  object name, OBJNAME, indicates the object number used in mask
  design, which is used to construct the filename of the corresponding
  one-dimensional spectrum FITS file (i.e. the object's {\it spec1d}
  file).}  There is one entry in the catalog for each observation of
an object (e.g., if an object was observed on two different slitmasks,
it is listed in the catalog twice); for convenience, we also
distribute a version of the {\it zcat} in which only a single redshift
estimate per object is included\footnote{A few objects which were
  observed on DEEP2 masks (e.g. supernova hosts) are not included in
  the {\it zcat}, as they were not selected according to DEEP2 target
  selection criteria.}.

The redshifts stored in a {\it zspec} file are first compared to those
provided in the corresponding mask's {\it zresult} file (i.e., the
original output from the {\it spec1d} redshift pipeline), which
contains information on 10 candidate redshifts,selected based on
reduced $\chi^2$, for each object. For all objects to which the {\it
  zspec} user assigned redshift quality code $Q=3$ or 4 (i.e.,
obtained a secure redshift measurement), we compare the redshift
identified by the zspecer to the $\chi^2$ minima compiled in the {\it
  zresult} file. If there is a $\chi^2$ minimum within $\Delta_z =
0.01$ of the user's selected $z$, then we use the redshift stored in
the {\it zresult} file, rather than the {\it zspec} result. This is
done to accommodate objects which went through the {\it zspec}
inspection process before the {\it spec1d} and {\it spec2d} code was
finalized; in such cases, redshifts may change slightly between
software versions, and we wish to use the best possible estimates of
object attributes.

For those objects which were assigned $Q=3$ or 4 but for which
$\Delta_z > 0.01$ for all minima in the {\it zresult} file, we attempt
to re-fit for the redshift, again using the {\it spec1d}
redshift-fitting code, but restricting the redshift range considered
to a limited window (half-width = 0.03 in $z$) centered about the
redshift chosen in the {\it zspec} process. If the re-fit redshift
agrees within $\Delta_z = 0.01$ of the {\it zspec} result, then the
properties corresponding to that $\chi^2$ minimum (its redshift,
template coefficients, velocity dispersion, etc.) are assigned to that
object in the {\it zcat}. However, if there is no $\chi^2$ minimum
within $\Delta_z=0.01$, the object is assigned redshift quality $Q=2$,
as we are unable to provide standard pipeline quantities (such as
velocity dispersion) in the same way as for a redshift with
well-defined minimum.  There are 138 objects assigned $Q=2$ for that
reason.

Next, we read in a set of ASCII-format tables that provide a list of
all redshift corrections that have been compiled outside of the usual
{\it zspec} process (q.v.\ below); in these cases, we override the
{\it zspec} results.  In cases where the the object is reassigned a
redshift with a quality code of $Q=-2$, $-1$, 1, or 2, we simply
update the redshift, redshift quality, and comment according to the
override information. For those assigned a secure redshift ($Q=3$ or
4), we again utilize the {\it spec1d} redshift-fitting code and search
for a $\chi^2$ minimum in a $\Delta_z = \pm 0.03$ redshift window
centered about the corrected redshift given in the override file. If
this fit fails (again a $\Delta_z = 0.01$ criteria is applied), which
occurs for 105 objects, we update the redshift and comment, but we set
the redshift quality to $Q=2$. When the fit succeeds, we update the
{\it zcat} with the full set of corrected parameters from {\it
  spec1d}, as well as the override-provided redshift quality and
comment.

The measurement of DEEP2 redshifts for difficult cases is still a work
in progress; in particular, objects with $Q = 2$ may still be possible
to extract redshifts from.  A good example are the superimposed
serendips ({\it supser}'s; 377 objects); i.e., cases where the spectra
of two separate galaxies are superimposed in the same spectrum (cf.\
\S\ref{datatables}).  These have not yet been disentangled (and
indeed, it is ambiguous which redshift should be assigned in such
cases).  Another class are the spectra that exhibit only one reliable
feature ({\it sngls}'s; 1161 objects).  Many such cases can be
resolved using $BRI$ photometry alone (Kirby et al.~2007), and many
more will eventually yield to high-accuracy photo$z$'s that are being
prepared using multi-band data from $u$ to 8 $\mu$m (Huang et al., in
prep).  These revised redshifts are not included in DR4.

\subsubsection{Redshift reinspections}

A few of the objects for which redshift overrides are provided
resulted from scattered anomalies discovered by the DEEP2 team in the
course of performing survey science.  A greater number resulted from
two focused efforts: an investigation of all objects in the Extended
Groth Strip with multiple {\it zspec} inspections or secure redshift
codes, conducted before DEEP2 Data Release 3; and a reinspection of
all $Q=2$ objects by experienced zspecers in order to recover
additional redshifts.

In preparation for the previous major data release (DR3), we undertook
a detailed review of the data quality in Field 1, in order to provide
the community with the most dependable redshift catalog in the EGS
region, as well as to ascertain the rates of error in our full
redshift catalog. This quality review was carried out in two stages:
1) A comparison and resolution of differences between {\it zspec}
efforts from different reviewers;
and 2) a review of all $Q\gteq 3$ redshifts in EGS.\\

By the summer of 2007 (the time of the review), most masks observed in
Field 1 had gone through multiple {\it zspec} checks. The zspec
results from all reviewers were combined together into a database with
14,509 independent entries, where an entry corresponded to a single
observation of a single target; i.e, multiple observations (on
different slitmasks or at very different times) of the same target
were treated as separate entries in the database. Out of all entries
in the database, 10,763 entries (75\%) had gone through the {\it
  zspec} process more than once.  Several criteria were applied to
flag cases of possibly erroneous redshifts or redshift quality among
the multiply-checked targets in the Extended Groth Strip.
The selected objects (1102 in total) were:\\

A) Targets with redshift quality $Q \gteq 3$ in all reviews, but with
differences in assigned redshift greater than 0.001. There were 63
such cases, most of which had complex spectra that were misidentified
by novice reviewers;\\ 

B) Targets assigned $Q \gteq 3$ by one or more reviewers and $Q<3$ by
at least one reviewer, the bulk of the sample; or\\ 

C) Targets classified as a 'supser' (cf.\ \S\ref{datatables}) by any
reviewer. In certain cases, rare and complex spectra, e.g. from AGNs,
were assigned this quality class in error.  There were roughly 100
such objects.

Using {\it zspec}, the flagged targets were subsequently reviewed
again carefully by expert team members and a final judgement was made
on their best-fitting redshift and redshift quality. From a total of
1102 flagged targets (10.2\% of multiply-checked spectra), 544 were
verified to have good redshifts ($Q\gteq 3$ or
spectroscopically-confirmed stars), 523 were assigned $Q=2$ and 30
were assigned $Q=1$. The objects assigned $Q=2$ were primarily either
supser's (which are assigned $Q=2$ due to their ambiguous redshifts)
or objects to which {\it zspec} users with limited experience had
assigned inappropriately high redshift confidence.

As a result of this process, we developed a procedure to visually
examine the spectra of all $Q\gteq 3$ objects in Field 1 in a rapid
manner, to ensure that egregious redshift errors would be caught.
Using custom-purpose software to display multiple spectra
simultaneously on a common rest-frame wavelength scale, we examined
approximately 10,000 spectra, 2,600 of which had only had a single
zspec review at the time. This was done in two passes. In the first
pass, sections of the two-dimensional spectra were examined in regions
around major emission lines to verify emission line redshifts and
catch mismatches. Objects which lacked strong emission lines were then
further reexamined in a second pass using their one-dimensional
spectra. At the end of both passes, objects for which the redshift
could not be verified by this quick-inspection procedure, 739 in
total, were subjected to another round of full {\it zspec} review by a
set of expert team members. This guaranteed that every
putatively-secure redshift in Field 1 was reviewed at least twice by a
human eye.

As a result of this process, $\sim 250$ objects (out of the 10,000
inspected) were downgraded to $Q=2$, and 8 were downgraded to $Q=1$.
As a check on the quality of the redshift catalogs we distribute, we
compared the results of our review with the redshift catalog from
which DEEP2 Data Release 2 was constructed; it contained the results
from 14175 spectra in the Extended Groth Strip. We found that out of
those objects, 31 galaxies had incorrect redshifts (0.3\% of all
$Q\gteq 3$ redshifts), while 249 new, secure redshifts had been
obtained for objects that previously had $Q< 3$(largely due to
reobservations of masks with problematic signal-to-noise), while there
were 493 galaxies with $Q \gteq 3$ that we downgraded to $Q<3$ during
reinspection (i.e., in 3\% of cases, the original {\it zspec} user was
judged to be overoptimistic on whether the assigned redshift was
secure). The results of these redshift reinspections were included in
the DEEP2 Data Release 3 (DR3) catalog, yielding a uniform, internally
consistent, high quality catalog for objects in the Extended Groth
Strip.

As a result of this inspection process, we have now required that all
DEEP2 masks were inspected by at least one expert {\it zspec} user,
and use their results in the current (Data Release 4) redshift
catalog.  Additionally, we have reinspected all objects assigned
redshift quality $Q=2$, searching for reliable redshifts that might
have been missed in more cursory examinations.  This was done by first
having a single expert user (Marc Davis) examine all spectra assigned
$Q=2$ and search for an improved redshift estimate.  Then, a second
expert user examined each spectrum assigned a new redshift and judged
whether they concurred with the result.  In cases where both experts
agreed that a redshift had been recovered, a new redshift, quality
code, etc. were assigned and the results incorporated into the {\it
  zcat}.  As a result of this process, more than 1000 $Q>3$ redshifts
were recovered out of more than 5000 spectra inspected.

\subsection{Redshift Results}
\label{redshiftresults}

\begin{figure}
\vspace{76mm}
\ifpdffig
\vspace*{-3.0in}
\begin{center}
\includepdf[scale=0.38]{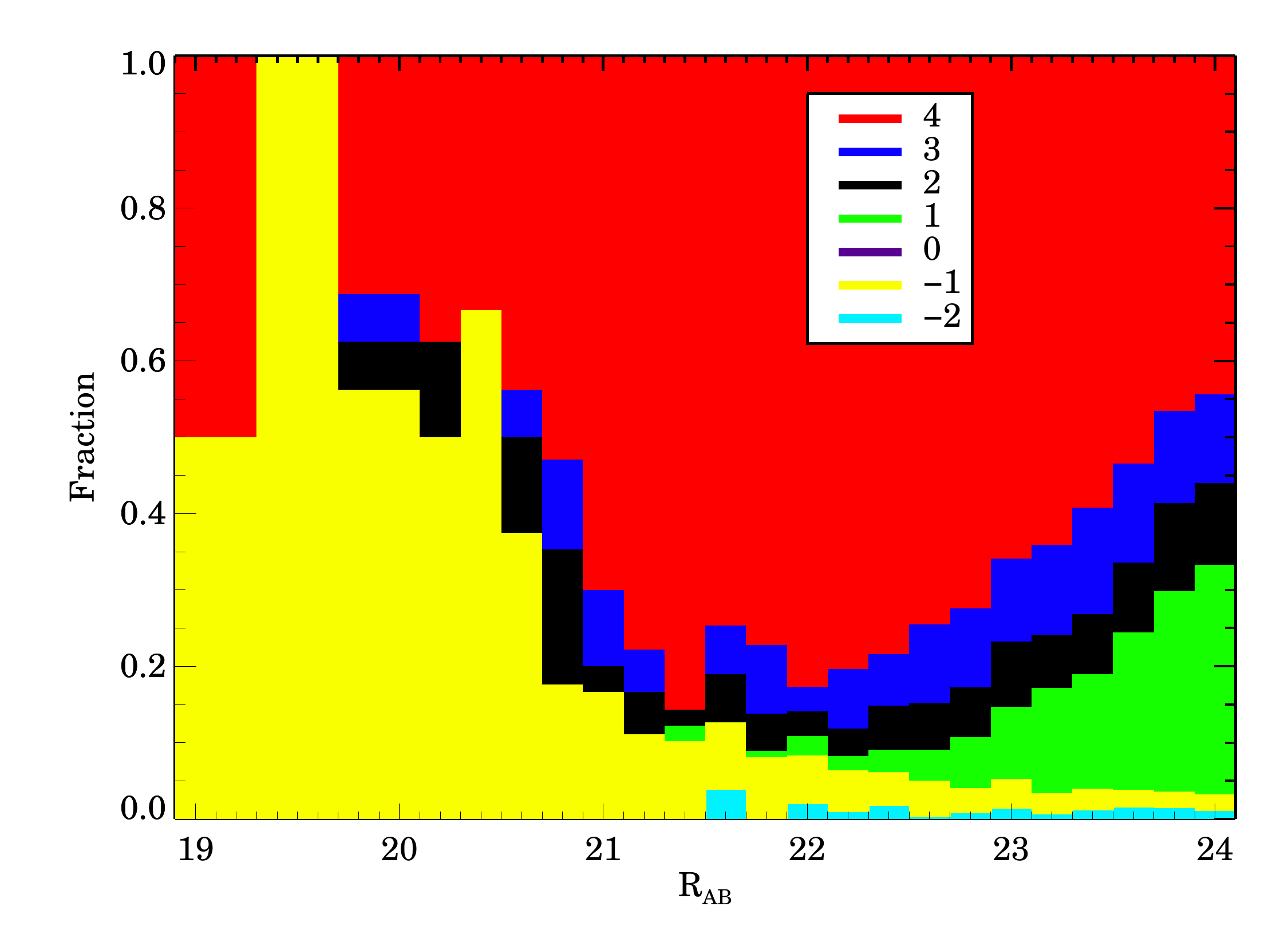}
\vspace*{-0.2in}
\end{center}
\else
\vspace*{-3.0in}
\begin{center}
\includegraphics[scale=0.38]{zqual_mag2.eps}
\end{center}
\fi
\caption{Histogram of fractions of DEEP2 spectra in Field 1 yielding
  redshifts with various quality codes in bins of apparent
  magnitude. Aside from the larger number of bright stars in EGS due
  to differences in sample design, this figure is similar to that for
  Fields 2, 3, and 4.}
\label{zqual_mag.egs.eps}
\end{figure}

Redshift results are summarized in Table~\ref{table.qualitycodes},
which gives the number of galaxies assigned each quality code in a set
of bins of apparent magnitude.  These are totals over all spectra in
the {\it zcat} redshift catalog; hence, a galaxy observed on multiple
slitmasks will be counted multiple times, once per observation.  A
total of 52,989 spectra were obtained.  1.0\% are so severely
compromised by instrumental issues that the object was effectively
never observed ($Q = -2$); 1.8\% are stars ($Q = -1$); 16.8\% are of
such poor quality/signal-to-noise that they will likely never yield
useful redshifts ($Q = 1$; many of these are faint blue galaxies
beyond our redshift range); 9.8\% contain information that could be
used to determine a redshift but no definitive determination was made
(or no single redshift could be assigned); 11.7\% yield a redshift
estimated by the {\it zspec} user to be reliable at the 95\% level ($Q
= 3$); and 58.8\% have redshifts estimated to be reliable at the
99.5\% level ($Q = 4$).  These reliability estimates appear to be
accurate (see below).  This information is also shown in
Figure~\ref{zqual_mag.eps} and Figure~\ref{zqual_mag.egs.eps}, which
show the fraction of spectra with different quality codes vs.~$R$
magnitude in Fields 2, 3, 4 and in Field 1 (EGS).  The trends are as
expected: in Fields 2, 3, 4, while the fraction with $Q = 4$ falls off
near the magnitude limit, the fraction with ($Q = 1$ rises greatly
(owing mainly to the onset of the faint blue galaxies), while the
fraction with $Q = 3$ is low among the brightest objects but remains
fairly constant fainter than $R_{\rm AB} = 21$.  The fraction of
potentially recoverable redshifts ($Q = 2$) is rather constant at
about 15 at all magnitudes, while the number of stars ($Q = -1$) and
catastrophic instrumental failures ($Q = -2$) is generally very small.
The behavior in Field 1 is similar except for the much larger fraction
of stars at the brightest magnitudes (where there are few total
targets), which is likely due to differences in target selection
strategy in that field.

\begin{figure}
\ifpdffig
\begin{center}
\includepdf[scale=0.38]{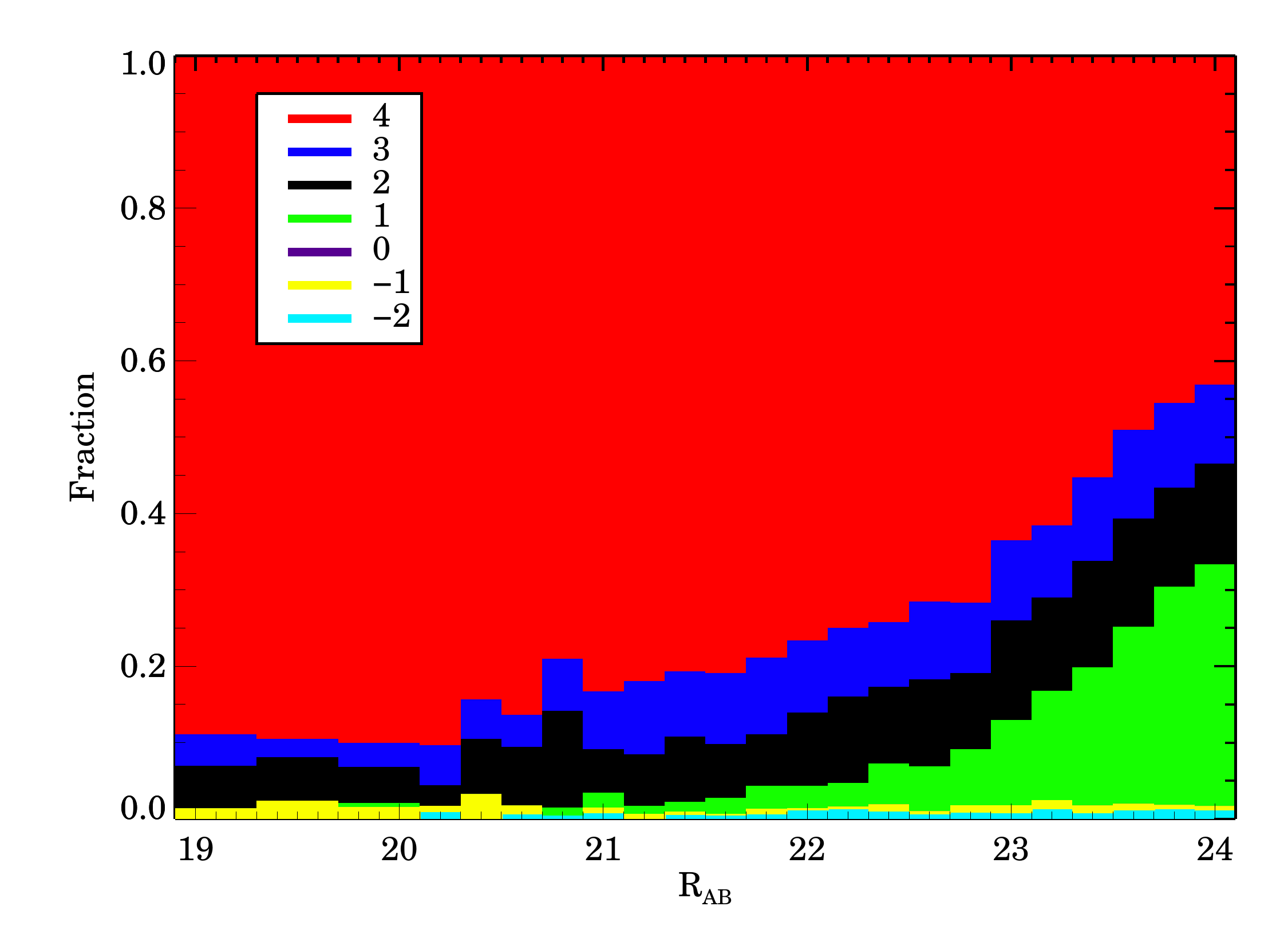}
\vspace*{-0.2in}
\end{center}
\else
\begin{center}
\includegraphics[scale=0.38]{zqual_mag1.eps}
\end{center}
%
\fi
\caption{Histogram of fractions of DEEP2 spectra in Fields 2, 3, and 4
  yielding redshifts with various quality codes in bins of apparent
  magnitude. Quality code $Q=4$ corresponds to redshifts that are
  $\ge$99\% secure; $Q=3$ redshifts are assessed to be $\ge$95\%
  secure; $Q=2$ indicates spectra with a low $S/N$ or a known problem
  as noted in comment codes, but for which a redshift may be
  recoverable; $Q=1$ indicates cases with low $S/N$, that are probably
  not recoverable; $Q=0$ indicates objects whose redshifts were never
  measured in $zspec$; $Q=-1$ indicates stars; and $Q=-2$ corresponds
  to instrumental problems so severe that the object was effectively
  never observed.  }
\label{zqual_mag.eps}
\end{figure}

\subsubsection{Reliability of DEEP2 Redshifts}

The accuracy of the redshifts and estimated errors can be assessed by
comparing multiple observations of the same object.  Duplicate data
come from the overlapping areas between the top and bottom rows of
masks, from the several overlapping pointings in Field 4, and from a
few other scattered regions which have been covered repeatedly (see
Figure~\ref{cooper.deep2.wfn.eps}).  Due to the relatively large
fraction of DEEP2 objects observed twice (roughly 2.5\%), the
reliability of our redshift measurements can be tested robustly.

We first consider whether the claimed reliability rates for quality
codes 3 (95\%) and 4 (99.5\%) are borne out by real data (using only
observations outside the EGS, as in that field discrepant redshifts
were investigated and reconciled).  Since the claimed error rate for
$Q = 4$ is 0.5\%, if we compare two independent observations of the
same galaxy both of which were assigned $Q = 4$, they should agree (to
within, say, 500 \kms) 99.0\% of the time and disagree 1\% of the
time.  In fact, two $Q = 4$ redshifts disagree only 0.29\% of the time
(the 95\% upper limit on the mismatch rate is 0.60\%), considerably
better than the claimed quality.\footnote{We note that, when assessing
  the probability of an event from a Poisson process which was
  observed to occur $N$ times, the typical assumption that the
  uncertainty in the number of events, $\sigma(N)$ is equal to $N$ is
  quite inaccurate when $N \ltsim 5$.  If we assume a flat prior on
  the true expected number of events, $\mu$, at 68\% / 95\% confidence
  we can conclude that $\mu < 1.14 / 3.00 $ if we observe $N=0$; or
  similarly we obtain upper limits $\mu < 2.35 / 4.73$ if we observe
  $N=1$, 3.49/6.23 for $N=2$, 4.59 / 7.48 for $N=3$, 5.61 / 8.42 for
  $N=4$, or 6.52 / 9.032 for $N=5$ (as opposed to 68\%/95\% upper
  limits of 6.05/8.68 for $N=5$ for the Gaussian approximation).  We
  present here the best estimate of the mismatch rate (=$N/N_{dup}$,
  where $N$ is the number of cases of mismatches and $N_{dup}$ is the
  number of objects with duplicate observations of the requisite
  qualities), as well as the 95\% upper limit on this rate.}
Similarly, the mismatch rate for $Q = 3$ vs.~$Q = 4$ pairs is only
0.68\% (upper limit 1.60\%), compared to a predicted rate of 5.5\%,
and the mismatch rate for $Q = 3$ vs.~$Q = 3$ pairs is 1.5\% (upper
limit 4.4\%), compared to a predicted rate of 90\%.  We conclude that
the actual catastrophic error rates in {\it zcat} redshifts are
roughly half as large as claimed; i.e., the claimed rates are very
conservative.  

\begin{figure*}
\ifpdffig
\vspace*{-0.5in}
\begin{center}
\includepdf[scale=0.85]{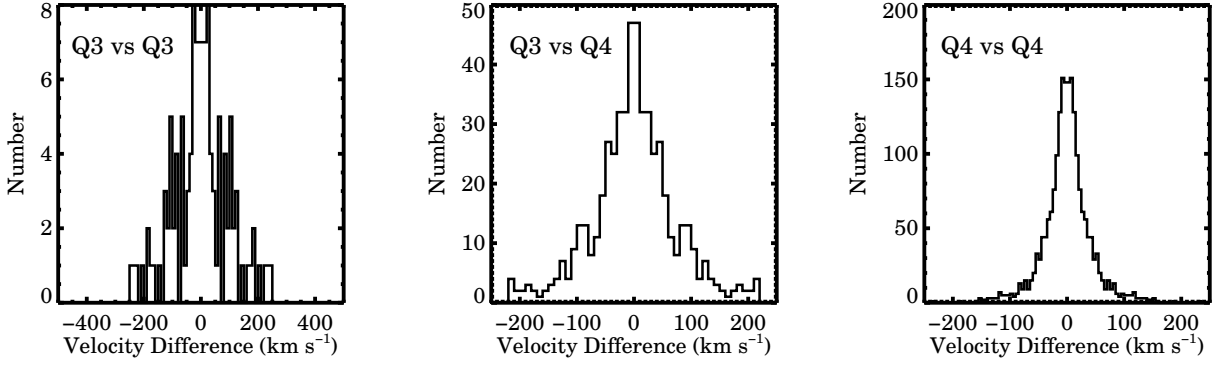}
\end{center}
\vspace*{-6.75in}
\else
\begin{center}
\includegraphics[scale=0.85]{zdifferences.eps}
\end{center}
\fi
\caption{Histograms of heliocentric recession velocity differences for
  multiple observations of the same galaxy. Comparisons are separated
  according to the by quality class of each redshift.  There are 70
  pairs in the left histogram ($Q=3/Q=3$), 306 pairs in the middle
  histogram ($Q=3/Q=4$), and 1195 pairs in the right histogram
  ($Q=4/Q=4$).  The RMS velocity difference for $Q = 3$ vs.~$Q = 3$ is
  62 \kms (with outliers removed).  The RMS for $Q = 3$ vs.~$Q = 4$ is
  51 km s$^{-1}$\, while for the most common $Q = 4$ vs.~$Q = 4$ case
  the RMS is 22 \kms.  }
\label{zdifferences.eps}
\end{figure*}

To assess the accuracy of our redshift error estimates, we consider
the set of all objects that have been observed exactly twice and for
which $Q \ge 3$ for both observations.\footnote{Note: all velocity
  errors and differences in this paper are given in the restframe of
  the object, which means they have been divided by $(1+z)$.  That is,
  $\Delta v = c \times \Delta z/(1+z)$, where $\Delta z$ is the
  redshift difference.}  Figure~\ref{zdifferences.eps} shows
histograms of actual velocity differences for pairs of objects with $Q
= 3$ vs.~$Q = 3$, $Q = 3$ vs.~$Q = 4$, and $Q = 4$ vs.~$Q = 4$.  The
observed RMS difference when comparing two redshifts assigned quality
3 is 62 \kms\ (with outliers removed), compared to a mean predicted
RMS difference of 21.5 \kms\ based on the pipeline errors of the same
duplicates, or 21 \kms\ based on the mean pipeline $Q = 3$ error for
the sample as a whole.  The observed RMS difference for cases where
one spectrum was assigned $Q = 3$ while the other received $Q = 4$ is
51 \kms, versus 16 \kms\ expected based on the pipeline errors of the
same duplicates, or 17 \kms\ based on the mean pipeline errors for $Q
= 3$ and $Q = 4$ spectra.  Finally, the observed RMS difference for
cases where both objects received $Q = 4$ is 22 \kms, compared to a
mean predicted RMS difference of 10 \kms\ based on either the objects'
or the full-sample mean pipeline error.  Though small number
statistics may distort these histograms, it is true in general that
the observed errors are larger than either set of pipeline-predicted
errors, and the latter need to be corrected upward.  From this test,
we find that the pipeline $Q = 4$ RMS errors can be made to match the
observed $Q=4$ RMS errors if 14 \kms\ is added in quadrature to each
error estimate, while the pipeline $Q = 3$ RMS errors can be made to
match the observed $Q = 3$ RMS errors if 41 \kms\ is added in
quadrature.

\begin{figure}
\ifpdffig
\includegraphics[scale=0.4]{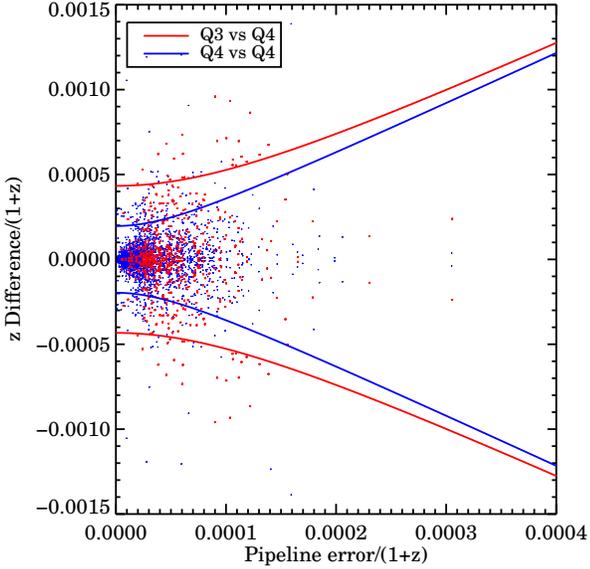}
\vspace*{-1.0in}
\else
\includegraphics[scale=0.4]{errortest_qual.eps}
%
\fi
\caption{Velocity differences between observations of doubly observed
  galaxies in which blue points represent cases where both
  observations yielded redshift quality 4, while red points represent
  cases where one observation yielded redshift quality ($Q$) of 3 and
  the other redshift quality 4.  The horizontal axis is the pipeline
  redshift error estimate for the velocity difference (combining the
  errors estimated for each observation in quadrature); the vertical
  axis is the observed redshift difference.  The blue and red solid
  lines are the expected 3-$\sigma$ errors if the estimated pipeline
  errors for the two measurements are combined in quadrature with 20
  \kms\ or 41 \kms, respectively; the corresponding excess for the
  smaller number of cases which yielded $Q$=3/3 is 58 \kms.  The
  quantities added in quadrature were determined by calculating the
  RMS excess error over the pipeline prediction for $Q$=4/4 and
  $Q$=3/4 cases, respectively; adding this to the pipeline predictions
  in quadrature corresponds reasonably well to the 3-$\sigma$ range of
  the data points in this plot.  The pipeline-predicted redshift
  errors appear to be modestly overoptimistic; it is likely the excess
  error is due to variation between observations (e.g., slightly
  different slit placement on rotating galaxies will lead to variation
  in the measured redshift at this level), rather than due to photon
  statistics.
}
\label{errortest_qual.eps}
\end{figure}

Figure~\ref{errortest_qual.eps} sheds more light by plotting the
individual measured velocity differences for each of the above
duplicate galaxy pairs versus the predicted velocity differences based
on the pipeline errors of the two galaxies.  The predicted pipeline
error is $\sqrt((\sigma_{z,1})^2+(\sigma_{z,2})^2)/(1+z)$, where
$\sigma_{z,1}$ is the pipeline redshift error of the first
observation, $\sigma_{z,2}$ is the pipeline redshift error of the
second observation, and $z$ is the mean of the two redshift estimates.
Points are color-coded by quality code (blue for $Q = 4$ vs.~$Q=4$
pairs and red for $Q=4$ vs.~$Q=3$).  We again confirm that the
pipeline-estimated errors are too small; i.e.  the measured
differences for galaxies with small pipeline errors are larger than
would have been predicted.  This can be cured by adding an extra error
in quadrature to the pipeline values.  The blue lines correspond to
the predicted $3-\sigma$ limits if an extra error of 14 \kms\ is
added in quadrature to the pipeline errors of each {\it individual}
galaxy, while the red lines have 43 \kms\ {\it total} added in
quadrature, values based on the excess errors determined above.  These
additions make the lines fit the observed 3-$\sigma$ excursions of the
two populations fairly well.

\begin{figure}
\ifpdffig
\includepdf[scale=0.4]{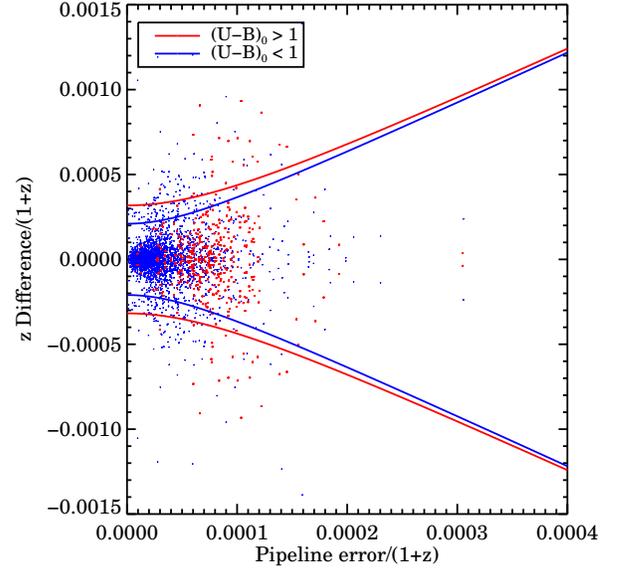}
\vspace*{-1.0in}
\else
\includegraphics[scale=0.4]{errortest.eps}
%
\fi
\caption{As in Figure~\ref{errortest_qual.eps} but with samples now
  split by color rather than redshift quality.  Blue points represent
  blue (restframe $(U-B) < 1$) star-forming galaxies whose redshifts
  are based predominantly on emission lines.  Red points represent
  red-sequence galaxies, whose redshifts are based mainly on
  absorption lines.  The blue and red solid lines are the 3-$\sigma$
  errors that are produced by adding the estimated pipeline error for
  each measurement in quadrature with an extra 17 \kms\ or 32 km
  s$^{-1}$, respectively, to roughly match the observed 3-$\sigma$
  points of the observed distributions.  These quantities correspond
  to the median excess error over the pipeline prediction for blue and
  red galaxies respectively.  Pipeline-predicted redshift errors again
  appear to be modestly overoptimistic, by an amount similar to found
  in Figure~\ref{errortest_qual.eps}.  }
\label{errortest.eps}
\end{figure}

A similar test is shown in Figure~\ref{errortest.eps}, which compares
errors for red and blue galaxies rather than dividing up objects
according to their quality codes (though blue, emission-line galaxies
predominantly yield $Q=4$ redshifts, with a greater $Q=3$ fraction for
red galaxies). By adding an extra 17 \kms\ error in quadrature for
blue galaxies, or 32 \kms\ for red galaxies, we can match the observed
results reasonably well.  These tests suggest that an effective method
for correcting DEEP2 pipeline redshift uncertainties to match
empirical errors is to make the replacement $\sigma_v = \sqrt
{\sigma_{v_{pipeline}}^2 + \sigma_Q^2}$, where $\sigma_v$ is the
predicted empirical error, $\sigma_{v_{pipeline}}$ is the error in $v$
predicted by the pipeline, and $\sigma_Q = 25$ \kms\ if $Q=3$, or 11
\kms\ if $Q=4$.  We note that the fact that the pipeline
underestimates observed errors is not necessarily a failing of the
{\it spec2d} or {\it spec1d} error model; it could instead reflect
miscentering of objects in slits or the relative rotational velocities
of the dominant star-forming regions in target galaxies.

The final way in which we can assess the accuracy of information in
the {\it zcat} is by checking the consistency with which the various
quality codes have been applied.
Table~\ref{table.comparequalitycodes} shows agreement among quality
codes for duplicate pairs.  Star identification ($Q = -1$) seems
highly reliable; $\sim 90\%$ of objects assigned $Q=-1$ in one
observation received the same quality code in the other.  The
assignment of $Q = 4$ codes is also quite repeatable: galaxies
receiving $Q=4$ in one observation received it again 68\% of the time
and $Q = 3$ another 19\% of the time, and hence received a
``reliable'' rating 87\% of the time.  The $Q=1$ quality code, too, is
a fairly consistent indicator of poor quality, being paired with
itself or $Q = 2$ 78\% of the time.

The $Q = 2$ codes are more highly scattered, being paired frequently
with other codes either worse or better, as we might expect for this
marginal category.  Nearly half of the $Q=2$ objects observed twice
were assigned $Q = 4$ on the second observation; this could happen
because a bad-data problem such as bad sky-subtraction or inconsistent
continuum levels ({\tt bsky} or {\tt bcont}, cf.\ \S\ref{datatables})
was not present in the $Q = 4$ observation.  $Q = 3$ codes also seem
to repeat relatively rarely; the second observation of a $Q=3$ object
is classified with $Q=4$ some 66\% of the time.

A final way to look at these issues is to group all $Q = 1$ and $Q =
2$ together to form the ``unreliable'' class, and all $Q = 3$ and $Q =
4$ together as the ``reliable'' class.  Problematic pairs are those
that have one member in each class.  According to
Table~\ref{table.comparequalitycodes}, this happens $14$\% of the
time. Of the discrepant cases, roughly half are due to technical
problems in one observation but not the other: i.e., these are
galaxies that would normally yield reliable redshifts if the data were
good.  The remainder appear to be dominated by objects which received
$Q=1$ or $Q=2$ in one observation due to poor signal-to-noise or a
single visible emission line.

\subsection{Target Selection and Redshift Success as a Function of Color and Magnitude}
\label{redshiftresults2}

\begin{figure}
\ifpdffig
\vspace*{-0.25in}
\includepdf[scale=0.4]{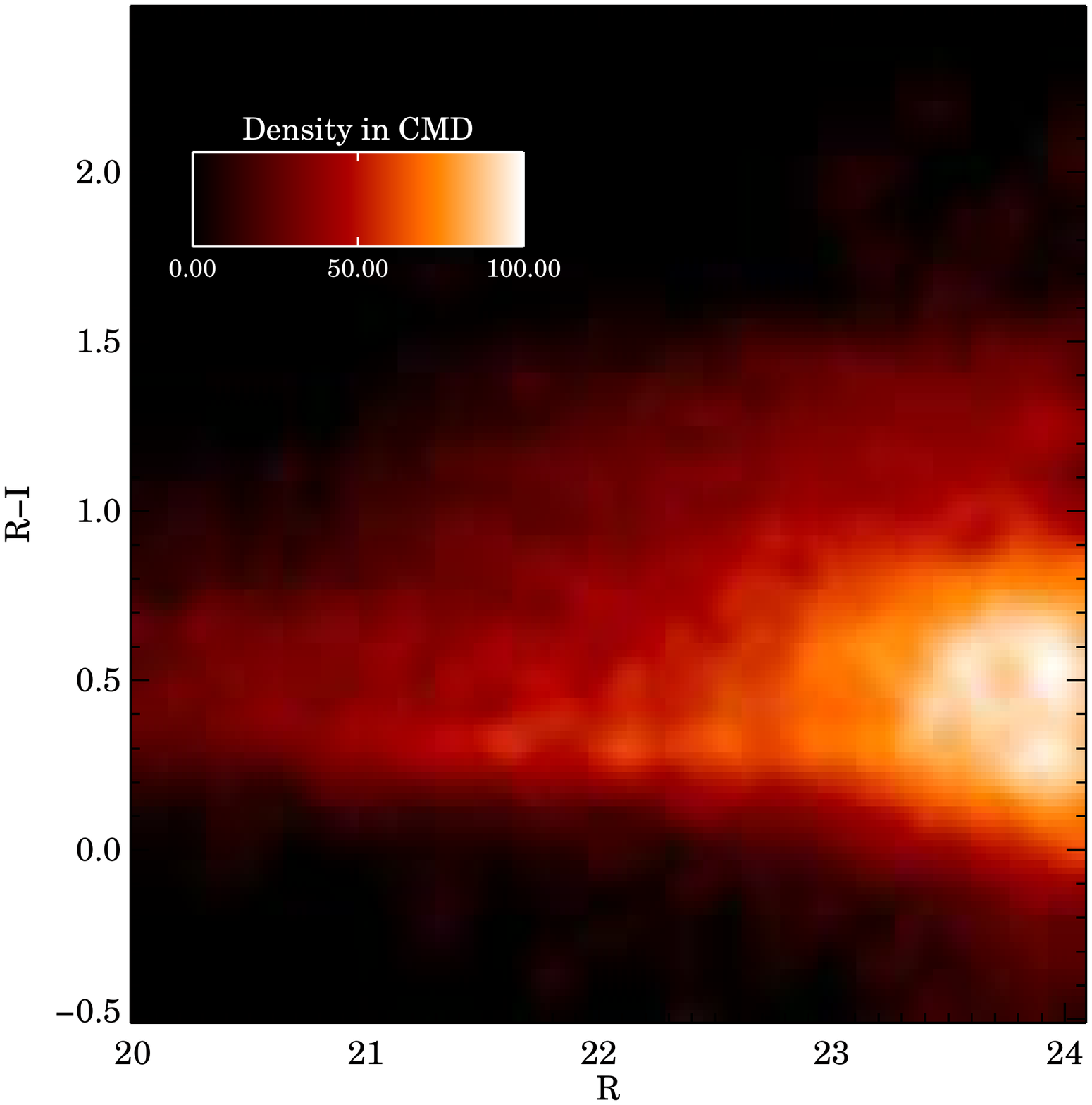}
\vspace*{-1.0in}
\else
\includegraphics[scale=0.4]{egscontourdistr.eps}
\fi
\caption{ Map of the density of DEEP2 galaxy candidate objects in
  Fields 1,2, 3, and 4 in apparent $R$ vs.~$R-I$ color-magnitude
  space.  Likely stars (i.e.\ objects with $p_{gal}<0.2$) are not
  included.  The color at each position is proportional to the square
  root of the number of objects in the catalog with that color and
  magnitude, on an arbitrary scale from 0 to 100.  }
\label{egscontourdistr.eps}
\end{figure}

\begin{figure}
\ifpdffig
\includepdf[scale=0.4]{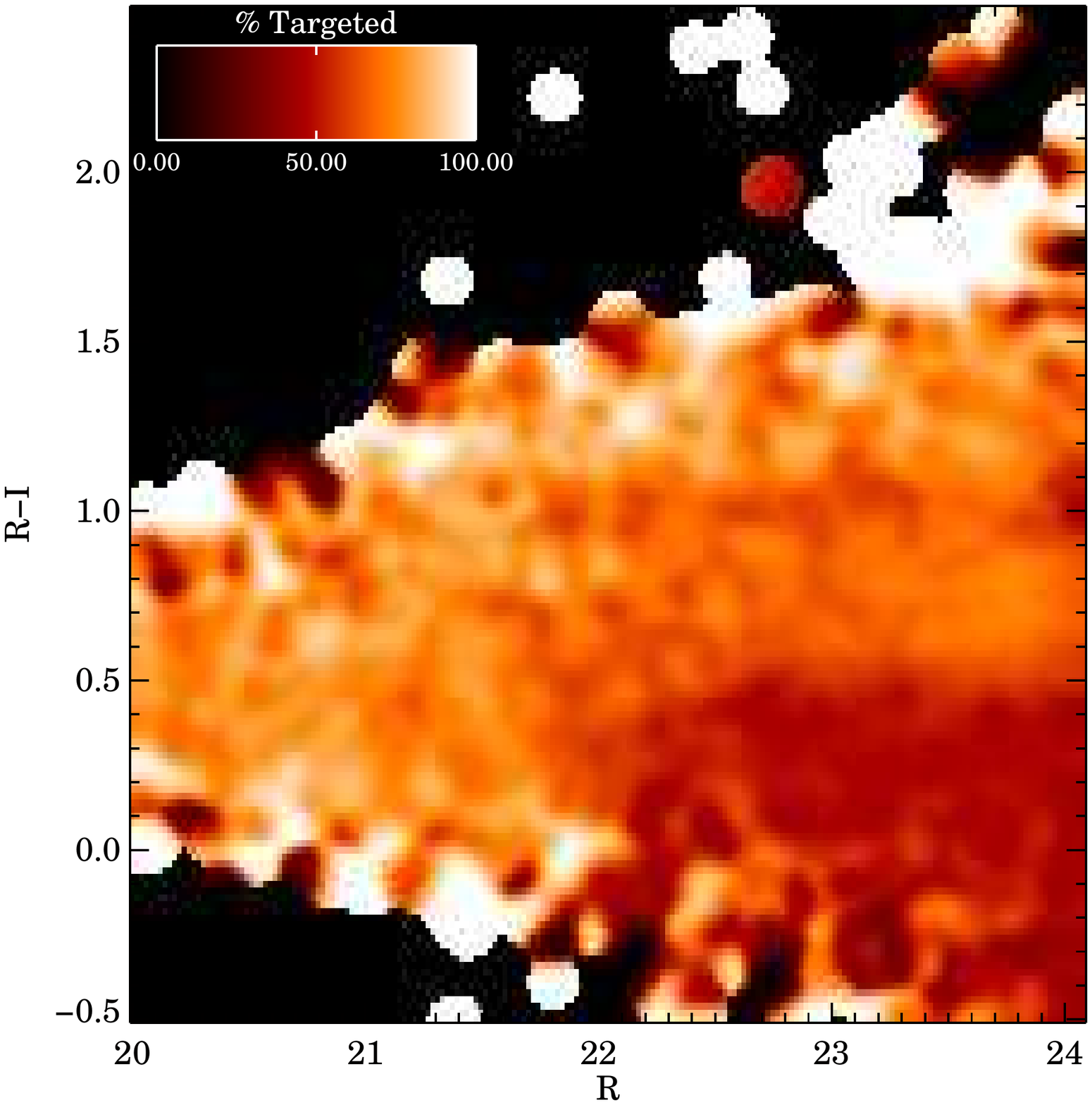}
\vspace*{-1.0in}
\else
\includegraphics[scale=0.4]{targ1_r.eps}
%
\fi
\caption{ The fraction of the candidate targets shown in
  Fig.~\ref{egscontourdistr.eps} which were placed on slitmasks in
  Field 1 (EGS).  }
\label{targ1_r.eps}
\end{figure}

\begin{figure}
\ifpdffig
\vspace*{-0.25in}
\includepdf[scale=0.4]{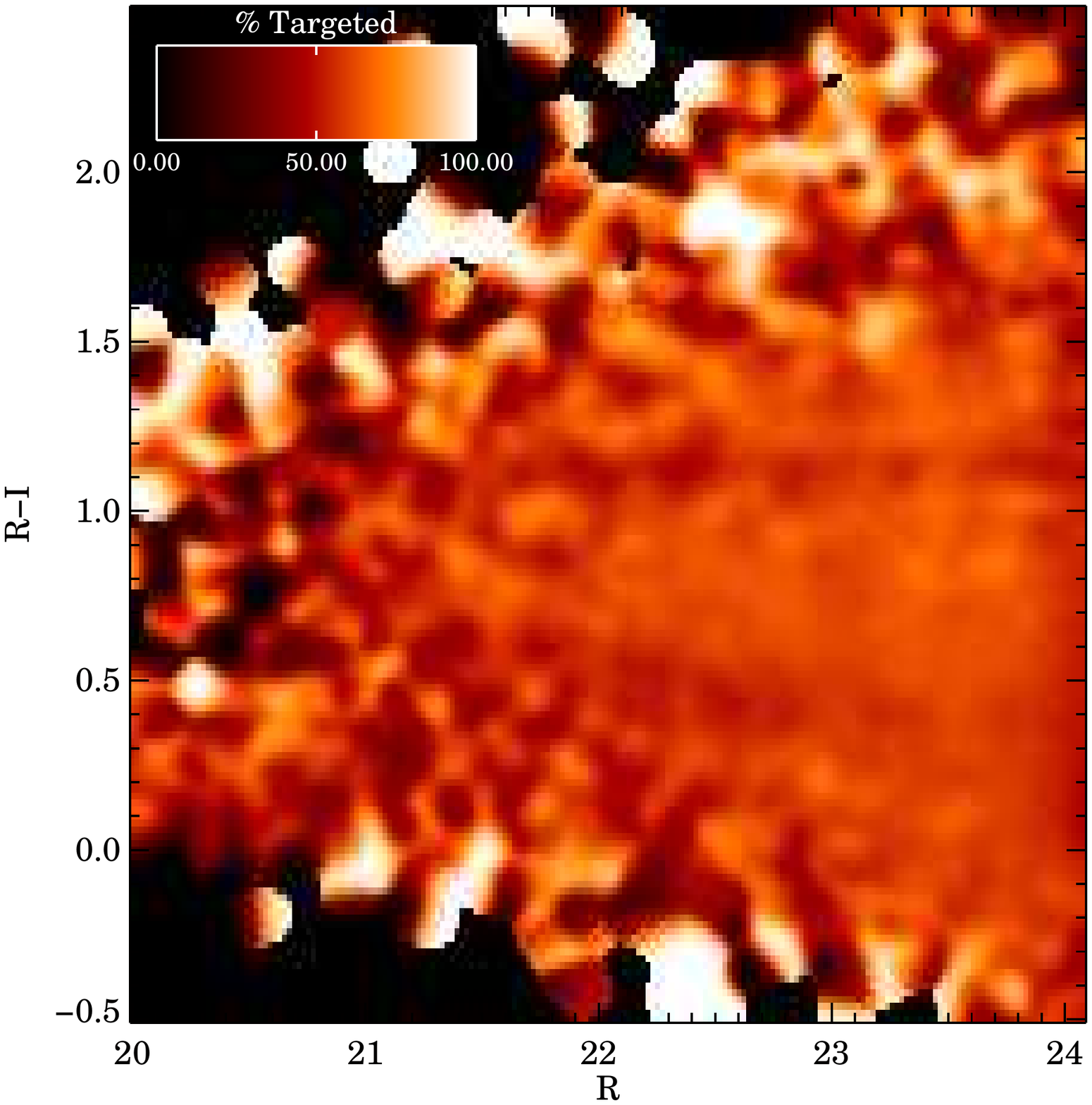}
\vspace*{-1.0in}
\else
\includegraphics[scale=0.4]{targ234_r.eps}
\fi
\caption{ The fraction of the candidate targets shown in
  Fig.~\ref{egscontourdistr.eps} which were placed on slitmasks in
  Fields 2, 3, and 4.  }
\label{targ234_r.eps}
\end{figure}

We conclude this section by showing two series of diagrams that
illustrate the density of objects in the original {\it pcat} galaxy
candidate catalog in color or color-magnitude space; then the fraction
that received slitlets; then the fraction that yielded either $Q = 3$
or $Q = 4$ redshifts (summed); and finally the fraction of $Q = 4$
redshifts alone.  The first series plots the density of galaxies in
the apparent $R-I$ vs.~$R_{\rm AB}$ plane.
Figure~\ref{egscontourdistr.eps} demonstrates that relatively blue
galaxies dominate a pure $R<24.1$ sample at faint magnitudes.
However, only half of them are targeted in EGS, as may be seen in
Figure~\ref{targ1_r.eps}, due to the fact that we de-weight faint
nearby galaxies (most of which are very blue) in target selection.
The same plot for Fields 2,3,4 (Figure~\ref{targ234_r.eps}) is much
more uniform, reflecting the mild 25\% roll-off of weights over the
faintest 0.3 mag in these fields.

\begin{figure}
\ifpdffig
\includegraphics[scale=0.4]{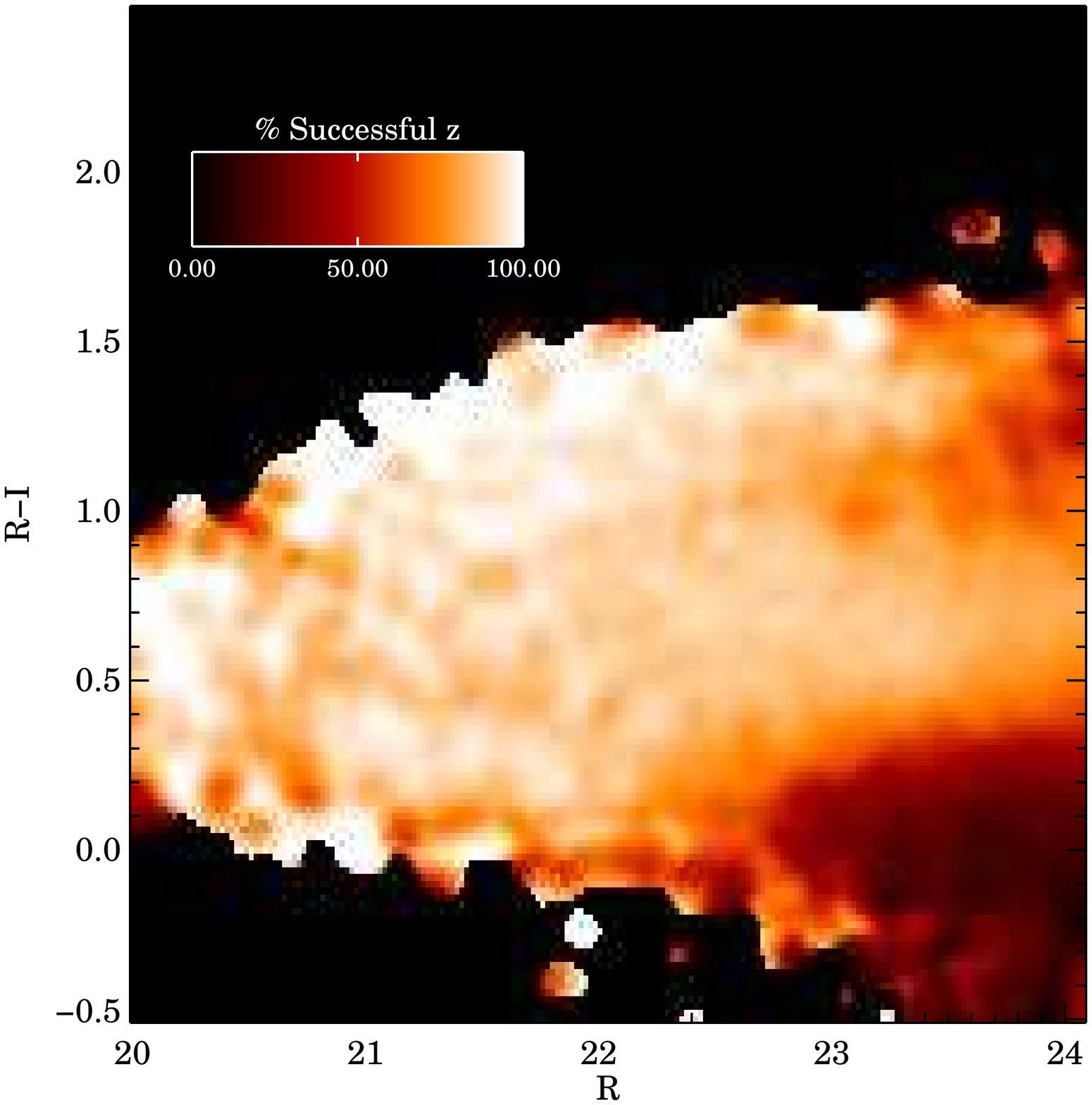}
\vspace*{-1.0in}
\else
\includegraphics[scale=0.4]{egscontourzsuccessr.eps}
%
\fi
\caption{ Fraction of the above objects placed on slitmasks that
  yielded either $Q=3$ ($>95\%$ confidence) or $Q = 4$ ($>99\%$
  confidence) redshifts.  Highly-problematic data ($Q=-2$) and stars
  are not counted as either successes or failures, so this is
  $(Q4+Q3)/(Q1+Q2+Q3+Q4)$, where $Q1$ is the number of objects
  assigned quality $Q=1$, $Q2$ is the number given $Q=2$, etc.  Hence,
  this is the fraction of galaxies which were targeted for a useful
  spectrum for which we obtained a secure redshift.  The lack of
  redshifts for faint, very blue galaxies reflects the fact that most
  of them are beyond our redshift limit of $z \sim 1.4$.}
\label{egscontourzsuccessr.eps}
\end{figure}

\begin{figure}
\ifpdffig
\vspace*{-0.25in}
\includepdf[scale=0.4]{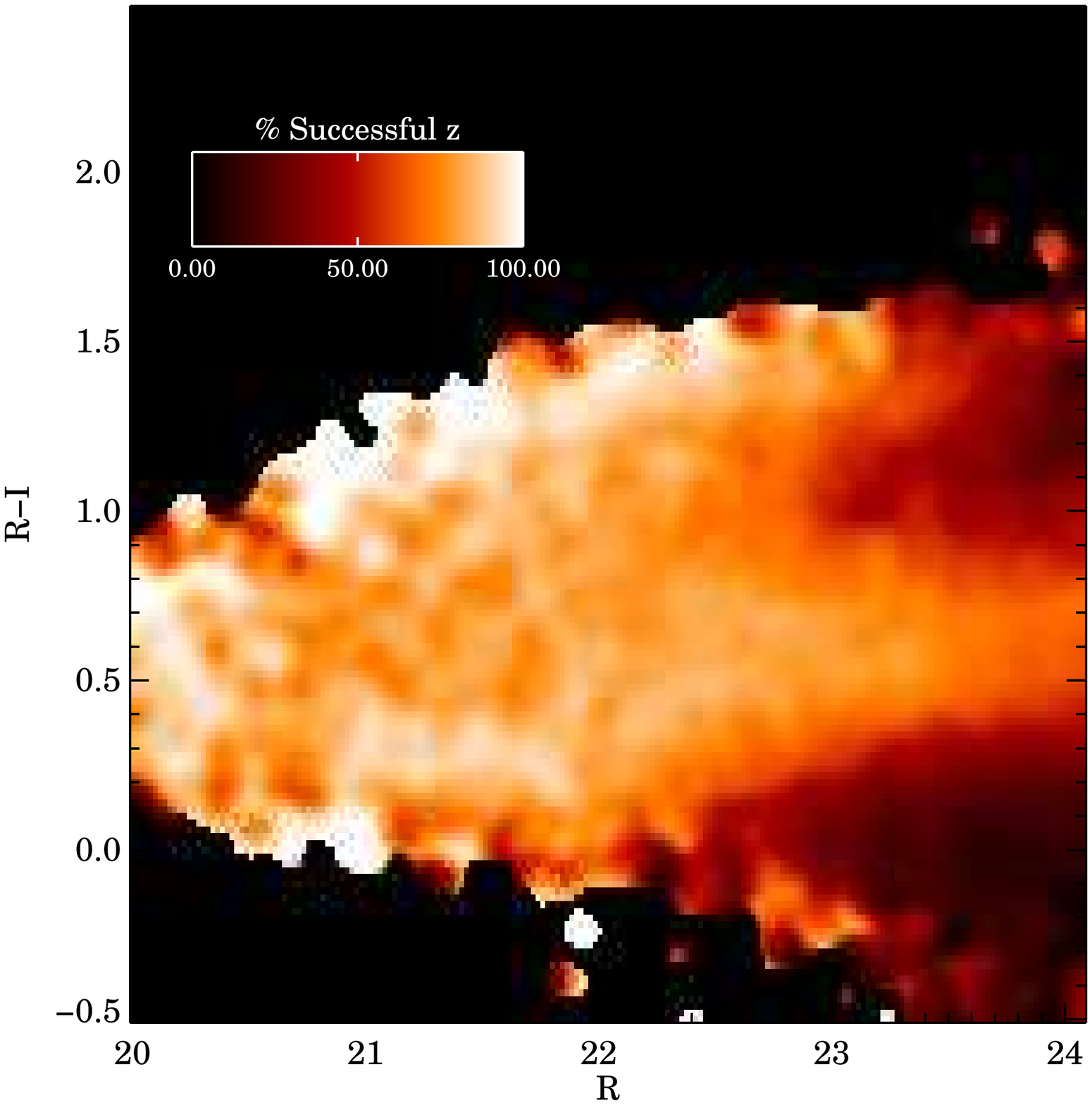}
\vspace*{-1.0in}
\else
\includegraphics[scale=0.4]{egscontourzsuccessr_q4.eps}
\fi
\caption{ As Fig.~\ref{egscontourzsuccessr.eps} but including only
  $Q=4$ ($\ge 99\%$ confidence redshifts) in the numerator.  The lack
  of high quality redshifts for very faint/blue galaxies continues; in
  addition there is a deficit of very faint red galaxies in the
  highly-secure category, due to their poor continuum $S/N$.  }
\label{egscontourzsuccessr_q4.eps}
\end{figure}

Figure~\ref{egscontourzsuccessr.eps} shows the fraction of targeted
galaxies (combining data from all fields) that yield either $Q=3$ or
$Q=4$ redshifts as a function of color and magnitude (i.e. over the
same plane as Figs.~\ref{egscontourdistr.eps} -- \ref{targ234_r.eps}.
The steep loss for faint blue galaxies is due to the fact that many
are beyond our $z \sim 1.4$ redshift limit.  There is also a slight
loss in the last 0.3 mag for red galaxies, reflecting the difficulty
of measuring $z$'s for objects with weak emission.  Finally,
Figure~\ref{egscontourzsuccessr_q4.eps} repeats this figure but
including $Q=4$ redshifts only.  There is relatively little change in
the faint blue corner but a large change for faint red galaxies,
showing that the latter tend to have $Q=3$ redshifts, in keeping with
the difficulty of measuring their redshifts.
 
\begin{figure}
\ifpdffig
\includegraphics[scale=0.4]{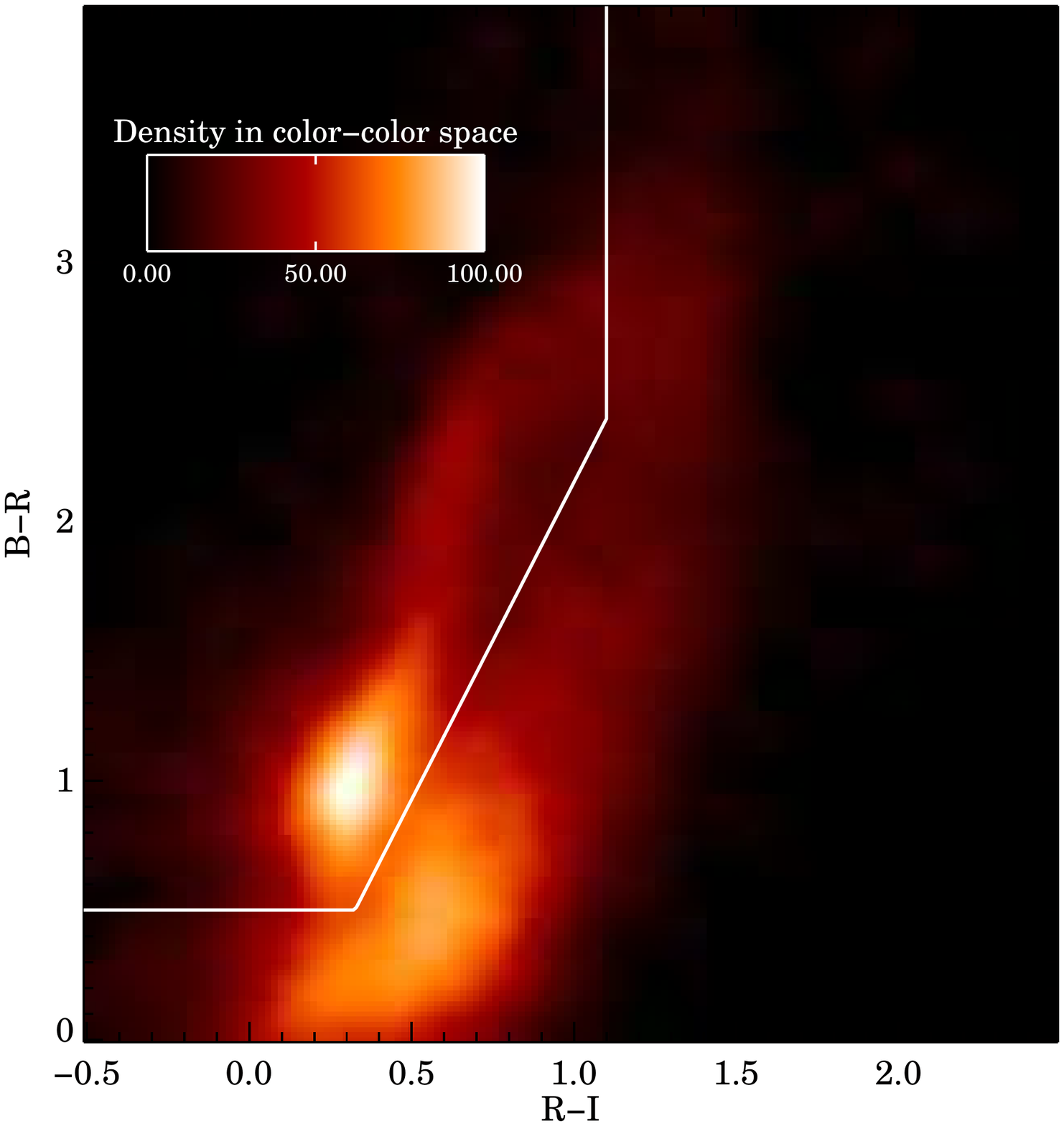}
\vspace*{-1.0in}
\else
\includegraphics[scale=0.4]{egscontourdist.eps}
\fi
\caption{ Map of the density of DEEP2 galaxy candidate objects in
  Fields 1, 2, 3, and 4 in apparent $B-R$ vs.~$R-I$ color-color space.
  Objects rejected as stars ($p_{gal}<0.2$) are not included.  The
  intensity at each position is proportional to the square root of the
  number of objects in the catalog with that color and magnitude on an
  arbitrary scale from 0 to 100.  The white line indicates the DEEP2
  color cut used to select high redshift objects; it corresponds to
  the dot-dashed line in Fig.~\ref{egsstars.eps}}
\label{egscontourdist.eps}
\end{figure}

\begin{figure}
\ifpdffig
\vspace*{-0.25in}
\includepdf[scale=0.4]{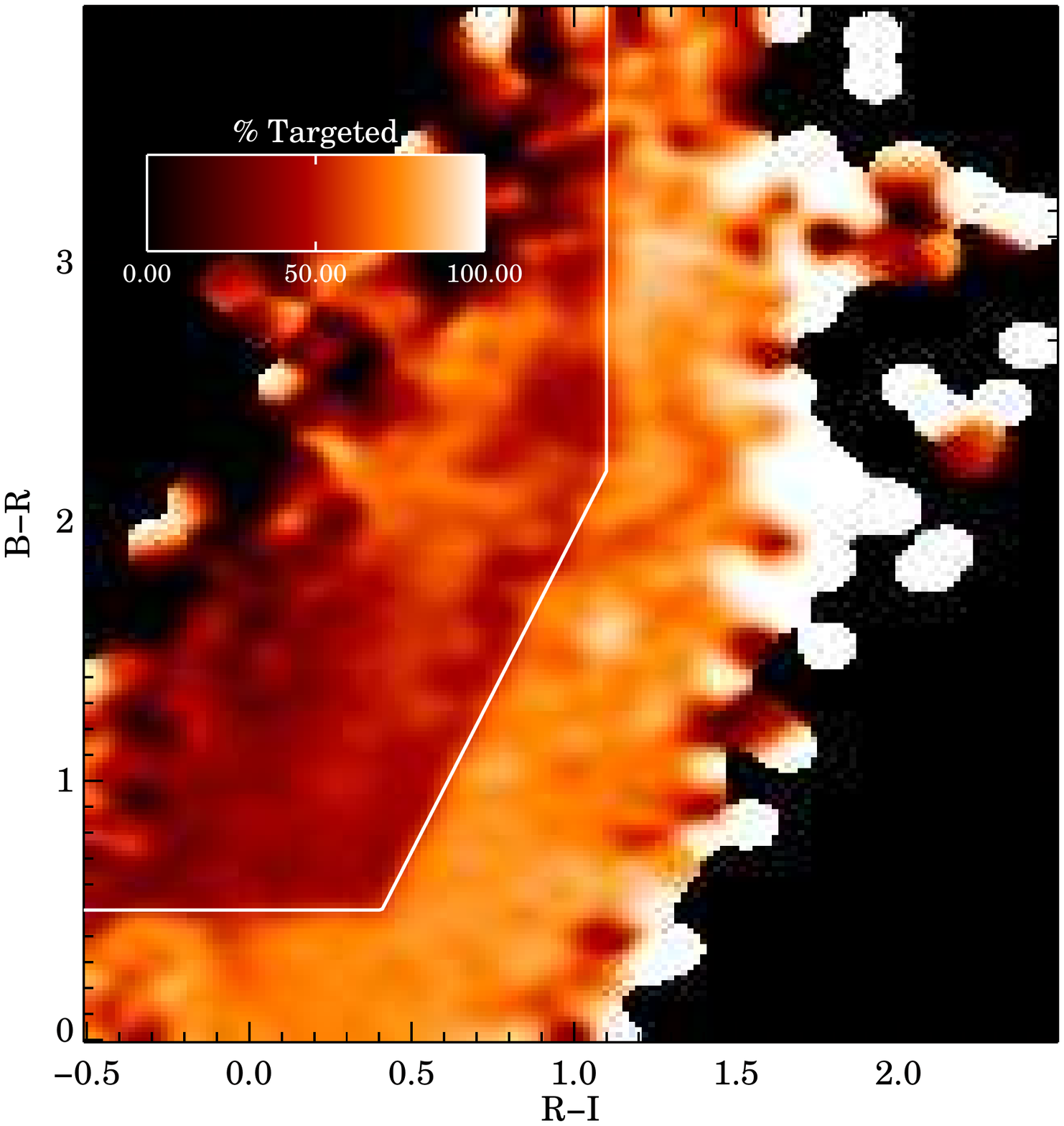}
\vspace*{-1.0in}
\else
\includegraphics[scale=0.4]{targ1.eps}
%
\fi
\caption{ The fraction of the candidate targets shown in
  Fig. ~\ref{egscontourdistr.eps} which were placed on slitmasks in
  Field 1 (EGS).  }
\label{targ1.eps}
\end{figure}

The second series of plots is similar to the first, but now plots
objects in apparent $B-R$ vs.~$R-I$ color-color space.
Figure~\ref{egscontourdist.eps} plots the density of the full target
sample in this diagram, while Figure~\ref{targ1.eps} shows the target
sampling density in Field 1 (EGS).  Several factors are evident here.
First, the de-weighting of nearby faint galaxies reduces the sampling
rate for low-redshift blue galaxies, which may be found in the
lower-left corner of the $z<0.7$ region (above the color cut line).
Second, the de-weighting of all nearby galaxies in EGS reduces the
sampling rate of objects to the upper left of the color pre-selection
boundary compared to those below and to the right.  Third, the high
sampling rate at the outer edges of the distribution reflects the high
priority given to objects with peculiar colors, which are outliers in
this diagram.

\begin{figure}
\ifpdffig
\includepdf[scale=0.4]{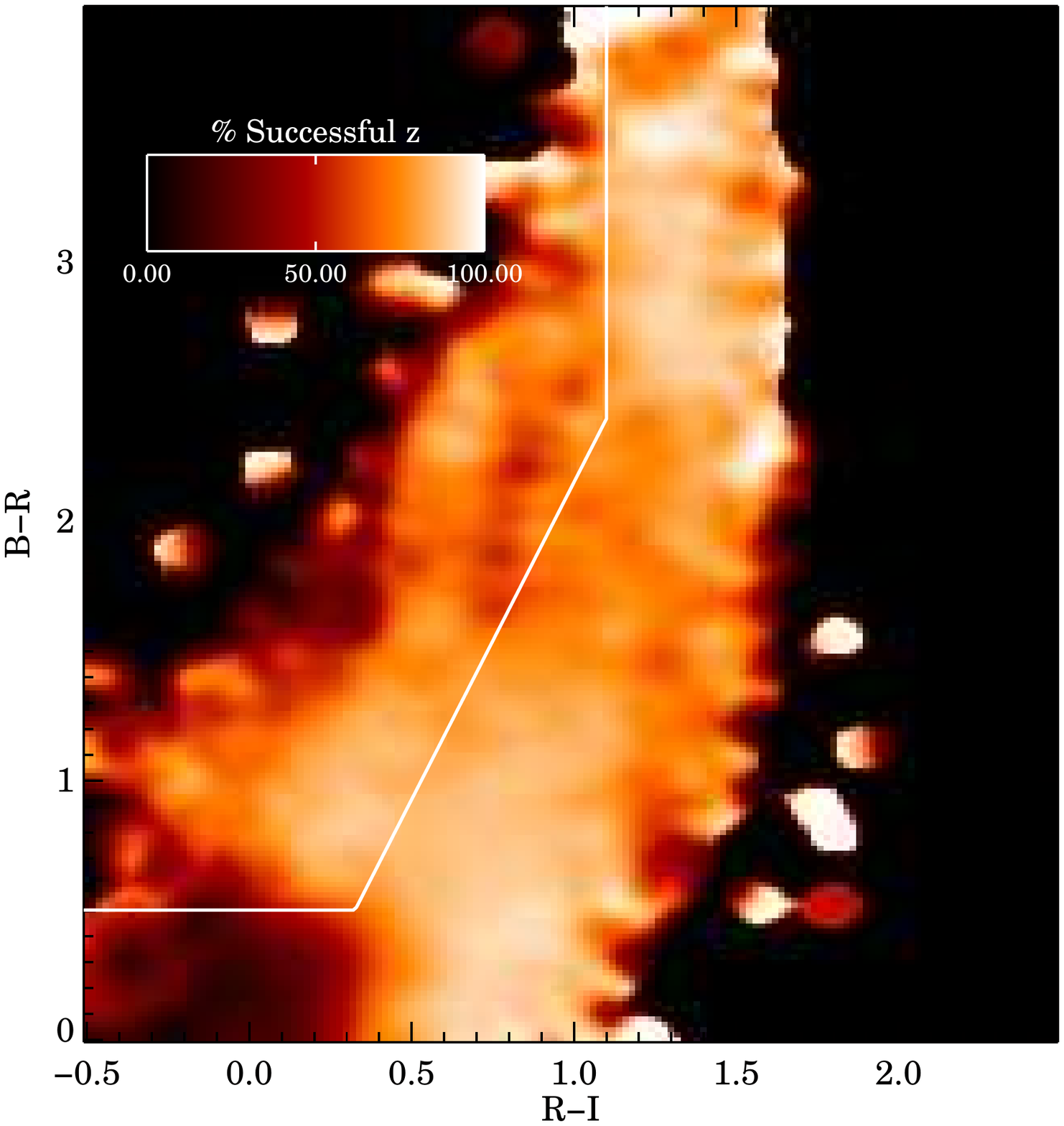}
\vspace*{-1.0in}
\else
\includegraphics[scale=0.4]{egscontourzsuccess.eps}
\fi
\caption{ Fraction of the objects placed on slitmasks that yielded
  either $Q=3$ or $Q = 4$ redshifts, over the same plane as
  Fig.~\ref{egscontourdist.eps}.  Instrumentally-compromised data and
  stars are not counted as either successes or failures, so the
  plotted quantity is $(Q4+Q3)/(Q1+Q2+Q3+Q4)$.  }
\label{egscontourzsuccess.eps}
\end{figure}

\begin{figure}
\ifpdffig
\vspace*{-0.25in}
\includepdf[scale=0.4]{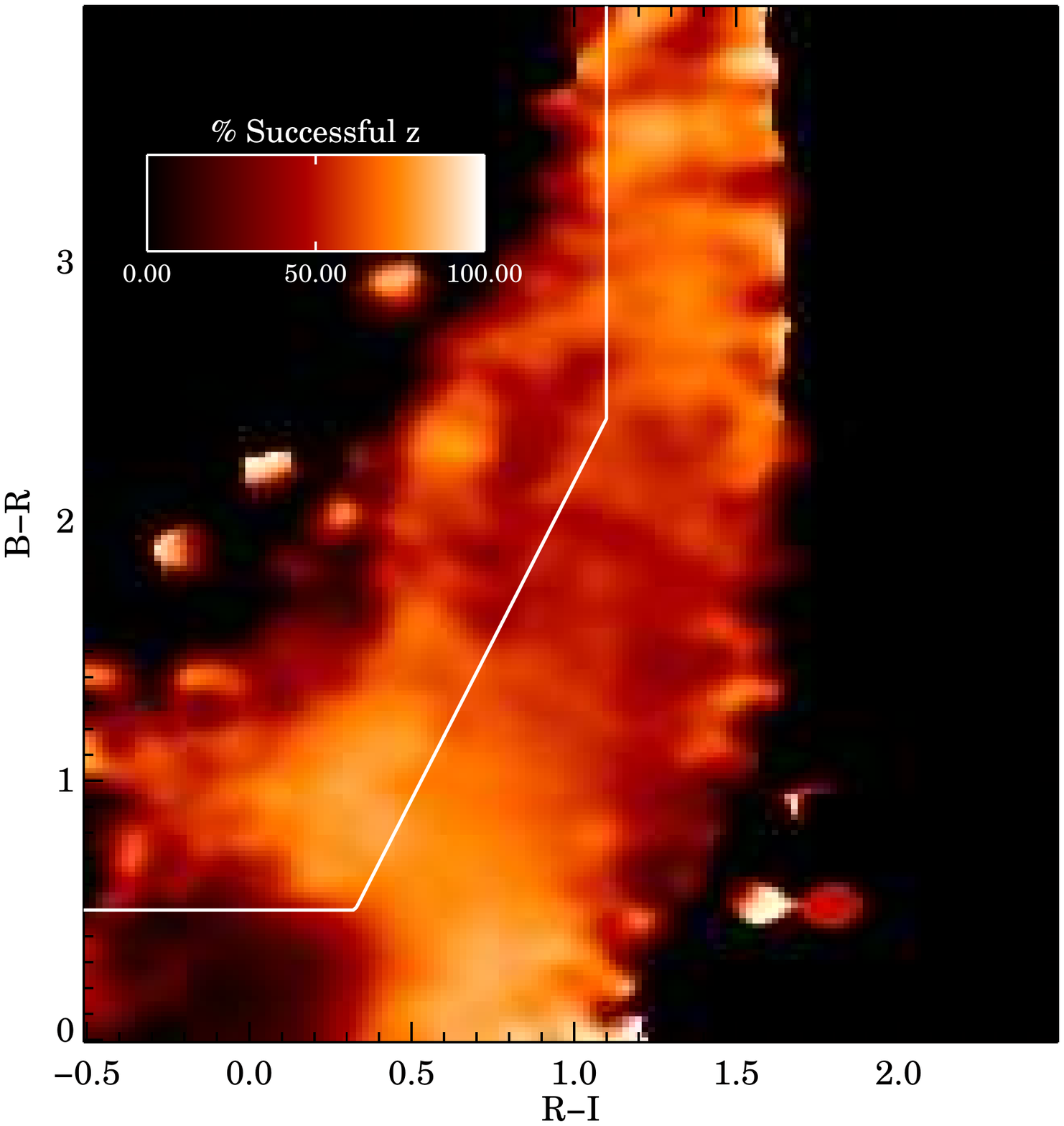}
\vspace*{-1.0in}
\else
\includegraphics[scale=0.4]{egscontourzsuccess_q4.eps}
%
\fi
\caption{ As Fig.~\ref{egscontourzsuccess.eps}, but counting only
  redshift quality $Q=4$ as successful.  }
\label{egscontourzsuccess_q4.eps}
\end{figure}

Figure~\ref{targ234.eps} repeats this diagram for Fields 2, 3, 4.  The
main features evident are the strong color pre-selection, the slow
roll-off in the selection function near this boundary due to
``pre-whitening,'' and, again, the strong preference given to objects
with peculiar colors in the outer parts of the distribution.  Redshift
success rates are plotted in Figure~\ref{egscontourzsuccess.eps} and
Figure~\ref{egscontourzsuccess_q4.eps}.  When considering quality
codes $Q>3$, the redshift success rate is a relatively flat function
for galaxies of all types save in the bluest corner of color-color
space, where $z>1.4$ galaxies may be found.  It is worth noting that
even in the most favorable regions of color space the redshift success
rate is approximately 90\%, not 100\%, however.  Since $Q=4$ redshifts
are predominantly assigned to blue galaxies, it is little surprise
that the objects assigned $Q=4$ are more localized in color space;
regions dominated by intrinsically red galaxies will generally yield a
lower rate of $Q=4$ redshifts.

The net result of all of these selection and success functions
multiplied together is summarized in Figure~\ref{neigh_figure.epsi},
which illustrates the fraction of galaxies that meet the DEEP2
selection criteria that receive slitlets and the total fraction that
yield $Q \ge3$ redshifts, as a function of distance to the
third-nearest neighbor.  Overall, 61\% of qualifying targets in Fields
2, 3, 4 are placed on slitlets (59\% in Field 1, EGS); and 43\% out of
all qualifying targets are in fact observed and yield $Q \ge3$
redshifts (43\% in Field 1, EGS).

\section{Multiple Galaxies}
\label{multiplicity}

Section~\ref{targetbiases} listed blending in the CFHT ground-based
$BRI$ imaging as a potential source of biases in the DEEP2 target
sample.  Such blends have two possible bad effects.  First, galaxies
that have ostensibly reliably-determined redshifts may in fact consist
of two superposed but separate galaxies, in which case the photometry,
as well as derived properties dependent on continuum strength such as
equivalent width, reflect the properties of the blend rather than the
particular galaxy providing a redshift.  Second, it may happen that
the combination of two galaxies, each of which is just below the
$R_{\rm AB} = 24.1$ magnitude limit, is bright enough that their
combined light brings them above that limit.  These pairs may or may
not contaminate the ``reliable'' ($Q>2$) redshift sample, as they may
be identified as {\tt supser}s in the {\it zspec} process; however,
even if they do not, their presence still falsely enlarges the target
sample and could distort its statistical properties.  However, the
presence of multiple subclumps within a single galaxy is not a
significant problem from either of these standpoints.  Similarly, in
cases where the light from a pair of galaxies with sufficiently high
brightness ratio (e.g., 5:1) is combined, one galaxy dominates and few
conclusions would be altered.

The number of superpositions (hereafter called ``multiple galaxies'')
has been estimated by comparing the low-resolution ground-based DEEP2
{\it pcat} catalog to a high-resolution Sextractor catalog based on
the {\it HST} ACS mosaic in EGS by Lotz et al.~(2008).  This latter
catalog consists of all objects detected in either the $V$ or $I$ ACS
images.  Four percent of DEEP2 candidate objects with $R_{\rm AB} <
24.1$ prove to have multiple Sextractor matches within a 0\s.75
radius, and 10\% have multiple matches within a 1\s.0 radius.
Inspection by eye confirmed that objects within either of these
separations frequently look single in the CFHT images, depending on
seeing.

\begin{figure}
\vspace{85mm}
\ifpdffig
\vspace*{-3.5in}
\begin{center}
\includepdf[scale=0.4]{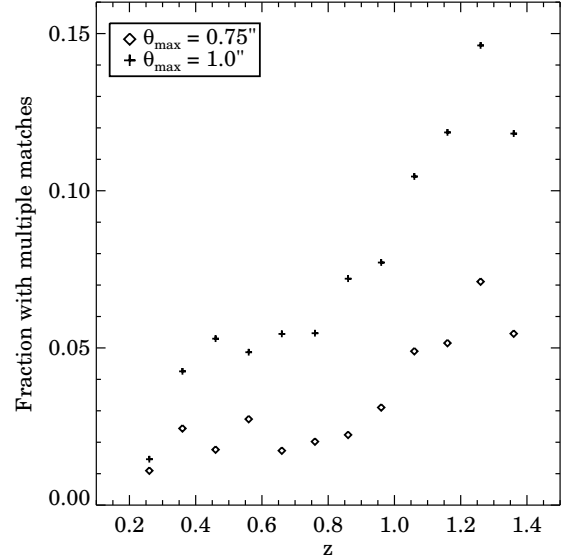}
\vspace*{-1.2in}
\end{center}
\else
\vspace*{-3.65in}
\begin{center}
\includegraphics[scale=0.4]{multi_z.eps}
\end{center}
%
\fi
\caption{ The fraction of multiple galaxies vs.~redshift for galaxies
  with secure $z$ measurements ($Q \ge 3$).  Multiples are found by
  matching DEEP2 {\it pcat} galaxies to objects detected in $V$ and
  $I$ images in the EGS {\it HST} mosaic. A multiple galaxy is defined
  as having two or more Sextractor matches within a radius of 0\s.75
  or 1\s.0 of the {\it pcat} position.  Below $z \sim 1$, only about
  40\% of the multiples found here are serious -- the remainder are
  due either to data, subclumps within single objects, or very
  high-brightness-ratio pairs.  If these are set aside, the remaining
  fraction of multiples (1-2\%) is consistent with the expected rate
  of random superpositions of physically unassociated galaxies along
  the line of sight.  The fraction of multiples rises significantly
  beyond $z \sim 1$.  Though some of the increase is due to subclumps
  (as seen in the {\it HST} images), much of the rise must be due to
  an increase in the number of separate but physically associated
  companion galaxies, which appear to be more prevalent at high
  redshifts.  }
\label{multi_z.eps}
\end{figure}

To understand these objects in more detail, we start with the quarter
that were targeted for spectroscopy and yielded reliable redshifts ($Q
\ge 3$). Note that this fraction is only half as large as the 50\%
targeting + redshift rate obtained for {\it pcat} objects overall in
Figure~\ref{neigh_figure.epsi}; we will return to that point later.
The fraction of multiples as a percent of all galaxies with redshifts
is plotted as a function of $z$ in Figure~\ref{multi_z.eps}.

The number of objects found in the larger aperture vs.~the smaller
remains constant at about 2.5 at all redshifts.  This number is larger
than the ratio of the areas of the two apertures (1.8).  One possible
explanation is that pairs with separations below $\sim 0.5$\asec\ are
treated as single objects by Sextractor, so the ratio of the two
numbers is really the ratio of the areas of two {\it annuli}, the
inner one relatively narrow.  Alternatively, to the degree to which
close pairs are likely to be associated with each other, it could
represent the impact of physical processes: 1\asec\ corresponds to
11$h^{-1}$ kpc comoving, or 5.6 $h^{-1}$ kpc physical, at $z=1$.
Close pairs could have merged, preventing us from seeing them;
alternatively, the background galaxy in a pair might be hidden from
view by dust in the foreground object.

Figure~\ref{multi_z.eps} shows a clear increase in the rate of
multiple galaxies near $z \sim 1$, apparent with either aperture size.
To explore this, we have inspected all {\it HST} images of the
multiple galaxies with redshifts by eye to see if Sextractor is
reliably identifying the same kind of galaxies at all redshifts. It
appears that this is not the case.  At low redshifts (where the
multiplicity fractions are low), we find that only 40\% of the
multiple galaxies consist of well-separated, distinct galaxies of
comparable brightness -- the remainder are either subclumps within the
same galaxy, pairs with high brightness ratios, or regions with bad
ACS data.

Since these latter cases are not a problem for DEEP2 data, we can
multiply the fractions of multiples below $z \sim 1$ in
Figure~\ref{multi_z.eps} by 0.4 to find the true numbers of
problematic galaxies.  We find that $\sim$1\% of DEEP2 galaxies at
$z<1$ have a contaminating galaxy within 0.75\asec, while $\sim$2\%
have a potential contaminant within 1\asec.  These numbers are quite
close to those expected if the companion objects consist of
$R<\sim25_{AB}$ mag galaxies (i.e., roughly those down to the
brightness limit of the Sextractor catalog) distributed randomly on
the sky.  We therefore estimate that at redshifts $z < 1$, 1-2\% of
DEEP2 galaxies are seriously contaminated by unrelated superimposed
foreground or background galaxies; we further conclude that most of
these superpositions are not caused by neighboring galaxies, but
rather by physically unassociated objects at widely separated
distances along the line of sight.

The situation changes above $z \sim 1$, where the fraction of
multiples in Figure~\ref{multi_z.eps} increases.  We would not expect
the number of random superpositions within a fixed angular separation
to depend on redshift.  As a consequence, the analysis we have done
for $z<1$ galaxies demonstrates that randomly superimposed galaxies
can make up at most a small fraction of of the multiple companions at
$z \ge 1$. Based on our inspection of ACS postage stamp images, the
fraction of multiples which are in fact single objects with several
subclumps doubles near $z \sim 1$; this is reasonable, since more
distant galaxies tend to look clumpier (both because we are observing
them at shorter rest-frame wavelengths and due to their intrinsic
properties; cf. Papovich et al. (2003).  However, this increase cannot
account for the full rise, and we are forced to conclude that the
number of physically associated {\it but distinct} companions also
rises beyond $z = 1$.

\begin{figure}
\ifpdffig
\vspace*{-0.3in}
\includepdf[scale=0.4]{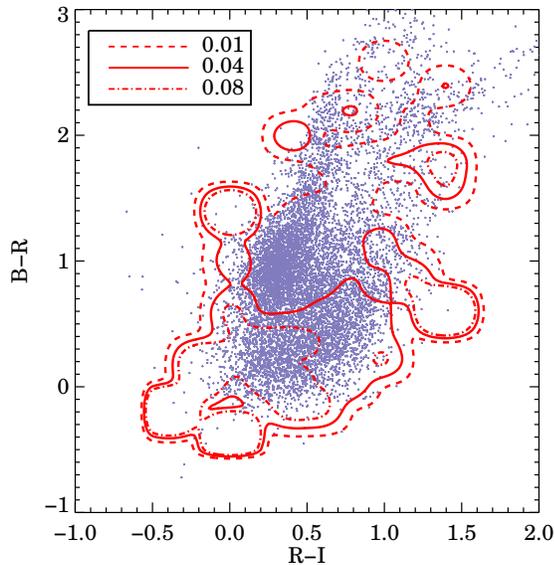}
\vspace*{-1.0in}
\else
\includegraphics[scale=0.4]{p_smooth.eps}
\fi
\caption{Smoothed contours showing the frequency of multiple objects:
  i.e. DEEP2 {\it pcat} objects having multiple companions within
  0.\s.75 radius in the Sextractor catalog based on the {\it HST}
  mosaic in EGS (see Figure~\ref{multi_z.eps}). Here we consider all
  objects in the {\it pcat} photometric catalog with errors on both
  $(B-R)$ and $(R-I)$ lower than 0.2, not only those for which
  spectroscopy was obtained.  The highest peaks lie off the main body
  of points and are dominated by just a few galaxies -- their colors
  are likely compromised by bad photometry due to multiplicity.  The
  remaining multiples lie mostly in the extreme bottom and lower-left
  of the disitribution, which is dominated by distant blue galaxies,
  most at $z>1.4$.  Inspection of the {\it HST} images reveals many
  subclumps within single objects but an even larger fraction of
  apparently distinct companion galaxies (see \S\ref{multiplicity}).
  Based on their location in color-color space, many of these
  multiples must lie near $z \sim 2$.  }
\label{smoothedmultiples.eps}
\end{figure}

These trends can be followed to even greater $z$ by using the location
of multiple galaxies in the color-color diagram as a crude redshift
guide.  Figure~\ref{smoothedmultiples.eps} plots the distribution of
the fraction of {\it pcat} objects which have multiple components
within 0\s.75 in the {\it HST} imaging over color-color space.  The
highest peaks in multiplicity rate lie away from the bulk of the
galaxy color distribution; they tend to be regions containing only one
or two galaxies.  Inspection suggests that the apertures used to
measure CFHT 12K photometry for these objects were incorrect due to
the multiple components, so that magnitudes and colors were
compromised, yielding peculiar colors.  Setting aside these outliers,
the main body of multiples cluster strongly in the lower-left part of
the diagram, where $z \gtsim 1.5$ blue galaxies may be found (cf.\
Figure~\ref{allcolor.eps}).  Evidently most multiples are quite
distant, beyond the $z\sim1.4$ limit where [O {\scriptsize II}] passes
beyond our spectral coverage.  This explains why relatively few
multiples were both targeted for spectroscopy and yielded reliable
redshifts (as mentioned above).

The increase in multiplicity rate at higher redshifts inferred from
Figure~\ref{smoothedmultiples.eps} agrees with the trend found
previously in Figure~\ref{multi_z.eps}, which showed a large increase
setting in around $z \sim 1$.  Inspection of the {\it HST} images of
the bluest and presumably most distant multiples in
Figure~\ref{smoothedmultiples.eps} shows a similar mix as at $z \sim
1$: a small subset of pairs with large brightness ratios, a moderate
number of galaxies consisting of multiple subclumps, and a large
fraction of apparently well separated, distinct companions which are
blended in ground-based photometry.

To sum up, the fraction of DEEP2 galaxies down to $R_{\rm AB} = 24.1$
whose ground-based CFHT photometry is blended due to purely random
superposition of physically unrelated objects along the line of sight
is of order 1-2\%.  These objects dominate the population of
significantly blended galaxies below $z\sim1$.  At higher redshifts,
the rate of single galaxies which are broken up into multiple clumps
in the {\it HST} data by Sextractor goes up; at the same time, the
occurrence of physically-associated pairs of galaxies that are
well-separated in {\it HST} images also increases.  This population
appears to extend to higher redshift ($z\sim 2$.  It is intriguing to
speculate that a different phase of galaxy formation has been detected
at these higher redshifts, in which a larger fraction of the mass
accretion occurs in major mergers rather than in minor mergers or
smooth flows.  A further inference is that ground-based photometry of
blue galaxies beyond $z \sim 1$ is significantly contaminated by
companions as much as 5-10\% of the time.

This investigation into multiplicity is qualitative and has barely
touched the surface of this subject.  It does, however, give some
rough indications for the number of contaminating blends expected in
ground-based data and their likely effects.  These issues are likely
to be important for deep weak lensing studies, as close pairs may
preferentially align along filamentary large-scale structure and their
photometric redshift estimates will be contaminated; photometric
redshift calibration requirements for next-generation projects such as
LSST are extremely tight (Ma, Hu \& Huterer 2006).  Future work will
benefit from comprehensive spectroscopic redshift surveys extending
beyond $z = 1.4$, better photometric redshifts for distant galaxies,
and near-IR {\it HST} images, which will establish whether the
separate-appearing blue clumps in the {\it HST} optical images are in
fact distinct galaxies or merely blue star-forming clumps within
larger potential wells.  The new CANDELS survey, which includes the
Extended Groth Strip amongst its fields, will provide an ideal testbed
for this study (Grogin et al.~2011; Koekemoer et al.~2011).

\section{Data Tables and DR4 Reference Guide}
\label{datatables}

The two tables in this section contain the main data from DEEP2 Data
Release 4 (DR4), which is the first release of the complete set of
spectra from the DEEP2 survey. Future data releases are planned as
redshifts are further improved and additional spectral quantities
(e.g., EWs, velocity widths) are added. Only the first few lines of
each table are presented here; the complete tables are available
electronically at \url{http://deep.berkeley.edu/DR4/}. In addition to the
tables summarized herein, DR4 includes a variety of other survey data
products, such as the design parameters for each DEIMOS slitmask,
sky-subtracted 1-d and 2-d DEIMOS spectra, and the CFHT 12K $BRI$
photometric catalogs from which the targets were selected. Below, we
provide a brief list of the primary DEEP2 DR4 data products along with
the relevant URLs: 

\vskip 0.1in
\indent \mbox{{\bf DEEP2 Redshift Catalog (Table 10):}} \linebreak
\indent \indent \indent \url{http://deep.berkeley.edu/DR4/zcatalog.html} \\
\vskip 0.1in
\indent \mbox{{\bf Summary of Slitmask Observations (Table 9):}} \linebreak
\indent  \indent \indent \url{http://deep.berkeley.edu/DR4/masktable.html} \\
\vskip 0.1in
\indent \mbox{{\bf Slitmask Design Parameters:}} \linebreak
\indent  \indent \indent \url{http://deep.berkeley.edu/DR4/maskdesign.html} \\
\vskip 0.1in
\indent \mbox{{\bf 1-d and 2-d DEIMOS Spectra:}} \linebreak
\indent  \indent \indent \url{http://deep.berkeley.edu/DR4/spectra.html} \\
\vskip 0.1in
\indent \mbox{{\bf 2-d Completeness Maps (Figs.\ 1 and 2):}} \linebreak
\indent  \indent \indent \url{http://deep.berkeley.edu/DR4/completeness.html} \\
\vskip 0.1in
\indent \mbox{{\bf CFHT 12K $BRI$ Photometric Catalogs:}} \linebreak
\indent  \indent \indent \url{http://deep.berkeley.edu/DR4/photo.html} \\
\vskip 0.1in

Table~\ref{table.maskdata} lists data on the individual slitmasks, one
line per observation.  In some cases, multiple masks with the same
mask number but covering different objects were observed due to the
mask centers being adjusted early in the survey; each mask/date
combination in the table corresponds to a unique slitmask.  For each
mask, this table records the four-digit mask ID number; the
observation date; a nominal RA and Dec near the center of the mask;
the position angle of the long axis of this mask; the number of
objects from this mask in the DR4 {\it zcat} data release (and hence
the number in Table~\ref{table.galaxydata1} below); the estimated
signal-to-noise ratio per pixel ($S/N$) for an $R=23.5$ galaxy
spectrum on the mask; the estimated seeing; the redshift completeness
obtained on this mask for all galaxies targeted; and the redshift
completeness obtained for red galaxies only ($R-I > 0.5$).

The first two digits of the mask ID number indicate the CFHT pointing,
using the same numbering scheme as in
Figures~\ref{cooper.deep2.wfn.eps}a and \ref{cooper.deep2.wfn.eps}b,
while the last two digits are a position code within each
pointing.\footnote{In Fields 2,3,4, the DEEP2 survey as designed
  consisted of three CFHT 12K pointings per field (see
  Figure~\ref{cooper.deep2.wfn.eps}.  The position code in each
  pointing has values 0-39 in the lower-declination row and 40-79 in
  the upper row, both in order of increasing RA.  Vertical masks in
  the ``fishtail'' at low RA are numbered 80 and 81.  This numbering
  scheme permits a total of 243 masks per field, not 120, but not all
  numbers are filled.  Field 1 (EGS) is covered by four overlapping
  pointings divided into eight blocks (see
  Figure~\ref{cooper.deep2.egs.wfn.eps}.  In each pointing, there are
  16 masks that run perpendicular to the strip, numbered 0-39 and
  ordered by increasing RA.  There are 14 masks, seven in each block,
  running parallel to the strip, numbered 40-79, and again ordered by
  increasing RA.  This allows for a total of 320 masks in EGS, but
  again not all numbers are filled.}  Completenesses are based on
reliable redshifts only, which are defined as having quality code
$Q=3$ or 4, and seeing is the FWHM measured from stars in the slitmask
alignment boxes.  The $S/N$ estimate used is the continuum
signal-to-noise per 1-d pixel in the extracted spectrum, computed from
skyline-free regions near 6900 \AA \ with 3$\sigma$ outliers
rejected. The value reported is based on a median-median line fit to
the S/N vs. magnitude for objects with $23.1 < R_{AB} < 24.1$,
evaluated at $R = 23.5$.

\begin{figure*}
\ifpdffig
\begin{center}
\includepdf[scale=0.75]{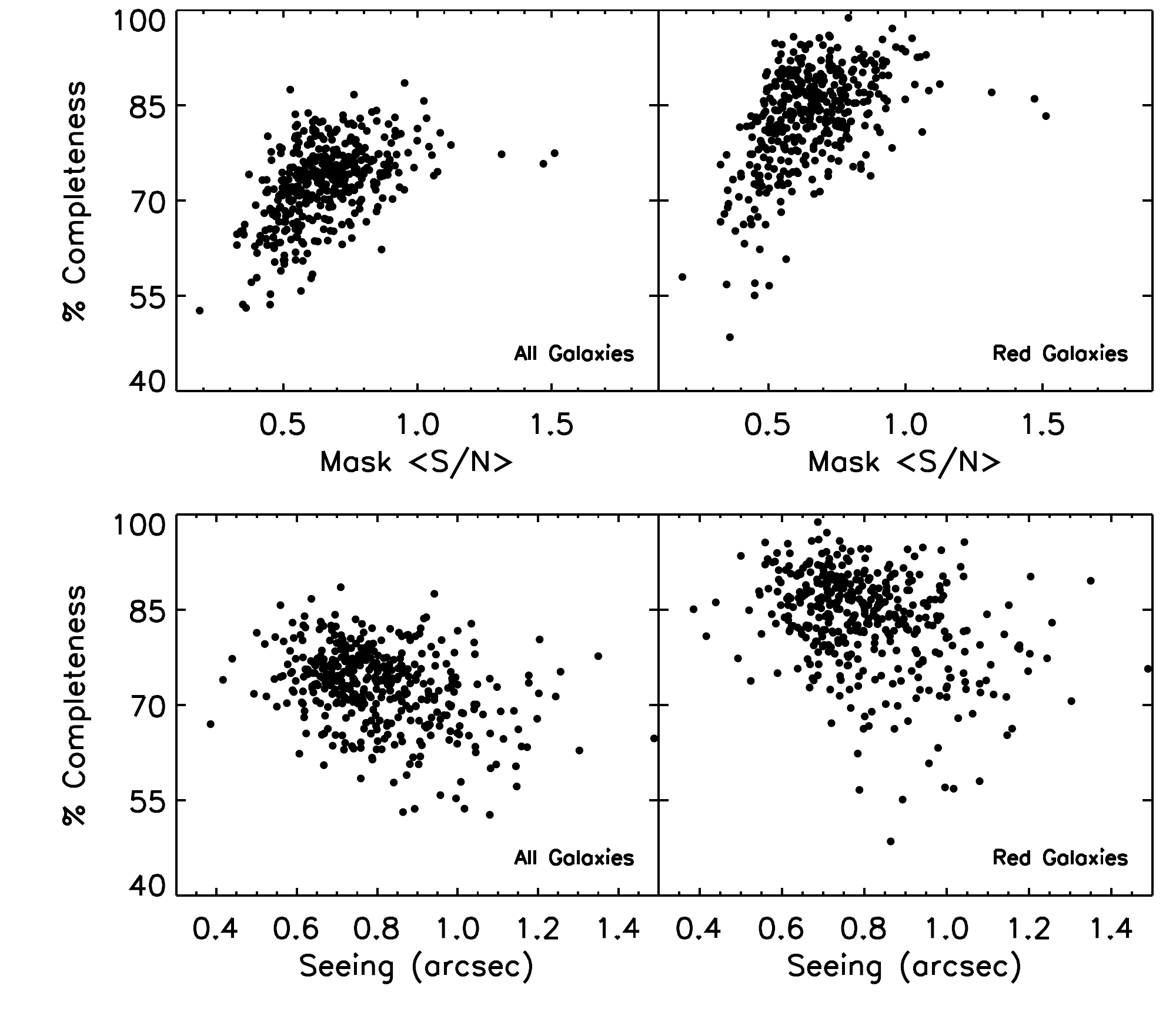}
\end{center}

\vspace*{-0.25in}
\else
\begin{center}
\includegraphics[scale=0.55]{sn_plots.eps}
\end{center}
\fi
\caption{ Four panels comparing various indicators of the quality of
  observations in each slitmask.  In all cases the y-axis is the
  percentage of reliable redshifts with $Q \ge 3$.  The left two
  panels evaluate redshift success using all galaxy targets on a mask,
  while the ones on the right include only "red" galaxies with $R-I >
  0.5$.  The measure of signal-to-noise ($S/N$) used here is the
  average continuum signal-to-noise value per 1-d pixel (throwing out
  3$\sigma$ outliers) computed in a window around 6900 \AA \ in the
  extracted 1-d spectra, based upon a fit to S/N vs. magnitude for a
  given mask evaluated at $R=23.5$.  "Seeing" indicates the average
  FWHM of the spectra of alignment stars.  The fractional redshift
  success for the full set of galaxies is lower than when we restrict
  to ``red'' galaxies because the former sample contains large numbers
  of blue objects which lie beyond $z = 1.4$.  The excluded blue
  galaxies influence the all-galaxy redshift success due to variations
  in their number from mask to mask from clustering; they fail to
  yield redshifts even in good conditions because of their lack of
  features in the DEEP2 spectral window.  As would be expected, for
  red galaxies our redshift success rate correlates with both
  signal-to-noise and seeing (which itself strongly influences S/N).
}
\label{sn_plots.epsi}
\end{figure*}

Both redshift completeness fractions are useful indicators of mask
quality and are plotted along with seeing and $S/N$ in
Figure~\ref{sn_plots.epsi}.  We report both as, for galaxies bluer
than $(R-I)=0.5$, the redshift completeness obtained does not
correlate with S/N or seeing; instead, it likely reflects only cosmic
variance in the number of bright $z>1.4$ galaxies in a given mask
region.  The trends are as expected, with the fractions of reliable
redshifts increasing strongly with both $S/N$ and seeing, which are
closely correlated.  Probably the single most useful mask-quality
indicator is the seeing FWHM, which ranges between 0\s.4 and 1\s.5
with a median value of 0\s.76. 

Table~\ref{table.galaxydata1} is the master data table for DEEP2,
presenting redshifts and associated data for each individual candidate
galaxy targeted for spectroscopy.  Each line corresponds to a single
observed slitlet spectrum.  Multiply-observed objects have multiple
entries, though a $z_{best}$ for each object is provided based on
information from all spectra (see below).  Data are given for targeted
objects only; many serendipitous objects are visible on the 2-d
spectra, but they have been uniformly analyzed.  Their presence is
signaled by the comment code {\it offser} (``offset serendip'') in the
comment field of the targeted galaxy (see below).

Table~\ref{table.galaxydata1} is similar to the DEEP2 {\it zcat} table
at \url{http://deep.berkeley.edu/DR4/}, but certain confusing or
rarely-used quantities have been omitted (we describe those quantities
in \S\ref{extraparams}), while some other quantities derived from mask
design files have been added (the dropped quantities continue to be
available in the {\it zcat} available online.)
Table~\ref{table.galaxydata1} is thus a melding of data in the DEEP2
DR4 spectral database, the previously-released {\it pcat} photometry
catalogs (Coil et al.~2004b)\footnote{See also
  \url{http://deep.berkeley.edu/DR4/photoprimer.html}}, and data from
the slitmask design files, collected together for easy reference.

Objects in Table~\ref{table.galaxydata1} are ordered by their object
number in the DEEP2 {\it pcat} catalogs.  The variable names in caps
are defined here or, if pre-existing, agree with published catalogs
and webpages.  Sources of all entries are given in footnotes to the
table.  The columns are as follows:

{\it Col. (1) --} OBJNO: A unique 8-digit DEEP2 object number drawn from the 
{\it pcat} photometric catalogs.     The object number format is {\it XXyyyyyy}, where 
{\it XX} indicates the CFHT field/pointing used to derive the photometry
(e.g., ``23'' for DEEP2 Field 2/Pointing 3) and 
{\it yyyyyy} is a unique identifier within the pointing. The fields and
pointings are shown in Figure~\ref{cooper.deep2.wfn.eps}
and Figure~\ref{cooper.deep2.egs.wfn.eps}.

{\it Col. (2) --} RA: Right ascension of the object, in decimal degrees.

{\it Col. (3) --} DEC: Declination of the object, in decimal degrees.

{\it Col. (4) --} MAGB: $B$-band apparent AB magnitude $m_B$ from the
{\it pcat} CFHT photometry.  This is defined as $m_B = m_R +
(B-R)_{1\asec}$, where $(B-R)_{1\asec}$ is the apparent color measured
through a 1\asec-radius aperture and $m_R$ is the apparent $R$
magnitude (next column).  Using a small fixed aperture for the color
yielded the most robust colors and total magnitude corrections from
$R$ to $B$ and $I$ (Coil et al.~2004b).

{\it Col. (5) --} MAGR: $R$-band apparent AB magnitude $m_R$ from the
{\it pcat} photometry.  This is measured through a circular aperture
whose radius is three times the Gaussian radius, $r_g$, defined in
column (10).  If $3r_g < 1$\asec, the magnitude is measured through a
1\asec-radius aperture (Coil et al.~2004b).  All magnitudes provided
have been corrected for Galactic extinction.

{\it Col. (6) --} MAGI: $I$-band apparent AB magnitude $m_I$, defined
as $m_I = m_R - (R-I)_{1\asec}$,where $(R-I)_{1\asec}$ is measured
through a 1\asec-radius aperture.  See column (4).

{\it Cols. (7), (8), (9) --} MAGBERR / MAGRERR / MAGIERR: RMS
$B$-band/$R$-band/$I$-band magnitude error ($\\sigma_B$ / $\sigma_R$ /
$\sigma_I$).  These error estimates include sky and photon noise only.
Based on the analyses presented in Coil et al.~(2004b), there may be
additional systematic zeropoint errors internal to each pointing of
order 0.04 mag in $B$ and 0.02 mag in $R$ and $I$, but mean zeropoints
should differ from pointing to pointing by $<0.01$ mag.

{\it Col. (10) --} RG: Gaussian radius $r_g$ of a circular 2-d
Gaussian fit to the $R$-band image, expressed in units of CFHT pixels
(one pixel is 0\s.207).

{\it Cols. (11), (12) --} EL / PA: Ellipticity $e_2$ (EL) and position
angle (PA) of the object's CFHT $R$-band image, derived from an
analysis of image moments.  The ellipticity parameter $e_2$ is defined
as $e_2=(1-b/a)=2\epsilon/(1+\epsilon)$, where $\epsilon$ is the
conventional eccentricity, and PA is relative to north.  Note that
position angles determined from ground-based images do not necessarily
match those from high-resolution Hubble images; e.g., see Wirth et
al.~(2004).

{\it Col. (13) --} PGAL: Probability of being a galaxy $p_{gal}$,
based on the $R$-band image.  A value $p_{gal} = 2$ indicates a source
with unusual $BRI$ colors; $p_{gal} = 3$ indicates an extended source.
All sources with $p_{gal} = 2$ or 3 are treated as galaxies in target
selection.  A value of $p_{gal}$ in the range 0.0-1.0 is the Bayesian
probability of a compact object's being a galaxy (not a star).
Objects with $p_{gal} < 0.2$ are treated as stars and excluded from
selection for spectroscopy; objects with $p_{gal}$ between 0.2 and 1.0
are included with a selection weight proportional to their value of
$p_{gal}$; and those with $p_{gal}$>1 are given the same weight as an
object with $p_{gal}=1$.  For further explanation see \S\ref{sample}
and Coil et al.~(2004b).

{\it Col. (14) --} SFD\_EBV: Galactic reddening $E(B-V)$ from
Schlegel, Finkbeiner \& Davis (1998).  This value was used to correct
the photometry for a given object for Galactic extinction.

{\it Col. (15) and (16) --} M\_B / UB\_0: Absolute CFHT 12K $B$-band
magnitude $M_B$ and restframe $U-B$ color corrected for Galactic
reddening, ${U-B}_0$ , both computed from the $BRI$ photometry as
described in Willmer et al.~(2006).  We use the subscript 0 here to
indicate that the calculations are done with $z=0$ passbands; no
correction for dust internal to DEEP2 galaxies has been applied.
$M_B$ is calculated assuming an LCDM cosmology with $h = 1$; i.e., the
listed value is $M_B - 5 log_{10} h$.

{\it Col. (17) --} OBJNO: repeated from {\it Col. (1)}.

{\it Col. (18) --} MASK: Slitmask number on 
which the object was observed, from Table~\ref{table.maskdata}.  

{\it Col. (19) --} SLIT: Slitlet number on which the object
was placed.  Slitlets are numbered in order of position along
the long axis of the mask, starting with Slitlet 0 lying to the far 
left on Figure~\ref{deimosfocalplane.thar.epsi}.

{\it Col. (20) --} DATE: UT date of observation in YYYY-MM-DD format. 
 
{\it Col. (21) --} MJD: Modified Julian date of observation. 

{\it Col. (22) --} SLITRA: RA of slitlet center.

{\it Col. (23) --} SLITDEC: Declination of slitlet center.

{\it Col. (24) --} SLITLEN: Slitlet length in arc seconds.

{\it Col. (25) --} SLITPA: Slitlet PA relative to north.

{\it Col. (26) --} Z: Observed best-fitting redshift $z$ from this
spectrum, in the geocentric reference frame.  This is the value
obtained from the {\it zspec} process.

{\it Col. (27) --} ZBEST: Best heliocentric-reference-frame
redshift combining information from all spectra obtained for an
object, $z_{best}^{hel}$.  Uses the highest-$Q$ value or, when there
are multiple choices of equivalent quality, uses the first value
observed.  A value of $z_{best}^{hel}$ is given only if there is at
least one redshift with $Q \ge3$.

{\it Col. (28) --} ZERR: Redshift error $\sigma_z$ from this observation, as
explained in \S\ref{redshiftmeasurements}.

{\it Col. (29) --} ZQUALITY: Redshift quality code $Q$ for this
observation.  Codes are as follows: $Q = -2$ (data so poor that object
was effectively never observed; tracked for statistical purposes);
$Q=-1$ (star); $Q = 1$ (probable galaxy but very low $S/N$; data are
not likely ever to yield a redshift); $Q = 2$ (objects with low $S/N$
and/or for which data are in some way compromised, but for which a
redshift may be obtainable with extra effort. The reason for assigning
$Q=2$ is generally listed in the COMMENT column (below) -- all {\it
  supser}'s and {\it sngls}'s automatically get $Q = 2$), $Q = 3$
(reliable redshift with probability of accuracy $\ge 95$\%; $Q=4$
(reliable redshift with probability of accuracy $\ge 99$\%).  Further
details are given in \S\ref{redshiftmeasurements}, while the relevant
comment codes are explained below.

{\it Col. (30) --} COMMENT: A comment provided in the course of the
{\it zspec} and catalog compilation process.  The comment can be used
to identify a variety of problems or issues with the data (see also
\url{http://deep.berkeley.edu/DR3/comments.html}).  As noted in
\S\ref{redshiftmeasurements}, their application has varied, and they
should be used with caution; many are a matter of judgment and may
have been applied differently by different users.  See further remarks
below.

There are two separate categories of problems that are given comments,
those that have a high probability of impairing redshift quality and
those that may or may not do so.  All $Q=2$ redshifts should have at
least one comment from the first category; where none is assigned, the
"{\it marg}" keyword (marginal S/N) may be assumed.  The first
category consists of:

{\tt bsky:} Bad sky subtraction. This comment highlights badly
subtracted sky continuum or sky lines or both.  An incorrect sky
spectrum shape results in incorrectly subtracted sky, and consequent
errors in the extracted object spectrum.  Examples are shown in
Figure~\ref{bad_data.epsi}.

{\tt bcol:} Bad column(s). One or more bad CCD columns spoil a
significant part of the extracted spectrum and/or the neighboring sky
region.  Note that ``columns'' here run parallel to the dispersion
direction; i.e., the name of the comment code refers to columns on the
detector, rather than columns within the extracted 2d spectrum.  This
is the opposite convention from
Figure~\ref{deimosfocalplane.thar.epsi}, where they run parallel to
the slitlet direction.  The pipeline interpolates across bad columns
and manages to repair a single bad column fairly well, but three or
more adjoining columns seriously degrade data quality.  {\tt bcont}
(and {\tt bs}) may also be assigned when a piece of spectrum that is
curved due to optical distortion falls onto a vignetted region columns
or onto a gap between CCDs.  An example is shown in
Figure~\ref{bad_data.epsi}.

{\tt bcont:} Bad continuum shape. This is a catchall phrase expressing
the fact that the continuum shape does not look right.  It is often
associated with {\tt bsky} or {\tt bcol}.
  
{\tt bext}: Bad extraction window.  The extraction window used by the
pipeline for the spectrum is displaced from the actual object, or is
too wide or too narrow, or is contaminated by a companion.

{\tt edge:} Edge. The object is too near the end of the slitlet and
consequently has poor sky subtraction and/or extraction.

In the case of these problems ({\tt bcol, bcont, bext, bs and edge}),
whether an object is assigned $Q=2$ (redshift possibly measurable),
$Q=1$ (redshift not measurable due to the object properties) or $Q=-2$
(object was effectively not observed) will sometimes involve judgement
calls, which may differ from one {\it zspec} user to another.

{\tt disc:} Discontinuity. There is an unphysical jump in the
continuum level between the blue and red-side spectra.  This is
typically caused by mismatched extraction windows on the red and blue
sides or by bad sky subtraction on one side.  An example is shown in
Figure~\ref{bad_data.epsi}.

{\tt marg:} Marginal : The $S/N$ is 
low and the spectrum barely meets standards for its assigned
quality code.

{\tt sngl:} Single. Only one feature is visible (we count a resolved
[O {\scriptsize II}] doublet as two features).  All {\it sngl}'s
automatically get $Q = 2$.

{\tt supser:} Superimposed serendipitous object ("serendip").  There
are two redshifts measurable from the spectrum, and hence it is
ambiguous what redshift should be assigned to the target.  If one or
both spectra have only one line, {\tt supser sngl} is used.  Either
$z$ can be entered into the {\it zspec} file, and the second $z$ is
not necessarily recorded. All {\it supser}'s get $Q = 2$.  We note
that {\it supser}'s are among the most difficult phenomena to spot,
and many have probably been missed.

The second category of comments represents conditions that
can often be present without impairing redshift quality:

{\tt fix:} Fix. Some element of the reductions went wrong and should
be redone.  The redshift is not necessarily affected.

{\tt iffy:} Iffy. The extracted 1-d spectrum may be adequate for
determining a redshift (as indicated by the $Q$ code), but there is
something wrong that makes the observation unreliable as an {\it
  integrated spectrum} of the galaxy, and therefore unsuitable for
other analyses such as measurement of equivalent widths, rotation
curves, linewidths, or SEDs.  Examples include: (1) a bad extraction
window that gets only part of a galaxy or is contaminated by a
companion, (2) emission extending beyond the extraction window and
therefore under-summed in the 1-d spectrum (this may also result in
oversubtracted sky, as shown in Figure~\ref{bad_data.epsi}.), (3) a
severe flux discontinuity across the blue/red CCD boundary {\tt disc}.
With rare exceptions, {\tt iffy} was used only for objects assigned
$Q=3$ or $Q=4$.  It may be further qualified by adding {\tt bs}, {\tt
  bcont}, {\tt bcol}, {\tt disc}, {\tt bext}, or {\tt edge} to explain
the nature of the problem.

{\tt offser:} Offset serendip. A second spectrum is present in the
slitlet but is well enough separated or dim enough that the extraction
window and sky-subtraction of the target object are not seriously
affected.  If only one line is visible, {\tt offser sngl} is used. The
redshift of the serendip is not necessarily recorded.  Serendip
spectra may be obtained by downloading the 2-d slitlet spectra at
\url{http://deep.berkeley.edu/DR4/}.  {\it Offser}'s are easier to spot than
supser's but, again, some have probably been missed.

{\tt fill-gap failed:} Identifies those instances where a redshift was
not able to be identified because the {\it spec2d} data reduction
pipeline failed to construct a robust one-dimensional spectrum for the
object. In the vast majority of cases, this resulted from an
exceptionally poor wavelength solution, by which the blue and red
halves of the object spectrum (see \S\ref{maskdesign} and
\S\ref{pipeline}) overlapped in wavelength space rather than being
separated by a $\sim \! 5$\AA corresponding to the gap between the two
DEIMOS CCDs.

{\tt ZREVISED:} Identifies those objects for which the redshift was
``fixed'' based on information obtained outside the first {\it zspec}
analysis of the mask. This includes objects whose redshifts were
assigned following the $Q=2$ rechecking process; for more details see
\S\ref{redshiftmeasurements}.

{\tt ZREVISED: zfix failed:} Identifies those cases where a redshift
revision was attempted (see {\tt ZREVISED}), but failed. This failure
generally resulted from the inability of the \emph{spec1d} redshift
pipeline to find a suitable fit that matched the revised redshift
value (e.g., within $\Delta z = 0.01$).

{\tt ZMATCH-FAIL:} Identifies those cases where a redshift that was
previously identified as $Q = 3,4$ was refit, but the \emph{spec1d}
redshift pipeline failed to yield a good fit that matched the previous
redshift value (e.g., within $\Delta z = 0.01$). One motivation for
this refitting process was to account for changes in the spectral
reductions after a given slitmask had been visually inspected using
the \emph{zspec} tool.

\subsubsection{Additional Parameters in Online DR4 Catalogs}
\label{extraparams}

In addition to the quantities described above, the {\it zcat} redshift
catalog distributed at the DR4 website includes several additional
parameters which are of more limited utility or as yet only partially
vetted. These are:

{\tt OBJNAME:} The DEEP2 object number in string format, used
to define filenames. This can differ from OBJNO in overlap regions
between CFHT pointings, where OBJNAME was defined from the object
number used for object selection, whereas OBJNO indicates the instance
of an object with the best photometry.

{\tt CLASS:} Indicates the class of template which yielded the best
redshift fit in the {\it zspec} process; possible values are 'STAR',
'GALAXY', or [broad-line] 'AGN' (see \S\ref{redshiftpipeline}).  It
should be noted that this only indicates the best-fitting template;
many AGN (both broad- and narrow-line) in the DEEP2 sample will have
'GALAXY' as their class.

{\tt SUBCLASS:} For objects fit with stellar templates, indicates the
spectral type of the best-fitting template.  Note that only a sparse
set of spectral types was used in the fits, so this does not
necessarily match the actual spectral type of the target.

{\tt RCHI2:} Reduced chi-squared for the best fit template compared to
the observed spectrum.  Based on results of the {\it zspec} process,
this value is not necessarily indicative that a given redshift is good
or not.

{\tt DOF:} The number of degrees of freedom used to determine RCHI2.

{\tt VDISP:} The estimated velocity dispersion of the given galaxy (in
\kms), determined as described in \S\ref{redshiftpipeline}.  VDISP
values appear to be a good estimate of velocity dispersion, but may be
subject to small systematics, especially at low VDISP.

{\tt VDISPERR:} The estimated uncertainty in the VDISP value for a given galaxy.

\section{Two-dimensional Selection Functions}
\label{selectionfunction}

As part of DEEP2 Data Release 4, we are releasing a set of files
describing the two-dimensional selection function of the DEEP2 Galaxy
Redshift Survey: i.e., the probability (as a function of position on
the sky) that an object meeting the DEEP2 target criteria (magnitude
limit and color cut, as applicable) is selected for observation and
successfully yielded a redshift.  The calculation of these selection
function (or, as they are sometimes labelled,``window function'') maps
has been described in Coil et al.~(2008) and references therein; we
review here.

The two-dimensional selection function maps take into account the
actual placement and geometry of the overlapping slit masks used for
DEEP2, as well as vignetting in the DEIMOS camera and the locations of
gaps between the DEIMOS CCDs.  In producing these maps, we account for
the multipass nature of DEEP2 targeting and the fact that some
overlapping DEEP2 slitmasks were designed simultaneously and some were
not (as the mask design code will not place an object on multiple
masks designed at the same time and in the same row of masks; cf.\
\S\ref{maskdesign}) in computing selection probabilities.  The
redshift completeness for red ($R-I >0.5$) galaxies is used to
determine the probability a targeted object yields a redshift; the
redshift success rate for bluer objects is both low and not correlated
with observing statistics such as seeing (rather it primarily reflects
the density of $z \sim 2$ galaxies at a given location on the sky), so
the red galaxy completeness provides the best estimate of the
probability that a targeted galaxy at a redshift where DEEP2 can
obtain a $z$ will actually yield a reliable redshift measurement.

In these maps, we also mask out all regions where the photometric data
are affected by either saturated stars or CCD defects and hence no
galaxies were targeted. The region where these maps are nonzero
provides a geometrical outline of the regions where DEEP2 targeted on
the sky, while the actual value at a given position represents the
probability that we both targeted a galaxy (which depends on how many
masks overlap a given area, whether the object is in the first or
second-pass region of those masks, and whether they were designed
simultaneously) and then successfully measured its redshift.  Note
that the provided files do not include pointing 14 in the northernmost
region of the EGS. Due to poorer photometry, the mask design differed
in that pointing from the rest of the survey, such that the spatial
selection is not uniform between it and the rest of DEEP2.  No DEEP2
clustering measurements use data from that pointing.  In an upcoming
paper (Matthews et al.~2012) we will provide improved photometry for
this region based on imaging obtained as part of the CFHT Legacy
Survey (Gwyn 2012).

The selection function maps are distributed at the DEEP2 DR4 website
in the form of FITS-format images, with World Coordinate System
headers describing the mapping from right ascension and declination to
pixel, and the value at a given pixel (ranging from 0 to 1) being the
combined selection and redshift success probability for a DEEP2 target
galaxy at that position. Outside of the EGS, where the mask-making was
done independently for each CFHT pointing, we provide one file per
pointing.  For the EGS, we provide one file describing the selection
function over the entire field.  In order to produce random catalogs
for calculating correlation functions, one should use the value of the
selection function at a given position as the probability of keeping
an object (between 0 and 1) at that location within the catalog.  We
note that for some measurements, it may also be necessary to correct
for the dependence of selection probability on source density at small
scales (cf.\ \S\ref{seleffects}); in DEEP2 science papers, this has
been done with mock galaxy catalogs (e.g.\ Yan, White, \& Coil 2004).

\section{Trends with Redshift}
\label{redshifttrends}

We conclude this overview of the DEEP2 survey by illustrating certain
major trends in the data as a function of redshift, summarizing
properties of the DEEP2 dataset.

\begin{figure*}
\ifpdffig
\begin{center}
\includepdf[scale=0.75]{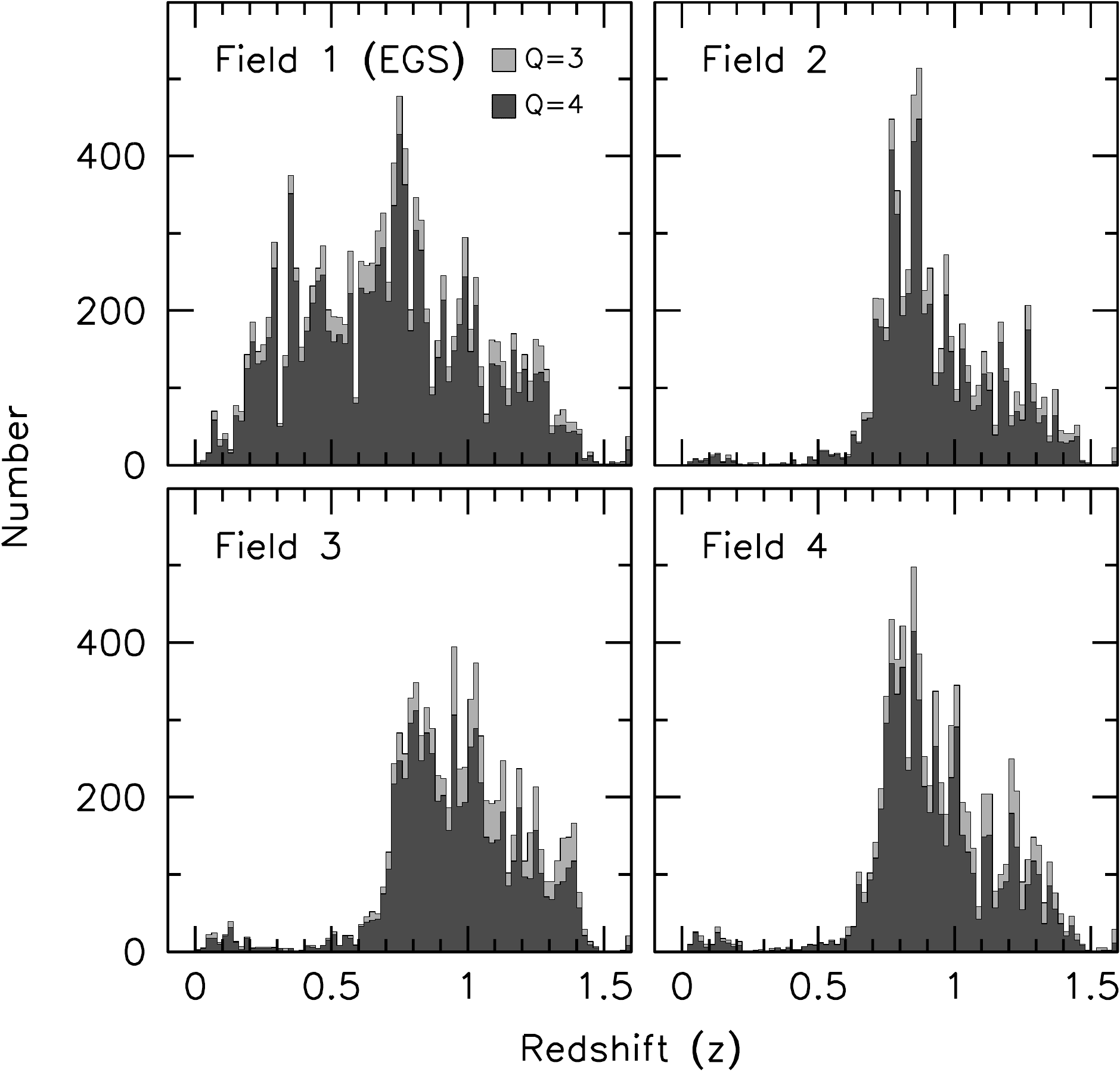}
\end{center}
\else
\begin{center}
\includegraphics[scale=0.6]{zhist_v5.eps}
\end{center}
%
\fi
\caption{The histogram of reliable redshifts in each field.  The
  histogram including only $Q = 4$ redshifts is shown in dark grey; $Q
  = 3$ redshifts are shown in light grey.  }
\label{zhist_0.02.epsi}
\end{figure*}

Figure~\ref{zhist_0.02.epsi} shows redshift histograms for each of the
four fields, using only reliable ($Q = 3$ and $Q = 4$) redshifts.
These figures illustrate graphically that, fractionally, the
contribution of $Q = 3$ ($>95\%$ reliable) redshifts is rather small;
the great majority (83\%) of secure redshifts have $Q = 4$ ($>99\%$
reliable).  Field 1 (EGS) is well sampled at all redshifts, whereas
our $BRI$ color pre-selection has efficiently eliminated foreground
galaxies in Fields 2, 3, and 4.  As expected, strong peaks due to
large-scale structure are evident, which differ in detail from field
to field, demonstrating the need for statistically independent samples
to beat down cosmic variance.

Figure~~\ref{zhist_0.02.epsi} also sheds light on whether a
significant number of of redshifts are lost when important spectral
features fall on atmospheric absorption bands (e.g.\ the A band of
O$_2$ at 7620 \AA).  The [O {\scriptsize II}] $\lambda$3727
emission line should fall on the A band for objects at $z = 1.045$.
There is no depression in the diagrams at that redshift, indicating
that any loss is small.
  
\begin{figure*}
\ifpdffig
\vspace*{0.25in}
\begin{center}
\includepdf[scale=0.825]{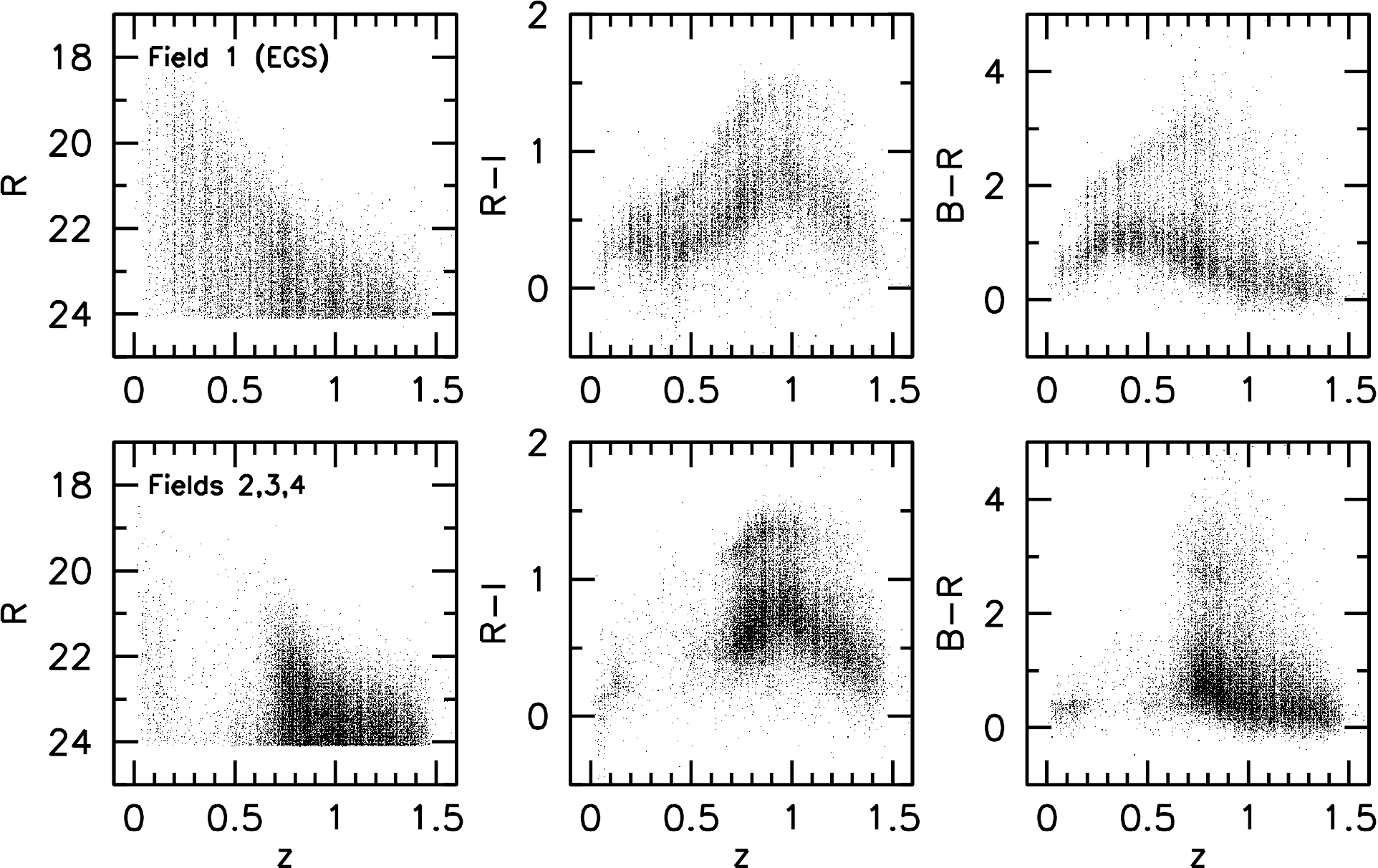}
\end{center}
\else
\begin{center}
\includegraphics[scale=0.825]{colz3_v3.eps}
\end{center}
%
\fi
\caption{Apparent magnitudes and colors of galaxies versus redshift
  for the completed DEEP2 survey.  Only galaxies with reliable
  (i.e. $Q = 3$ and $Q = 4$) redshifts are shown.  The upper row shows
  objects from the EGS, while the lower row consists of objects from
  Fields 2, 3, and 4.  Color bimodality is clearly visible even in
  apparent colors.  }
\label{apparentdatavsredshift.eps}
\end{figure*}

Figure~\ref{apparentdatavsredshift.eps} plots apparent magnitudes and
colors versus redshift.  The top row is for Field 1 (EGS); the bottom
row combines all objects in Fields 2--4. The most striking feature of
these diagrams is the clear bimodality visible in the color plots.
The lower, densely populated sequence consists of blue-cloud galaxies,
while the more thinly populated upper trace is the red sequence.  The
latter fades out near $z\sim1$, due partly to a real reduction in
numbers there (Bell et al.~2004, Willmer et al.~2006, Faber et
al.~2007) and partly because the survey begins to lose red galaxies
rapidly at that redshift.  This is a combined effect of such objects
having difficulty meeting the DEEP2 magnitude limit (as the CFHT $R$
band is shortward of the 4000 \AA\ break past $z=1$) and poorer
redshift success for faint red objects without emission lines (cf.\
the discussion of color-magnitude diagrams below).

Also notable are the two small islands of very blue, $z \sim 0.15$
galaxies visible in Fields 2, 3, and 4.  These reside in the blue
corner of the color-color diagram where the colors of low-redshift,
high star-formation rate galaxies overlap those of more distant blue
galaxies (cf. Figure~\ref{allcolor.eps}); as a result, a $BRI$ color
pre-selection cannot reject them cleanly without also rejecting real
high-redshift galaxies.  Some of these are exceedingly low in
luminosity, with $M_B$ as low as -14 (see below), and likely present an interesting
population of objects for follow-up studies.

The vertical striping visible in these figures (especially for Field
1) is a consequence of cosmic variance: redshifts where the density in
our fields are higher than average will have more objects of all
colors and magnitudes, and those with low density will have few.  The
stripes correspond directly to the features visible in the redshift
histograms in Fig.~\ref{zhist_0.02.epsi}.  In Fields 2-4, the impact
of large-scale structure is much reduced, as we are effectively
averaging the properties of three statistically independent regions of
the Universe.  A significant dearth of galaxies near $z = 1.15$ is
nevertheless still visible, despite combining roughly 22,000 galaxies
in three widely separated fields, illustrating how difficult it is to
reduce cosmic variance to a small error at every redshift (at
$z=1.15$, [O {\scriptsize II}] $\lambda$3727 overlaps with one of the
cleanest regions in the night sky spectrum, eliminating sky
subtraction errors as a possible culprit).  The impact of cosmic
variance increases the smaller the redshift bins ($\delta z$)
considered.  As a result, we find no detectable correlation between
the DEEP2 redshift histogram and the DEIMOS night sky spectrum
(evaluated either at the central wavelength of [O {\scriptsize II}] or
at the wavelength of either doublet component).  Variations in the
DEEP2 redshift success rate on the scale of the width of a night
skyline are completely swamped by the large cosmic variance on those
scales, if they exist at all.

\begin{figure*}
\ifpdffig
\begin{center}
\includepdf[scale=1.0]{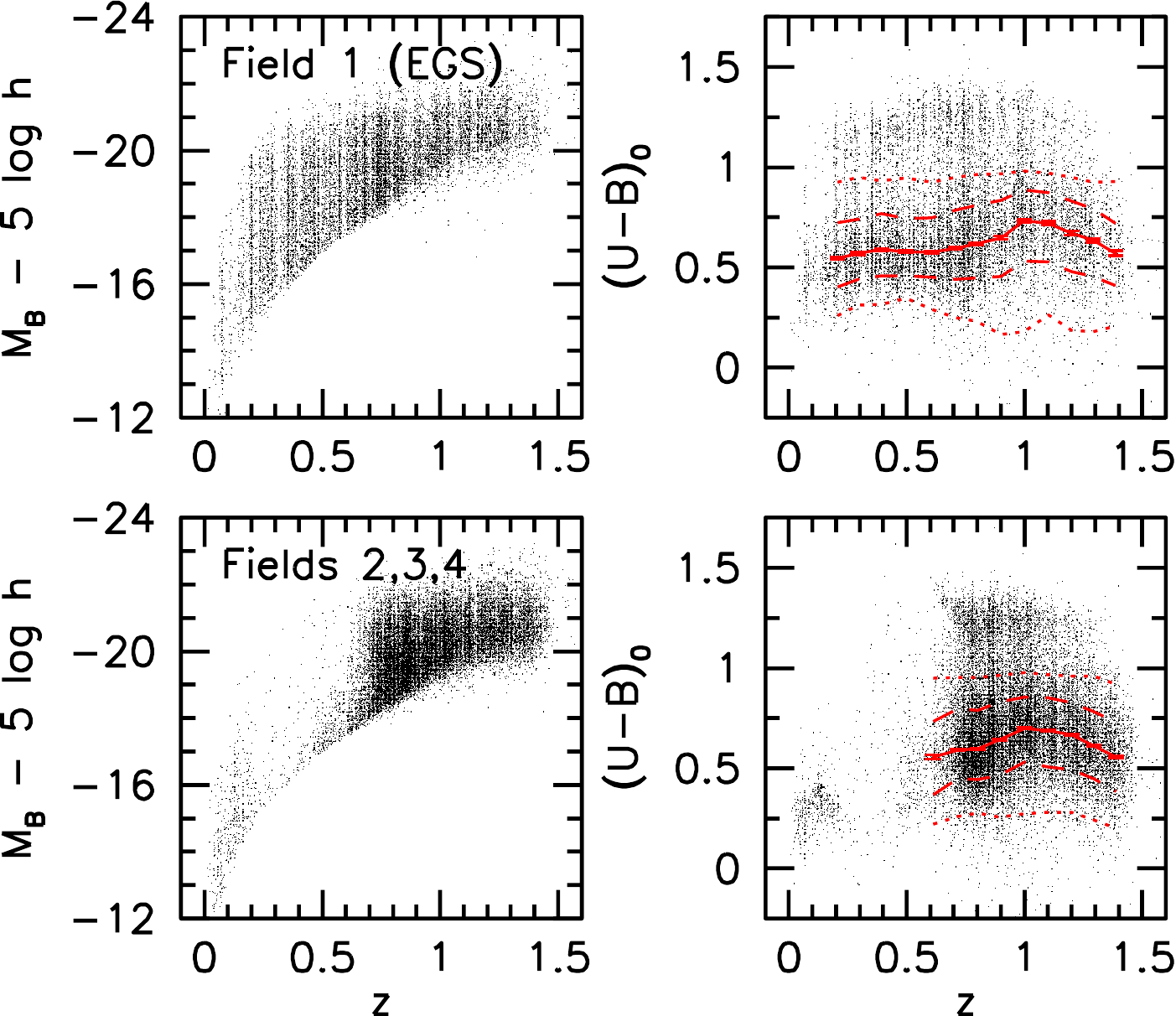}
\end{center}
\else
\begin{center}
\includegraphics[scale=1.0]{colz4_v8.eps}
\end{center}
\fi
\caption{Absolute magnitudes and colors of galaxies versus redshift
  for the completed DEEP2 survey.  Galaxies are the same as in
  Figure~\ref{apparentdatavsredshift.eps}.  The upper row represents
  EGS, the lower row Fields 2, 3, and 4.  Lines in the right panels
  represent the 1- and 2-$\sigma$ ranges for blue-cloud galaxies only.
  We argue that most of the complicated trend evident in the plot of
  restframe $U-B$ color ($(U-B)_0$) vs.~redshift is created by the
  apparent $R$-band magnitude selection limit beating against the
  intrinsic distribution of blue-cloud galaxies in the color-magnitude
  diagram at each redshift.  See discussion in \S\ref{redshifttrends}
  and Figure~\ref{cmdiagvsredshift.eps}.  }
\label{absdatavsredshift.eps}
\end{figure*}

\begin{figure*}
\ifpdffig
\begin{center}
\includepdf[scale=0.8]{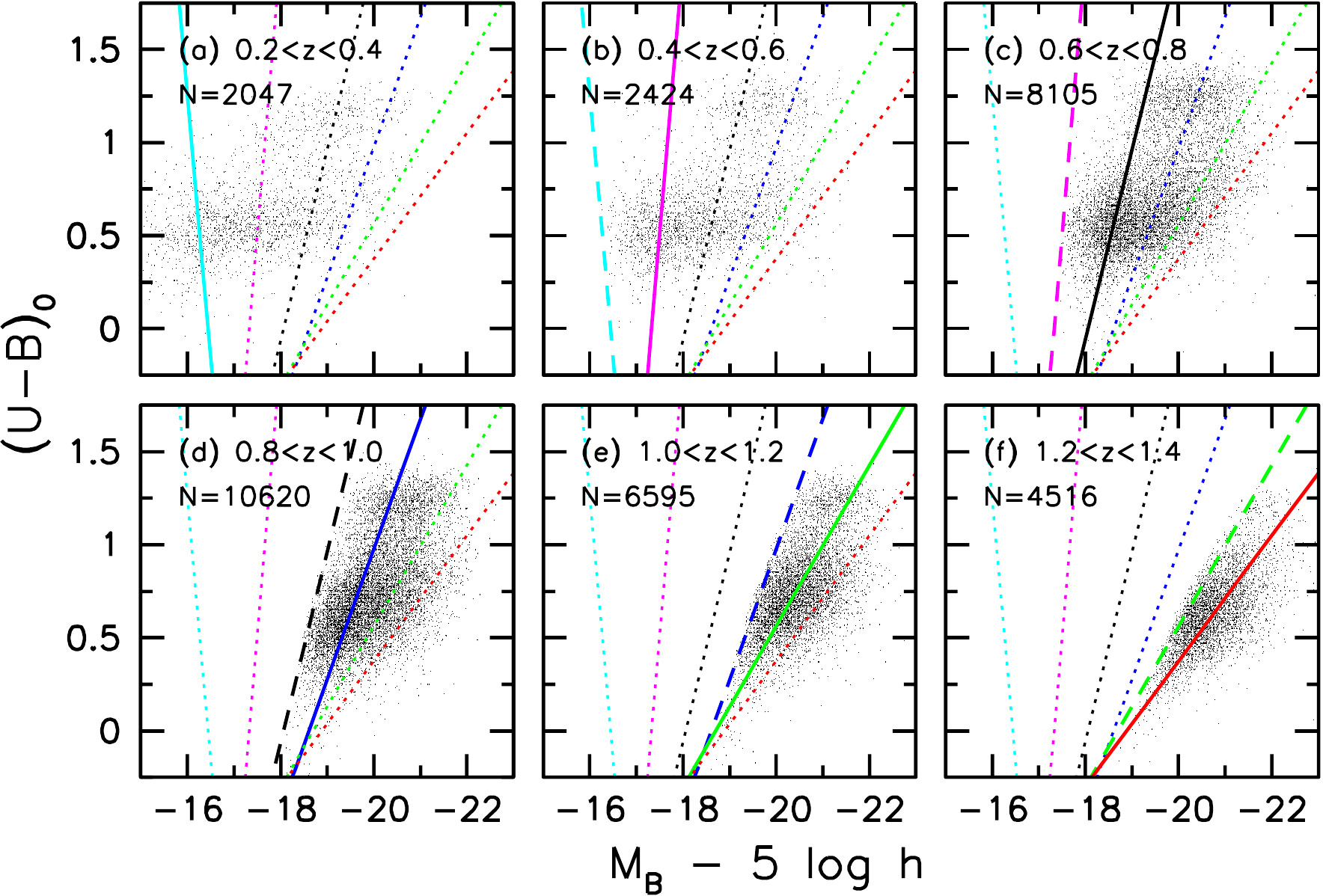}
\end{center}
\else
\begin{center}
\includegraphics[scale=0.8]{colmagz3_v6.eps}
\end{center}
\fi
\caption{Color-magnitude diagrams for all fields, binned by
  redshift. The straight lines approximate how the apparent $R_{\rm
    AB} = 24.1$ limit of the survey translates into restframe color
  and absolute magnitude limits at each redshift.  The solid line in
  each panel is the completeness limit for the far side of
  (i.e. highest redshift within) that bin and represents the absolute
  completeness limit for that bin; the dashed line indicates the limit
  at the near side of a bin.  Dotted lines repeat the limits from
  other redshift bins, using the same colors in all panels.  For
  blue-cloud objects, the magnitude limit favors rather blue galaxies
  at the highest redshifts, redder galaxies at intermediate redshifts,
  and finally bluer galaxies again at low redshifts.  These trends are
  a major reason why the median $(U-B)_0$ line is curved for
  blue-cloud galaxies in Figure~\ref{absdatavsredshift.eps}.  }
\label{cmdiagvsredshift.eps}
\end{figure*}

Figure~\ref{absdatavsredshift.eps} plots absolute $B$ magnitude and
restframe $(U-B)_0$ color versus redshift.  These are calculated from
the CFHT 12K $BRI$ photometry using the methods of Willmer et
al.~(2006); we use the subscript 0 to indicate that restframe
passbands are used (no correction for dust internal to DEEP2 galaxies
has been applied).  The color bimodality and the low-redshift blue
galaxies are again evident.  The most interesting aspect of this
figure is the curved locus in $(U-B)_0$ vs. $z$ traced out by
blue-cloud galaxies.  The solid line shows the median color of blue
cloud galaxies (defined as having ($(U-B)_0 < 1.0$) as a function of
$z$, while the dotted and dashed lines shows quintiles of the
distribution.  Figure~\ref{cmdiagvsredshift.eps} suggests that this
curvature is not due to errors in the K-correction procedure for
restframe $(U-B)_0$ but is instead caused by the interaction of the
fixed $R_{\rm AB} = 24.1$ magnitude limit with the color-magnitude
distribution of galaxies in the blue cloud (see below), combined with
the impact of any color evolution (Blanton 2006).

Figure~\ref{cmdiagvsredshift.eps} plots restframe color-magnitude
diagrams binned by redshift.  The well-known color trends within both
the red sequence and blue cloud are clearly evident, with brighter
galaxies being redder in both cases.  Straight lines approximate the
boundaries imposed by the fixed $R_{\rm AB} = 24.1$ survey limit at
different redshifts.  In each panel, the heavy solid line shows this
limit for the {\it far} side of the indicated redshift bin while the
dashed line is the corresponding limit for the {\it near} side of that
bin. Light dotted lines repeat the lines from all other bins to guide
the eye.

The aggregate trends of $(U-B)_0$ with $z$ seen in
Figure~\ref{absdatavsredshift.eps} can be explained by what is seen
here.  At the highest redshifts, the survey tends to include bluer
blue-cloud galaxies but lose redder ones (at fixed $M_B$), as the
limit lines become highly slanted in restframe color-absolute
magnitude space.  The mean color of the most distant detected
blue-cloud galaxies is therefore by necessity bluer than for
lower-distance samples.  At lower $z$, the $R=24.1$ limit line is more
vertical, permitting redder blue-cloud galaxies to enter the survey
across a wider range in $M_B$.  At even lower redshifts, the survey is
able to pick up abundant, very faint blue galaxies, which causes the
median color of blue-cloud galaxies selected to fall again.  We have
not modeled these effects quantitatively, but it is clear that the
most obvious trends in color plots like
Figure~\ref{absdatavsredshift.eps} need careful interpretation.

\section{Conclusions and Acknowledgements}
\label{conclusions}

This paper has presented DEEP2 Data Release 4, the first spectroscopic
database resulting from a complete DEEP2 Galaxy Redshift Survey
sample, containing 52,989 spectra and 38,348 reliable redshift measurements for
objects as faint as  $R_{\rm AB} = 24.1$.  The information in
this paper is intended as a handbook for users of the survey, as well
as for those using the DEEP2 data reduction pipelines to reduce DEIMOS
data.  Both raw and reduced spectra, photometry, and all redshifts are
available for download from the Data Release 4 website
(\url{http://deep.berkeley.edu/DR4/}) as well as other websites listed
in \S\ref{reduceddata}.  We do not expect this to be the final data
release from the DEEP2 survey.  Future releases may include improved
photometry, revised redshift estimates (e.g. from incorporating
photometric redshift information for $Q=2$ objects), improved data
reductions and extractions, and/or additional derived parameter
measurements not included here.

The discoveries from DEEP2 have led to the development of the DEEP3
Galaxy Redshift Survey (Cooper et al.~2011, 2012a).  This project has
obtained more than 8,000 spectra, primarily of objects in the Extended
Groth Strip. The primary goals of DEEP3 were to obtain redshifts for
three different samples:
\begin{itemize}
 
 \item Objects of interest identified through AEGIS multiwavelength observations;
 
 \item $R_{\rm AB} < 24.1$ galaxies that were not previously targeted by DEEP2 to
   avoid slit overlaps, enhancing the sample of objects which can
   be used to trace galaxy environments in the EGS, and providing an
   enlarged DEEP2-like sample of galaxies all of which have deep
   multiwavelength imaging; and
  
 \item Faint, highly star-forming galaxies down to $R_{\rm AB} = 25.5$,
   allowing us to test whether they continue to follow the same
   scaling relations as brighter galaxies.
 
 \end{itemize} 
 In order to maximize complementarity to DEEP2 data, DEEP3
 observations were conducted with a 600-line/mm ($R \sim 3000$)
 grating on DEIMOS, with spectra typically covering the wavelength
 range from 4600 \AA\ to 9800 \AA.  Hence, DEEP3 provides worse
 kinematic information than DEEP2, but covers more spectral lines in a
 given galaxy.  We expect the first data release from DEEP3 to occur
 in the next year, and that the legacy of DEEP2 will continue into the
 future.

\acknowledgements

Thanks are due to the many institutions and individuals who have made
the DEEP2 survey possible.  First thanks go to the W.~M.~Keck
Foundation, the University of California, and NASA for providing funds
to construct and operate the Keck telescopes.  Second, we wish to
thanks the technical teams in the UCO/Lick Shops and at Keck
Observatory for their role in building and commissioning the DEIMOS
spectrograph and for their superb support during many observing runs.
Funds for the spectrograph were provided by instrumentation grant ARI
92-14621 from the National Science Foundation and instrument funds
from the California Association for Research in Astronomy (Keck
Observatory) and from the University of California/Lick Observatory.

The DEIMOS data-reduction pipeline is based in part on code from the
Princeton Sloan Digital Sky Survey spectrograph pipeline, and we thank
Scott Burles and David Schlegel for their help in adapting this
software for DEIMOS.  Particularly vital components were the SDSS code
developed for slitlet extractions, B-spline sky subtraction, and
redshift determination.  Patrik Jonsson's study of fringing in the
LRIS spectrograph led to the basic understanding of fringing needed to
establish image-stability specifications for DEIMOS.  The ATV image
display tool created by Aaron Barth, SPLOT created by David Schlegel,
the IDL Astronomy User's Library supported by the Goddard Space Flight
Center, and the {\it idlutils} package were all instrumental in this
work.  The {\it submit\_prepare} perl script by Marshall Perrin was of
great assistance in the submission of this paper.

We benefitted
greatly from advice and input from Jeremy Mould, Charles Steidel, Mark
Metzger, Richard Ellis, Chris Conselice, Harlan Epps, Edward Groth,
Garth Illingworth, Richard Kron, Alex Szalay, Roc Cutri, Charles
Beichman, Peter Eisenhardt, Giovanni Fazio, and Kevin Bundy in
conceiving, designing, and executing the survey.  We also wish to
thank the entire DEEP1 team for their pioneering efforts which helped
greatly in designing DEEP2.

The DEEP2 survey was initiated under the auspices of the NSF Center
for Particle Astrophysics. Major grant support was provided by
National Science Foundation grants AST 95-29098, 00-711098, 05-07483,
and 08-08133 to UCSC, AST 00-71048, 05-07428, and 08-07630 to UCB, and
08-06732 to the University of Pittsburgh.  Computing hardware used to
analyze DEEP2 data was provided by Sun Microsystems. The {\it HST} ACS
imaging mosaic in EGS was constructed by Anton Koekemoer and Jennifer
Lotz and was funded by grant HST-AR-01947 from NASA.  NASA imaging of
the original Groth Strip was planned and executed by Ed Groth and
Jason Rhodes with support from NAS5-1661 and NAG5-6279 to the WFPC1
IDT. Support for this work was provided by NASA through Hubble
Fellowship grants 51256.01 and 51269.01 awarded to ENK and MCC,
respectively, by the Space Telescope Science Institute, which is
operated by the Association of Universities for Research in Astronomy,
Inc., for NASA, under contract NAS 5-26555. Sandra Faber would like to
thank CARA for a generous research grant and the Miller Institute at
UC Berkeley for a Visiting Miller Professorship, during which much of
this paper was written.  Jeffrey Newman and Alison Coil acknowledge
support from Hubble Fellowships during their DEEP2 work, and Michael
Cooper acknowledges support from both Hubble and Spitzer Fellowships.

Marc Davis acknowledges support from the entire DEEP2 team who filled
in while he was recovering from a stroke.

Finally, we recognize and acknowledge the highly significant
cultural role and reverence that the summit of Mauna Kea has always 
had within the indigenous Hawaiian community; it has been a privilege to be
given the opportunity to conduct observations from this mountain.

\appendix
 
\section{APPENDIX}

\subsection{Sky Subtraction} 
\label{appendix}

This appendix describes the steps we have taken to allow
photon-limited subtraction of night sky emission, even in the limit of
bright OH lines.  There are two major requirements necessary to
achieve this goal.

First, we must ensure that the photometric response of all pixels
remains constant to high accuracy between afternoon flat-fielding and
evening observations.  If this response varies, the flux calibration
between object and sky pixels will differ and sky subtraction will be
imperfect.  As explained below, the response in each pixel varies
periodically with wavelength at red wavelengths, a phenomenon commonly
known as ``fringing.''  Hence, in order to ensure accurate sky
subtraction we must keep the wavelength of light falling on a given
pixel constant to high accuracy.  We can attain this by keeping the
spectral image of a given mask accurately centered on exactly the same
pixels at all temperatures and spectrograph position angles.  Since
natural flexure in large spectrographs is larger than this tolerance,
we have incorporated a ``flexure compensation system'' in DEIMOS to
keep the images accurately centered.  This system has essentially
solved the problem of fringing by ensuring that the fringe pattern is
identical in flat-field and science frames.

The second key requirement is that we be able to employ brightness
information in the sky spectrum offset from a target object to infer
sky brightness at the same wavelength on the object spectrum
accurately.  This requires a highly accurate determination of the
wavelength of light falling on each pixel compared to its neighbors.
The challenge is very high because of the extremely steep gradient in
brightness on the shoulders of the OH sky lines.  For the 1200-line
grating used for DEEP2, the slit image has a FWHM of 3.9 px, but the
native FWHM of the spectrograph is only 2.3 px
(Table~\ref{table.instrumentparams}).  It is this latter quantity that
sets the gradient in the measured sky line flux. Since the native PSF
is quite sharp, the brightness falls off extremely fast, and shifts of
only 0.015 px between the sky spectrum and the object spectrum produce
detectable systematic residuals.  We therefore need a wavelength
solution that is better than this over all pixels of the 2-d spectrum,
plus a model sky spectrum that tracks the steep gradients with
wavelength in the sky spectrum to high accuracy.

The next section discusses the basic physics of fringing and derives
an image-motion tolerance needed to keep flat-field errors due to
fringing negligible.  We then describe the flexure compensation system
and its performance in light of this tolerance.  Finally, we describe
the methods used to determine wavelength solutions, the b-spline
modeling of the sky spectrum, and the sky-subtraction process.

\subsection{Fringing Physics and Image Motion Tolerance}
\label{fringing}

The peak OH sky-line brightness in a one-hour exposure is roughly 3000
photons per pixel with the 1200-line grating.  Hence, photon-limited
sky subtraction requires a flatfielding accuracy much better than
$1/\sqrt{3000}$, or 1.8\%.  Lower-dispersion gratings have higher peak
counts and need even higher accuracy, as do multiple stacked
exposures.  In practice we chose the target of $\pm$0.25\% for DEIMOS.
A major contributor to flat-fielding errors is CCD ``fringing,'' as
explained above.  The CCD silicon layer is a resonant cavity that
supports constructive and destructive interference through internal
light reflections (Newton's rings), and the pixel sensitivity varies
according to whether the interference is constructive or destructive.
Tests on the Kast spectrograph detector at Lick and the LRIS
spectrograph detector at Keck show that the fractional sensitivity
variation, $F$, is sinusoidal as a function of the wavelength shift in
\AA, $s$.  Hence, it can be described by the function $F(s) = A
\sin{{2\pi s} \over {P}}$, where $A$ is the fractional fringe
amplitude on a pixel and $P$ is the phase wrap period in \AA.  The
latter is given by $P={\lambda \over N}$, where $N$ is the
interference order $N = { 2tn \over \lambda}$, $n$ is the refractive
index of silicon (3.6 at these wavelengths), $\lambda$ is the
wavelength of light falling on a given pixel, and $t$ is the detector
thickness, which for DEIMOS is 40$\mu$m. Thus, $N \sim 300$ at 9000
\AA, so $P \sim 30$ \AA.  The interference is extremely high-order
because the effective thickness of the detector is much larger than
the wavelength of light.

The greatest sensitivity to wavelength shifts occurs where the
derivative of $F$ is at its maximum value, ${dF \over ds}_{max} =
{2A\pi \over P}$.  The fringe amplitude $A$ in the thick CCDs used for
DEIMOS is only $\pm$2\% at 9000 \AA, which minimizes fringing.
Keeping changes in $F$ below 0.25\% therefore requires keeping $s
\ltsim 0.6$ \AA, corresponding to an image shift of roughly 1.8 pixels
with the 1200-line grating.  However, the tolerance is tighter for
lower-dispersion gratings because their dispersion is lower and
because the OH sky brightness per pixel is higher.  Coadding many
spectra for the same object also requires better flat-fielding for
flat-field errors to remain negligible compared to uncertainties in
sky subtraction from photon statistics.  Allowing for the use of other
gratings and/or longer exposure times motivated a tighter design
tolerance of 0.6 pixels RMS flexure in the wavelength direction.
Errors in the direction along the slit are less problematic; their
design tolerance was set at 1.0 pixels RMS.

\subsection{The DEIMOS Flexure Compensation System}
\label{fcsystem}

In order to ensure such a high degree of stability, a closed-loop
flexure compensation system (FCS) was built into the basic design of
the DEIMOS spectrograph.  The system contains two movable elements:
the grating tilt mechanism, which is used to steer the image in the
dispersion direction, and a stage in the dewar, which moves the CCD
perpendicular to the dispersion.  Two pairs of optical fibers pipe
light into the spectrograph at opposite ends of the slitmasks.  The
light in spectroscopic mode comes from a ThAr arc lamp (an LED is used
for direct imaging mode), which produces arc-line images (spots) on
the two separate flexure-compensation CCD detectors, which are shown
in Figure~\ref{deimosfocalplane.thar.epsi}).

Despite the small size of these detectors (600 $\times$ 1200 15-$\mu$
pixels), the ThAr spectrum is so rich that spots are available with
nearly all possible gratings and tilts.  The FCS spectra are
determined from $\sim 1$-$3$ second exposures on the FCS CCDs, and
then the locations of the ThAr spots are processed.  Any shifts needed
to recenter the spots in their proper locations are then fed back into
the grating tilt and CCD actuator mechanisms.  The loop takes roughly
10 sec to complete.  A sensing system turns FCS tracking off and on
automatically whenever the shutter is opened or closed.  An automatic
computer script is used to ensure that the afternoon calibrations and
evening observations are taken at the same spot locations.

Several issues had to be taken into account when designing the DEIMOS
flexure compensation system.  The capture range of the actuators had
to be large enough to accommodate the maximum flexure expected.  The
lens element mounts in the camera turned out to be softer than planned
and other mechanical flexure also cropped up, so the initial design
capture range, $\pm$15 pixels, proved to be barely adequate.  Image
rotation is also not corrected by the system, necessitating rigid
mounts for all fold mirrors and gratings, which can rotate the image.

The most important limitation is that all parts of the image do not
move uniformly when certain optical elements are moved, owing to
higher-order optical distortions.  Some of these distortions were
anticipated, such as the distortion due to slit curvature, but a worse
effect is that the central parts of the image move slightly
differentially from the edges due to differential sags among the
heavily-curved camera elements.  Since the FC spots are at the edges
of the focal plane, their motions do not precisely represent image
motions near the center of the FOV.  This last factor limits the
ultimate accuracy of the system, causing image motions of 0.30 pixels
RMS along the dispersion direction and 0.50 pixels RMS perpendicular
to the dispersion.  However, both motions more than meet their specs;
the 0.30 pixel motions along the dispersion direction are 6 times
smaller than the tolerance of 1.8 pixels that must be achieved to
flat-field the 1200-line grating used for DEEP2.

\subsection{Pipeline Sky Subtraction}
\label{skysubtractcode}

Our method for modeling the sky spectrum and subtracting it depends on
the fact that the flexure compensation system will reliably place a
given wavelength on a given pixel, and thus that the afternoon
wavelength calibration closely matches the wavelength calibration of
the evening observation. We use the technique of b-spline fitting to
produce a very precise model of the night sky spectrum.  The code used
is based on similar code implemented by Scott Burles and David
Schlegel for a data reduction pipeline designed for the Sloan Digital
Sky Survey fiber spectrograph.  \footnote{See {\tt
    http://spectro.princeton.edu/idlutils\_doc.html} and {\tt
    http://spectro.princeton.edu/idlspec2d\_doc.html} for details.}
Related techniques are discussed in detail in Bolton \& Burles (2007).

B-splines (short for "basis splines") are smooth curves that can
accurately represent an arbitrary, continuous and continuously
differentiable function given a sufficiently dense set of control
points.  Essentially, b-splines provide the interpolation power of
cubic splines, but unlike cubic splines, they need not pass through
the control points used to determine the interpolating curve.  The
technique has found wide application in smooth modeling of irregular
high-dimensional surfaces and can filter out noise and reject
outliers, which is very important for sky modeling. The cubic ($n=4$)
b-splines used in the DEEP2 DEIMOS reduction pipelines are comprised
of localized cubic basis functions that are least-squares fitted to
the measured intensity in each 2D sky pixel of a spectrum as a
function of the wavelength of that pixel (for a general introduction,
see de Boor 2001); these basis functions are continuous at the
``breakpoints'' between successive segments through the second
derivative.

For example, suppose that we have a slitlet with five arc seconds of
slit-length free of any objects, and hence suitable for estimation of
the sky spectrum. The DEIMOS plate scale corresponds to eight pixels
per arc second, so there are 40 separate sky spectra, each consisting
of 4096 pixels per chip (i.e., 4096 on the blue CCD and the red CCD
separately).  We tilt each slitlet by at least 5 degrees relative to
the detector rows so that the 40 different measures of the sky
brightness each sample the sky spectrum at slightly different
wavelength. This tilting is critical, as it results in a strongly
oversampled sky spectrum free of significant gaps in wavelength
coverage.  By means of KrArNeXe arc spectra taken during the
afternoon, we fit a 2-d wavelength solution to each slitlet with a
fifth-order polynomial describing the variation of wavelength with
pixel number in the dispersion direction (i.e. along a row of the
extracted spectrum), a second-order polynomial describing the average
derivative of wavelength with pixel number along the slit direction as
a function of pixel number in the dispersion direction, and a constant
offset for each row compared to the value predicted from the
polynomial terms (e.g. to correct for any shifts in the average
wavelength at some position along the slit due to dust grains
partially blocking the slit).  In order to achieve a robust model for
the local wavelength solution on each pixel, the program takes the
second and higher Legendre-polynomial coefficients from the 2-d
wavelength calibration done each afternoon, but supplements them with
a linear term and zeropoint shift determined from fits to the
cross-correlation shift between the observed sky spectrum and a
template sky spectrum from the Keck HIRES spectrograph (Osterbrock,
Fulbright, \& Bida 1997) in 100 \AA\ windows, measured as a function
of wavelength.

We use 6144 breakpoints evenly spaced in wavelength for the 1-d
b-spline, which is simultaneously fit with outlier rejection to data
from all sky rows on the slitlet (after rectification to remove the
spectral curvature).  The b-spline minimizes the deviations of a
locally cubic polynomial from all measurements of sky flux as a
function of wavelength (40*4096 points in our example).  The FWHM of
the DEIMOS PSF of $\sim$2.3 pixels means that the image of the slitlet
is critically sampled or better (it is the width of the native PSF
that is important in determining gradients in the measured sky
brightness, not the slitwidth itself, which is 3.9 pixels FWHM).  The
number of breakpoints used (1.5 times the number of pixels along a row
of the extracted, rectified spectrum) is therefore more than adequate
to ensure full sampling.\footnote{In the case of slitlets which were
  not been tilted, as was done early in the survey, we use only one
  breakpoint per pixel, placed at the wavelengths corresponding to the
  pixels along the center of the slitlet, to minimize ringing.} After
an initial fit, the code then discards points that are outliers by
more than 20-$\sigma$, and the process runs again for a maximum of
three iterations.

We use individual b-splines to estimate the 1-d sky spectrum of each
frame of spectroscopic data for each DEIMOS slitlet.  Almost all of
the time, the b-spline technique produces a sky spectrum that is
extremely clean and allows us to reach the Poisson noise limit, as
shown in \S\ref{dataquality}.  As a result, it was not necessary for
us to dither objects up and down along slits to ensure good sky
subtraction, and doing so would have been a waste of detector real
estate.  The method breaks down when major pieces of the 2-d spectrum
are missing due to CCD gaps or bad columns, or in cases where large
amounts of skyline light is scattered in from neighboring
alignment-star boxes, but overall the system works very well, as
demonstrated by Figures~\ref{hist_onsky.epsi} and
\ref{hist_offsky.epsi}.


\begin{table}
\tablenum{1}
\scriptsize
\begin{center}
\caption{Selected Publications Describing the DEEP2 Survey \& Science Highlights}
\begin{tabular} {ll}
\tableline\tableline
\noalign{\smallskip}
\small{Paper}   & \small{Content}    \cr
\noalign{\smallskip} 
\tableline
\noalign{\smallskip}
\footnotesize{Survey design, infrastructure:} \cr
\xx Davis et al.~(2003)     &  Science objectives and early results  \cr
\xx Faber et al.~(2003)     &  DEIMOS integration and testing  \cr
\xx Coil et al.~(2004b)     &  CFHT $BRI$ photometry; {\it pcat} catalog \cr
\xx Yan et al.~(2004)       &  DEEP2 mock catalogs \cr
\xx Davis et al.~(2005)     &  Survey description; early group and cluster counts   \cr
\xx Davis et al.~(2007)     &  AEGIS data sets \cr
\noalign{\smallskip}
\footnotesize{Bimodality, red sequence:} \cr
\xx Willmer et al.~(2006)   &  Luminosity functions; growth of red sequence since $z\sim 1$ \cr
\xx Faber et al.~(2007)     &  Lum. funcs for DEEP2+COMBO-17; quenching of blue galaxies feeds RS \cr
\xx Bundy et al.~(2006)     &  Stellar masses and mass functions; down-sizing of quenching mass \cr
\xx Gerke et al.~(2007)     &  Quenching of red galaxies stronger in groups; onset at $z \sim 1.3$ \cr
\xx Schiavon et al.~(2006)  &  Red sequence contains recently quenched populations from $z\sim1$ to now \cr
\xx Harker et al.~(2006)    &  $U-V$ color of RS evolves slowly due to continuing new arrivals \cr
\noalign{\smallskip}
\footnotesize{Groups and clusters:} \cr
\xx Newman et al.~(2002)    &  Measuring $w$ with group and cluster counts  \cr
\xx Gerke et al.~(2005)     &  Group and cluster catalog, group properties \cr
\xx Gerke et al.~(2012)     &  Final group and cluster catalog \cr
\noalign{\smallskip}
\footnotesize{Interactions, morphologies:} \cr
\xx Lin et al.~(2004)       &  Pair counts; galaxy merger rate is not as fast in past as thought \cr
\xx Lin et al.~(2008)       &  Pair counts by type; separate merger rates for ``wet,'' ``dry,'' and ``mixed''   \cr
\xx Lotz et al.~(2008)      &  Gini/M$_{20}$ morphologies; galaxy merger rate is not as fast in past as thought \cr
\noalign{\smallskip}
\footnotesize{Environments:} \cr
\xx Cooper et al.~(2006)    &  Color-environment relation at $z\sim1$; bright blue galaxies in dense regions  \cr
\xx Cooper et al.~(2007)    &  Red galaxies in dense environments increase continuously after $z \sim 1.3$  \cr
\xx Cooper et al.~(2008)    &  Sloan-DEEP2 comparison; disappearance of massive blue galaxies  at $z \sim 0$ \cr     
\xx Lin et al.~(2010)       & Dry and mixed merger rates increase rapidly with density, wet merger rates do not \cr
\xx Cooper et al.~(2010)    &  Color-density relation existed but was weaker at $z\sim1$, even for constant-mass samples \cr     
\xx Cooper et al.~(2012)    & The impact of environment on the
size-stellar mass relation at $z \sim 1$ \cr     

\noalign{\smallskip}
\footnotesize{Clustering, satellites:} \cr
\xx Coil et al.~(2004a)     &  First $\xi(r)$ at $z \sim 1$ for large sample; clustering trends resemble local ones \cr  
\xx Coil et al.~(2006a)     &  First $\xi(r)$ for groups and clusters at $z\sim1$; galaxies do not follow NFW profiles \cr  
\xx Coil et al.~(2007)      &  Galaxy-QSO clustering; QSOs cluster like blue galaxies, similar halo masses \cr 
\xx Conroy et al.~(2007)    &  Satellite motions; halo masses, mass-light ratios, stellar mass growth \cr
\xx Coil et al.~(2008)      &  Definitive DEEP2 $\xi(r)$ vs.~color and luminosity  \cr
\xx Coil et al.~(2009)      &  Galaxy-AGN clustering; AGN cluster similarly to their (predominantly red) hosts \cr 
\noalign{\smallskip}
\footnotesize{Star formation, feedback:} \cr
\xx Yan et al.~(2006)       &  [O {\scriptsize II}] in SDSS and DEEP2 red sequence galaxies is due to LINERS, not SFR \cr
\xx Lin et al.~(2007)       &  SFR is enhanced by $\times2$ in close pairs \cr
\xx Noeske et al.~(2007ab)  &  Star-forming ``main sequence'', staged SFR model, downsizing in SFR  \cr
\xx Weiner et al.~(2007)    &  Extinction, SFR, and AGN tracers based on emission-line indices  \cr 
\xx Harker (2008)           &  Optical spectra at $z\sim 0.75$ consistent with the ``main sequence'' SFR model \cr    
\xx Konidaris (2008)        &  [O {\scriptsize II}] in DEEP2 red sequence galaxies is due to LINERs, not SFR \cr
\xx Yan et al.~(2009)       &  Post-starburst galaxies; ``absolute'' environments constant with redshift \cr
\xx Salim et al.~(2009)     &  Most mid-IR (24 $\mu$) dust emission comes from older stars, not current SFR \cr
\xx Weiner et al.~(2009)    &  Ubiquitous outflows from star-forming galaxies at $z \sim 1.3$ \cr
\xx Sato et al.~(2009)      &  Outflows at low $z$ more often in IR-luminous or post-starbursts \cr
\noalign{\smallskip}
\footnotesize{Structural properties:} \cr
\noalign{\smallskip}
\xx Kassin et al.~(2007)    &  $S_{0.5}$ linewidths; TF relation already in place at $z \sim 1$ for normals and mergers \cr 
\xx Dutton et al.~(2011)    &  Observed Tully-Fisher evolution agrees with LCDM-based models \cr 
\noalign{\smallskip}
\footnotesize{AGNs and black holes:} \cr
\xx Nandra et al.~(2007)    &  AGNs populate RS, green valley, upper blue cloud; persist after quenching \cr 
\xx Pierce et al.~(2007)    &  AGNs are found mainly in E/S0/Sa galaxies, rare in mergers \cr   
\xx Gerke et al.~(2007)     &  Dual merging supermassive black hole in early-type galaxy \cr
\xx Georgakakis et al.~(2008)   &  Population of deeply buried, persistent, dust-obscured AGNs on the RS \cr
\xx Bundy et al.~(2008)     &  AGN ``trigger rate'' matches red sequence quenching rate after $z \sim 1$ \cr
\xx Montero-Dorta et al.~(2009) & Numbers and environments of red sequence Seyferts and LINERS evolve after $z \sim 1$ \cr
\xx Comerford et al.~(2009) &  Thirty-two candidate in-spiraling supermassive black holes  \cr 
\xx Yan et al.~(2011) &  Fractions of obscured and Compton-thick AGN at $z \sim 0.6$ were at least as large as today  \cr 

\tableline
\end{tabular}
\end{center}
\label{table.previouspapers}
\end{table}

\begin{table}
\tablenum{2}
\footnotesize
\begin{center}
\caption{Instrument and Exposure Parameters}
\begin{tabular}{ll}
\noalign{\smallskip}
\noalign{\smallskip}
\tableline
\tableline
\noalign{\smallskip}
Parameter       &    Value           \\
\noalign{\smallskip} \tableline \noalign{\smallskip}
Slitmask parameters:                             &  {}        \cr
\xx Total slit length                            &  16\m.7    \cr
\xx Usable slit length                           &  16\m.3    \cr
\xx Maximum mask width                           &  5\m.3     \cr
\xx Typical number of slitlets per mask          &  140       \cr
\xx Slit width                                   &  1\s.0     \cr
\xx Slitlet tilts                                &  Up to 30\deg  \cr
\xx FWHM along slit for point source             &  0\s.5-1\s.2 (best/worst seeing) \cr 
Spectral parameters:                             &  {}   \cr
\xx Grating                                      &  1200 l mm$^{-1}$, gold-coated, 7500 \AA  \ blaze \cr
\xx Typical spectral range                       &  6500-9300 \AA   \cr
\xx Dispersion                                   &  0.33 \AA \ px$^{-1}$ \cr
\xx FWHM of sky line in spectral direction       &  1.29-1.39 \AA; center best, worse at detector edges  \cr
\xx Spectrograph image quality, point source     &  2.1 px (field center), 2.7 px (corners) [FWHM]  \cr
\xx Spectral resolution$^a$                      &  $R \equiv \lambda/\Delta\lambda = $5900 at 7800 \AA  \cr
\xx Equivalent velocity $\sigma$                 &  24 km s$^{-1}$ \cr
\xx Order blocking filter                        &  GG495 \cr
\xx Peak throughput, telescope+spectrograph      &  29\% with fresh coatings \cr
\xx RMS image stability$^b$                      &  0.3 px along spectrum, 0.5 px perpendicular \cr
Detector parameters:                             &  {} \cr
\xx Detector mosaic                              &  Eight 2K$\times$4K CCDs in 2 $\times$ 4 array \cr
\xx Pixel size                                   &  15 $\mu$   \cr
\xx Total array size                             &  8192 px $\times$ 8192 px; 12.6 cm $\times$ 12.6 cm  \cr
\xx Blue-red gap in spectral direction           &  7 px \cr
\xx Gap between CCDs, spatial direction          &  Approx. 8\asec \cr
\xx Detector scale                               &  0\s.1185 px$^{-1}$ \cr
\xx Gain                                         &  1.25 e$^-$ per DN \cr
\xx Read noise                                   &  2.55 e$^-$   \cr
Exposure parameters:                             &  {} \cr
\xx Nominal exposure                             &  3 $\times$ 20 min. \cr
\xx Typical sky level between OH lines           &  13 photon px$^{-1}$ in 20 min. near 8000 \AA \cr
\xx Brightest OH sky line                        &  $\sim$2000 photons px$^{-1}$ in 20 min. \cr
\noalign{\smallskip} \tableline
\end{tabular}
\end{center}

Notes: \\
$^a$ $\Delta\lambda$ is the FWHM of an OH night-sky line. \\
$^b$ Daily, spanning evening observations and matching afternoon calibrations. \\

\label{table.instrumentparams}
\end{table}

\begin{table}
\tablenum{3}
\footnotesize
\begin{center}
\caption{DEEP2 Field Information}
\begin{tabular} {lrrrrcccccrrr}
\tableline
\tableline
\noalign{\smallskip}
Fld   & RA     & Dec    & $l$    &   $b$  & Red.$^a$ & Area    & Area     & Pre-    &Masks   & Masks    & Targ.    & $Q$$\ge3$  \cr
No    & {}     & {}     & {}     &   {}   &  {}     & nom.$^b$& done$^c$ & sel?$^d$&nom.$^e$& done$^f$ & done$^g$ & $z$'s$^h$  \cr
\noalign{\smallskip} \tableline  
\noalign{\smallskip}                                                                                
1       & 14:19 & 52:50 &  96.52 & +59.57 & 08-11 & 0.60    & 0.60  & N & 120 & 104 & 17,745 & 12,051 \cr
2       & 16:52 & 34:55 &  57.38 & +38.63 & 15-24 & 0.93    & 0.62  & Y & 120 &  85 & 10,201 &  6,703 \cr
3       & 23:30 & 00:00 &  83.79 & -56.55 & 34-49 & 0.93    & 0.90  & Y & 120 & 103 & 12,472 &  8,126 \cr
4       & 02:30 & 00:00 & 168.10 & -53.99 & 22-40 & 0.93    & 0.66  & Y & 120 & 103 & 12,494 &  7,943 \cr
\noalign{\smallskip} \tableline
\end{tabular}
\end{center}

Notes: \\
$^a$ $E(B-V)$ range within nominal field area in units of 0.001 mag, from Schlegel, Finkbeiner \& Davis (1998). \\
$^b$ Nominal area of field as originally planned, in square degrees.   \\
$^c$ Area actually covered, in square degrees.   \\
$^d$ Indicates whether targets are pre-selected using $BRI$ color cuts.   \\
$^e$ Nominal number of masks in field as originally planned.   \\
$^f$ Number of successful masks actually obtained.   \\
$^g$ Number of targeted candidate galaxies on slitmasks.  These are the objects listed in Table 10; duplicate 
observations are included.  Counts objects eventually found to be stars but not stars in alignment boxes. \\
$^h$ Number of reliable galaxy redshifts in this field with $Q = 3$ or 4, not including stars or
duplicate redshifts.\\

\label{table.fields}
\end{table}

\begin{table}
\tablenum{4}
\scriptsize
\begin{center}
\caption{Existing and Planned Data in DEEP2 Field 1 (Extended Groth Strip)$^a$}
\begin{tabular} {lllll}
\noalign{\smallskip}
\noalign{\smallskip}
\tableline\tableline
\noalign{\smallskip}
\footnotesize{Data} & \footnotesize{Wavelength} & \footnotesize{Depth$^b$} & \footnotesize{Area$^c$} & \footnotesize{Contact}  \cr  
\noalign{\smallskip} \tableline
\noalign{\smallskip}
\noalign{\smallskip} 
\xx Keck DEIMOS (DEEP2)            & 6500-9100 \AA    & $R = 24.1$           &   16'$\times$125'  & M. Davis  \cr
\xx  \xx $z$ = 0-1.4               &     {}           &   {}                 &    {}              &    {}  \cr
\xx  {\it Keck DEIMOS (DEEP3)}$^d$ & 4800-9600 \AA    & $R = 25.5$           &   16'$\times$60'   & S. Faber  \cr
\xx  \xx $z$ = 0-1.4               &     {}           &   {}                 &    {}              &    {}   \cr
\xx  MMT Hectospec                 & 4500-9000 \AA    & $R = 22.5$           &   17'$\times$120'  & C. Willmer, A. Coil  \cr
\xx  {\it GTC-EMIR GOYA}           & 1-2.5$\mu$       & $K = 24.5$           &   $\sim$0.3 deg$^2$  &  R. Guzman  \cr
\noalign{\medskip} 
\xx  Chandra ACIS                  &  1-10 keV        &  200 ksec            & 17'$\times$120'    & P. Nandra  \cr
\xx  Chandra ACIS (AEGIS-X)        &  1-10 keV        &  800 ksec            & 17'$\times$40'     & P. Nandra  \cr
\xx  XMM EPIC                      &  0.1-15 keV      &  70-82 ksec          & 30' diam           &  -----  \cr
\noalign{\medskip} 
\xx  GALEX UDeep imaging           &  FUV,NUV         &  154, 270 ksec       & 1.25\deg\ diam      & S. Salim   \cr
\xx  GALEX grism                   &  FUV,NUV         &  291, 291 ksec       & 1.25\deg\ diam      & C. Martin  \cr
\noalign{\medskip} 
\xx  Hubble ACS                    &  $F606W,F814W$            &  28.7,28.1           & 10'$\times$67'     & A. Koekemoer, J. Lotz  \cr
\xx  Hubble ACS (CANDELS) &  $F606W,F814W$            &  28.9,28.8           & 7'$\times$26'     & A. Koekemoer  \cr
\xx  Hubble NICMOS                 &  $JH$            &  25.0,24.8           & 0.013 deg$^2$$^e$   & S. Kassin  \cr
\xx  Hubble WFC3 (CANDELS) &  $F125W,F160W$            &  27.2,27.3          & 7'$\times$26'     & A. Koekemoer  \cr
\noalign{\medskip}  
\xx  CFHT 8K$\times$12K ({\it pcat}) &  $BRI$         &  24.5,24.5,23.5$^f$  & 4$\times$28'$\times$42' & J. Newman, A. Coil  \cr
\xx  CFHT Megacam (CFHTLS)$^g$     &  $ugriz$         &  $ugri\sim$27, $z\sim$25.5  & 1 deg$^2$   & S. Gwyn        \cr
\xx  MMT Megacam                   &  $u'g'i'z'$      &  26.2,27.2,26.0,26.0 & 1 deg$^2$          & M. Ashby  \cr
\xx  {\it LBT}                     &  $UY$            &  $\sim25$,23.9            & 17'$\times$110'                 & B. Weiner        \cr   
\xx  Subaru                        &  $R$             &  27.0                & 1 deg$^2$          & M. Ashby  \cr
\xx  KPNO~4m NDWFS$^h$              &  $B_{W}RI$        &  27.1,26.2,25.8      & 1.4 deg$^2$        & A. Dey  \cr
\noalign{\medskip} 
\xx  Palomar~5m (POWIR)$^i$         &  $JK_s$          &  23.9,21.7-22.5       & 0.2,0.7 deg$^2$    & K. Bundy  \cr
\xx  {\it CFHT WIRCAM}             &  $YJHK$          &  $\sim25$             & 20'$\times$20'     & R.  Pello  \cr
\xx  {\it CFHT WIRCAM (WIRDS)}$^j$ &  $JHK$           &  24.8,24.6,24.5       & 3$\times$20'$\times$20'  & C. Willott  \cr
\xx  Subaru                        &  $K$             &  24.5                 & 7'$\times$40' & T. Yamada  \cr
\xx  KPNO4~m NEWFIRM                &  $JK$            &  24.4,23.9            & 1.4 deg$^2$        & M. Dickinson  \cr
\xx  {\it KPNO~4m NEWFIRM (NMBS)}$^k$ &  $JH$(med.)$^j$ &  $K$ = 23.4          & 28'$\times$28'     & P. van Dokkum    \cr
\noalign{\bigskip} 
\xx  IRAC GTO                      &  3.6,4.5,5.8,8.0 & 0.9,0.9,6.3,5.8$\mu$Jy  & 10'$\times$120'   & P. Barmby  \cr
\xx  IRAC GTO (``Handle")          &  3.6,4.5,5.8,8.0 & 1.0,1.5,9.3,12.0$\mu$Jy & 60'$\times$25'    & J. Huang  \cr
\xx  {\it IRAC (SEDS)}$^l$         &  3.6,4.5         & 25.7,25.7 $\mu$Jy    & 12'$\times$75'     & G. Fazio  \cr
\noalign{\medskip} 
\xx  AKARI$^m$                     &  15$\mu$         & 115-150 $\mu$J$^l$   & 10'$\times$70'       & M. Im  \cr
\noalign{\medskip} 
\xx  MIPS GTO                      &  24,70$\mu$      & 77$\mu$J,10.3mJy     & 10'$\times$120'      & J. Huang, R. Hickox    \cr
\xx  MIPS FIDEL$^n$                &  24,70,160$\mu$  & 30$\mu$J,3mJ,20mJy   & 10'$\times$90'       & M. Dickinson  \cr
\noalign{\medskip} 
\xx  {\it Herschel PACS (HERMES)}$^o$  &  110,170$\mu$ & 5.2,7.4mJy          &  10'$\times$67'                & D. Lutz  \cr
\xx  {\it Herschel SPIRE (HERMES)}$^o$ &  250,350,450$\mu$ & 11.1,15.2,12.9mJy   &  10'$\times$67'           & S. Oliver  \cr
\noalign{\medskip} 
\xx  {\it Scuba2 Legacy Deep}      &  850 $\mu$       & 3.5mJy               & 1 deg$^2$            & R. Ivison  \cr
\noalign{\medskip} 
\xx  VLA                           &  6cm             & 0.6mJ$^p$            & 30'$\times$80'       & S. Willner  \cr
\xx  VLA                           &  20cm            & 100$\mu$J            & 30'$\times$80'       & R. Ivison  \cr
\noalign{\medskip} 
\xx  GMRT                          &  50cm            & 75mJ                 & 10'$\times$90'      & A. Biggs  \cr      
\noalign{\smallskip}
\tableline
\end{tabular}
\end{center}
Notes: \\
$^a$  Planned or in progress are in italics.  For a general overview, see \url{http://aegis.ucolick.org/astronomers.html}. \\
$^b$  All magnitudes are AB mags.  Limiting magnitudes are 5 $\sigma$ unless otherwise stated.  \\
$^c$  Areas are approximate.  \\
$^d$  DEEP3 will acquire $\sim$8,000 new spectra and double the redshift sampling density in the Hubble ACS mosaic region. \\
$^e$  NICMOS: 63 pointings. \\
$^f$  CFHT {\it pcat}: 8 $\sigma$. \\
$^g$  CFHTLS = CFHT Legacy Survey (\url{http://www.cfht.hawaii.edu/Science/CFHTLS}).  \\
$^h$  NDWFS = NOAO Wide-Deep Field survey; EGS is an extension of the main NDWFS. \\
$^i$  POWIR = Palomar Observatory Wide Infrared Survey (Conselice et al.~2008).   \\
$^j$  WIRDS = WIRCAM Infrared Deep Survey (\url{http://terapix.iap.fr/rubrique.php?id\_rubrique=256}) \\
$^k$  NMBS = Newfirm Medium Band Survey; five medium-band filters from $J$ through $H$ (van Dokkum et al.~2009). \\
$^l$  SEDS = {\it Spitzer} Extragalctic Deep Survey (G.~Fazio, priv.~comm.).  \\
$^m$  AKARI: 50\% of the field is at the two quoted depths (5 $\sigma$; M.~Im, priv.~comm.).  \\
$^n$  FIDEL = Far-Infrared Deep Extragalactic Legacy Survey (\url{http://irsa.ipac.caltech.edu/DATA/SPITZER/FIDEL}) \\
$^o$  HERMES = Herschel Multi-Tiered Extragalactic Survey (\url{http://astronomy.sussex.ac.uk/~sjo/Hermes}). \\
$^p$  VLA, 6 cm: 10-$\sigma$ (S.~Willner, priv.~comm.).  \\  

\label{table.otherdata}
\end{table}

\begin{table}
\tablenum{5}
\scriptsize
\begin{center}
\caption{Existing Data in DEEP2 Fields 2,3,4}
\begin{tabular} {lllll}
\noalign{\smallskip}
\noalign{\smallskip}
\tableline\tableline
\noalign{\smallskip}
                Data               &  Wavelength      &  Depth$^a$           &   Area$^b$           &   Contact  \cr  
\noalign{\smallskip} \tableline
\noalign{\smallskip}
\noalign{\smallskip}
Field 2 (16:52, 34:55):            &     {}           &   {}                 &   {}                 &   {}  \cr
\noalign{\smallskip}
\xx Keck DEIMOS/DEEP2              & 6500-9100 \AA    & $R = 24.1 $          & 28'$\times$84'       & M. Davis  \cr
\xx  \xx $z$ = 0.75-1.4            &     {}           &   {}                 &   {}                 &   {}  \cr
\xx  Chandra ACIS$^c$              &  1-10 keV        & 9 ksec               & 30'$\times$100'      & S. Murray  \cr
\xx  CFHT Megacam                  &  $i,z$           &  24.6,23.2           & 1 deg$^2$            & L. Lin   \cr
\xx  Palomar/KPNO4m (POWIR)$^d$    &  $JK$            &  22.4,21.5 (variable)     & 0.2 deg$^2$          & K. Bundy  \cr
\xx  IRAC$^f$                      &  3.6,4.5,5.8,8.0 &  1.9,2.9,17,21       & 45'$\times$100'      & J. Huang  \cr
\xx  MIPS$^g$                      &  24,70,160$\mu$      &  0.227,40,200 mJy
& 0.7 deg$^2$                  & B. Weiner  \cr
\noalign{\smallskip}  
\noalign{\smallskip}      
Field 3 (23:30, 00:00):            &     {}           &   {}                 &   {}                 &   {}  \cr
\noalign{\smallskip}
\xx  Keck, DEIMOS/DEEP2            &  6500-9100 \AA   &  $R = 24.1$          & 28'$\times$126'      & M. Davis  \cr
\xx  \xx $z$ = 0.75-1.4            &     {}           &   {}                 &   {}                 &   {}  \cr
\xx  Chandra ACIS$^c$              &  1-10 keV        & 9 ksec               & 30'$\times$100'      & S. Murray  \cr
\xx  CFHT Megacam                  &  $i,z$           & 24.6,23.7            & 2 deg$^2$            & L. Lin   \cr
\xx  Sloan Equatorial Stripe       & $ugriz$          & $\sim21$ AB          & All                  & -----  \cr  
\xx  Palomar/KPNO4m (POWIR)$^d$    &  $JK$            & 22.4,21.5 (variable)      & 0.3 deg$^2$          & K . Bundy  \cr
\noalign{\smallskip}  
\noalign{\smallskip}      
Field 4 (02:30, 00:00):            &     {}           &   {}                 &   {}                 &   {}        \cr      
\noalign{\smallskip}
\xx  Keck, DEIMOS/DEEP2            & 6500-9100 \AA    & $R = 24.1$           & 28'$\times$90'       & M. Davis  \cr
\xx  \xx $z$ = 0.75-1.4            &     {}           &   {}                 &   {}                 &   {}  \cr
\xx  Chandra ACIS$^c$              &  1-10 keV        & 7   ksec             & 30'$\times$100'      & S. Murray  \cr
\xx  CFHT Megacam                  &  $i,z$           & 24.5,23.7            & 2 deg$^2$            & L. Lin   \cr
\xx  Sloan Equatorial Stripe       & $ugriz$          & $\sim21$ AB          & All                  & -----  \cr  
\xx  Palomar/KPNO4m (POWIR)$^d$    &  $JK$            & 22.4,21.5 (variable)      & $\sim$0.25 deg$^2$    & K. Bundy  \cr
\xx  CFHT WIRCAM    &  $J$            & 24.0 AB      & 0.45 deg$^2$    & L. Lin  \cr
\xx  IRAC$^e$                      &  3.6,4.5,5.8,8.0 & 4,6,25,25 $\mu$Jy    & 0.95 deg$^2$         & R. Hickox  \cr
\noalign{\smallskip}
\tableline
\end{tabular}
\end{center}

Notes: \\
$^a$  All magnitudes are AB mags.  Limiting magnitudes are 5 $\sigma$ unless otherwise stated.  \\
$^b$  Areas are approximate.  \\
$^c$  Chandra ACIS program \#9900045 (Murray et al.~2008). \\
$^d$  POWIR = Palomar Observatory Wide Infrared Survey (Conselice et al.~2008).   \\
$^e$  Spitzer program \#50660 (Jones et al.~2008). \\
$^f$  Spitzer program \#40689.  \\
$^g$  Spitzer program \#40455. \\

\label{table.otherdata234}
\end{table}

\begin{table}
\tablenum{6}
\footnotesize
\begin{center}
\caption{DEEP2 Compared to Other $z\sim$ 1 Surveys}
\begin{tabular} {llllllll}
\tableline\tableline
\noalign{\smallskip}
{}                 & DEEP2           & DEEP2-         & TKRS$^a$           & VVDS-           & VVDS-           &  zCOSMOS-    & PRIMUS$^e$\cr
{}                 & {}              & EGS            & {}                 & deep$^b$        & wide$^c$        &  bright$^d$  & {}        \cr        
\noalign{\smallskip} \tableline
\noalign{\smallskip}
Fld name(s)        & Fields 2-4      & Field 1         & GOODS-N           & F02,CDFS        & F02,10,14,22   & COSMOS         & 10 fields  \cr
Mag lim$^f$        & 24.1 ($R$)      & 24.1 ($R$)      & 24.4 ($R$)        & 24.0 ($I$)      & 22.5 ($I$)     & 22.5 ($I$)     &  23 ($I$)   \cr
Spectral res, $R$  & 5900            & 5900            & 2000              & 230             & 230            &  600           & 7-100    \cr
Nom targs$^g$      & 45,000          & 17,775          & 2,018             & 35,000          & $\sim$100K     & 20,000         & 270,000   \cr
Compl targs$^h$    & 35,214          & 17,775          & 1,987             & 10,949          & 31,196         &  10,643        & 270,300   \cr
Reliab $z$'s$^i$   & 24,785          & 12,617          & 1,536             & 4,506           & 8,961          &   ~900         & 79,419$^e$  \cr
$z$ err, \kms      & 30              & 30              & $\ltsim$100       & 275             & 275            &  55            & 1800 -- 3300      \cr
Nom area$^j$       & 2.80            & 0.60            & 0.046             & 2.0             & 16.0           & 1.7            & 10.0 \cr
Compl area$^k$     & 2.18            & 0.60            & 0.046             & 0.61            &  6.1           & ---            & 9.1 \cr
$z$'s deg$^2$$^l$  & 11,400          & 21,000          & 21,300            & 9,400           & 2,300          & 11,800         & $ 14,300$   \cr 
$zeffic_1^m$       & 0.711           & 0.717           & 0.725             & 0.399           & 0.280          & 0.565          & 0.45     \cr  
$zeffic_2^n$       & 0.734           & 0.725          & 0.762             & 0.432           & 0.418          & 0.592          & 0.48$^e$    \cr
Environ merit$^o$  &  503            & ---            & 5.1               & 33.3            & 13.2           & 115            & ---       \cr
\noalign{\smallskip} 
\tableline
\end{tabular}
\end{center}

Notes: \\
$^a$ Wirth et al.~(2004).  $^b$ Le Fevre et al.~(2005).  $^c$ Garilli et al.~(2008).  $^d$ Lilly et al.~(2007,2009). \\
$^e$ A.~Coil (priv.~comm.). \\
$^f$ Magnitude limit in AB mags. \\
$^g$ Number of target slitlets in survey as originally planned.   \\
$^h$ Number of target slitlets completed to date.  Includes stars but not serendips or (for DEEP2)
     objects without usable spectra ($Q$ = -2).  Roughly 1.1\% of DEEP2 targets have $Q$ = -2.\\
$^i$ Reliable galaxy redshifts to date with probability of correctness being $\ge$95\%, stars
     not included. For all surveys but PRIMUS, this means quality codes 3 and 4 or equivalent.  
     VVDS-wide number includes bright galaxies that are also part of VVDS-deep.     \\
$^j$ Total field area as originally planned, in square degrees.  \\
$^k$ Total field area covered to date (single-pass only in parts of VVDS-deep and VVDS-wide).  \\
$^l$ Reliable galaxy redshifts deg$^2$. Includes only galaxies with quality codes 3 and 4 
     (no serendips or stars) and attempts to be representative for non-uniformly covered 
     surveys (VVDS-wide and VVDS-deep).  DEEP2 numbers use galaxies in this paper and actual field areas  
     minus regions lost to stars, CCD gaps, and incomplete mask coverage.  TKRS is similar. 
     VVDS-deep estimate uses existing redshifts in Field F02 with an assumed field size of 0.41 deg$^2$; 
     VVDS-wide estimate uses numbers of published 
     galaxies in F02, F10, F14, and F22, which have mostly single pass; does not include
     the smaller area in F02, which is mostly double-pass.
     zCOSMOS-bright estimate uses the target 
     density and star rate from Lilly et al.~(2007) together with the 
     reliable $z$-success rate measured from the zCOSMOS DR2 sample (Lilly et al.~2009) 
     to predict a final density for $Q3$+$Q4$ galaxies of 10,400 galaxies deg$^2$.  \\
$^m$ {\it Overall} redshift efficiency, defined as the fraction of all targeted slitlets 
     that yield reliable galaxy and/or QSO redshifts ($\ge95$\%). 
     Efficiency for VVDS-deep is calculated from the released set of 8981 redshifts, 
     of which 35\% have $I<22.5$ and may also be in the VVDS-wide sample. \\
$^n$ {\it Galaxy} redshift efficiency, defined as the fraction of targeted galaxies and/or QSOs that 
     yield reliable redshifts ($\ge95$\%). Total failures are allocated amongst stars and galaxies 
     proportional to their numbers amongst the objects with successful redshifts; if all failures 
     are in fact galaxies, the redshift efficiency would be lower than listed here.   \\
$^o$ Figure of merit indicating a survey's weight for environmental and clustering measures, given by  
     $N^2/A$, where $N$ is the number of reliable redshifts (in thousands) and $A$ is field area in      
     degrees.  Values used are the existing numbers for all surveys except zCOSMOS-bright, for which the 
     design numbers are used.  Combined numbers for DEEP2 and EGS are used.  No entry is given for
     PRIMUS since its redshifts are not accurate enough to localize individual objects within the web of large-scale structure (which requires redshift errors significantly smaller than the correlation length, $<\sim 500$ km/s; cf. Cooper et al.~2005).  However, PRIMUS should provide accurate measurements of larger-scale overdensity around individual objects, and can provide information on the relationship between galaxies and dark matter on smaller scales via projected cross-correlation functions.\\ 
\label{table.othersurveys}
\end{table}

\begin{table}
\tablenum{7}
\begin{center}
\footnotesize
\caption{Redshift Quality Codes Assigned to all Targets}
\begin{tabular} {llrrrrrrr}
\tableline\tableline
\noalign{\smallskip}
{$R_{AB}$ mag range}  & Color$^a$   & $Q$=-2$^b$  & $Q$=-1$^c$  & $Q$=1$^d$  & $Q$=2$^e$  & $Q$=3$^f$  & $Q$=4$^g$ & Total \cr
\noalign{\smallskip} \tableline
\noalign{\smallskip}
$<$20.0         & Blue    &     0 &    11 &     2 &    13 &     6 &   168 &    200 \cr
    {}               & Red     &    2 &    29 &     0 &    10 &     8 &   215 &    264 \cr
    {}               & Total   &      2 &    40 &     2 &    23 &    14 &   383 &    464 \cr
\noalign{\smallskip}
20.0-21.0            & Blue    &     6 &    16 &     5 &    63 &    23 &   401 &    514 \cr
    {}               & Red     &      3 &    48 &     5 &    45 &    34 &   416 &    551 \cr
    {}               & Total   &    9 &    64 &    10 &   108 &    57 &   817 &   1065 \cr
\noalign{\smallskip}
21.0-22.0            & Blue    &    9 &     9 &    29 &   104 &    64 &  1020 &   1235 \cr
    {}               & Red     &   14 &    89 &    30 &   122 &   213 &  1658 &   2126 \cr
    {}               & Total   &   23 &    98 &    59 &   226 &   277 &  2678 &   3361 \cr
\noalign{\smallskip}
22.0-23.0            & Blue    &    29 &    12 &   346 &   396 &   217 &  2203 &   3203 \cr
    {}               & Red     &   70 &   211 &   306 &   571 &   954 &  6158 &   8270 \cr
    {}               & Total   &  99 &   223 &   652 &   967 &  1171 &  8361 &  11473 \cr
\noalign{\smallskip}
$>23.0$            & Blue    184 &    17 &  5871 &  2050 &  1071 &  6033 &  15226 \cr
    {}               & Red     &   217 &   504 &  2327 &  1812 &  3634 & 12906 &  21400 \cr
        {}               & Total   401 &   521 &  8198 &  3862 &  4705 & 18939 &  36626 \cr
    \noalign{\smallskip}
Full sample            & Blue    228 &    65 &  6253 &  2626 &  1381 &  9825 &  20378 \cr
    {}               & Red     306 &   881 &  2668 &  2560 &  4843 & 21353 &  32611 \cr
    {}               & Total   &   534 &   946 &  8921 &  5186 &  6224 & 31178 &  52989 \cr
 \cr
\noalign{\smallskip}
\tableline
\end{tabular}
\end{center}
Notes: \\
$^a$ Color: Blue = $R-I < 0.5$, Red = $R-I > 0.5$. \\
$^b$ $Q$ = -2: data so poor that object was effectively never observed. \\
$^c$ $Q$ = -1: star. \\
$^d$ $Q$ = 1: probable galaxy but very low S/N; data not likely to yield redshift.  Many
of these are distant galaxies beyond the nominal redshift limit of $z = 1.4$. \\
$^e$ $Q$ = 2: low S/N or data are somehow compromised for reliable redshift; reason given in comments. \\
$^f$ $Q$ = 3: reliable redshift with probability of accuracy $\ge$95\%. \\
$^g$ $Q$ = 4: reliable redshift with probability of accuracy $\ge$99\%. \\

\label{table.qualitycodes}
\end{table}

\begin{table}
\tablenum{8}
\begin{center}
\footnotesize
\caption{Quality Codes for Duplicate Observations}
\begin{tabular} {rrrrrr}
\noalign{\smallskip}
\noalign{\smallskip}
\tableline\tableline
\noalign{\smallskip}
 $Q$  &  -1   &   1   &     2   &    3    &     4      \cr
\noalign{\smallskip} \tableline 
\noalign{\smallskip}
-1  &  43  &   {}   &    {}  &    {}   &     {}      \cr
 1  &    4  &   264  &     {}  &    {}   &     {}      \cr
 2  &    2   &   186  &     52  &    {}   &     {}     \cr
 3  &    0   &   48  &     54  &    68   &     {}      \cr
 4  &    0   &   82  &     114 &    295  &   1038     \cr
\noalign{\smallskip} \tableline
\end{tabular}
\end{center}
Note: Entries represent the number of duplicate pairs with the 
indicated combination of quality codes. \\
\label{table.comparequalitycodes}
\end{table}

\begin{deluxetable}{cccccccccc}
\tabletypesize{\scriptsize}
\tablenum{9}
\tablecaption{Mask Data (Sample)}
\tablewidth{0pt}
\tablehead{
\colhead{Mask No.} & \colhead{Obs. Date} & \colhead{RA} & \colhead{Dec} & \colhead{PA} & \colhead{$N_{obj}$} & \colhead{$<S/N>$} & \colhead{Seeing} & \colhead{\%(tot)} & \colhead{\%(red)} \\
\colhead{(1)} & \colhead{(2)} & \colhead{(3)} & \colhead{(4)} & \colhead{(5)} & \colhead{(6)} & \colhead{(7)} & \colhead{(8)} & \colhead{(9)} & \colhead{(10)} 
}
\startdata
1100 & 2003-05-03 & 213.74732 & 52.07412 & -48.86 & 152 & 0.718 & 0.677 & 66.23
 & 73.85\\
1101 & 2003-05-06 & 213.79093 & 52.09807 & -48.86 & 155 & 0.348 & 1.017 & 53.64
 & 56.79\\
1102 & 2003-06-30 & 213.82309 & 52.12490 & -48.86 & 158 & 0.897 & 0.737 & 76.47
 & 81.54\\
1103 & 2003-05-03 & 213.86179 & 52.14789 & -48.86 & 150 & 0.746 & 0.705 & 71.62
 & 81.82\\
1104 & 2003-05-30 & 213.89094 & 52.17260 & -48.86 & 162 & 0.355 & 0.767 & 66.24
 & 69.51\\
1105 & 2003-07-01 & 213.92597 & 52.19302 & -48.86 & 147 & 0.555 & 1.177 & 73.47
 & 78.82\\
1106 & 2003-05-05 & 213.95854 & 52.22014 & -48.86 & 151 & 0.449 & 0.999 & 65.56
 & 72.62\\
1107 & 2004-05-22 & 213.99695 & 52.24292 & -48.86 & 156 & 0.340 & 1.028 & 65.16
 & 67.90\\
1108 & 2004-05-21 & 214.02546 & 52.26719 & -48.86 & 153 & 0.515 & 0.928 & 71.71
 & 78.12\\
1109 & 2003-05-06 & 214.06047 & 52.28759 & -48.86 & 154 & 0.187 & 1.080 & 52.67
 & 57.97\\
1110 & 2003-05-04 & 214.08909 & 52.31194 & -48.86 & 159 & 0.439 & 0.798 & 62.99
 & 66.23\\
1111 & 2003-05-28 & 214.12444 & 52.33257 & -48.86 & 155 & 0.734 & 0.823 & 75.48
 & 84.81\\
1112 & 2003-05-04 & 214.15082 & 52.35534 & -48.86 & 157 & 0.413 & 0.979 & 64.47
 & 63.22\\
1113 & 2003-05-30 & 214.18373 & 52.37425 & -48.86 & 155 & 0.477 & 0.957 & 67.74
 & 72.29\\
1114 & 2003-06-01 & 214.20993 & 52.39691 & -48.86 & 156 & 0.688 & 0.740 & 71.61
 & 71.43\\
1115 & 2003-05-05 & 214.24393 & 52.41659 & -48.86 & 155 & 0.451 & 0.996 & 55.26
 & 56.99\\
1140 & 2003-05-05 & 213.76523 & 52.12802 &  41.14 & 150 & 0.360 & 0.864 & 53.10
 & 48.48\\
1141 & 2003-05-06 & 213.80431 & 52.10401 &  41.14 & 142 & 0.450 & 0.893 & 53.62
 & 55.07\\
1142 & 2004-04-21 & 213.84704 & 52.08417 &  41.14 & 150 & 0.535 & 1.047 & 70.07
 & 72.46\\
1143 & 2003-05-05 & 213.88612 & 52.06016 &  41.14 & 143 & 0.400 & 1.008 & 57.86
 & 74.29\\
1144 & 2003-05-30 & 213.92886 & 52.04032 &  41.14 & 143 & 0.552 & 0.670 & 74.29
 & 84.62\\
1145 & 2003-05-30 & 213.96794 & 52.01631 &  41.14 & 156 & 0.435 & 0.720 & 64.00
 & 67.11\\
1146 & 2003-05-06 & 214.01067 & 51.99647 &  41.14 & 151 & 0.503 & 0.788 & 61.38
 & 56.60\\
1150 & 2003-05-04 & 214.04399 & 52.32691 &  41.14 & 147 & 0.554 & 0.764 & 70.83
 & 72.41\\
1151 & 2003-05-28 & 214.08325 & 52.30289 &  41.14 & 146 & 0.575 & 0.674 & 74.83
 & 74.19\\
1152 & 2003-05-04 & 214.12617 & 52.28306 &  41.14 & 145 & 0.469 & 0.784 & 65.49
 & 62.32\\
1153 & 2003-06-01 & 214.16543 & 52.25904 &  41.14 & 141 & 0.873 & 0.743 & 73.91
 & 73.91\\
1154 & 2004-05-22 & 214.20835 & 52.23921 &  41.14 & 153 & 0.610 & 0.863 & 75.00
 & 80.00\\
1155 & 2003-06-30 & 214.24761 & 52.21519 &  41.14 & 152 & 0.848 & 0.616 & 82.55
 & 87.06\\
1156 & 2004-04-19 & 214.29054 & 52.19536 &  41.14 & 150 & 0.410 & 1.159 & 63.45
 & 66.25\\
\enddata
\tablecomments{
(1) Mask number: the first two digits represent the CFHT pointing and the last two digits 
are a position code within the pointing (see Section~\ref{datatables}); (2) UT observing date in YYYY-MM-DD for the night contributing the majority of a mask's data; 
(3)-(4) RA/Dec (2000.00) of mask center; (5) PA of long axis of mask; (6) number of target objects on 
mask not counting alignment stars; (7) median FWHM of alignment star spectra in arcsec (seeing
measurement); (8) continuum 
$S/N$ per pixel near 6900 \AA (based on median flux and measured variance, omitting the atmospheric B-band region); (9) percentage yield of reliable redshifts for overall target sample on mask 
($Q \ge 3$); (10) percentage yield of reliable redshifts for redder galaxies ($R-I > 0.5$) only.
}
\label{table.maskdata}
\end{deluxetable}

\clearpage

\begin{deluxetable}{cccccccccccrcccc}
\tablenum{10a}
\tablecaption{Galaxy Data, Part 1 (Sample)}
\tabletypesize{\scriptsize}
\tablewidth{0pt}
\setlength{\footskip}{3.0in}
\tablehead{
\colhead{ID} & \colhead{RA} & \colhead{Dec} & \colhead{$B$} & \colhead{$R$} & \colhead{$I$} & \colhead{$\sigma_B$} & \colhead{$\sigma_R$} & \colhead{$\sigma_I$} & \colhead{$r_g$} & \colhead{$e_2$} & \colhead{PA} & \colhead{$p_{gal}$} & \colhead{$E(B-V)$} & \colhead{$M_{B}$} & \colhead{$U-B$} \\
\colhead{(1)} & \colhead{(2)} & \colhead{(3)} & \colhead{(4)} & \colhead{(5)} & \colhead{(6)} & \colhead{(7)} & \colhead{(8)} & \colhead{(9)} & \colhead{(10)} & \colhead{(11)} & \colhead{(12)} & \colhead{(13)} & \colhead{(14)} & \colhead{(15)} & \colhead{(16)} }
\startdata
11001673 & 213.86870 & 51.95644 &  23.49 &  23.14 &  22.58 &  0.026 &  0.022 & 
 0.029 &  1.370 & 0.046 &  -46.95 & 1.000 &  0.011 & -999 & -999 \\
11001699 & 213.81047 & 51.94232 &  22.07 &  20.03 &  19.55 &  0.009 &  0.001 & 
 0.002 &  1.957 & 0.026 &   -0.40 & 3.000 &  0.011 & -19.23 &   1.130 \\
11001770 & 213.84843 & 51.94888 &  24.14 &  24.10 &  24.02 &  0.068 &  0.077 & 
 0.152 &  2.184 & 0.332 &   -9.43 & 3.000 &  0.011 & -999 & -999 \\
11001800 & 213.83176 & 51.95255 &  25.34 &  23.51 &  23.08 &  0.169 &  0.033 & 
 0.050 &  1.796 & 0.069 &   -2.61 & 0.510 &  0.011 & -999 & -999 \\
11001860 & 213.83255 & 51.95417 &  24.38 &  23.40 &  22.57 &  0.079 &  0.034 & 
 0.035 &  2.023 & 0.048 &  -12.92 & 3.000 &  0.011 & -20.08 &   0.752 \\
11001861 & 213.81732 & 51.95325 &  23.25 &  22.82 &  22.21 &  0.035 &  0.028 & 
 0.035 &  2.823 & 0.114 &  -38.11 & 3.000 &  0.011 & -20.22 &   0.532 \\
11001878 & 213.81459 & 51.93851 &  23.97 &  23.42 &  23.13 &  0.056 &  0.032 & 
 0.053 &  1.844 & 0.077 &   62.49 & 0.663 &  0.011 & -17.67 &   0.395 \\
11001898 & 213.81851 & 51.96013 &  24.18 &  23.14 &  22.60 &  0.090 &  0.040 & 
 0.054 &  3.049 & 0.359 &  -81.78 & 3.000 &  0.011 & -18.75 &   0.646 \\
11001909 & 213.82000 & 51.94717 &  23.10 &  21.18 &  20.52 &  0.025 &  0.004 & 
 0.005 &  2.059 & 0.037 &  -24.73 & 3.000 &  0.011 & -21.02 &   0.813 \\
11001922 & 213.82240 & 51.93072 &  22.01 &  21.26 &  21.06 &  0.014 &  0.008 & 
 0.016 &  2.415 & 0.089 &   21.47 & 3.000 &  0.011 & -14.92 &   0.563 \\
11001927 & 213.82467 & 51.96038 &  25.29 &  22.05 &  20.96 &  0.208 &  0.012 & 
 0.010 &  2.547 & 0.145 &  -28.47 & 3.000 &  0.011 & -20.43 &   1.253 \\
11001934 & 213.82196 & 51.95469 &  22.89 &  22.14 &  21.83 &  0.022 &  0.013 & 
 0.021 &  2.430 & 0.108 &   43.00 & 3.000 &  0.011 & -19.66 &   0.417 \\
11001950 & 213.83030 & 51.94160 &  23.60 &  22.41 &  22.12 &  0.052 &  0.017 & 
 0.028 &  2.445 & 0.295 &  -75.92 & 3.000 &  0.011 & -999 & -999 \\
11001974 & 213.80832 & 51.94306 &  23.05 &  22.47 &  21.92 &  0.019 &  0.012 & 
 0.016 &  0.799 & 0.053 &   -2.44 & 0.999 &  0.011 & -20.59 &   0.464 \\
11001978 & 213.78782 & 51.95398 &  24.46 &  23.85 &  23.55 &  0.064 &  0.042 & 
 0.070 &  1.339 & 0.096 &   36.55 & 0.769 &  0.011 & -999 & -999 \\
11002016 & 213.79909 & 51.95631 &  24.10 &  23.91 &  23.44 &  0.046 &  0.044 & 
 0.064 &  1.664 & 0.042 &   78.09 & 1.000 &  0.011 & -19.49 &   0.460 \\
11002019 & 213.78484 & 51.94730 &  23.99 &  22.75 &  22.26 &  0.052 &  0.018 & 
 0.025 &  1.909 & 0.197 &  -43.56 & 0.429 &  0.011 & -19.15 &   0.639 \\
11002024 & 213.75619 & 51.93770 &  24.12 &  23.60 &  23.07 &  0.069 &  0.040 & 
 0.053 &  1.979 & 0.395 &  -19.89 & 3.000 &  0.012 & -18.69 &   0.516 \\
11002039 & 213.78086 & 51.95369 &  25.60 &  24.09 &  23.36 &  0.183 &  0.058 & 
 0.059 &  1.644 & 0.042 &   72.13 & 0.980 &  0.011 & -999 & -999 \\
11002051 & 213.77227 & 51.94480 &  24.61 &  23.20 &  22.40 &  0.110 &  0.032 & 
 0.033 &  2.262 & 0.192 &   87.64 & 3.000 &  0.011 & -999 & -999 \\
11002064 & 213.73228 & 51.93605 &  24.63 &  23.52 &  23.22 &  0.131 &  0.048 & 
 0.077 &  2.336 & 0.437 &  -23.33 & 3.000 &  0.012 & -999 & -999 \\
11002085 & 213.76673 & 51.95897 &  23.55 &  23.33 &  23.05 &  0.032 &  0.030 & 
 0.051 &  1.898 & 0.062 &   61.91 & 0.941 &  0.011 & -999 & -999 \\

\enddata
\tablecomments{See Section~\ref{datatables} for details.  Columns are:
  (1) Unique DEEP2 object ID from {\it pcat}; (2)-(3) object RA and
  Dec (2000.00); (4)-(9) total $BRI$ magnitudes and errors from {\it
    pcat}; (10) Gaussian radius (11)-(12) ellipticity and PA from
  proprietary team {\it pcat}; (13) probability of being a galaxy
  determined from photometry; (14) Galactic $E(B-V)$ from Schlegel,
  Finkbeiner \& Davis (1998); (15)-(16) absolute $B$ magnitude ($M_B -
  5 \cdot \log_{10} h$) and rest-frame $(U-B)$ from Willmer et
  al.~(2006) K-corrections (sometimes labeled $(U-B)_0$ in DEEP2 papers).  }
\label{table.galaxydata1}
\end{deluxetable}

\clearpage
\begin{landscape}
\setlength{\footskip}{3.0in}
\begin{deluxetable}{cccccccrcccccl}
\tabletypesize{\scriptsize}
\tablecaption{Galaxy Data, Part 2 (Sample)}
\tablenum{10b}
\tablewidth{0pt}
\tablehead{
\colhead{ID} & \colhead{Mask} & \colhead{Slit} & \colhead{Date} & \colhead{MJD} & \colhead{Slit RA} & \colhead{Slit Dec} & \colhead{Slit Length} & \colhead{Slit PA} & \colhead{$z$} & \colhead{$z_{best}^{hel}$} & \colhead{$\sigma_z$} & \colhead{$Q$} & \colhead{Comment} \\
\colhead{(17)} & \colhead{18)} & \colhead{(19)} & \colhead{(20)} & \colhead{(21)} & \colhead{(22)} & \colhead{(23)} & \colhead{(24)} & \colhead{(25)} & \colhead{(26)} & \colhead{(27)} & \colhead{(28)} & \colhead{(29)} & \colhead{(30)}   }
\startdata
11001673 & 1101 & 0 & 2003-05-06 & 52765.46 & 213.86906 & 51.95637 &   4.86 & 
 -48.86 & -1 & -1 & 999.9 & 2 & 
bcol; bext                                      \\
11001699 & 1100 & 7 & 2003-05-03 & 52762.44 & 213.81044 & 51.94244 &   4.20 & 
 -48.86 &  0.29064 &  0.29061 & 4.26e-05 & 4 & 
                                                \\
11001770 & 1101 & 3 & 2003-05-06 & 52765.46 & 213.84874 & 51.94867 &  10.27 & 
 -18.86 & -1 & -1 & -5 & 1 &                                                 \\
11001800 & 1100 & 4 & 2003-05-03 & 52762.44 & 213.83186 & 51.95259 &   3.77 & 
 -48.86 & -1 & -1 & -5 & 1 &                                                 \\
11001860 & 1101 & 10 & 2003-05-06 & 52765.46 & 213.83284 & 51.95412 &   4.98 & 
 -48.86 &  0.95771 &  0.95767 & 9.87e-05 & 4 & 
                                                \\
11001861 & 1101 & 14 & 2003-05-06 & 52765.46 & 213.81763 & 51.95319 &   6.33 & 
 -48.86 &  0.90908 &  0.90904 & 2.95e-05 & 4 & 
                                                \\
11001878 & 1100 & 3 & 2003-05-03 & 52762.44 & 213.81496 & 51.93842 &   4.57 & 
 -48.86 &  0.50301 &  0.50297 & 5.04e-06 & 4 & 
bsky; iffy                                      \\
11001898 & 1101 & 17 & 2003-05-06 & 52765.46 & 213.81873 & 51.96013 &   5.01 & 
 -78.86 &  0.64987 &  0.64984 & 5.21e-05 & 3 & 
                                                \\
11001909 & 1101 & 11 & 2003-05-06 & 52765.46 & 213.82013 & 51.94721 &   4.64 & 
 -48.86 &  0.67948 &  0.67944 & 6.97e-05 & 3 & 
ZREVISED                                        \\
11001922 & 1100 & 0 & 2003-05-03 & 52762.44 & 213.82199 & 51.93105 &  10.01 & 
 -48.86 &  0.07411 &  0.07408 & 1.01e-05 & 4 & 
fix; bext; ZREVISED                             \\
11001927 & 1146 & 15 & 2003-05-06 & 52765.30 & 213.82497 & 51.96045 &   7.18 & 
  41.14 &  0.68392 &  0.68388 & 6.57e-05 & 4 & 
                                                \\
11001934 & 1100 & 9 & 2003-05-03 & 52762.44 & 213.82227 & 51.95463 &   3.56 & 
 -48.86 &  0.65117 &  0.65114 & 1.21e-05 & 4 & 
                                                \\
11001950 & 1101 & 5 & 2003-05-06 & 52765.46 & 213.83038 & 51.94163 &   4.89 & 
 -75.92 & -1 & -1 & 2.54e-05 & 2 & 
sngl; ZREVISED                                  \\
11001974 & 1100 & 8 & 2003-05-03 & 52762.44 & 213.80847 & 51.94308 &   3.05 & 
 -48.86 &  0.93205 &  0.93201 & 1.34e-05 & 4 & 
                                                \\
11001978 & 1146 & 3 & 2003-05-06 & 52765.30 & 213.78787 & 51.95391 &   7.38 & 
  41.14 & -1 & -1 & 1.42e-04 & 1 & 
offser                                          \\
11002016 & 1146 & 7 & 2003-05-06 & 52765.30 & 213.79964 & 51.95657 &   5.57 & 
  41.14 &  1.05800 &  1.05796 & 2.03e-05 & 4 & 
                                                \\
11002019 & 1100 & 18 & 2003-05-03 & 52762.44 & 213.78477 & 51.94747 &   3.80 & 
 -43.56 &  0.64948 &  0.64945 & 2.48e-05 & 4 & 
                                                \\
11002024 & 1100 & 23 & 2003-05-03 & 52762.44 & 213.75636 & 51.93768 &   4.02 & 
 -19.89 &  0.74553 &  0.74550 & 3.40e-05 & 4 & 
                                                \\
11002039 & 1100 & 22 & 2003-05-03 & 52762.44 & 213.78119 & 51.95361 &   4.63 & 
 -48.86 & -1 & -1 & -5 & 1 &                                                 \\
11002051 & 1100 & 21 & 2003-05-03 & 52762.44 & 213.77234 & 51.94483 &   6.49 & 
 -78.86 & -1 & -1 & 7.34e-05 & 2 & 
bcont; disc; ZREVISED                           \\
11002064 & 1100 & 30 & 2003-05-03 & 52762.44 & 213.73241 & 51.93608 &   4.74 & 
 -23.33 & -1 & -1 & -1 & 1 &                                                 \\
11002085 & 1100 & 29 & 2003-05-03 & 52762.44 & 213.76717 & 51.95884 &   5.37 & 
 -48.86 & -1 & -1 & -5 & 1 &                                                 \\

\enddata
\tablecomments{See Section~\ref{datatables} for details.  Columns are:
(17) Unique DEEP2 object ID from {\it pcat}, (18) slitmask number, (19) slitlet number 
on mask; (20) UT date of observation; (21) modified Julian date of observation; (22)-(23) RA and Dec of slitlet center;
(24)-(25) slitlet length in arcsec and PA; (26) redshift from this observation in geocentric 
reference frame; (27) best redshift from all observations (heliocentric reference frame); (28) error of this redshift observation (negative values indicate $\chi^2$ minima whose widths, and hence the resulting redshift errors, were not well determined); (29) redshift quality code: $Q$ = -2 (data 
so poor that object was effectively never observed); $Q$ = -1 (star); 1 (very low $S/N$; data are 
not likely to yield a redshift); $Q = 2$ (redshift information present, but definitive redshift not obtained); $Q = 3$ (reliable redshift with probability 
of accuracy $\ge 95$\%; $Q$ = 4 (reliable redshift with probability of accuracy $\ge 99$\%); (30) comment codes: see explanations in 
Section~\ref{datatables} (none of these sample objects had comments assigned).
} 
\label{table.galaxydata2}
\end{deluxetable}

\clearpage
\end{landscape}


\begin{references}

\reference {} Abazajian, K. et al.~2003, \aj, 126, 2081
\reference {} Aihara, H. et al.~2011, \apjs, 193, 29
\reference {} Arnouts, S., de Lapparent, V., Mathez, G., Mazure, G.,
  Mellier, Y., Bertin, E., \& Kruszewski, A.~1997, \aaas, 124, 163
\reference {} Balogh, M.~L. et al.~2004, \mnras, 348, 1355
\reference {} Barger, A. J., Cowie, L. L., Mushotzsky, R. F.,
  Yang, Y., Wang, W.-H., Steffen, A. T., \& Capak, P.~2005, \aj, 129, 578 
\reference {} Barger, A.~J., Cowie, L. L., \& Wang, W.-H.~2008, \apj, 689, 687
\reference {} Barth, A.~J. 2001, in ASP Conf.~Ser., Vol.~238,
Astronomical Data Analysis Software and Systems X, eds. F. R. Harnden,
Jr., F. A. Primini, \& H. E. Payne (San Francisco: ASP), 385
\reference {} Bell, E.~F. et al.~2004, \apj, 609, 752
\reference {} Bellanger, C., de Lapparent, V., Arnouts, S., 
  Mathez, G., Mazure, A., \& Mellier, Y.~1995, \aaas, 110, 159 
\reference {} Bernstein, G.~M. et al.~1994, \apj, 424, 569
\reference {} Blanton, M.~R.~2006, \apj, 648, 268 
\reference {} Bolton, A.~S. \& Burles, S.~2007,{\em N. J. Ph.}, 9, 443
\reference {} Bruzual, G. \& Charlot, S.~2003, \mnras, 344, 1000
\reference {} Bundy, K. et al.~2006, \apj, 651, 120
\reference {} Chen, Y.-M. et al.~2009, \mnras, 393, 406
\reference {} Cohen, J. G. et al.~2000, \apj, 538, 29
\reference {} Cohen, J. G.~2002, \apj, 567, 672
\reference {} Coil. A. L. Davis, M., \& Szapudi, I.~2001, \pasp, 113, 1312  
\reference {} Coil, A. L. et al.~2004a, \apj, 609, 525   
\reference {} Coil, A. L. et al.~2004b, \apj, 617, 765 
\reference {} Coil, A. L. et al.~2006a, \apj, 638, 668 
\reference {} Coil, A. L. et al.~2006b, \apj, 644, 671   
\reference {} Coil, A. L. et al.~2007, \apj, 654, 115   
\reference {} Coil, A. L. et al.~2008, \apj, 672, 153   
\reference {} Coil, A. L. et al.~2009, \apj, 701, 1484 
\reference {} Coil, A. L. et al.~2011, \apj, 741, 8   
\reference {} Coil, A. L., Davis, M., \& Szapudi, I.~2001, \pasp, 113, 1312 
\reference {} Coleman, G. D., Wu, C.-C., \& Weedman, D. W.~1980, \apjs, 43, 393 
\reference {} Colless, M., Ellis, R. S., Taylor, K., \& Hook,
  R. N.~1990, \mnras, 244, 408 
\reference {} Colless, M. et al.~2001, \mnras, 328, 1039
\reference {} Comerford, J.~M. et al.~2009, \apj, 698, 956
\reference {} Conroy, C. et al.~2005, \apj, 635, 982
\reference {} Conroy, C. et al.~2007, \apj, 654, 153
\reference {} Conselice, C., Bershady, M. A., Dickinson, M., \&
Papovich, C.~2003, \aj, 126, 1183 
\reference {} Conselice, C., Bundy, K., Vivian, U., Eisenhardt, P., Lotz, J.,
   \& Newman, J. 2007, \mnras, 383, 1366
\reference {} Cooper, M. C., Newman, J. A., Madgwick, D. S., 
Gerke, B. F., Yan, R. \& Davis, M.~2005, \apj, 634, 833
\reference {} Cooper, M. C. et al.~2006, \mnras, 370, 198
\reference {} Cooper, M. C. et al.~2007, \mnras, 376, 1445
\reference {} Cooper, M. C. et al.~2008, \mnras, 383, 1058
\reference {} Cooper, M. C. et al.~2010, \mnras, 409, 337
\reference {} Cooper, M. C. et al.~2011, \apjs, 193, 14
\reference {} Cooper, M. C. et al.~2012a, \mnras, 419, 3018
\reference {} Cooper, M. C. et al.~2012b, {\it spec2d}, ASCL, 1203.003 
\reference {} Covinginton, M. D. et al.~2009, [ArXiv:902.0566]
\reference {} Cowie, L. L., Songaila, A., Hu, E. M., \& Cohen, 
  J. G.~1996, \aj, 112, 839
\reference {} Croton, D. J. et al.~2006, \mnras, 365, 11 
\reference {} Cuillandre, J.-C., Luppino, G., Starr, B., \& Isani,
S. 2001, in {\it Proceedings of Semaine de l'Astrophysique Fran\c{c}aise}, 
eds. F. Combes, D. Barret, F. Th\'evenin (Les Ulis: EdP-Sciences), 605  
\reference {} Davis, M. et al.~2003, SPIE, 4834, 161
\reference {} Davis, M., Gerke, B., \& Newman, J.~2005, in {\it Observing Dark Energy}, Astron.
Soc. Pacific Conf. Series, Vol.~339, eds. S. Wolff \& T. Lauer, 
San Francisco: Astronomical Society of the Pacific, p. 128
\reference {} Davis, M. et al.~2007, \apj, 660, L1
\reference {} de Boor, C. 2001, {\it A Practical Guide to Splines,} Applied
  Mathematical Sciences, New York: Springer 
\reference {} Dekel, A. \& Birnboim, Y.~2006, \mnras, 368, 2
\reference {} Dickinson, M. \& the FIDEL team, 2007, {\em AAS}, 211, 5216
\reference {} Drinkwater, M. J. et al.~2010, \mnras, 401, 1429
\reference {} Drory, N., Bender, R., Feulner, G., Hopp, U., Maraston,
C., Snigula, J., \& Hill, G. J.~2003, \apj, 595, 698
\reference {} Dutton, A. A. et al. 2011, \mnras, 410, 1660
\reference {} Eisenstein, D. J. et al. 2003, \apj, 585, 694
\reference {} Faber, S. M. et al.~2003,  {\em S.P.I.E.}, 4841, 1657
\reference {} Faber, S. M. et al.~2007, \apj, 665, 265
\reference {} Garilli, B. et al.~2008, A\&A, 486, 683
\reference {} Gebhardt, K. et al.~2003, \apj, 597, 239
\reference {} Georgakakis, A. et al.~2008, \mnras, 385, 2049 
\reference {} Gerke, B. F. et al.~2005, \apj, 625, 6 
\reference {} Gerke, B. F. et al.~2007a, \mnras, 376, 1425
\reference {} Gerke, B. F. et al.~2007b, \apj, 660, 23 
\reference {} Gerke, B. F. et al.~2012, \apj, in press
\reference {} Grogin, N. A. et al.~2011, \apjs, 197, 35
\reference {} Groth, E. J., Kristian, J. A., Lynds, R., O'Neill,
E. J., Balsano, R., Rhodes, J., \& WFPC1 IDT, 1994, BAAS, 26, 1403
\reference {} Gwyn, S. D. J.~2012, \aj, 143, 38  
\reference {} Guzman, R. et al.~1997, \apj, 489, 559
\reference {} Haiman, Z. et al.~2001, \apj, 553, 545
\reference {} Harker, J. J., Schiavon, R. P., Weiner, B. J., \&
Faber, S. M.~2006, \apj, 647, 103
\reference {}Harker, J. J.~2008, PhD thesis, University of California, Santa Cruz
\reference {} Hopkins, P. F., Hernquist, L., Cox, T. J., 
  di Matteo, T., Robertson, B., \& Springel, Volker~2006, \apjs, 163, 1
\reference {} Hogg, D. W., Cohen J. G., \& Blanford, R.~2000, \apj, 545, 32
\reference {} Horne, K.~1986, \pasp, 98, 609
\reference {} Hoyos, C., Koo, D. C., Phillips, A. C., Willmer, C. N. A.,
\& Guhathakurta, P.~2005, \apj, 635, L21
\reference {} Im, M. et al.~2002, \apj, 571, 136
\reference {} Ivezic, Z. et al.~2007, \aj, 134, 973
\reference {} Jones, C. et al.~2008, Spitzer proposal \#50660
\reference {} Kaiser, N., Squires, G., \& Broadhurst, T.~1995, \apj,
449, 460  
\reference {} Kassin, S. A. et al.~2007, \apj, 660, L35
\reference {} Kinney, A. L., Calzetti, D., Bohlin, R. C., McQuade, K.,
Storchi-Bergmann, T., \& Schmitt, H. R.~1996, \apj, 467, 38
\reference {} Kirby, E. N., Guhathakurta, P., Faber, S. M., 
Koo, D. C., Weiner, B. J., and Cooper, M. C.~2007, \apj, 660, 62
\reference {} Kobulnicky, H. A. et al.~2003, \apj, 599, 1006
\reference {} Kochanek, C. S. et al.~2011, [ArXiv:1110.4371]
\reference {} Koekemoer, A.~M. et al.~2011, ApJS, 197, 36
\reference {} Konidaris, N. P.~2008, PhD thesis, University of
California, Santa Cruz
\reference {} Koo, D. N. C. et al.~2005, \apjs, 157, 175
\reference {} Laird, E.S. et al.~2009, \apjs, 180, 102
\reference {} Le Fevre, O. et al.~2005, A\&A, 439, 845
\reference {} Lilly, S. J., Le F\'evre, O., Crampton, D., Hammer, F.,
  \& Tresse, L.~1995, \apj, 455, 50
\reference {} Lilly, S. et al.~2007, \apjs, 172, 70
\reference {} Lilly, S. et al.~2009, zCOSMOS catalog DR2, ADS bib code 2009yCat..21720070L
\reference {} Lin, L. et al.~2004, \apj, 617, L9
\reference {} Lin, L. et al.~2007, \apj, 660, L51
\reference {} Lin, L. et al.~2008, \apj, 681, 232
\reference {} Lin, L. et al.~2010, \apj, 718, 1158
\reference {} Lowenthal, J. et al.~1997, \apj, 481, 673
\reference {} Lotz, J. M. et al.~2008, \apj, 672, 177
\reference {} Ma, Z., Hu, W., \& Huterer, D.~2006, \apj, 636, 21
\reference {} Madgwick, D. et al.~2003, \apj, 599, 997
\reference {} Matthews, D. J. et al.~2012, \apjs, in prep.
\reference {} Montero-Dorta, A. D. et al.~2009, \mnras, 392, 125 
\reference {} Morton, D.~C.~1991, \apjs, 77, 119
\reference {} Murray, S., Forman, W. R., Jones, C., Hickox, R., Kenter, A., Willner, S.
  2008, American Astronomical Society HEAD Meeting \#10, \#32.03
\reference {} Newman, J. A. \& Davis, M.~2000, \apj, 534, 11
\reference {} Newman, J. A. \& Davis, M.~2002, \apj, 564, 567
\reference {} Noeske, K. G. et al.~2007a, \apj, 660, L43
\reference {} Noeske, K. G. et al.~2007b, \apj, 660, L47
\reference {} Osterbrock, D. E., Fulbright, J. P., \& Bida, T. A.~1997, \pasp, 109, 614
\reference {} Papovich, C. et al.~2003, \apj, 598, 827
\reference {} Phillips, A. C. et al.~1997, \apj, 489, 543 
\reference {} Pierce, C. M. et al.~2007, \apj, 660, L19
\reference {} Rix, H.-W. et al.~2004, \apjs, 152, 163
\reference {} Salim, S. et al.~2009, \apj, 700, 161
\reference {} Sato, T., Martin, C. L., Noeske, K. G., Koo, D. C., \& Lotz, J. L.~2009, \apj, 696, 214
\reference {} Sawicki, M. et al.~2008, \apj, 687, 884
\reference {} Schiavon, R. P. et al.~2006, \apj, 651, L93
\reference {} Schlegel, D.~\& Burles, S.~2012, \url{http://www.astro.princeton.edu/$\sim$schlegel/code.html}
\reference {} Schlegel, D. J., Finkbeiner, D. P., \& Davis, M.~1998,\apj, 500, 525
\reference {} Shectman, S.~A., Landy, S.~D., Oemler, A., Tucker,
D.~L., Lin, H., Kirshner, R.~P., Schechter, P.~L.~1996, \apj, 470, 172
\reference {} Simard, L., et al.~2002, \apjs, 142, 1
\reference {} Tonry, J. \& Davis, M. 1979, \aj, 84, 1511
\reference {} van Dokkum, P. G., et al.~2009, \pasp, 121, 2 
\reference {} Weiner, B. J. et al.~2005, \apj, 620, 595
\reference {} Weiner, B. J. et al.~2006a, \apj, 653, 1027
\reference {} Weiner, B. J. et al.~2006b, \apj, 653, 1049
\reference {} Weiner, B. J. et al.~2007, \apj, 660, L39
\reference {} Weiner, B. J. et al.~2009, \apj, 692, 187
\reference {} Willmer, C. N. A. et al.~2006, \apj, 647, 853 
\reference {} Wirth, G. D. et al.~2004, \aj, 127, 3121 
\reference {} van den Bergh, S., Cohen, J. G., Hogg, D. W., Blanford, R.~2000, \aj, 120, 2190
\reference {} van Dokkum, P. G. et al.~2009, \pasp, 121, 2
\reference {} Vogt, N. P. et al.~1996, \apj, 465, L15
\reference {} Vogt, N. P. et al.~1997, \apj, 479, L121
\reference {} Vogt, N. P. et al.~2005, \apjs, 159, 41 
\reference {} Yan, R., White, M., \& Coil, A. L.~2004, \apj, 607, 739
\reference {} Yan, R. et al.~2006, \apj, 648, 281
\reference {} Yan, R. et al.~2009, \mnras, 398, 735
\reference {} Yan, R. et al.~2011, \apj, 728, 38
\reference {} Yee, H. K. C. et al.~1996 \apjs, 102, 289
\reference {} Yee, H. K. C. et al.~2000, \apjs, 129, 475
\reference {} Yoon, J., Peterson, D. M., Kurucz, R. L., \& Zagarello, R. J.~2010, \apj, 708, 71
\reference {} York, D.~G. et al.~2000, \aj, 120, 1579

\end{references}
\end{document}